\documentclass[twocolumn,preprintnumbers, amsmath, amssymb, floatfix]{revtex4}
\usepackage{graphicx}
\usepackage{dcolumn}
\usepackage{bm}

\def\be{\begin{equation}}
\def\ee{\end{equation}}
\def\beq{\begin{eqnarray}}
\def\eeq{\end{eqnarray}}
\def\n{\nonumber}
\def\bay{\begin{array}}
\def\eay{\end{array}}

\def\kakueq{{\raisebox{-2.0ex}{$\sim$} \atop \raisebox{0.8ex}{$_{\mathrm{K}}$}}}
\begin{document}
\preprint{CIRI/03-smw05}
\title{Foundations of a Universal Theory  of Relativity}

\author{Sanjay M. Wagh}
\affiliation{Central India Research Institute, \\ Post Box 606,
Laxminagar, Nagpur 440 022, India\\
E-mail:cirinag\underline{\phantom{n}}ngp@sancharnet.in}

\date{May 8, 2005}
\begin{abstract}
Earlier, we had presented \cite{heuristic} heuristic arguments to
show that a {\em natural unification\/} of the ideas of the
quantum theory and those underlying the general principle of
relativity is achievable by way of the measure theory and the
theory of dynamical systems. Here, in Part I, we provide the
complete physical foundations for this, to be called, the {\em
Universal Theory of Relativity}. Newton's theory and the special
theory of relativity arise, situationally, in this Universal
Relativity. Explanations of quantum indeterminacy are also shown
to arise in it. Part II provides its mathematical foundations. One
experimental test is also discussed before concluding remarks. \\

\centerline{To be submitted to: }
\end{abstract}
\maketitle

\newpage
The organization of this paper is such that physical foundations
of the proposed Universal Theory of Relativity are to be found in
\S\, \ref{pf}, its mathematical foundations in \S\, \ref{mf}, one
experimental test in \S\,\ref{results} and concluding remarks in
\S\,\ref{conclude}.

In \S\, \ref{introduction}, we discuss the general background for
the present considerations. It should help distinguish the
proposed Universal Theory of Relativity from various other
theories, including, Quantum Geometry \cite{ashtekar1} and String
Theory \cite{strings}.

In \S\, \ref{idea-basis}, we discuss the physical basis of some of
the newtonian concepts, in particular, inertia and force. In this
discussion, we mainly stress that the Galilean concept of the
inertia of a material body is, undoubtedly, more fundamental, more
general, than the newtonian concept of force. Therefore, we may
expect the concept of inertia to necessarily find a place in any
future physical theory, but not the concept of force.

Newton's theory does not explain the origin of either the inertia
or the electric charge of material bodies. For any theory that
attempts to explain the origin of inertia and the electric charge,
it then becomes necessary to replace the newtonian concept of
force with some suitable other. The concept of the material body
as a {\em source\/} of force is, consequently, to be {\em
completely abandoned\/} in any such theoretical framework.
Decisively, this must, {\em simultaneously}, hold for all the
forces that need to be {\em postulated\/} to describe the motions
of material bodies in Newton's and other theories.

In other words, it is decisive to recognize that the mathematical
framework of any theory which ``explains'' origins of newtonian
{\em source properties\/} of the physical matter must, {\em
necessarily}, be also applicable, {\em simultaneously}, to {\em
all\/} the ``fundamental'' forces that are needed in Newton's and
other theories to describe the motions of material bodies. This is
\cite{heuristic} the key to the ``new'' theory.

Clearly, a formulation that replaces the concept of only a {\em
single\/} source property of material bodies cannot then be
physically satisfactory as well as conceptually consistent.

In this section, we then stress the importance of the physical
construction of the reference frames or the coordinate systems. We
also stress that the motions of material bodies must, in general,
affect the constructions of coordinate frames.

In \S\, \ref{gpr}, we then discuss the status of the general
principle of relativity. This important principle states that the
laws of physics must be applicable to all the frames of reference.
Consequently, the universal theory of relativity, a theory
consistent with the general principle of relativity, will
necessarily have to incorporate the physical construction of the
coordinate systems.

In the context of this above discussion, we also consider
Einstein's equivalence principle and stress that the equivalence
principle essentially establishes {\em only\/} the {\em
consistency\/} of the phenomenon of gravitation with the general
principle of relativity. It then needs to be emphasized here that
the equivalence principle is {\em not\/} logically equivalent to
the general principle of relativity.

In section \S\, \ref{gent}, we discuss various general
expectations from a theory consistent with the general principle
of relativity. This section essentially sets the conceptual
background for the sections to follow. But, a reader is requested
to go through even the earlier sections.

Then, in \S\, \ref{qamfnt}, we consider quantum aspects
\cite{heisenberg} vis-$\acute{a}$-vis the general principle of
relativity and the requirements of the mathematical formalism
implied by such considerations.

In \S\, \ref{mf}, we provide the mathematical foundations for the
proposed unified theory. Specifically, methods of measure theory
and dynamical systems are reviewed in this section.

Then, in \S \,\ref{results}, we analyze a torsion balance
experiment. Any theory using the concept of force always predicts
a non-null effect for this experiment. However, a ``null effect''
is \underline{also} obtainable in the Universal Theory of
Relativity apart from possible {\em non-null\/} effects (of the
theories using the concept of force). We therefore suggest that
the existence of such {\em anomalous null-results\/} be searched
for in torsion balance experiments, preferably involving dynamic
measurements.

[Such anomalous null results will be some of the main features of
the proposed Universal Theory of Relativity. This is because a
transformation of the underlying space providing null result for
any given experimental situation is, thinkably, permissible. But,
physical space of Universal Relativity changes with changes in
matter. Care is therefore needed in establishing null-results.]

Finally, \S\, \ref{conclude}, contains some concluding remarks
about the proposed theory.

\section{PHYSICAL FOUNDATIONS} \label{pf}

\subsection{General Background} \label{introduction} For the sake of
completeness, we recall here the discussion from \cite{heuristic}.
Its purpose here is to contrast the present approach with some
other approaches \cite{ashtekar1, strings} to the unification of
the ideas of the quantum theory and Einstein's general relativity.

In an essence, Newton's deliberations define \cite{cartan} {\em
specific mathematical structures or fields\/} (scalar, vector,
tensor functions) over the {\em metrically flat 3-continuum\/} and
consider the laws of their transformations. The 3-continuum admits
the {\em same Euclidean metric structure\/} before and after the
coordinate transformations. The ever-flat 3-continuum is, in this
sense, an {\em absolute space\/} and, in Newton's theory,
accelerations of material bodies can refer only to this absolute
space.

Furthermore, in Newton's theory, physical laws for these
quantities are the mathematical statements form-invariant under
the galilean coordinate transformations which are basic to the
newtonian formulation of mechanics.

Mathematical methods \cite{class-mech} for these {\em newtonian\/}
fields are, evidently, required to be consistent with the
underlying flat 3-continuum admitting the same metric structure
before and after the transformations of these fields. {\em This
is, truly, the sense of any theory being newtonian}.

Then, the galilean transformations under which the newtonian laws
are form-invariant are, as opposed to general, {\em specific\/}
transformations of the coordinates of the continuum
$\mathbb{R}^3$.

Newton's theory also attaches {\em physical\/} meaning to the
space coordinates and to the time coordinate. In this theory, the
space coordinate describes the ``physical distance'' separating
physical bodies and the time coordinate describes the reading of a
``physical'' clock. Euclidean space is then also the {\em
physical\/} space of Newton's theory.

In addition, the (newtonian) temporal coordinate has {\em
universally\/} the same value for all the spatial locations, {\em
ie}, all synchronized clocks at different spatial locations show
and maintain the same time. In other words, the newtonian time
coordinate is the {\em absolute\/} physical time.

Basically, Newton's theory imagines a material body as a
point-mass endowed with the inertia of that material body. It is a
primary {\em physical\/} conception of this theory. Necessarily, a
point-mass moves along a one-dimensional curve of the unchanging
Euclidean 3-continuum.

In {\em physical\/} associations of Newton's theory, it is then
tacitly assumed that the interaction of a measuring instrument
(observer) and the object (a particle whose physical parameters
are being measured) is negligibly small or that the effects of
this interaction can be eliminated from the results of
observations to obtain, as accurately as desired, the values of
these parameters \cite{bohr1}.

An issue closely related to the above one is that of the
causality. Given initial data, Newton's theory predicts the values
of its variables of the point-mass exactly and, hence, assumes
strictly causal development of {\em its\/} physical world.

Conceptually, in Newton's theory, force is the cause behind
motions of material bodies. Next to inertia, force is the second
most important of the conceptions of Newton's theory.

Furthermore, as Lorentz had first realized very clearly, the
sources of the newtonian {\em forces\/} are the singularities of
the corresponding fields defined on the flat 3-continuum. Although
unsatisfactory, this nature of the newtonian framework causes no
problems of mathematical nature since this distinction is
maintainable within the formalism, {\em ie}, well-defined
mathematical procedures for handling this distinction are
possible.

Here, one could imagine bodies of vanishing inertia moving with
the {\em same speed\/} relative to all the inertial observers.
But, acceleration (relative to absolute space) in Newton's second
law of motion has no meaning for vanishing inertia. This inability
is a certain indication of the limitations of Newton's theoretical
framework.

Then, if a zero rest-mass object were to exist in reality, and
nothing in Newton's theory prevents this, it is clear that we need
to ``extend'' various newtonian conceptions. Only experiments can
tell us about the existence of such bodies.

Now, light displays phenomena such as diffraction, interference,
polarization etc. But, as is well known, Newton's corpuscular
theory needs {\em unnatural, non-universal}, inter-particle forces
to explain these phenomena. That light displays phenomena needing
{\em unnatural\/} explanations in Newton's theory could, with
hindsight, then be interpreted \cite{100-yrs} to mean that light
needs to be treated as a zero rest-mass particle. Then, the speed
of Light is the {\em same\/} for all the inertial observers.

Revisions of newtonian concepts were necessary by the beginning of
the 20th century. Firstly, efforts to reconcile some experimental
results with newtonian concepts failed and associated as well as
independent conceptions led Einstein to Special Theory of
Relativity \cite{ein-pop}. Secondly, other experiments related to
the wave-particle duality, of radiation and matter, both,
ultimately led to non-relativistic quantum theory
\cite{heisenberg}.

The methods of Non-Relativistic Quantum Field Theory \cite{qt} are
also similar of nature to the above newtonian methods in that
these consider {\em quantum fields\/} definable on the metrically
flat 3-continuum. For these fields of quantum character, we are of
course required to modify the newtonian mathematical methods. The
Schr\"{o}dinger-Heisenberg formalism achieves precisely this.

Quantum considerations only change the nature of the mathematical
(field) structure definable on the underlying metrically flat
3-continuum. That is, differences in the newtonian and the quantum
fields are mathematically entirely describable as such. But, the
metrically flat 3-continuum is also, in the above sense, an
absolute space in these non-relativistic quantum considerations.

Now, importantly, the ``newtonian source properties'' of physical
matter are differently treated in the non-relativistic quantum
field theory than in Newton's theory. The mass and the electric
charge of a physical body appear as pure numbers, {\em to be
prescribed by hand\/} for a point of the metrically flat
3-continuum, in Schr\"{o}dinger's equation or, equivalently, in
Heisenberg's operators.

Quantum theory then provides the probability of the {\em
location\/} of the mass and the charge values in certain specific
region of the underlying metrically flat 3-continuum. The
mathematical formalism of the quantum theory provides only
probability and it is basically a set of mathematical rules to
calculate the probability of a physical event.

However, certain physical variables of the newtonian mass-point
acquire {\em discrete values\/} in the mathematical formalism of
the quantum theory. This discreteness of certain variables is the
genuine characteristic of the quantum theory and is a significant
departure from their continuous values in Newton's theory.

This quantum theory is fundamentally a theory that divides the
physical world into two parts, a part that is a system being
observed and a part that does the observation. Therefore, quantum
theory always refers to an {\em observer\/} who is {\em
external\/} to the system under observation. The results of the
observation, of course, depend in detail on just how this division
is made.

But, it must be recognized that classical concepts are not
completely expelled from the physical considerations in the
quantum theory. On the contrary, in Bohr's words \cite{bohr2}:

$\spadesuit$ {\em ... it is decisive to recognize that, however
far the phenomena transcend the scope of classical physical
explanation, the account of all evidence must be expressed in
classical terms}. $\spadesuit$

\noindent This applies in spite of the fact that classical
(newtonian) mechanics does not account for the observations of the
microphysical world. (Bohr offered ``complementarity of
(classical) concepts'' as an explanation for this.)

We also note that it is not possible to treat zero rest mass
particles in the non-relativistic quantum theory. As is well known
\cite{qt}, Schr\"{o}dinger's equation or Heisenberg's operators of
this theory are meaningful only when mass of the considered
particle is non-vanishing. Essentially, it is the same limitation
as that of Newton's theory. Non-relativistic quantum field theory
cannot then describe the phenomena displayed by light.

But, in these non-relativistic quantum considerations, a physical
body is described as a non-singular point-particle, not as an
extended object. That is, mass and electric charge appearing
herein are non-singularly defined only for a point of the
metrically always-flat 3-continuum.

Now, special relativity \cite{jackson} implies that the {\em
particle\/} of electromagnetic radiation has zero rest-mass -
follows from the mass-variation with velocity. Special relativity
{\em enlarges\/} the {\em galilean group\/} of transformations of
the metrically flat 3-continuum and time to the {\em Lorentz
group\/} of transformations of the metrically flat 3-continuum and
time, also treatable as a {\em metrically flat 4-dimensional
Minkowski-continuum} \footnote{Nothing special about
4-dimensionality. It also existed with Newton's theory.
Differences between these two theories, Newton's theory and the
Theory of Special Relativity, arise from only the kind of
transformations that are being used by them. In fact, the
4-dimensional methods were discovered by Minkowski much after
Einstein formulated the special theopy of relativity.}.

Lorentz transformations keep Minkowski metric the same. Then,
special relativistic laws for electromagnetic fields (mathematical
structures on the metrically flat 4-continuum), Maxwell's
equations, are mathematical statements form-invariant under
Lorentz transformations.

Then, ``special relativistic laws of motion'' exist for the
sources and Maxwell's equations exist for the fields. So long as
we treat the sources and the fields separately, mathematical
problems do not arise since well-defined mathematical procedures
exist to handle these concepts.

Standard mathematical methods then permit us again considerations
of classical fields on the Minkowski-continuum \cite{class-mech}.
The ``newtonian'' mathematical methods hold also for them, now in
4-dimensions, and are consistent with the fact that the flat
4-continuum admits the same metric structure before and after the
Lorentz transformations of these fields. This is, now, the sense
of any theory being classical. The Minkowski-spacetime is then an
{\em absolute 4-space}.

To describe motions of zero rest mass particles, we ascribe
vanishing rest mass to a point of the space. A point of the space
then has $m=0$ when $E=\pm\, p$ and such a point necessarily moves
with the speed of light.

Notably, Lorentz transformations under which special relativistic
laws are form-invariant are {\em specific\/} coordinate
transformations.

Further, since the involved transformations are very different
than those of Newton's theory, concepts of a {\em measuring rod\/}
and a {\em clock\/} are subject to critical examination and it
then becomes clear that the ordinary newtonian these concepts
involve the tacit assumption that there exist, in principle,
signals that are propagated with an infinite speed. Then, as was
shown by Einstein \cite{ein-dover}, the absolute character of time
is lost completely: initially synchronized clocks at different
spatial locations do not keep the same time-value.

However, like with Newton's theory, coordinates have a direct
physical meaning in special theory of relativity. Although it is
the same association of physical character, the Lorentz
transformations constitute significant departure from the
newtonian concepts since time is no longer the absolute time in
special relativity.

But, classical considerations of special relativity, like with
Newton's theory, assume exact measurability as well as strict
causality.

Now, quantum fields require suitable equations that are
form-invariant under Lorentz transformations to describe quanta
moving close to the speed of light in vacuum. These quantum fields
are, once again, suitable mathematical structures definable on the
metrically ever-flat 4-dimensional (Minkowski) spacetime.

Methods of the special relativistic quantum field theory
\cite{dirac} then handle such {\em quantum fields\/} defined on
the metrically ever-flat 4-continuum admitting a Minkowskian
metric. Next, the quantum mathematical methods are appropriate
generalizations of the mathematical methods of
Schr\"{o}dinger-Heisenberg formalism. This, the
Dirac-Schwinger-Tomonaga formalism \cite{dirac}, achieves for the
metrically flat minkowskian 4-continuum that which the
Schr\"{o}dinger-Heisenberg formalism achieves for the newtonian
3-space and time.

Then, the differences in the (special-relativistic) classical and
quantum fields are mathematically entirely describable as such.
Non-relativistic results are recoverable when the velocities are
small compared to the speed of light.

However, the underlying Minkowski spacetime does not change under
the (Lorentz) transformations keeping the quantum equations
form-invariant and is also, in the earlier sense, an {\em absolute
4-space\/} here.

Likewise with non-relativistic theory, a body is represented in
these special relativistic quantum considerations by ascribing in
{\em non-singular\/} sense the mass and the charge as pure numbers
to points of the ever-flat Minkowski 4-continuum in the
corresponding operators.

This special relativistic quantum field theory then provides us
the probability of the {\em spatial location\/} and the {\em
temporal instant\/} of the mass and the charge values in a region
of the Minkowski 4-continuum, for all velocities limited by the
speed of light in vacuum.

Other massless particles, {\it eg}, neutrinos, are also allowed in
the special relativistic quantum field theory due to the group
enlargement from that of the galilean group to the Lorentz group
of transformations.  This group enlargement permits form-invariant
Dirac equation \cite{dirac} and also the theory of massive spin
$\frac{1}{2}$ fermions.

But, there is no possibility of explaining the origin of ``mass''
as well as of ``charge'' in, quantum or not, special relativistic
theories. It is only after we have {\em specified\/} the values of
mass and charge for a source particle that we can {\em obtain},
from the mathematical formalisms of these theories, its further
dynamics based on the given (appropriate) initial data. Hence, the
values of mass and charge are not {\em obtainable\/} in these
theories.

Clearly, the newtonian and the special relativistic frameworks,
both, are not sufficiently general to form the basis for the
entire physics. Therefore, some new developments are needed to
account for the ``origins'' of inertia and electric charge. We
recall here that these are the physical properties by which we
``identify'' or ``characterize individual material or physical
bodies.

Next, Lorentz had recognized \cite{subtle} (p. 155) the notion of
the {\em inertia of the electromagnetic field}. He then had a
clear conception that inertia (opposition of a physical body to a
change in its state of motion) could possess {\em origin\/} in the
field conception. Just as a person in a moving crowd experiences
opposition to a change in motion, a particle (region of
concentrated field) moving in a surrounding field experiences
opposition to a change in its state of motion. This is Lorentz's
conception of the field-origin of inertia.

Now, firstly, the distinction between the source and the field
must necessarily be obliterated in any formulation of this
conception. In other words, a field is the only basic concept and
a particle is a derived concept here. Secondly, the mathematical
formulation of this conception is also required to be {\em
intrinsically nonlinear}.

Solutions of linear equations obey superposition principle, and
required number of solutions can be superposed to obtain the
solution for any assumed field configuration. But, sources
generating the assumed field configuration continue to be the
singularities of the field. Hence, the distinction between source
and field cannot be obliterated.

Solutions of some (non-linear) field equations would not obey the
superposition principle. Then, one could hope that non-singular
solutions of non-linear equations for the field would permit
appropriate treatment of sources as singularity-free regions of
concentrated field energy.

An important question is now that of the appropriate (non-linear)
field equations, of obtaining these equations without venturing
into meaningless arbitrariness. In fact, this question is of some
appropriate non-linear mathematical formalism that need not even
possess the character of non-linear (partial) differential
equations for the field as a mathematical structure on the
underlying continuum. (It is also the issue of whether the most
fundamental formalism of physics could have a mathematical
structure other than that of the (partial) differential
equations.)

Historically, the very difficult and lengthy path to appropriate
non-linear equations was developed by Einstein alone.

The pivotal point of Einstein's formulation of the relevant ideas
is the equivalence of {\em inertial\/} and {\em gravitational\/}
mass of a physical body, a fact known since Newton's times but
which remained only an assumption of Newton's theory.

The above {\em equivalence principle\/} implies that the Lorentz
transformations are not sufficient to incorporate the explanation
of this equivalence of inertial and gravitational mass of a
material body. It then follows that general transformations of
coordinates are required and the physical basis is that of the
{\em general principle of relativity}.

On the basis of the {\em equivalence principle}, Einstein then
provided us the ``curved 4-geometry'' as a ``physically
realizable'' entity.

To arrive at his formulation of general relativity, Einstein
raised \cite{schlipp} (p. 69) the following questions:

$\spadesuit$ {\em Of which mathematical type are the variables
(functions of the coordinates) which permit the expression of the
physical properties of the space (``structure'')? Only after that:
Which equations are satisfied by those variables?} $\spadesuit$

He then proceeded to develop this theory in two stages, namely,
those dealing with
\begin{description} \item{(a)} pure gravitational field, and
\item{(b)} general field (in which quantities corresponding
somehow to the electromagnetic field occur, too).
\end{description}

The situation (a), the pure gravitational field, is characterized
by a symmetric (Riemannian) metric (tensor of rank two) for which
the Riemann curvature tensor does not vanish.

For the case (b), Einstein \cite{schlipp} (p. 73) then set up the
``preliminary equations'' to investigate \footnote{Einstein
expressed \cite{schlipp} (p. 75) his judgement of his preliminary
equations in the following words:  $\spadesuit$ {\em The right
side\/} (the matter part) {\em is a formal condensation of all
things whose comprehension in the sense of a field theory is still
problematic. Not for a moment, of course, did I doubt that this
formulation was merely a makeshift in order to give the general
principle of relativity a preliminary closed expression. For it
was essentially not anything {\em more\/} than a theory of the
gravitational field, which was somewhat artificially isolated from
a total field of as yet unknown structure}.  $\spadesuit$} the
usefulness of the basic ideas of General Relativity. His
(makeshift) field equations of this formulation of General
Relativity are form-invariant under general (spacetime) coordinate
transformations. The form-invariance of field equations under
general coordinate transformations is known as the principle of
general covariance \cite{std-texts} \footnote{The exact statement
used by Einstein
\cite{ein-dover} for this purpose is the following:  \\
$\spadesuit$ ``So there is nothing for it but to regard all
imaginable systems of coordinates, on principle, as equally
suitable for the description of nature. This comes to requiring
that: - \\  {\em The general laws of nature are to be expressed by
equations which hold good for all systems of coordinates, that is,
are covariant with respect to any substitutions whatever
(generally covariant)}. \\ It is clear that a physical theory
which satisfies this postulate will also be suitable for the
general postulate of relativity.'' $\spadesuit$  \\ Notice the
word ``suitable'' above. The strength of the requirement of
covariance depends upon the {\em a priori\/} selection of
geometrical quantities and can be relaxed by adding more
geometrical quantities to the theory. It is therefore based on the
Principle of Simplicity, rather than being any fundamental demand
of the general principle of relativity. De Witt in \cite{dewitt}
(1967, Vol. 160) remarks ``General relativity is concerned with
those attributes of physical reality which are
coordinate-independent and is the rock on which present day
emphasis on invariance principles will ultimately stand or
fall.''}.

Through these equations, geometric properties of the spacetime are
supposed to be determined by the physical matter. In turn, the
spacetime geometry is supposed to tell the physical matter how to
move. That is, the geodesics of the spacetime geometry are
supposed to provide the law of motion of the physical matter.

The ideas of general relativity essentially free Physics from the
association of physical meaning to coordinates and coordinate
differences, an assumption implicit in Newton's theory and in
special relativity. The formulation of Einstein's (makeshift)
field equations however attaches physical meaning to the invariant
distance of the curved spacetime geometry and considers it to be a
physically exactly measurable quantity.

Now, we may imagine \cite{std-texts} a {\em small\/} perturbation
of the {\em background\/} spacetime geometry and obtain equations
governing these perturbations. We may also consider
\cite{std-texts} quantum fields on the {\em unchanging
background\/} spacetime geometry.

Then, such methods (of perturbative analysis and also of the
Quantum Field Theory in Curved Spacetime) are quite similar of
nature to methods adopted for either the flat 3-continuum or the
flat 4-continuum in that these consider ``mathematical fields''
definable on the fixed and metrically curved, absolute,
4-continuum.

But, as far as Lorentz's or Einstein's ideas are concerned, these
above considerations of quantum field theory in curved spacetime
or perturbations of a curved spacetime geometry are, evidently,
{\em not\/} self-consistent since matter fields must affect the
background spacetime geometry. However, these are {\em not\/} the
real issues here.

Importantly, Einstein's approach to his field equations is beset
with internal contradictions of serious physical nature
\cite{smw-field}. These contradictions originate in the fact that
gravity is given preferential treatment in it. (See later.)

Firstly, Einstein's vacuum field equations  \footnote{To quote
Pais \cite{subtle} (p. 287) on this issue: \\  $\spadesuit$
``Einstein never said so explicitly, but it seems reasonable to
assume that he had in mind that the correct equations should have
no solutions at all in the absence of matter.''  $\spadesuit$ }
are entirely unsatisfactory \cite{smw-issues, smw-field} since
these are field equations for the pure gravitational field {\em
without\/} even a possibility of the equations of motion for the
sources of that field.

Certainly, matter cannot be any part of the theory of the vacuum
or the pure gravitational field. Then, there cannot be physical
objects in considerations of the pure gravitational field, except
as sources of such fields.

Now, a material particle is necessarily a spacetime singularity of
the pure gravitational field and, hence, mathematically, no
equations of motion for it are possible. Then, we have only
equations for the pure field but no equations of motion for the
sources creating those fields.

But, the vacuum field equations alone are {\em not enough\/} to
draw any conclusions of physical nature. Without the laws for the
motions of sources generating the (vacuum) fields, we have no
means of ascertaining or establishing the ``causes'' of motions of
sources. No conclusions of physical nature are therefore
permissible in this situation and, thus, the vacuum field
equations cannot lead us to physically verifiable predictions.

[Note that this above situation is markedly different from that
with special relativity. In special relativity, the background
geometry does not possess any {\em geometric singularity\/} at any
location, but only the (mathematical) fields defined on this
geometry can be singular. Then, similar to Newton's theory,
situations in special relativity lead us to physically testable
predictions.]

Secondly, recall that the energy-momentum tensor deals with the
density and fluxes of particles. Then, unless a definition of what
constitutes a particle is, a-priori, available to us, we cannot
construct the energy-momentum tensor.

Now, various relevant solutions of Einstein's field equations
represent a point particle as a spacetime singularity for which no
laws of motion are possible. Consequently, no acceptable
description of a particle is available in Einstein's approach to
the General Theory of Relativity.

Therefore, the concept of a particle is {\em not\/} clearly
defined to begin with and, hence, is {\em not\/} a-priori
available in Einstein's approach to his (makeshift) field
equations. Thus, Einstein's preliminary field equations are
ill-posed \cite{smw-issues, smw-field}.

Notably, this above does not, however, invalidate or question the
General Principle of Relativity in any manner whatsoever. (See
later.)

Next, recall that the quantum theory based on Schr\"{o}dinger's
$\Psi$-function provides us, essentially, the means of calculating
the probability of a physical event. It presupposes that we have
specified, say, the lagrangian or, equivalently, certain physical
characteristics of the problem under consideration. Evidently,
this is necessary to determine the $\Psi$-function using which we
then make (probabilistic) predictions regarding that physical
phenomenon under consideration.

At this stage, we then note the following fundamental limitation
of any theory that uses probabilistic considerations. (This
limitation is clearly recognizable for statistical mechanics in
relation to the newtonian theory.)

Importantly, the method of obtaining the probability of the
outcome of its toss is {\em irrelevant\/} to {\em intrinsic\/}
properties of the coin \footnote{Clearly, any unbiased coin has
the same probability of toss as that for another unbiased coin.
Then, the spatial extension and the material of the coin do not
determine the probability for the toss of an unbiased coin. In
turn, the laws leading us to only this probability of toss cannot
determine these {\em intrinsic\/} properties of the coin.}.

Therefore, methods of quantum theory, these leading us to the
probability of the outcome of a physical experiment about a chosen
physical object, cannot provide us the means of ``specifying''
certain intrinsic properties of that physical body. This fact,
precisely, appears to be the reason as to why we had to {\em
specify by hand\/} the values of the mass and the charge in
various operators of the non-relativistic as well as relativistic
versions of the quantum theory.

Therefore, quantum theory {\em presupposes\/} that we have {\em
specified\/} intrinsic properties of physical object(s) under
consideration. Hence, origins of such properties are to be sought
``elsewhere'' and not within the quantum theory.

Hence, we have that the formulation of general relativity as only
a theory of gravitation, Einstein's 1916 (makeshift) field
equations \cite{ein-dover}, is entirely unsatisfactory. We also
have that the probabilistic quantum theory cannot hope to explain
the origins of inertia and electric charge.

But, even when Einstein's field equations are physically
ill-posed, the underlying conceptions of the geometry being
indistinguishable from the physical matter need not be so. {\em
The General Principle of Relativity makes sense even without
Einstein's equations}. (See later.) Proper recognition of this
issue is then important.

A question therefore arises of some satisfactory mathematical
formulation of not only the fundamental conceptions underlying the
general principle of relativity but also of unifying them with the
fundamental conceptions of the quantum theory in an appropriate
manner.

But, for the ``new'' theory, we need the conceptual framework of
only the General Principle of Relativity or only that of the
Probabilistic Quantum Theory, and not the both. Let us then turn
to the issues related to this choice.

\subsection*{Other approaches to unification} \label{others}

Now, the two ``most successful'' theories of the 20th century,
namely, the Quantum Theory and Einstein's Theory of Gravity,
possess profoundly different conceptual frameworks and have led us
to adopt ``separate'' approaches to various problems of the micro
and the macro world \footnote{In this section, we will generally
refer to books or reviews wherein references to associated
original works can be found.}.

As far as the theories of the micro-world are concerned, these are
based on the principles of the Quantum Theory. QED, QCD etc.\ have
been experimentally justified by way of the verification of their
predictions, some to remarkable accuracies. These successes
\cite{feynman} lead us to accept the conceptual basis of the
Quantum Theory.

But, these theories of the micro physical world are certainly {\em
incomplete\/} without the incorporation of gravitation of the
micro-objects.

Now, the General Principle of Relativity has the appropriate
conceptual framework for gravitation. Einstein's equivalence
principle provides us the appropriate basis to formulate a theory
of gravitation. Einstein, in 1916, had followed exactly this path
to propose his preliminary equations for the field theory of
gravitation.

Einstein's formulation of General Relativity as {\em only\/} a
theory of gravitation leads us to classic tests of this theory of
gravity such as the precession of the perihelion of Mercury, the
bending of light, the gravitational red-shift etc.

These classic tests of General Relativity, though not as accurate
as those of the theories of the micro world, provide us adequate
reasons to also accept, simultaneously, the conceptual framework
of the General Principle of Relativity.

As an early recognition of the diverse conceptual frameworks of
these two aforementioned physical theories and also as an early
warning about the involved issues, Einstein wrote in 1916
(Preussische Akademie Sitzungsberichte) that:

$\spadesuit$ {\em Nevertheless, due to the interatomic movement of
electrons, atoms would have to radiate not only electromagnetic
but also gravitational energy, if only in tiny amounts. As this is
hardly true in Nature, it appears that quantum theory would have
to modify not only Maxwellian electrodynamics but also the new
theory of gravitation.} $\spadesuit$

Surely, Einstein's formulation deals only with the phenomenon of
gravitation and, consequently, does not incorporate
electromagnetism as well as other aspects of various known
micro-particles on the same footing as gravity. It is therefore
quite natural to expect that aspects related to quantum nature of
(gravitating) matter would necessitate fundamental changes to, the
then new, Einstein's theory of gravitation \footnote{For the same
reasons, ``explanations'' of the classic tests of Einstein's
theory of gravity can also be expected to be ``different'' when
these fundamental changes are taken into account.}.

Equally surely, an appropriate synthesis of the quantum theory and
the general principle of relativity is also necessary as their
diverse conceptual frameworks force on us a ``schizophrenic'' view
\cite{ashtekar1} of the physical world in which we treat macro
world as per Einstein's theory of gravity and the micro world as
per the quantum theory.

A question then arises of the ``final correctness'' of the
conceptual basis. Einstein, as is well known \cite{subtle}, chose
the General Principle of Relativity while most like Bohr,
Heisenberg, Dirac, Pauli chose the probabilistic Quantum Theory.

Einstein's attempts at the Unified Field Theory led him and
others, like Schr\"{o}dinger, de Broglie \cite{qmalter}, to
nowhere. But, Einstein sang his ``solitary song'' in favor of the
conceptual basis of General Relativity till the end \cite{subtle}.

Learning, perhaps, from the failures of Einstein's numerous
attempts at the formulation of a satisfactory Unified Field Theory
and keeping thereby ``faith'' in probabilistic methods of the
quantum theory, some like Bronstein, Rosenfeld, Pauli, then
attempted to {\em quantize} \footnote{See \cite{ashtekar1} for an
excellent historical account of related conceptual developments.}
Einstein's gravity in the same manner as was followed for other
fields such as the electromagnetic field.

But, such an approach to the ``quantum theory of gravity'' was
slated to face serious mathematical difficulties. The foremost of
these difficulties is that the metric of the spacetime geometry is
not just an inert arena but also the primary dynamical quantity in
Einstein's theory of gravity which has {\em no\/} background
metric.

The known procedures of quantum theory were geared to the
existence of a background metric such as the Minkowski metric.
Therefore, by giving up Einstein's most cherished dream,
inseparability of geometry and matter, the 4-metric was treated as
a perturbative tensor field over the (usually) flat background.
This gave us the covariant formalism of quantum gravity.

For this formalism, Feynman then extended perturbative methods of
QED to Einstein's gravity. Then, De Witt formulated \cite{dewitt}
the Feynman rules for covariantly quantized Einstein's gravity.
This all then led us to the notion of a massless spin-$2$
graviton. But, this perturbative quantum gravity turned out to be
non-renormalizable.

The non-renormalizable nature of perturbative quantization of
Einstein's gravity was interpreted to mean that important high
energy processes (at the Planck energy scale) were being ignored
by these perturbative methods.

A cure for this problem was sought by coupling Einstein's gravity
to other fields, as it must be. In particular, ``super-gravity''
imagined cancellation of bosonic infinities of the gravity by
those of the suitable fermionic fields \cite{strings}.

It was soon realized that super-gravity will be non-renormalizable
at the fifth and at higher order loops. In the mean while, an
innovative idea of replacing point particles by a 1-dimensional
(Nambu-Goto) string, an extended object, was invoked for the
theory of strong interactions.

Originally, the ``Duality Hypothesis'' that the $s$- and
$t$-channel diagrams provide ``dual'' descriptions of the same
physics, where $s$ and $t$ are the Mandelstam variables, was tried
\cite{strings} for the strong interactions. However, models based
on the above duality hypothesis predicted a variety of massless
particles which do not exist in the hadron world. Then, this
failure of the duality theories eventually yielded the way for
QCD.

But, duality theories could accommodate high spin particles
without ultraviolet anomalies and, in ``quantized'' general
relativity, the gravitational field is to be a massless spin-2
graviton. Hence, the idea that some ``duality theory'' could be a
``theory of {\em all\/} interactions'' soon caught attention.
Then, the Veneziano duality model was also shown to be a
relativistic string.

In the String Theory approach, different modes of oscillations of
the string correspond to particle-like states. Then, it turns out
that, in addition to the spin-1 mode, there also exists in String
Theory a spin-2 mode. A boon in disguise, the spin-2 mode could
then represent gravity.

Within the theoretical framework of the String Theory, only one
fundamental quantity, the string tension, needs to be specified
a-priori. Then, it is tempting indeed to think that a built-in
unification of all interactions by way of the modes of vibrations
of the string is possible. This expectation led to a flurry of
theoretical activity.

As many implications of String Theory were being developed,
usefulness of its ideas was also explored in the context of
cosmological conceptions. Such studies explored mainly the
``cosmological '' implications of higher dimensions necessarily
required for the String Theory.

The string theory \cite{strings} necessarily uses dimensions
higher than the usual four (10 for the super-string and 26 for the
bosonic string for which quantum anomalies do not occur in the
theory). It also uses the ideas of super-symmetry and works with
background fields as essential ingredients. The overall thrust of
the String Theory is then certainly on the unification of all the
four interactions, including Einstein's gravity by way of the
spin-2 mode of the string oscillations.

Still, it needs to be adequately realized that the String Theory
{\em cannot\/} hope to explain the origins of either the inertia
or the electrostatic charge on the basis of only the string
tension which is an arbitrary constant of this theory.

But, a ``theory of everything'' must provide these aforementioned
explanations \footnote{We could always question any chosen value
of the string tension. Why not any other value?}. If not anything
else, this aforementioned inability of the String Theory alone
forces us to look ``elsewhere'' for the explanations of properties
of matter.

Next, another approach to quantum theory of gravity also evolved
simultaneously to the String Theory. It was shown by Dirac
\cite{dirac01} that the hamiltonian of Einstein's theory of
gravity is a mathematically well-defined quantity. Motions
generated by this hamiltonian are then evolutions in time of the
initial spatial section, the Cauchy surface of the Einstein field
equations.

These theoretical developments led to the canonical approach to
Einstein's gravity which is then to be viewed as the dynamical
theory of the 3-geometries - the geometrodynamics \footnote{The
formalism of geometrodynamics is a conceptually consistent,
rigorous, mathematical description of the ``evolution'' of a
3-geometry to a 4-geometry of the spacetime. An appropriate
``quantization'' of geometrodynamics therefore leads us to a
rigorous mathematical formalism for the
corresponding quantum theory, that is, to the quantum theory of geometry. \\
However, as we shall see later, the curvature of geometry is {\em
not\/} a sufficiently general mathematical concept that can
substitute the physical, the newtonian, notion of force in its
entirety. This fact then severely limits the ``physical
usefulness'' of these approaches.}.

The ADM-formalism then led to further developments in canonical
approach. The 3-metric and the extrinsic curvature of the
3-geometry are the canonically conjugate variables of the
geometrodynamics. Notably, Einstein's field equations for gravity
then reduce to two types of equations: constraints and evolution
equations. One could then think of using (generalizations of)
Dirac's methods for quantization of ``constrained systems'' for
these sets of equations of gravity.

These developments led to a definite (Wheeler's) program of
ambitious nature to quantize Einstein's gravity. However, this
proposal remained mostly formal and quite separate from quantum
theories of the micro world.

In this last context, deserving special mention are the recent
developments related to Quantum Geometry \cite{ashtekar1}.
Notably, the Ashtekar phase space of Einstein's gravity is the
same as that of the gauge theories of the micro world.

The basis of these developments is a canonical transformation of
the ADM variables of gravity that yields, at the most, polynomial
constraints. ``Spin connection'' and ``triad'' achieve together
this simplification. The 3-metric, obtainable from Ashtekar's
spinorial variables, is nowhere needed in the ``metric-free''
formalism.

Canonical gravity being non-perturbative, these achievements were
quite important for quantum gravity. The quantization of the
Einstein-Ashtekar gravity leads to ``loop-states,'' 1-dimensional
excitations, from which the continuum arises only as a
coarse-grained approximation over the ``weave'' states of quantum
geometry.

This ``quantized'' Einstein-Ashtekar gravity, the Theory of the
Quantum Geometry, then appears to be the ``ultimate'' logical end
of the program of canonical gravity. But, it has not provided yet
any principle or procedure for incorporating other three
interactions.

However, this formalism of Quantum Geometry is basically a
rigorous mathematical theory, like the Euclidean geometry, in
which one needs to ``insert by hand'' physical qualifications of
matter to connect it to the physical world.

In the context of this above issue, one is then bound to recall
Einstein's theorem \cite{schlipp} (p.63) that:

$\spadesuit$ {\em ... nature is so constituted that it is possible
logically to lay down such strongly determined laws that within
these laws only rationally completely determined constants occur
(not constants, therefore, whose numerical values could be changed
without destroying the theory). -\,-\,-} $\spadesuit$

Then, an additional ``physical difficulty'' of the Quantum Theory
of Geometry is that physical constants (such as Planck's constant,
Newton's constant of gravitation etc.) also do not arise in it
from various permissible mutual relationships of physical bodies,
just exactly as we obtain them experimentally out of mutual
relationships of the involved physical objects.

But, physical constants have to be specified by hand not only in
Quantum Geometry but also in String Theory. Consequently, these
theories are, physically speaking, quite limited.
\footnote{Quantum Theory of Geometry, String Theory etc.\ could,
however, provide useful tools, just exactly as the Euclidean
geometry is for our ordinary, day-to-day, purposes. But,
Einstein's vacuum as well as (preliminary) field equations with
matter, these being based on physically inconsistent pictures,
cannot be ``trusted'' in any such sense.}. The same limitations
apply to other highly original and motivating approaches such as
the Euclidean quantum gravity \cite{eqg}, twistor theory
\cite{twist}, non-commutative geometry \cite{ncg}, the theory of
H-spaces \cite{newman} etc., although these approaches are not
discussed here for want of space and purpose.

Now, as seen earlier, Einstein's formulation of his field
equations is itself beset with problems of serious physical
concerns. Moreover, as also seen earlier, methods of quantum
theory, leading us to the calculation of only the probability of a
physical event, cannot provide us the ``origins'' of intrinsic
properties of physical objects.

Consequently, it is necessary to ``look'' beyond the mathematical
formalism of either of these theories to reach to some
appropriate, theoretically satisfactory, explanations of the
origins of the properties of physical matter.

Now, as will be discussed in the next sections, the General
Principle of Relativity still holds. It therefore seems advisable
to follow the conservative path of developing appropriate
mathematical formulation based on the general principle of
relativity and basic conceptions of the quantum theory, and to let
it suggest to us the explanations of physical phenomena.

Hence, even at the cost of being elementary and pedantic, it
appears to be certainly worthwhile to recall here as to what the
``phenomenon of gravitation'' is all about and how exactly
Newton's and Einstein's theories attempt to explain it.

\subsection{Conceptual Preliminaries} \label{idea-basis}

To begin with, let us note that the foremost of the concepts
behind Newton's theory is, undoubtedly, (Galileo's) concept of the
inertia of a material body. We postulate that every material body
has this inertia for motion.

The association of the inertia of a material body with the points
of the Euclidean space is the first primary physical conception
that is necessary for Newton's theory to describe motions of
physical bodies. Then, with this association, the Euclidean
distance becomes the {\em physical distance\/} separating material
bodies and the Euclidean space becomes the {\em physical space\/}
for any further considerations of Newton's theoretical scheme.

Next, a {\em physical clock\/} is a material body undergoing
periodic motion or a periodic phenomenon. Essentially, in Newton's
theory, a physical clock is a set of points of the Euclidean space
exhibiting periodic motion under a {\em periodic\/}
transformation. Mathematically, in Newton's theory, let $A$ be the
set of all points $x_{_A}$ of the Euclidean space making up the
clock. Let $T$ be the {\em periodic\/} transformation such that
$\forall \;x_{_A}\in A,\,T^n\,x_{_A}=x_{_A}$, where $n$ is the
period of the transformation $T$.

In Newton's theory, an observer can {\em observe\/} the {\em
entire periodic motion\/} of the material body of the clock (under
the use of the transformation $T$) without disturbing the clock in
any manner whatsoever. Then, the {\em known\/} state or the
reading of the clock {\em represents\/} the {\em physical time}.
Any ``measurement'' of the physical time gives the period or the
part of the period $n$ of the transformation $T$. It is tacitly
assumed in these considerations that the involved quantities are
{\em exactly\/} measurable.

Now, consider a material point with an initial location
$\vec{x}_o$. In Newton's theory, the trajectory of this material
point is a (continuous) sequence of points of the Euclidean space.
It is then a ``curve'' traced by the point $\vec{x}_o$ under some
transformation $\tilde{T}_t$ of the Euclidean space where
parameter $t$ labels points of the sequence. Of course, the
transformation $\tilde{T}_t$ need not be periodic.

The label parameter $t$ can then be made to correspond to the
physical time in a one-one correspondence. This is theoretically
permissible as the measurement of the physical time does not
disturb the clock in any manner whatsoever in Newton's theory.
This correspondence is the physical meaning of the labelling
parameter $t$.

An observer can thus check the position of another material point
against the state of a {\em physical clock}. Without this
``correspondence,'' the geometric curve of the Euclidean space has
no physical sense for the path of a material point.

When such physical associations are carried out, we say that the
material point is at ``this'' location given by the three space
coordinates and the physical clock is {\em simultaneously\/}
showing ``this'' time. This {\em simultaneity\/} is inherent in
the physical associations of Newton's theory.

Also, as a consequence of the fact that Newton's theory treats
material bodies as {\em existing\/} independently of the space,
the state of a physical clock or its reading is {\em assumed\/} to
be independent of the motion of another material point or points
separately under considerations. Then, physical time is {\em
independent\/} also of the coordination of the metrically-flat
Euclidean space.

But, then, the motion of a material point in the ``physical''
space does not produce any change in that space. Clearly, this
fact applies also to the periodic motion or the periodic
phenomenon making up the physical clock.

The ``physical'' construction of the coordinate axes and clocks
must also be using the material bodies, for example, coordinate
axes could be constructed using ``sufficiently long'' material
rods, say, of wood. Then, any material object, a road-roller, say,
crossing the coordinate axis must ``affect'' the corresponding
wooden rod.

But, in Newton's theory, the coordinate axes of the Euclidean
space do not get affected by the motions of other material bodies.
Clearly, use of non-cartesian coordinates does not change this
state of affairs with Newton's theory.

Now, any difference of coordinates in the Euclidean space is a
``measuring stick or rod'' that can be used to ``measure'' the
physical separation of material bodies.

Furthermore, in Newton's theory, each observer has a coordinate
system of such measuring rods and clocks. Then, when one observer
is in motion (relative to another one), the {\em entire\/} system
of coordinate axes and clocks is also carried with that observer
in motion.

Conceptually, the aforementioned physical situation is
``acceptable'' except in one case. Surely, we cannot have a
material rod with one observer and {\em simultaneously}, another
material rod, {\em at the same place}, with other observer in
(uniform or not) motion relative to the first one.

But, in Newton's theory, measuring stick of one observer does not
collide with that of another observer in motion even when both
these sticks arrive at the same place. Unacceptably, any two such
measuring sticks just pass through each other without even
colliding on their first contact.

The same situation does not arise for other material bodies which
are supposed to collide on their first contact. Then, in Newton's
theory, the measuring rods of the physical space - Euclidean space
- are treated {\em separately\/} than other material objects. But,
measuring rods must also be made up of material objects. Then,
their separate treatment is, theoretically, not appropriate one.
Surely, this problem with Newton's theory is, undeniably, of
serious theoretical concern.

(This above issue would not have been relevant if the Euclidean
distance were also not, simultaneously, the physical distance
separating material bodies. Mathematically, the continuum
$\mathbb{R}\times\mathbb{R}\times \mathbb{R}$ can be assumed.)

Hence, Newton's theory attempts to {\em explain all phenomena as
relations \footnote{The constants of Nature then arise in Newton's
theory from such relationships ``postulated'' to be existing
between physical objects.} between objects existing in Euclidean
space and time}. It achieves this by attributing {\em absolute\/}
properties to the space and the time, thereby totally separating
them from the properties and motions of matter.

Thus, limitations of Newton's theoretical scheme (providing his
famous three laws of motion) originate in its use of the Cartesian
concepts related to the Euclidean space and the associations of
properties of material bodies with the points of this
metrically-flat space.

Now, the entire physical structure of Newton's theory is woven
around only two basic concepts, namely, those of the inertia and
the force.

Clearly, the force, as a cause of motion, is another pivotal
concept of Newton's theory. Hence, consider the status of the
concept of force, the cause behind motions of material bodies,
within Newton's theory of mechanics.

Firstly, we could ask: What is the {\em cause\/} of this force?
Within Newton's overall theoretical scheme, only a material point
can be the source of force. A material point ``here'' {\em acts\/}
on a material point located ``there'' with the specified force.
Newton's theoretical scheme is therefore an {\em action at a
distance\/} framework.

Then, in Newton's theory, we can consider a physical body as one
material point and also other physical bodies as other material
points. We vectorially add the forces exerted by each one on the
first physical body to obtain the total force acting on it. It is
this {\em total force\/} that is used by Newton's second law of
motion to provide the means of establishing the path followed by
that physical body under the action of that total force.

In Newton's second law of motion, we {\em must\/} first {\em
specify\/} the {\em force\/} acting on a material body. Only then
can we solve the corresponding differential equation(s) and
obtain, subject to the given initial data, the path of the
material point representing that material body.

Then, without the {\em Law of Force}, it is clear that the {\em
Law of Motion\/} is {\em empty\/} of contents in Newton's
theoretical framework. This is an extremely important issue for a
physical theory.

From our ordinary, day-to-day, observations, we notice that
various objects fall to the earth when left ``free'' in the air.
We then say that objects {\em gravitate\/} towards the Earth. This
is, in a nutshell, the {\em phenomenon of gravitation}.

We then need to explain as to why the objects ``ordinarily''
gravitate to Earth, {\em ie}, why they have a tendency to come
together or why the distance between them decreases with time.

In Newton's theory, only {\em forces\/} ``cause'' motions of
material bodies. Then, the gravitating behavior of objects is
``explainable'' only by {\em postulating\/} a suitable {\em force
of gravity\/} that makes material bodies fall to the Earth.

But, in Newton's theory, a force acts between {\em any two\/}
separate material points possessing the {\em required source
property\/} by virtue of which the force in question is generated.
Furthermore, for the internal consistency of Newton's scheme, the
force so generated by one material point on the second material
point must also be equal in amplitude but opposite in direction to
that generated by the second material point on the first material
point. This is Newton's third law of motion. This law also has the
status of a postulate within the overall scheme of the newtonian
mechanics.

Newton had {\em assumed\/} that the force of gravity is
proportional to the {\em inertias\/} of the two material points
under consideration because, following Galileo, he had postulated
that inertia ``characterizes'' a physical body.

Such a force of gravity must then be generated by a chosen body
(Earth) on {\em all\/} the other material bodies because, by
postulate, every material body possessed inertia. The force of
gravity must then be {\em universal\/} in character.

To explain many of the day-to-day observations involving the
terrestrial bodies as well as planetary motions that were already
known in details, Newton was therefore compelled to state a Law of
Force - Newton's Law of Gravity - to explain the phenomenon of
gravitation.

In fact, in Newton's theory, a material body has two {\em
independent\/} attributes: the first, its {\em inertial mass}, is
a measure of the opposition it offers to a change in its state of
motion, and the second, its {\em gravitational mass}, is a measure
of the property by virtue of which it produces the force of
gravity on another material point.

Various observations, since Galileo's times, then indicate
\cite{0411052} that the inertial and the gravitational masses of a
material body are equal to a high degree of accuracy. However,
this equality becomes an assumption of Newton's theory.

Inverse-square dependence of the gravitational force on the
distance separating two bodies is also an assumption of Newton's
theory.

In relation to the inverse-square dependence of Newton's force of
gravity on distance separating two bodies, we could then always
raise questions: Why not any other power of distance? Why should
this force not contain time-derivatives of the space coordinates?
Clearly, Newton's theory offers no explanation for even the
inverse-square dependence of the force of gravitation.

Hence, in Newton's scheme, his law of gravitation has the status
of a {\em postulate\/} about the {\em force\/} acting between two
material particles separated by some spatial distance.

Also, Coulomb's law from the electrostatics provides another,
postulated, fundamental force. It is also assumed to exist {\em
universally\/} between any two {\em charged\/} material bodies. It
is an ``additional'' force, over and above that of gravity, which
Newton's theory postulates to explain the motions of charged
material points.

Now, every object does not fall to the Earth. So, ``something''
opposes the attractive force of gravity. That ``something'' must
also be another force. Thus, in a nutshell, a force can oppose
another. But, {\em every\/} force is an assumption here.

Then, if we find that some physical body, for example, a star, is
stable, we could, in Newton's theory, explain its stability by
postulating another suitable {\em force \/} which counterchecks
the force of self-gravity of the star. On the other hand, if the
star were unstable, existence of ``unbalanced'' forces in the star
is implied.

In Newton's theory, there are no fundamentally important issues
involved here than those related to finding the nature of the
force opposing the self-gravity of the star. It is essential to
recognize this important aspect of Newton's theory.

Nonetheless, in spite of it being an assumption of Newton's
theory, Newton's inverse-square law of gravitation does possess
certain experimental justification - it is this inverse-square
dependence that is known to be consistent with various
observations and experiments.

Still, it cannot be denied that Newton's law of gravitation is an
important {\em assumption\/} of the newtonian mechanics.

To reemphasize the status of the laws of the force in Newton's
theory, we note that every force is an assumption of this theory.
Some forces are assumed to exist {\em universally\/} between any
two material bodies. In particular, ``the force of gravitation''
is postulated by this theory.

In Newton's theory, every ``fundamental'' notion of the force
necessarily requires a source property to be attributed to
material bodies. Then, the action-at-a-distance force has this
important characteristic always.

Obviously, Newton's theory cannot hope to ``explain'' the origin
of any of such source attributes, each of these source attributes
being an assumption of that theory. Evidently, the same applies to
other action-at-a-distance theories. It is important to recognize
this fact at this early stage of our present considerations.

Perhaps, we would have been satisfied even with these assumptions
of Newton's theory if it were not for the fact that Newton's
theory does not explain the phenomena displayed by Light.
Moreover, various observations related to the wave-particle
duality of light as well as matter are also unexplainable within
the newtonian scheme.

Apart from various fundamental reasons of theoretical nature as
discussed earlier, it is also for such experiments or observations
which cannot be explained by Newton's theory, that some suitable
``new'' theory becomes a necessity.

Of course, different results of Newton's theory which successfully
describe motions of material bodies must be obtainable within the
new theory in some suitable way.

Then, to formulate ``new'' theory, we need to, not just modify
but, abandon some newtonian concepts at a fundamental level.

Now, let us also note the newtonian {\em principle of
relativity\/} which states that: {\em If a coordinate system $K$
is chosen so that Newton's laws of motion hold good without the
introduction of any pseudo-forces with respect to this frame then,
the {\em same\/} laws also hold good in relation to other
coordinate system $K'$ moving in {\em uniform\/} translation
relatively to $K$}. This principle is a direct consequence of the
experiments conducted by Galileo.

These issues then bring us to the question of the status of the
principle of relativity in a theoretical framework that abandons
the newtonian concept of the force. It is to this and other
related issues that we now turn to.

\subsection{The General Principle of Relativity} \label{gpr}

To incorporate the physical description of the phenomena displayed
by Light, zero rest-mass object, Einstein modified the newtonian
principle of relativity as: {\em If a coordinate system $K$ is
chosen so that physical laws hold good in their simplest form with
respect to this frame then, the {\em same\/} laws also hold good
in relation to other coordinate system $K'$ moving in {\em
uniform\/} translation relatively to $K$}. This is the {\em
special principle of relativity}. As is well known, together with
the principle of the constancy of the speed of light in vacuo, it
leads to the special theory of relativity.

Einstein's this special principle of relativity is essentially the
{\em same\/} as the principle of relativity of Newton's theory.
The word ``special'' indicates here that the principle is
restricted to the case of uniform translational motion of $K'$
relative to $K$ and does not extend to non-uniform motion of $K'$
in relation to the system $K$.

But, even the special theory of relativity is not sufficiently
general to offer explanations for various physical phenomena as
observed. Not only gravity, but, as should be amply clear, the
origin of inertia as well as the origin of electrostatic charge
are also not explainable in special relativity.

Primarily, the special theory of relativity is an extension of
{\em only\/} the newtonian laws to accommodate properties of
motions of material bodies with vanishing inertia \cite{100-yrs}.
It achieves this extension by acknowledging the fact that, in our
day-to-day experiences, we use Light to observe.

But, the special theory of relativity also rests on the metrically
flat continuum and is, thereby, beset with the problems of
treating the measuring rods and clocks separately from all other
objects. There is therefore the need to extend the special
principle of relativity.

Then, Einstein extended \cite{ein-dover} this principle on the
basis of Mach's reasoning as follows.

Mach's reasoning concerns the following situation. Consider two
{\em identical\/} fluid bodies so far from each other and from
other material bodies that only the {\em self-gravity\/} of each
one needs to be considered. Let the distance between them be
invariable, and in neither of them let there be ``internal
motions'' with respect to each other. Also, let either body, as
judged by an observer at rest relative to the other body, {\em
rotate\/} with constant angular velocity about the line joining
them. This is, importantly, a {\em verifiable\/} relative motion
of the two identical fluid bodies.

Now, using surveyor's instruments, let an observer at rest
relative to each body make measurements of the surface of that
body. Let the revealed surface of one body be {\em spherical\/}
and of the other body be an {\em ellipsoid of revolution}.

The question then arises of the {\em reason\/} behind this
difference in these two bodies. Of course, no answer is to be
considered satisfactory unless the given reason is {\em
observable}. This is so because the Law of Causality has the
``genuine scientific'' significance only when observable effects
ultimately appear as causes and effects.

As is well known, Newton's theory as well as the special theory of
relativity require the introduction of {\em fictitious or the
pseudo\/} forces to provide an answer to this issue. The reason
given by these two theories is, obviously, entirely unsatisfactory
since the pseudo-forces are unobservable.

Any cause within the system of these two bodies alone will not be
sufficient as it would have to refer to the absolute space only.
But, the absolute space is necessarily unobservable and,
consequently, any such ``internal'' cause will not be in
conformity with the law of causality.

The only satisfactory answer is that the cause must be outside of
this physical system, and that must be referred to the {\em real
difference\/} in motions of {\em distant\/} material bodies
relative to each fluid body under consideration.

Then, the frame of reference of one fluid body is {\em
equivalent\/} to that of the other body for a description of the
``motions'' of other bodies. As Mach had ``concluded''
\cite{mach}, no observable significance can be attached to the
cause of the difference in their shapes without this equivalence.

{\em The laws of physics must then be such that they apply to
systems of reference in any kind of motion (without the
introduction of any fictitious causes or forces)}. This is then
the extended or the {\em general principle of relativity}.

{\em Clearly, the reference frames must be constructed out of
material bodies and any motions of ``other'' material bodies must
affect the constructions of the reference frames. Therefore, the
general principle of relativity also means that the laws of
physics must be so general as to incorporate even these situations
in their entirety.}

Now, equally important is the fact that {\em the notion of the
{\em physical time\/} must undergo appropriate changes when the
above is implemented}. In particular, the correspondence of the
labelling parameter of the ``path'' of a physical body with the
time displayed by a physical clock must be different than that in
Newton's theory or in special relativity. {\em Notably, the
underlying continuum and the physical space are then different}.

Einstein connected the general principle of relativity with the
observation that a possible {\em uniform\/} gravitation imparts
the same acceleration to all bodies. This insight leads us to
Einstein's equivalence principle. It arises as follows.

Let $K$ be a Galilean frame of reference relative to which a
material body is moving with uniform rectilinear motion when far
removed from other material bodies. Let $K'$ be another frame of
reference which is moving relatively to $K$ in uniformly
accelerated translation. Then, relatively to $K'$, that same
material body would have an acceleration which is independent of
its material content as well as of its physical state.

The observer at rest in frame $K'$ can then raise the question of
determining whether frame $K'$ is ``really'' in an accelerated
motion. That is, whether this is the only cause for the
acceleration of bodies being independent of material content.

Now, let various bodies, of differing material contents and of
differing inertias, fall freely under the action of Earth's
gravity after being released from the same distance above the
ground and at the same instant of time. Galileo had, supposedly at
the leaning tower of Pisa, observed that these bodies reach the
ground at the same instant of time and had thereby concluded that
these bodies fall with the {\em same\/} accelerations.

Hence, the {\em decrement in distance\/} between material bodies
displaying only the phenomenon of gravitation is then {\em
uniquely\/} characterized by the fact that the acceleration
experienced by material bodies, occupying sufficiently small
region of space near another material body of large spatial
dimension, is independent of their material content and their
physical state. Here, the gravitational action of the larger
material body can then be treated as being that of {\em uniform\/}
gravitation.

Therefore, the answer to the question raised by the observer at
rest in the frame $K'$ is in the negative since there does exist
an analogous situation involving the phenomenon of uniform
gravitation in which material bodies can possess acceleration that
is independent of their material content and the physical state.

Thus, the observer at rest in the frame $K'$ can alternatively
explain the observation of the ``acceleration being independent of
the physical state or the material content of bodies'' on the
basis of the phenomenon of {\em uniform\/} gravitation.

The mechanical behavior of involved material bodies relative to
the frame $K'$ is then the same as that in the frame $K$, being
considered ``special'' as per the special principle of relativity.
We can therefore say that the two frames $K$ and $K'$ are {\em
equivalent\/} for the description of the facts under
consideration. Clearly, we can then extend the special principle
of relativity to incorporate even the ``accelerated'' frames.

Borrowing Einstein's words on this issue \cite{ein-dover}, this
above situation is then {\em suggestive\/} that:

$\spadesuit$ {\em the systems $K$ and $K'$ may both with equal
right be looked upon as ``stationary,'' that is to say, they have
an equal title as systems of reference for the physical
description of phenomena}. $\spadesuit$

\noindent [Note the word ``suggestive'' in this statement.]

Now, the equality of inertial and gravitational masses of a
material body refers to the ``equality'' of corresponding
qualities of a material body. But, this is permissible only in a
theory that assumes the concept of a force as an external cause of
motions of material bodies. The concept of the gravitational mass
is, however, {\em irrelevant\/} when the concept of force is
abandoned. Only the concept of the inertia of a material body is
then relevant to the motions of physical bodies.

What then is the status of the general principle of relativity in
a theory that completely abandons the concept of force? Does it
hold in the absence of the concept of force?

From the above, it should now be evident that the general
principle of relativity stands even when the concept of force is
abandoned because it only deals with the {\em observable\/}
concept of an acceleration of a material body. Specifically, in
the context of Einstein's equivalence principle, it rests only on
the observation that uniform gravity imparts the same acceleration
to all the bodies.

Now, it is crucial to recognize that the equivalence principle
establishes only the {\em consistency\/} of the phenomenon of
gravitation with the general principle of relativity. Clearly, the
equivalence principle {\em is not logically equivalent\/} to the
general principle of relativity.

As noted earlier, Einstein had, certainly, been quite careful to
use the word ``suggestive'' in stating the relation of these two
{\em different\/} principles. He further wrote in
\cite{ein-dover}:

$\spadesuit$ ... in pursuing the general theory of relativity we
shall be led to a theory of gravitation, since we are able to
``produce'' a gravitational field merely by changing the system of
coordinates. $\spadesuit$

But, in spite of the phenomenon of gravitation being consistent
with the general principle of relativity, there could, in
principle, exist other physical phenomena which could be
inconsistent with the general principle of relativity. This would
have been the situation if Newton's theory had shown us the
existence of some ``fundamental real force''  distinguishing
accelerated frames from the inertial frames but that force
``explaining'' the observed motions of material bodies.

This last is, of course, not the situation and we therefore have
the faith that the general principle of relativity should be the
basis of any ``satisfactory'' physical theory. Moreover, the
``verifiability'' of the motion in Mach's arguments considered
earlier is also indicative of the ``universality'' of the general
principle of relativity.

From the above, it should then be clear that the general principle
of relativity can be reached from more than one vantage issues.
Each such issue can then indicate only that some physical
phenomenon related to that issue is {\em consistent\/} with this
principle of relativity. The mutual consistency of the general
principle of relativity and various physical conceptions then
becomes the requirement of a satisfactory physical theory.

Therefore, physical construction of the frames of reference, the
{\em physical coordination\/} of the {\em physical space\/} using
measuring rods, is also one of the primary requirements of the
satisfactory theory based on the general principle of relativity.

Then, it should now be also clear that the {\em universal theory
of relativity}, a physical theory explicitly based on the general
principle of relativity, will not be just a theory of gravitation
but, of necessity, also the {\em theory of everything}. It is
certainly decisive to recognize this.

{\em Therefore, a theory which abandons the concept of force
completely can ``explain'' the phenomenon of gravitation by
demonstrating that the decrement of distance between material
bodies is, {\em in certain situations}, independent of their
material contents and physical state.} By showing this, a theory
of the aforementioned type can incorporate the phenomenon of
gravitation.

Why is this above mentioned demonstration expected to hold only in
{\em certain situations}?

To grasp the essentials here, let us recall that, in Newton's
theory, only the {\em total force\/} acting on a physical body is
used by Newton's second law of motion. We usually also {\em
decompose\/} this force into different parts in the well known
manner as the one arising due to gravity, the one arising due to
electrostatic force etc.

But, what matters in Newton's theory for the motion of any
physical body is the total force acting on it and not the
decomposition of this total force in parts, the decomposition,
strictly speaking, being quite {\em irrelevant}.

Thus, the phenomenon of gravitation is, then, ``displayed'' by
material bodies, essentially, only in {\em certain situations},
those for which the total force is that due to gravity.

This above is, in overall, the significance of the general
principle of relativity.

\subsection{General Expectations from the Universal Theory of Relativity} \label{gent}

Now, what are our general expectations from any ``new'' theory
then?

Clearly, the concept of the inertia of a material body is {\em
more fundamental\/} than that of the force because the conception
of gravitational force requires the introduction of gravitational
mass which is conceptually very different but ``equals'' the
inertia in value to a high degree of accuracy \cite{0411052}.
Hence, only the newtonian concept of force comes under scrutiny
for modifications.

Therefore, it must be adequately recognized that the newtonian
concept of ``force'' will have to be abandoned in the process of
developing the new theoretical framework. In other words, the
``cause'' behind the motions of material bodies will have to be
conceptually entirely different than has been considered by
Newton's theory.

Consequently, ``agreements'' of the results of the new theory with
the corresponding ones of Newton's theory can only be {\em
mathematical\/} of nature. The physical conceptions behind the
mathematical statements of the new theory will not be those of
Newton's theory.

Therefore, any explanation of the phenomenon of gravitation in the
new theory will only involve the demonstration of the ``decrement
of distance'' under certain situations involving material bodies.
It must, of course, be shown that this decrement in distance is,
for these situations, such that the ``acceleration'' of the bodies
is independent of their material content and the physical state.
It must also be shown that the ``known'' inverse-square dependence
of this phenomenon arises in the new theory in some mathematical
manner.

Any such ``new theory'' must then explain the ``origin'' of the
inertia of material bodies. It must also incorporate the
``physical'' construction of the coordinate system that must,
necessarily, change with the motions of material bodies in the
``physical'' space. Without the appropriate incorporation of these
two issues, no theory can be considered to be physically
satisfactory.

Any such ``new'' theory needs also ``explain'' the equality of the
{\em inertial\/} and the {\em gravitational\/} mass of a material
body. The equality of these two entirely different physical
conceptions, even with experimental uncertainties, is a sure
indication that the {\em same quality\/} of a material body
manifests itself, according to circumstances, as its inertia or as
its weight (heaviness).

But, it must be remembered that the concept of the gravitational
mass and that of the electrostatic charge owe their origins to the
newtonian concept of the force.

But, the ``source properties'' cannot be basic to the new theory
that abandons the concept of the force. Consequently, the
gravitational mass of the material body will not be fundamental to
the new theory, but the inertial mass will be. Some ``entity''
that replaces the electrostatic charge will also be basic to the
new theory.

Then, it must also follow from the mathematical framework of the
new theory that the inertial mass can also be ``naturally''
considered as the ``source'' in the mathematical quantity that can
be the newtonian gravitational force.

Similarly, the quantity that, in the new theory, replaces the
electrostatic charge must also naturally appear as the ``source''
in the mathematical quantity that can be considered to be
Coulomb's electrostatic force.

Furthermore, we need to demand that the ``new'' theory must also
not contain the law of motion which is ``independent'' of the law
of the force. That is to say, the force as an external quantity,
to be ``specified'' separately of the law of motion, must not
occur in this new theory. In it, we can only have the law of
motion.

Crucially, abandonment of the concept of force that is independent
of the law of motion applies, at the same time, to ``every kind of
(fundamental) force'' postulated to be acting between the material
particles by Newton's theory.

Therefore, the {\em conceptual framework as well as the
mathematical formalism or procedure\/} by which we ``replace'' the
concept of force (as an ``external cause of motion'' independent
of the law of motion) {\em will have to be applicable to every
kind of (fundamental) force that Newton's or any other theory has
to postulate or assume to explain the observed motions of material
bodies}.

The {\em Principle of the Simplicity\/} (of Theoretical
Construction) dictates that this above must be the case for a
satisfactory theory.

Replacing only the concept of the gravitational force is then
unacceptable not only from this point of view of the simplicity
but also because the resultant theory then cannot account for the
entirety of charged material bodies within its hybrid framework.
Charged material bodies will have to be the singularities of the
electrostatic force but not of the gravitational force in the
mathematical framework of such a hybrid theory. Any such hybrid
framework is then bound to be physically inconsistent and, hence,
unacceptable.

Then, the mathematical procedure by which we replace the notion
of, say, Newton's gravitational force {\em cannot\/} be expected
to be entirely {\em different\/} than the one adopted, say, for
replacing the notion of Coulomb's electrostatic force.

Now, Einstein attempted to replace the notion of {\em force\/}
with that of the {\em curvature\/} of the spacetime manifold.
Then, this replacement must also be applicable to every notion of
force in Newton's theory. In particular, it must apply to
Coulomb's electrostatic force.

Then, Coulomb's electrostatic force can be attractive as well as
repulsive depending on the ``relative'' signs of the involved
electric charges. An immediate implication of this above for
Einstein's aforementioned attempt is that the relative sign of
charges determines the curvature for the geometry ``experienced''
by them.

But, how this is to be achieved is unclear. Einstein's field
equations with matter do not implement this in any non-singular
manner. Moreover, what about origins of electric charges that are
to determine the curvature?

We may then conclude that the concept of the curvature of geometry
is \underline{not} \underline{sufficientl}y g\underline{eneral} to
replace the notion of force.

Now, the concept of force is, in a definite mathematical sense
\cite{dyn-sys}, {\em equivalent\/} to that of certain
transformation of the points of the (Euclidean) space in Newton's
theory. This then ``suggests'' that mathematical transformations
of points of the (suitable underlying) space can, quite generally
as well as naturally, ``replace'' the newtonian concept of force
as a cause of motion.

A transformation which ``brings'' two bodies ``together''
corresponds to an attractive force between them while the one
which ``pushes'' these bodies ``away'' from each other corresponds
to a repulsive force between them. Moreover, the action of a
transformation can ``naturally'' depend on some parameters (eg,
electric charges). We may then be able to incorporate Coulomb's
law (and other laws of force) in a framework based on
transformation of some suitable (mathematical) space as a possible
generalization of the newtonian conception of force as a cause of
motion.

Evidently therefore, the concept of transformation
\underline{a}pp\underline{ears} to be \underline{sufficientl}y
g\underline{eneral} to replace the (newtonian) notion of force.

It should then be also evident that the mathematical laws
obtainable using this replacement of force by the transformation
of suitable underlying space will be applicable to {\em every
physically constructed frame of reference}, and, hence, this
mathematical formalism will be in conformity with the general
principle of relativity.

It should then be equally clear that the phenomenon of gravitation
is incorporated in this framework as the concept of transformation
is ``applicable'' in all the relevant situations.

This then brings us to issues of the quantum considerations within
the theme of the universal theory of relativity whose certain
characteristics we have considered above.

\subsection{Quantum aspects and the related requirements of the
Mathematical Foundations for the Universal Theory of
Relativity}\label{qamfnt}

Now, Bohr had captured \cite{bohr1} the ``essence'' of the quantum
theory in the following succinct words:

$\spadesuit$ ... {\em The quantum theory is characterized by the
acknowledgement of a fundamental limitation in the classical
physical ideas when applied to atomic phenomena. ... the so-called
quantum postulate, which attributes to any atomic process an
essential discontinuity, or rather individuality, completely
foreign to the classical theories. ...}

{\em ... Strictly speaking, the idea of observation belongs to the
causal spacetime way of description. ... According to the quantum
theory, just the impossibility of neglecting the interaction with
the agency of measurement means that every observation introduces
a new uncontrollable element. ...}

{\em This postulate implies a renunciation as regards the causal
spacetime coordination of atomic processes. Indeed, our usual
description of physical phenomena is based entirely on the idea
that the phenomena concerned may be observed without disturbing
them appreciably. This appears, for example, clearly in the theory
of relativity, which has been so fruitful for the elucidation of
the classical theories. As emphasized by Einstein, every
observation or measurement ultimately rests on the coincidence of
two independent events at the same spacetime point. Just these
coincidences will not be affected by any differences which the
spacetime coordination of different observers may exhibit. Now,
the quantum postulate implies that any observation of atomic
phenomena will involve an interaction with the agency of
observation not to be neglected. Accordingly, an independent
reality in the ordinary physical sense can neither be ascribed to
the phenomena nor to the agencies of observation. ...}
$\spadesuit$

Indeed, the quantum theory is then supposed to acknowledge an
essential limitation of the classical newtonian ideas by
recognizing that any observation of a physical system involves,
necessarily, an uncontrollable disturbance of that physical
system. This is, in spirit, similar to the special theory of
relativity acknowledging a fundamental fact that we, ordinarily,
use electromagnetic radiation, Light, to observe material bodies
\cite{100-yrs}.

The quantum hypothesis, through Heisenberg's indeterminacy
relations \cite{heisenberg}, then renders the {\em exact\/}
measurability  of the coordination of the space and the time
questionable indeed. This is the decisive role of indeterminacy
relations.

Then, let us consider that an observer chooses a certain spatial
location as the origin of the coordinate system and intends to
attach {\em spatial labels\/} to various events in the vicinity
with respect to it. The observer needs to use a {\em measuring
rod}, made using physical matter, to achieve this.

The observer intends to also attach suitable {\em temporal
labels\/} to each event in the vicinity. For this purpose, the
observer then needs to place, at each suitable location in the
vicinity, near the ``physical events'' to be observed, physical
bodies executing periodic motions as {\em clocks}.

But, by the quantum postulate, these concepts must involve {\em
indeterminacies}. Neither the origin of the coordinate system nor
the coordinate labels can be determined in a {\em physical
measurement\/} any more accurately than permitted by Heisenberg's
relevant indeterminacy relation. The {\em physical\/} coordination
used by an observer {\em cannot\/} then be identical with the
(mathematical) coordination of the underlying continuum, if any.

But, this last is also the implication of the issues involved with
even the physical construction of the coordinate frames.

Recall from \S\,\ref{idea-basis} that the ``coordinate axes'' of
the reference frame must be using physical matter, and, in
general, motions of other material bodies must affect the
coordinate axes.

Now, to attach coordinate labels to various spatial locations, the
observer needs to ``move'' the ``measuring rod'' suitably and
measure distances to these locations. But, changes must, in
general, result to the coordinate axes due to the motion of the
measuring rod. Therefore, the observer will {\em not\/} be able to
measure {\em exactly\/} the distances necessary to label the
spatial locations.

The same also applies to the measurement of the physical time
since any physical clock must be constructed from physical matter
and any such measurement involves the measurement of the location
of the hands of the clock.

Once again, we see even here that the {\em physical\/}
coordination used by an observer {\em cannot\/} be identical with
the (mathematical) coordination of the underlying continuum, if
any.

It is then natural to expect that these issues of the physical
construction of the coordinate axes and those of the quantum
postulate are related to each other in some manner.

But, of course, the observer will be able to ``neglect'' these
changes to the coordinate axes in some situations. Then, in
approximation, the observer could peruse the classical equations.

Now, in Newton's theory, ordinary material bodies are considered
as some ``rigid'' or ``non-rigid'' collection of material points,
bound together by inter-particle forces.

``Spatial extensions and associated properties'' of material
bodies then arise in Newton's theory only from such conceptions.

In Newton's theory, a {\em physical\/} rod is then to be treated
as such a spatially extended physical object. Of course, we then
cannot determine the distances of locations from a chosen origin
any more accurately than the chosen unit - the spatial extension
of the chosen physical rod. To be able to do any better, we need
to ``divide'' the physical rod into smaller parts. Within Newton's
overall theoretical scheme, this is, evidently, the conceptual
origin of the involved experimental limitations of the distance
measurement. However, in principle, the distance measurements can
be as accurate as desired in Newton's theory.

But, when the concept of force is abandoned, as has to be the
situation with the ``universal theory of relativity,'' the
aforementioned newtonian conception of an ``extended physical
object'' too will have to be suitably replaced.

Now, Bohr's approach \cite{bohr2} {\em permits\/} us to postulate,
notably, an {\em exactly localizable material point}, a classical
newtonian particle, to represent a physical body but shifts the
onus of the indeterminateness of its location entirely on the
process of measurement, a physical process constituted
``externally'' to the system being observed.

The use of classical concepts to describe an experimental
arrangement is the basis of Bohr's approach. Recall from
\cite{bohr2}:

$\spadesuit$ ... by the word ``experiment'' we refer to a
situation where we can tell others what we have done and what we
have learned and that, therefore, the account of the experimental
arrangement and of the results of the observations must be
expressed in unambiguous language with suitable application of the
terminology of classical physics.

This crucial point, ..., implies the {\em impossibility of any
sharp separation between the behavior of atomic objects and the
interaction with the measuring instruments which serve to define
the conditions under which the phenomena appear}. ...
Consequently, evidence obtained under different experimental
conditions cannot be comprehended within a single picture, but
must be regarded as {\em complementary\/} in the sense that only
the totality of the phenomena exhausts the possible information
about the objects.

Under these circumstances an essential element of ambiguity is
involved in ascribing conventional physical attributes to atomic
objects, as is at once evident in the dilemma regarding the
corpuscular and wave properties of electrons and photons, where we
have to do with contrasting pictures, each referring to an
essential aspect of empirical evidence. ... $\spadesuit$

Bohr then insists on the use of ``classical ideas'' (including the
concept of force) for the ``description'' of the quantum
properties of matter. Only the ``classical concepts'' permit him
an {\em experimental arrangement to be constituted externally\/}
to the system being observed because these concepts involve
agencies, forces, fundamentally external to any conceivable
system.

But, the newtonian point-particle then cannot describe the
``evidence'' related to the wave phenomena displayed by electrons,
say. As shown by Heisenberg \cite{heisenberg}, the corpuscular
picture is limited by the indeterminacy relations, which are then
interpreted to mean a definite lack of sharp distinction between
the interaction of electron with the instrument and the observed
physical phenomenon involving electrons.

Bohr's is then a ``hybrid'' approach that uses classical concepts
without ``fundamentally changing'' those concepts. But, it
associates probability aspects with the classical concepts.

Recall from \cite{bohr2}:

$\spadesuit$ {\em ... any arrangement suited to study the exchange
of energy and momentum between the electron and the photon must
involve a latitude in the space-time description of the
interaction sufficient for the definition of wave-number and
frequency ... Conversely, any attempt of locating the collision
between the photon and the electron more accurately would, on
account of the unavoidable interaction with the fixed scales and
clocks defining the the space-time reference frame, exclude all
closer account as regards the balance of energy and momentum}.

 {\em ... an adequate tool for a complementary way of
description is offered precisely by the quantum mechanical
formalism which represents a purely symbolic scheme permitting
only predictions, on lines of the correspondence principle, as to
results obtainable under conditions specified by means of
classical concepts. ... Thus, a sentence like ``we cannot know
both the momentum and the position of an atomic object'' raises at
once questions as to the physical reality of two such attributes
of the object, which can be answered only by referring to the
conditions for the unambiguous use of the space-time concepts, on
the one hand, and dynamical conservation laws, on the other hand.
While the combination of these concepts into a single picture of a
causal chain of events is the essence of classical mechanics, room
for regularities beyond the grasp of such a description is just
afforded by the circumstance that the study of the complementary
phenomena demands mutually exclusive experimental arrangements.}
$\spadesuit$

When the classical concepts are retained without fundamental
modifications, the ``quantum postulate'' {\em imposes\/} the
viewpoint that certain classical concepts are ``complementary'' in
accounting for the physical experiences.

Bohr \cite{bohr1, bohr2} then explains the indeterminacy relations
as mathematical realizations of the complementarity of the
involved classical conceptions and supports their probabilistic
basis.

Recall, once again from \cite{bohr2}, that Bohr had been
``critical'' of even the conceptual foundations of the relativity
theory:

$\spadesuit$ {\em ...causal description is upheld in relativity
theory within any given frame of reference, but in quantum theory
the uncontrollable interaction between the objects and the
measuring instruments forces us to a renunciation even in such
respect.} $\spadesuit$ and went on to express optimism that:

$\spadesuit$ {\em ... the viewpoint of complementarity may be
regarded as a rational generalization of the very ideal of
causality.} $\spadesuit$

Whether this optimism can be upheld in a ``hybrid'' approach, as
is Bohr's, is doubtful.

Now, as seen before, certain newtonian concepts, in particular,
the concept of force, will have to be fundamentally abandoned to
accommodate the General Principle of Relativity.

Consequently, in the context of various implications of the
General Principle of Relativity, Bohr's complementarity hypothesis
will be ``limited'' as an explanation of the indeterminacy
relations whose basis must now be sought within only the context
of those implications.

Thence, in relation to the above discussion, we now recall from
\S\,\ref{gent} that, in the Universal Theory of Relativity,
transformations of the points of the underlying space are to
replace, naturally, the notion of force in Newton's theory. A
transformation of the points of the underlying space is the only
``cause'' of the motion of a chosen material body in the universal
theory of relativity.

Then, a transformation of the underlying space can be performed
which does {\em not\/} affect the material body whose location is
being measured, but ``moves'' only the measuring rod in use in the
manner desired by the observer for the involved measurement.
Notably, such a transformation is to represent then ``all'' the
actions, including that of the (electromagnetic) radiation, if
any, in the same measurement process.

Now, if a material body were to be representable as an ``exactly
localizable material point'' within this theoretical framework
then, it would be possible to measure the ``exact'' location of
that material body by employing a transformation of the
aforementioned type.

[In this context, we note that force in Newton's theory can be
looked upon as a transformation of the points of the Euclidean
space. Newton's theory represents a physical body as an exactly
localizable material point. Then, it is possible in Newton's
theory to measure the exact location of the material point because
a transformation that does not affect the material point but moves
only the measuring rod is permissible.]

In other words, we could locate, exactly, a material point by
moving a measuring rod by its side without affecting the location
of that material point since a transformation of the underlying
space that achieves this, including Light to ``see'' the process,
is, now, thinkable.

This would, however, violate the indeterminacy relations which
arise, naturally, from the quantum postulate that has the
impeccable support of the empirical evidence of great value.

{\em Then, in the context of the General Principle of Relativity,
the quantum postulate implies that it must be impossible in the
theory to hypothesize an exactly localizable material point to
represent a physical object.} Hence, an {\em intrinsic
indeterminacy\/} in the location of a material point representing
a physical body is implied here.

In the Universal Theory of Relativity, the origin of Heisenberg's
indeterminacy relations will then be a combination of this {\em
intrinsic indeterminacy\/} in the location of a material point
representing a material body and the transformation of the
underlying space that is the ``cause'' of the motion of that
material point.

An ``act of observation'' involves transformation of points of the
underlying space representing the observed physical system and
also of points of the space representing the used measuring
apparatus. Then, transformations of the underlying space and
intrinsic indeterminacies are the keys to quantum aspects of
matter.

Now, a transformation of the points of the underlying space is to
represent {\em unique evolution\/} of a physical system fixed {\em
deterministically\/} by the initial conditions. (See, for details,
\S\, \ref{mf}.) Then, an observation of a physical system may
involve the ``disturbance'' of that particular system, but not an
uncontrollable or unpredictable disturbance. The viewpoint that
``any act of observation of a physical system involves,
necessarily, an uncontrollable disturbance of that physical
system'' will have to be rested therefore.

{\em The quantum hypothesis thus acknowledges the existence of an
{\em intrinsic indeterminacy\/} in the very conceptualization of a
material point to represent a physical body}. The mathematical
formalism of the universal theory of relativity must then respect
this acknowledgement. This is, now, the ``true essence'' of the
quantum postulate.

Then, having renounced newtonian conceptions at a fundamental
level, we needed to ``reanalyze'' various such aspects as
discussed above.

Recapitulating, moving a material body from its given ``location''
should cause changes to the construction of the coordinate system
and, hence, to the ``physical geometry'' because the construction
of the coordinate system is the basis of the ``metric function''
of the geometry.

In turn, ``given the metric or the distance function of the
geometry'' we would know how the totality of {\em all\/} the
material bodies are ``located'' relative to each other.

Hence, the physical geometry is determined by material bodies
which, in turn, are also determined by the physical geometry.

However, the issue remains of physical characteristics of material
bodies such as, for example, their inertia, electrostatic charge
etc.

Recalling implications of the quantum postulate here, these
physical ``qualities'' of material bodies must, basically, be
``definable'' for various subsets, but {\em not\/} for any {\em
singleton subset}, of the underlying space. (A singleton subset
would otherwise be an ``exactly localizable'' material point.)
But, such aspects belong to the framework of the mathematical
theory of measures \cite{measure-theory, srivastava}.

The physical ``qualities'' of matter are then to be treated as
{\em measures\/} defined on certain subsets of the underlying
continuum. Such subsets are then the {\em basic\/} physical
objects for the universal theory of relativity and each physical
object can be viewed, necessarily, as a {\em measurable set\/}
\cite{measure-theory, srivastava} of the corresponding measure
space.

But, a measure can be averaged over a measurable set, the average
being a property of each point of that set. The averaged measure
then provides \cite{smw-indeterminacy, heuristic} the non-singular
notion of a point-object with the physical characteristics.

But, the ``location'' of the point-object so defined is {\em
indeterminate\/} within the measurable set. This mathematical
situation is precisely in accord with the requirement as imposed
by the quantum postulate on the mathematical formalism of the
universal theory of relativity.

Then, the {\em physical distance\/} separating two material
objects is the appropriate {\em mathematical distance\/} between
the corresponding measurable sets. ``Kinematical'' quantities such
as ``velocity'' and ``acceleration'' (for one measurable set
relatively to another measurable set) involve change in the
so-defined physical distance under the action of the
transformation $T_{_t}$ of the underlying space. The parameter $t$
of $T_{_t}$ ``defines'' time in an appropriate sense as will be
discussed later in \S\,\ref{mf}.

Various physical phenomena can then arise from the effects of
transformations of the underlying space on the measurable sets and
the measures defined on them. This is then the framework of the
theory of dynamical systems.

Then, these ideas as well as their mathematical renderings are,
evidently, fundamentally different from those of Newton's theory,
of special relativity and, of orthodox quantum theory.

Notably, there ``do not occur'' any ``physical constants'' to be
``specified by hand'' in this above framework. But, {\em all\/}
the physical constants can arise in this framework only from
``mutual relationships'' of involved physical objects, just
exactly as we determine them experimentally.

For example, consider the phenomenon of gravitation within the
present conceptual framework. It involves the action of $T_{_t}$
for which the ``acceleration'' experienced by one measurable set,
in relation to another reference measurable set, is {\em
independent\/} of the ``measure'' defined on that set, but is
proportional to the measure defined on the reference measurable
set, both measure classes being invariant under $T_{_t}$. (In
general, measures change under the action of $T_{_t}$.)

Newton's gravitational constant then ``arises'' when
``acceleration'' is expressed as the ``inverse-square'' of the
physical distance. (This demonstration is, of course, somewhat
involved, and will be the subject of an independent study.)
Clearly, the possibility of theoretically obtaining the ``value''
of Newton's constant of gravitation can be seen to arise in this
manner within the overall framework of the Universal Relativity.

Furthermore, it should also be equally clear that, within
universal relativity, other physical constants can similarly
``arise'' from different permissible situations, mutual
relationships of measurable sets and the effects of
transformations of the underlying space on them.

Clearly, the ``values'' of such physical constants cannot be
changed ``without'' destroying the theory and this situation is,
precisely, as per Einstein's theorem \cite{schlipp} (p. 63) quoted
earlier.

This above is recognizable as an extremely important issue for any
complete physical theory. Whether such theoretically obtained
values of the fundamental physical constants are also their
experimentally determined values is then another extremely
important issue for us.

Notably, an observer, possessing any consciousness or not, has
only the ``background'' role to play in the Universal Theory of
Relativity. Nowhere in this theory,  in its explanations of
physical phenomena, do we require the ``intervention'' by an
``observer.'' The problems of incorporating a conscious observer
within its conceptual framework do not therefore arise. Newton's
theory also had the same role for an observer.

Transformation of the underlying space is then a {\em unique
evolution\/} of the points of the space. It therefore represents
{\em unique evolution of a physical system ``fixed''
deterministically by the initial conditions}. Consequently, the
Universal Theory of Relativity provides therefore \cite{schlipp}
``the complete description of any individual real situation as it
supposedly exists irrespective of any act of observation or
substantiation.''

Importantly, it should also be clear at this stage of our
considerations that in pursuing the Universal Theory of Relativity
we shall be led to explanations of general physical phenomena that
are ``radically different'' from those of the newtonian ways as
well as from those of the ways of the orthodox (probabilistic)
quantum theory.

In this above context, it needs to be stressed, and re-stressed,
that \underline{all} \underline{the} \underline{known}
\underline{ex}p\underline{erimental} \underline{results} would be
\underline{ex}p\underline{lainable} within the overall framework
of the Universal Theory of Relativity by treating the ``standard
forces'' as corresponding transformations of some suitable space
underlying the Universal Theory of Relativity.

Needless to say, we will then have to analyze, in the framework of
Universal Relativity, each of the known physical phenomena case by
case to check if certain new predictions are permissible for the
case under study.

An example of such a type is provided in \S\,\ref{results} in the
form of an analysis of the torsion balance experiment. In
particular, the use of the newtonian gravitational force implies
that the torsion balance will always be torqued when some external
masses are moved around it.

As had been remarked earlier, the existence of torqued motion of
the torsion balance is explainable in Universal Relativity by
treating the newtonian total force on it as a corresponding
transformation of the space underlying the Universal Theory of
Relativity.

However, in Universal Relativity, there does occur a situation
when the transformation of the underlying space does not act on
the torsion balance to produce its oscillatory motion even when
the external masses are moved around it. In this situation, a null
result obtains irrespective of the speed of motion of external
masses.

Certainly, a very careful torsion balance experiment involving
\underline{d}y\underline{namic} \underline{measurements} needs to
be designed to verify this null effect situation. Perhaps, some
very carefully selected geometrical design of the torsion balance
experiment may also be needed for this purpose.

We conclude this section, \S\,\ref{pf}, with the following remarks
that highlight the spirit behind our endeavors of this discussion.

The approach followed in this discussion, of \S\,\ref{pf},
presents the Universal Theory of Relativity as an appealing, but
not as a simplest, system of thoughts. We essentially developed
the physical foundations of this theory as being psychologically
``natural'' or appealing.

One could however perceive this approach as also being a logically
compelling one. This is primarily because, apart from the concept
of transformation of suitable space, there appears to be no
another sufficiently general mathematical concept to replace the
newtonian notion of force, it also being in conformity with the
general principle of relativity and, simultaneously, allowing the
physical constructions of reference frames to be affected by
motions of other bodies.

Granted this above and having then laid the physical foundations
for the Universal Theory of Relativity in sufficient details, let
us now turn to mathematical aspects implied by these
considerations. Mathematical foundations for this theory then rest
on the mathematical theories of measures and dynamical systems,
both. This much is a certain conclusion of the above.

\section{Mathematical Foundations of the Universal Theory of Relativity} \label{mf}

In Einstein's (and Descartes's) conceptions \cite{ein-pop}, {\em
physical geometry\/} is not any inert stage for the physical
fields. This notion of {\em physical geometry\/} will be made
precise in the following.

As we aim to interest general physics community here, we provide
below basic mathematical notions \cite{srivastava, dyn-sys,
kdjoshi, measure-theory, trim6} to be used frequently. A
knowledgeable reader may wish to skip it.

\subsection{Preparatory Mathematical Notions}
\subsection*{Sets, Topologies, Groups, Measures ...}

Let $\mathbb{N}$ be the set of natural numbers, $\mathbb{Q}$ that
of rational numbers, $\mathbb{R}$ that of real numbers,
$\mathbf{2}$ the binary set $\{0,1\}$ and $\mathbb{Z}$ the set of
integers.

A collection of {\em all\/} subsets of a chosen set $X$ is a
\underline{Power} \underline{Set}, $\mathcal{P}(X)$, of $X$. A set
of all $k$-tuples, $(x_1, x_2, ..., x_k)$, of elements of $X$ is
$X^k$. The set of all {\em finite\/} sequences of elements of $X$,
including the empty sequence, is $X^{<\,\mathbb{N}}$.

A family $\mathcal{F}$ of nonempty sets is said to have the
\underline{finite intersection} p\underline{ro}p\underline{ert}y
if the intersection of every finite subfamily of $\mathcal{F}$ is
nonempty.

A \underline{cartesian} p\underline{roduct} of sets $X$ and $Y$ is
a set $X\times Y$ of all {\em ordered pairs\/} $(x,y)$ with $x\in
X$ and $y\in Y$. A cartesian product of a sequence of sets $X_1$,
$X_2$, ..., $X_n$ will be, usually, denoted by $\prod_{i=1}^nX_i$
or by $\times_{i=1}^nX_i$.

A \underline{relation} $R:X\to Y$ is any set of ordered pairs
$(x,y)$. Note that $R\subseteq X\times Y$. A set of all ordered
pairs $(y,x)$ whenever $(x,y) \in R$ is an {\em inverse
relation\/} $R^{-\,1}$. A \underline{com}p\underline{osition} of
relations $f$ and $g$ is a relation $g\circ f = \left\{ (x,z):
{\rm for\;some}\;y,\,(x,y) \in f \right.$ $\left.{\rm
and}\;(y,z)\in g\right\}$. The set $\triangle X=\left\{ (a,a):
a\in X\right\}$ is the \underline{dia}g\underline{onal}
(\underline{relation}) on set $X$.

The set $G_f=\left\{(x,y)\in X\times Y: y=f(x)\right\}$ is called
as the g\underline{ra}p\underline{h} of $f$. The set $X$ is called
the \underline{domain} and the set $Y$ is called the
\underline{co}-\underline{domain} of $f$. The element $x\in X$ is
called the p\underline{re}-\underline{ima}g\underline{e} of $y\in
Y$ if $y=f(x)$. The set $\{f(x):x\in X\}$ is called the
\underline{ran}g\underline{e} of $f$.

A single-valued relation is a \underline{function},
\underline{ma}p or a \underline{transformation}. It can be many to
one. A function $f:X\to \mathbb{R}$ is
\underline{sim}p\underline{le} if its range is finite.

A function $f:X\to Y$ is \underline{in}j\underline{ective} or {\em
1-1\/} if for all $x,y \in X,\,f(x)=f(y)\Rightarrow\,x=y$. A
function $f:X\to Y$ is \underline{sur}j\underline{ective} or
\underline{onto} if for each $y\in Y$ there is some $x\in X$ such
that $f(x)=y$. A 1-1 and onto function is
\underline{bi}j\underline{ective}. A bijection $p:A\to A$ is
called a p\underline{ermutation} of the set $A$.

The \underline{collection of all functions} from set $Y$ to set
$X$ forms a set which is denoted by $X^Y$.

An \underline{e}q\underline{uivalence} \underline{relation} on a
set $X$ is a relation $\sim\; \subset X\times X$  such that for
all $x,y,z\in X$ \begin{description} \item{(i)} $x\sim x$
(\underline{reflexive}) \item{(ii)} $x\sim y \Rightarrow y\sim x$
(\underline{s}y\underline{mmetric}) and \item{(iii)} $x\sim y$ and
$y\sim z$ $\Rightarrow x\sim z$ (\underline{transitive}).
\end{description} A set of all $y\in X$ such that $x\sim y$ is the
\underline{e}q\underline{uivalence} \underline{class} of $x$,
denoted by $R[x]$.

A family $\mathcal{D}$ of pairwise disjoint nonempty subsets of a
set $X$ such that $\bigcup_i D_i=X: D_i\in\mathcal{D}$ is a
\underline{decom}p\underline{osition} or p\underline{artition} of
$X$.

There exists a 1-1 correspondence, {\em ie}, bijection, from the
set of all equivalence relations of $X$ and the set of all
decompositions of $X$.

The decomposition of a set $X$ by an equivalence relation $R$ is
the q\underline{uotient set of} $X$ by $R$ or {\em set of quotient
classes modulo $R$}, denoted by $X\diagup R$. The function $p:X\to
X\diagup R,\;p(x)=R[x]$ for $x\in X$ is the
p\underline{ro}j\underline{ection} or q\underline{uotient
function}.

A \underline{strict order} on $X$ is a transitive relation, {\em
ie}, $(xRy$ and $yRz)$ $\Rightarrow\;xRz$, and $\forall\, a\in X$,
$(a,a)\notin R$. For any strict order $R$, $aRb$ and
$bRa\Rightarrow\;a=b$ for all $a,b \in X$,
\underline{anti}-\underline{s}y\underline{mmetr}y.

Two sets $A$ and $B$ are \underline{e}q\underline{uinumerous} or
of the same \underline{cardinalit}y if there exists a 1-1 map,
bijection, $f$ from $A$ to $B$. A set $A$ is \underline{finite} if
there is a bijection from $\{0, 1, ..., n\}$, $n \in \mathbb{N}$,
onto $A$. If $A$ is not finite, it is \underline{infinite}. A set
$A$ is \underline{countable} if it is finite or if there exists a
bijection from $\mathbb{N}$ onto $A$. An \underline{uncountable}
set is not countable.

To each set $X$ we can associate a symbol, $|X|$, its
\underline{cardinal number}, such that $X=Y \Longleftrightarrow
|X|$ and $|Y|$ are the same. Some cardinals are denoted by special
symbols, {\em eg}, $|\{0,1,...,n-1\}|=n, n\in \mathbb{N}$;
$|\mathbb{N}|= \aleph_o$ and $|\mathbb{R}|=\mathbf{c}$. We can
add, multiply as well as compare cardinal numbers by suitably
defining cardinal arithmetic.

Note that we have $\aleph_o<|\mathbf{2}^{\mathbb{N}}|=2^{\aleph_o}
=\mathbf{c}$, and $\aleph_o+\aleph_o = \aleph_o\,\cdot\,\aleph_o
=\aleph_o$, and $\mathbf{c}^n= \mathbf{c}^{\aleph_o} =\mathbf{c}
\;(n>1)$, and $|\mathbb{N}^{\mathbb{N}}|=\mathbf{c}$ etc.

If $A\subset X$, its \underline{characteristic}
\underline{function} is a many-one map $\chi_{_A}:\,X \to
\mathbf{2}$, where $\chi_{_A}(x) = 1$ if $x\in A$ and
$\chi_{_A}(x)=0$ otherwise. Then, $A\to \chi_{_A}$ defines a 1-1
map from $\mathcal{P}(X)$ onto the set $\mathbf{2}^X$. Note that
if a set $X$ has $n$ elements then, the power set $\mathcal{P}(X)$
has $2^n$ elements.

For an extended real-valued function $f:X\to \mathbb{R}$, the set
$\{x\in X\;|\; f(x)\neq 0\}=\mathfrak{Support}\,(f)$ is a
\underline{su}pp\underline{ort} of $f$ on $X$.

A real-valued function $u:E\to \mathbb{R}$ is said to
\underline{dominate} another function $v:E\to\mathbb{R}$ if
$v(\epsilon)\leq u(\epsilon)$ for all $\epsilon\in E$.

A p\underline{artial order} on a set $P$ is a binary relation $R$
which is reflexive, transitive and anti-symmetric. A set $P$ with
a partial order $R$ is a {\em partially ordered set\/} or a
p\underline{oset}. A \underline{linear} or \underline{total} or
\underline{sim}p\underline{le} \underline{order} on a set $X$ is a
partial order $R$ on $X$ such that for any $x,y \in X$ either
$xRy$ or $yRx$ holds.

For any two sets $X$ and $Y$, a p\underline{artial function} $f:
X\to Y$ is a function with domain a subset of $X$ and range
contained in $Y$. If $f$, $g$ are two partial functions from $X$
to $Y$ then, $g$ is \underline{extension} of $f$, or $f$ is a
\underline{restriction} of $g$, if ${\rm domain}(f) \subset \,{\rm
domain}(g)$ and $f(x)=g(x)$ for all $x\in \,{\rm domain}(f)$, and
we write $g \succeq f$ or $f \preceq g$. If $f$ is a restriction
of $g$ and ${\rm domain}(f)=A$ then, we write $f=g|A$.

A \underline{chain} in a set $P$ of a fixed poset $(P, R)$ is a
subset $C$ of $P$ such that $R$ restricted to $C$ is a linear
order. An \underline{u}pp\underline{er bound} for a set $A
\subseteq P$ is an $x\in P$ such that $yRx$ for all $y\in A$. An
$x \in P$ is called a \underline{maximal element} of $P$ if for no
$y\in P$ different from $x$, $xRy$ holds.

An element $x\in L$ of a linearly ordered set $(L,\leq)$ is the
\underline{first} (\underline{last}) \underline{element} of $L$ if
$x\leq y$ ($y\leq x$) for every $y\in L$. A linearly ordered set
$L$ is \underline{order}-\underline{dense} if for every $x<y$
there exists $z\in L$ such that $x<y<z$. Two linearly ordered sets
are \underline{order isomor}p\underline{hic} if there is a 1-1,
order-preserving map from one onto the other.

A \underline{well}-\underline{order} on a set $W$ is a linear
order $\leq$ on $W$ such that every nonempty subset $A$ of $W$ has
a first element. If $\leq$ is a well-order on $W$ then $(W,\leq)$,
or simply $W$, will be called a
\underline{well}-\underline{ordered set}. For $w,w'\in W$, we
write $w<w'$ if $w\leq w'$ and $w\neq w'$. A linearly ordered set
$(W,\leq)$ is well-ordered if and only if there is no descending
sequence $w_o>w_1>w_2> ...$ in $W$.

\underline{Zorn}'s \underline{Lemma}: If $P$ is a nonempty poset
with every chain in $P$ having an upper bound in $P$, then $P$ has
a maximal element. Equivalently, we have the \underline{Axiom of
Choice}: If $\{ A_i\}_{i \in I}$ is a family of nonempty sets,
then there is a \underline{choice} \underline{function} $f:\,I\to
\bigcup_iA_i$ such that $f(i)\in A_i$ for every $i\in I$.
Equivalently, we have the \underline{Well Orderin}g
\underline{Princi}p\underline{le}: every set can be well ordered.

If $W_1$ and $W_2$ are two well ordered sets and if some $f:W_1\to
W_2$ is an order preserving bijection, we call the sets $W_1$ and
$W_2$ as being \underline{order isomor}p\underline{hic} and $f$ as
an \underline{order isomor}p\underline{hism}. We write $W_1\sim
W_2$. Order isomorphic sets are of the same cardinality.

For well ordered set $W$, let $w\in W$ and $w^-\in W$ be such that
$w^-<w$ and suppose that there is no $v\in W$ satisfying
$w^-<v<w$. If existing, such $w^-$ is a unique member of $W$ and
is the \underline{immediate} p\underline{redecessor} of $w$ and
$w$ is the \underline{successor of} $w^-$. If a well ordered set
$W$ has an element $w$ other than the first element with no
immediate predecessor, such $w\in W$ is a \underline{limit
element} of $W$.

For $W$ being a well ordered set and $w\in W$, sets of the form
$W(w)=\left\{ u\in W: u<w\right\}$ are called as the
\underline{initial se}g\underline{ments} of $W$. Note that a well
ordered set $W$ cannot be order isomorphic to an initial segment
$W(u)$ of itself.

Principles of induction on natural numbers extend to generally
well ordered sets in the form of the so called
\underline{com}p\underline{lete induction} on well ordered sets -
proof by \underline{transfinite induction}.

For two well ordered sets $W_1$ and $W_2$, let $W_1\prec W_2$ if
$W_1$ is order isomorphic to an initial segment of $W_2$. Further,
let $W_1 \preceq W_2$ if either $W_1\prec W_2$ or $W_1\sim W_2$.
Then, the {\em Trichotomy Theorem of Well Ordered Sets\/} states
that for any two well ordered sets $W$ and $W'$, exactly one of
$W\prec W'$, $W\sim W;$ and $W'\prec W$ holds.

To each well ordered set $W$, we can associate a well ordered set
$t(W)$, called the \underline{t}yp\underline{e of} $W$, such that
$W\sim t(W)$ and if $W'$ is another well ordered set then, $W\sim
W' \Longleftrightarrow t(W)=t(W')$. The {\em fixed types\/} of
well ordered sets are called the \underline{ordinal numbers} and
the class of the ordinal numbers will be denoted by {\bf ON}.
Clearly, $|W|=|t(W)|$ and, hence, $\alpha=t(W)$. We say that an
ordinal $\alpha=t(W)$ is of cardinality $\kappa$ if $|W|=\kappa$.

Every ordinal $\alpha$ can be uniquely written as $\alpha=\beta+n$
where $\beta$ is a limit ordinal and $n$ finite. We call $\alpha$
{\em even or odd\/} if $n$ is even or odd.

An ordinal $\alpha$ will be called \underline{successor ordinal}
if $\alpha=\beta+1$ for some $\beta$, otherwise it will be called
a \underline{limit ordinal}. Note that $\alpha$ is a limit ordinal
if and only if any well ordered set $W$ such that $\alpha=t(W)$
has no last element.

A \underline{set of all countable ordinals} is an uncountable
well-ordered set, denoted by $\omega_1$, and the type of
$t(\omega_1)$ will also be denoted by $\omega_1$. Then, $\omega_1$
is the \underline{first} \underline{uncountable}
\underline{ordinal}. Cardinals are identified with initial
ordinals, and each is then denoted by the symbol $\aleph$.
\underline{Cantor}'s \underline{Continuum H}yp\underline{othesis}
states that $\mathbf{c}=\aleph_1$.

Ordinal numbers can be added, multiplied and compared by defining,
suitably, the ordinal arithmetic with ordinal addition and
multiplication as non-commutative operations in {\bf ON}.

Now, for $s \in A^{<\,\mathbb{N}}$, $A$ non-empty set, let $|s|$
be the length of $s$. Let $s=(a_o, a_1, ...,a_{n-1}) \in
A^{<\,\mathbb{N}}$ and $m<n$, we write $s|m = (a_o, a_1,
...,a_{m-1})$. If $t=s|m$, we say that $t$ is an
\underline{initial se}g\underline{ment} of $s$ or that $s$ is an
\underline{extension} of $t$, and we write $t\prec s$ or $s \succ
t$. We write $t \preceq s$ if either $t \prec s$ or $t=s$. We say
that $s$ and $t$ are {\em compatible\/} if one of them is an
extension of the other, otherwise they are said to be
\underline{incom}p\underline{atible}, written $s\bot t$. The
\underline{concatenation} $(a_o, a_1, ..., a_{n-1}, b_o, b_1, ...,
b_{m-1})$ of $s=(a_o, a_1, ..., a_{n-1})$ and $t=(b_o, b_1, ...,
b_{m-1})$ will be denoted by $\widehat{s\,t}$.

Then, a \underline{tree}, $T$, on a set $A$ is a nonempty subset
of $A^{<\,\mathbb{N}}$ such that if $s\in T$ and $t\prec s$ then
$t\in T$. Empty sequence $e$ belongs to all trees. Elements of $T$
are called \underline{nodes of} $T$. A node $u$ of $T$ is called
\underline{terminal} if for no $a\in A, \;\widehat{u\,a} \in T$. A
tree is called \underline{finitel}y
\underline{s}p\underline{littin}g if for every node $s$ of $T$,
the set $\left\{ a\in A: \widehat{s\,a}\in T\right\}$ is finite.
The \underline{bod}y \underline{of a tree} $T$ on a set $A$ is the
set $[T]= \left\{ \alpha\in A^{\mathbb{N}}: \forall\,k\;(\alpha|k
\in T) \right\}$. Members of $[T]$ are the \underline{infinite
branches} of $T$. A tree $T$ is called \underline{well}
-\underline{founded} if its body is empty and if $[T]\neq
\emptyset$, $T$ is called \underline{ill}-\underline{founded}.

A tree $T$ is well-founded if and only if there is no sequence
$\{s_n\}\in T$  such that $...\succ s_n \succ ... \succ s_1 \succ
s_o$. If $T$ is a tree and $u$ a node of $T$ then, the set
$T_u=\left\{ v\in A^{<\,\mathbb{N}}: \widehat{u\,v}\in T \right\}$
forms a tree.

\underline{K\"{o}ni}g's \underline{Infinit}y \underline{Lemma}
states and proves that a finitely splitting, infinite tree $T$ on
a set $A$ is ill-founded. For a tree $T$ on a finite set $A$,
$[T]\neq\emptyset$ $\Longleftrightarrow \left(\forall\,k\in
\mathbb{N}\right) \left(\exists \,u\in
T\right)\left(|u|=k\right)$. Sets $\{e\}$,
$\mathbb{N}^{<\,\mathbb{N}}$ etc.\ form trees on $\mathbb{N}$.

Consider a tree $T$ on a well ordered set $(A,\leq)$. Fix nodes
$s=(a_o, a_1, ..., a_{n-1})$ and $t=(b_o,b_1,$ $..., b_{m-1})$ of
$T$. Set $s <_{_{KB}}\,t$ if either $t \prec s$ or if there is an
$i=\min(m,n)$ such that $a_j=b_j$ for every $j<i$ and $a_i<b_i$.
Set $s\leq_{_{KB}}\,t$ if either $s <_{_{KB}}\,t$ or $s=t$. The
ordering $\leq_{_{KB}}$, the \underline{Kleene}-\underline{Brouwer
order}, is a linear order on $T$.

A tree $T$ on a well ordered set $A$ is well founded if and only
if $\leq_{_{KB}}$ is a well order on $T$. Transfinite induction
extends to well founded trees.

For a well founded tree $T$, define $\rho_{_T}:T\to {\rm {\bf
ON}}$ by $\rho_{_T}(u)=\sup\left\{ \rho_{_T}(v)+1:u\prec v, v\in T
\right\},\,u\in T$. Define $\rho_{_T}=\rho_{_T}(e)$ and call it
the \underline{rank} of $T$.

Note that $\rho_{_T}(u)=0$ if $u$ is terminal in $T$. Note also
that every well founded tree on the set $\mathbf{2}$ is of finite
rank.

Of importance is \underline{Cantor}'s \underline{Ternar}y
\underline{Set} $C$ defined as follows. Take $C_0=[0,1]$. Suppose
$C_n$ is defined and is a union of $2^n$ pairwise disjoint closed
intervals $\{I_j:1\leq j\leq2^n\}$ of length $1/3^n$ each. Obtain
$C_{n+1}$ by removing the open middle third of each $I_j$.
Finally, put $C=\bigcap_nC_n$.

Now, let $\mathcal{F}$ be some family of subsets of a set $X$.
Then, let $\mathcal{F}_{\sigma}=\left\{ \bigcup_{n\in\mathbb{N}}
A_n: \,A_n \in \mathcal{F} \right\}$ and
$\mathcal{F}_{\delta}=\left\{ \bigcap_{n\in\mathbb{N}} A_n: \,A_n
\in \mathcal{F} \right\}$. The family of finite unions
(intersections) of sets in $\mathcal{F}$ will be denoted by
$\mathcal{F}_s$ ($\mathcal{F}_d$). Also, let $ \neg\,\mathcal{F} =
\left\{A\subseteq X:\,X\setminus A \in \mathcal{F} \right\} $
where $X\setminus A=A^c$ denotes the
\underline{com}p\underline{lement} of $A$ in $X$. Then, evidently,
\[ \mathcal{F}_s \subseteq \mathcal{F}_{\sigma}, \;\mathcal{F}_d
\subseteq
\mathcal{F}_{\delta},\;\mathcal{F}_{\sigma}=\neg(\neg\mathcal{F})_{\delta},
\;\mathcal{F}_{\delta}=\neg(\neg\mathcal{F})_{\sigma} \] For a
non-empty set $X$, a family \[\left\{ A_s:\,s\in\,
A^{<\,\mathbb{N}} \right\} \] of subsets of $X$ is a
\underline{s}y\underline{stem of sets}, usually written $\{A_s\}$.
A system of sets $\{A_s\}$ is called
\underline{re}g\underline{ular} if $A_s\subseteq A_t$ whenever
$s\succ t$.

Now, define \[ \mathcal{A}_{_A}\left( \{A_s\}\right) =
\bigcup_{\alpha\in A^{\mathbb{N}}}\, \bigcap_nA_{\alpha|n} \] In
all the interesting cases $A$ is finite or $A=\mathbb{N}$. When
$A=\mathbb{N}$, we write $\mathcal{A}$ for $\mathcal{A}_{_A}$ and
call it the \underline{Souslin o}p\underline{eration}. The Souslin
operation is \underline{idem}p\underline{otent}:
$\mathcal{A}\left(\mathcal{A}(\mathcal{F})\right)=
\mathcal{A}\left(\mathcal{F}\right)$. The Souslin operation
involves uncountable unions. If $A=\mathbf{2}$, we write
$\mathcal{A}_2$ for $\mathcal{A}_{_A}$.

For the family $\mathcal{F}$ of subsets of a set $X$, let \[
\mathcal{A}_{_A}(\mathcal{F})=\left\{ \mathcal{A}_{_A}\left(\{
\mathcal{A}_s\}\right): \mathcal{A}_s\in \mathcal{F}; s\in
A^{<\,\mathbb{N}} \right\} \] be the family of sets obtained by
applying the Souslin operation on a system of sets in
$\mathcal{F}$. Then, for every family $\mathcal{F}$ of subsets of
$X$, we have \[ \mathcal{F}, \mathcal{F}_{\sigma},
\mathcal{F}_{\delta} \subseteq \mathcal{A}(\mathcal{F}) \]

For $s\in \mathbb{N}^{<\,\mathbb{N}}$, let $\Sigma(s)=\{\alpha\in
\mathbb{N}^{\mathbb{N}}: s \prec \alpha\}$ and $B=\bigcap_k
\bigcup_{|s|=k}\left[ A_s \times \Sigma(s)\right]$. Then, it is
seen that $\mathcal{A}(\{A_s\})=\pi_{_X}(B)$, where $\pi_{_X}: X
\times \mathbb{N}^{\mathbb{N}} \to X$ is the {\em projection\/}
map.

Now, any family $\mathcal{T}$ of subsets of $X$, with $X$ and
$\emptyset$ being its members and such that it is closed under
{\em arbitrary\/} unions and {\em finite\/} intersections, is
called a \underline{To}p\underline{olo}gy on $X$. A pair
$(X,\mathcal{T})$ is called a
\underline{to}p\underline{olo}g\underline{ical}
\underline{S}p\underline{ace}. Set $A\in\mathcal{T}$ is
\underline{o}p\underline{en}. Set $A\subseteq X$ is
\underline{closed} if $X\setminus A$ is open. Sets can be
simultaneously open and closed, {\em ie},
\underline{clo}p\underline{en}, {\em eg}, $X$ and $\emptyset$,
trivially. There can be non-trivial clopen sets in a topology on a
set $X$.

For $\mathcal{G}\subseteq\mathcal{P}(X)$, there exists topology
$\mathcal{T}$ on $X$ containing $\mathcal{G}$ such that if
$\mathcal{T}'$ is any topology containing $\mathcal{G}$, then
$\mathcal{T}\subseteq\mathcal{T}'$. The family $\mathcal{G}$ is
said to g\underline{enerate} the topology $\mathcal{T}$ on $X$ or
$\mathcal{G}$ is a \underline{subbase} for the topology
$\mathcal{T}$ on $X$. If $\mathcal{G}$ is countable, $\mathcal{T}$
is a \underline{countabl}y g\underline{enerated} topology.

A \underline{base} for a topology $\mathcal{T}$ on $X$ is a family
$\mathcal{B}$ of sets in $\mathcal{T}$ such that every $U\in
\mathcal{T}$ is a union of elements in $\mathcal{B}$. If
$\mathcal{G}$ is a subbase for $\mathcal{T}$, then the family of
finite intersections of elements of $\mathcal{G}$ is a base for
the topology $\mathcal{T}$. A topological space is said to be
\underline{second countable} if it has a countable base. A
subspace of a second countable topological space is second
countable.

A space $(X,\mathcal{T})$ is
\underline{zero}-\underline{dimensional} if its base consists of
clopen sets. Product of a family of zero-dimensional spaces is
zero-dimensional.

For any $A\subseteq X$, ${\rm cl}(A)$ denotes the intersection of
all closed sets containing $A$ and is called the
\underline{closure} of $A$. It is the smallest closed set
containing $A$. The largest open set contained in a set $A\subset
X$ is called the \underline{interior}, ${\rm int(A)}$, of $A$. Any
set $U$ with $x\in {\rm int}(U)$ is a
\underline{nei}g\underline{hborhood} of $x$.

An element $x\in X$ is an \underline{accumulation}
p\underline{oint} of $A \subseteq X$ if every neighborhood of $x$
contains a point of $A$ other than $x$. The set of all
accumulation points of $A$ is a \underline{derivative set} of $A$,
denoted by $A^{\prime}$. The elements of $A\setminus A'$ are the
\underline{isolated} p\underline{oints} of $A$. A set $A \subseteq
X$ is \underline{dense}-\underline{in}-\underline{itself} if it is
non-empty and has no isolated points.

A set $D \subseteq X$ is \underline{dense} in $X$ if $U\bigcap D
\neq \emptyset$ for every non empty open set $U$. A topological
space $X$ is \underline{se}p\underline{arable} if it has a
countable dense set. $\mathbb{R}^3$ is a separable topological
space.

For $A\subseteq X$ in $(X, \mathcal{T})$, a family $\mathcal{U}$
of sets whose union contains $A$ is called a \underline{cover of}
$A$. A subfamily of $\mathcal{U}$ that is a cover of $A$ is called
a \underline{subcover}. The set $A$ is called
\underline{com}p\underline{act} if every open cover of $A$ admits
a finite subcover.

If $X$ is a compact space, every closed subset of $X$ is compact.
Notably, any closed and bounded subset of $\mathbb{R}^3$ is
compact. Also, the Cantor ternary set is closed and bounded and,
hence, is compact in $I=[0,1]$.

A map $f:X\to Y$, $X$ and $Y$ being topological spaces, is called
\underline{continuous} if and only if $f^{-\,1}(V)$ is open
(closed) in $X$ for every open (closed) subset $V$ in $Y$.

A continuous image of a compact space or its compact subset is
compact.

A function $f: X \to Y$ is a \underline{homeomor}p\underline{hism}
if it is a bijection and both $f$ and $f^{-\,1}$ are continuous. A
homeomorphism $f$ from $X$ onto a subspace of $Y$ is called an
\underline{embeddin}g.

A subset $A$ of a topological space $X$ is called as a
\underline{retract} of $X$ if there is a continuous function
$f:X\to A$ such that $f|A$ is an identity map. In such a case, $f$
is called a \underline{retraction}. If $X$ is metrizable,
$A=\{x\in X:f(x)=x\}$ is closed when $A$ is a retract and $f$ a
retraction.

A \underline{metric s}p\underline{ace} is a pair $(X,d)$ where $X$
is a set, $d:X\times X \to \mathbb{R}$ is a (real-valued)
\underline{metric function} that satisfies, $\forall\; \,x,y,z \in
X$, \begin{description} \item{(${\rm \bf {a}}$)} $d(x,y)\geq 0$
and $d(x,y)=0$ iff $x=y$, \item{(${\rm \bf {b}}$)} $d(x,y)$
$=d(y,x)$ (Symmetry property), and \item{(${\rm \bf {c}}$)}
$d(x,z)\leq d(x,y)+d(y,z)$ (Triangle inequality).
\end{description}

A p\underline{seudo}-\underline{metric s}p\underline{ace} is a
pair $(X,\ell)$ where $X$ is a set, $\ell:X\times X\to \mathbb{R}$
is a p\underline{seudo}-\underline{metric function} that
satisfies, $\forall\; \,x,y,z \in X$, \begin{description}
\item{(${\rm \bf {a}}^{\prime}$)} $\ell(x,y)\geq 0$ and
$\ell(x,x)=0$\end{description} as well as the above properties
(${\rm \bf {b}}$) and (${\rm \bf {c}}$) of the metric function.

Define on set $X$ of a pseudo-metric space $(X,\ell)$ an
equivalence relation, $\sim$, such that $x\sim y$ iff
$\ell(x,y)=0$. Let $Y=\left\{\sim[x]:x\in X\right\}$. Further,
define for $A, B\in Y$ a \underline{canonical} \underline{metric}
\underline{function} on $Y$ as $e(A,B)=\ell(x,y)$ where $x\in A$
and $y\in B$. Now, let $\Pi: X\to Y$ be the natural projection,
{\em ie}, for $x\in X$, $\Pi(x)=\{ y\in X:\, x\sim y\}=\sim[x]$.
The function $\Pi$ is an \underline{isometr}y: it preserves the
canonical metric function $e$.

Define $B(x,r)=\{y\in X: d(x,y)<r\}$, where $x\in X$ and $r>0$, as
an \underline{o}p\underline{en ball} with center $x$ and radius
$r$. Then, defining the family $\mathcal{T}$ as the set of all
subsets $U$ of $X$ such that $U$ is the union of a family of open
balls in $X$, we obtain a \underline{to}p\underline{olo}gy
\underline{induced} \underline{b}y \underline{the}
\underline{metric} $d$ on $X$.

A topological space $(X,\mathcal{T})$ whose topology is induced by
the metric $d$ is a \underline{metrizable}
(\underline{to}p\underline{olo}g\underline{ical})
\underline{s}p\underline{ace}. (A pseudo-metric topology is
defined exactly as the metric topology.)

Two metrics $d_1$ and $d_2$ (or two pseudo-metrics $\ell_1$ and
$\ell_2$) on a set $X$ are said to be
\underline{to}p\underline{olo}g\underline{icall}y
\underline{e}q\underline{uivalent} to each other if they induce
the same topology $\mathcal{T}$ on $X$.

For any two points  $x=(x_1, x_2, x_3)$ and $y=(y_1, y_2, y_3)$ in
$\mathbb{R}^3$, $d(x, y) = \sqrt{\sum_{i=1}^3\,(x_i-y_i)^2}$ is
called the \underline{usual} (\underline{Euclidean})
\underline{metric} of $\mathbb{R}^3$. It induces a
\underline{usual to}p\underline{olo}gy on $\mathbb{R}^3$.

For any set $X$, function $d$: $d(x,y)=0$ if $x=y$ and $d(x,y)=1$
otherwise, defines a \underline{discrete} \underline{metric} on
$X$. Discrete metric induces a \underline{discrete
to}p\underline{olo}gy on $X$ consisting of all subsets of $X$.

Let $\{x_n\}$ be a \underline{se}q\underline{uence} of elements of
$X$ of $(X,d)$ and $x\in X$. $\{x_n\}$ is said to
\underline{conver}g\underline{e} to $x$, written $x_n\to x$ or
$\lim x_n=x$, if $d(x_n,x)\to 0$ as $n\to \infty$. Such an $x$ is
called the \underline{limit} of $\{x_n\}$. A sequence can have at
most one limit.

A function $f: (X,d)\to (Y,\rho)$ is called \underline{uniforml}y
\underline{continuous} on $X$ if for any $\epsilon > 0$ $\exists\,
\delta
> 0$ such that $d(x,y)<\delta\Rightarrow\rho\left( f(x),
f(y)\right) < \epsilon$ for any $x,y\in X$. A function $f:
(X,d)\to (Y,\rho)$ is an \underline{isometr}y if $\rho\left( f(x),
f(y) \right)=d(x,y)$ $\forall\, x,y\in X$.

A subset of a metrizable space is called a
$G_{\delta}$-\underline{set} if it is a countable intersection of
open sets. Hence, a closed subset of metrizable space is a
$G_{\delta}$ set. The complement of a $G_{\delta}$ set is called
as an $F_{\sigma}$-\underline{set}. It is a countable union of
closed sets. Every open set of a metrizable space is an
$F_{\sigma}$ set.

Let $f_n, f: (X,d)\to (Y,\rho)$. Then, the sequence, $(f_n)$, of
functions is said to \underline{conver}ge p\underline{ointwise} to
$f$ if for all $x$, $f_n(x)\to f(x)$ as $n\to\infty$. Furthermore,
we say that $f_n$ \underline{conver}g\underline{es}
\underline{uniforml}y to $f$ if for any $\epsilon > 0$, there
exists $N\in \mathbb{N}$ such that whenever $n\geq N$,
$\rho\left(f_n(x), f(x)\right)< \epsilon$ for all $x\in X$.

A map $f:X\to\mathbb{R}$, $X$ a metric space, is called as
\underline{u}pp\underline{er}-\underline{semicontinuous}\,\,
(\underline{lower}-\underline{semicontinuous}) if for every real
number $a$, the set $\{x\in X:f(x)\geq a\}$ ($\{x\in X: f(x)\leq
a\}$) is closed.

A sequence $\{x_n\}$ in a metric space $(X,d)$ is called a
\underline{Cauch}y \underline{se}q\underline{uence} if for every
$\epsilon > 0$, there is $N \in \mathbb{N}$ such that $d\left(x_n,
x_m\right) < \epsilon$ for all $m,n \geq N$. A metric $d$ on set
$X$ is called \underline{com}p\underline{lete metric} if every
Cauchy sequence in $(X, d)$ is convergent. A metric space is
called \underline{com}p\underline{lete} if $d$ is complete on $X$.
A metric that is topologically equivalent to a complete metric
need not be complete.

An arbitrary subspace of a complete metric space need not be
complete but a closed subspace is. $\mathbb{R}^3$ with the usual
metric is complete.

A topological space $X$ is called \underline{locall}y
\underline{com}p\underline{act} if every point of $X$ has a
compact neighborhood. $\mathbb{R}^3$ is a locally compact space.

If $\left\{X_i:i\in I\right\}$ is a family of topological spaces,
$X=\prod_{i\in I}X_i$, and $\pi_i:X\to X_i, \,i\in I$, then the
smallest topology on $X$ making each {\em projection map\/}
$\pi_i$ continuous is called the p\underline{roduct}
\underline{to}p\underline{olo}gy. The set $\left\{\pi^{-\,1}(U):
U\;{\rm open\;in\;}X_i,\,i\in I\right\}$ is a subbase for the
product topology.

For any set $A\subseteq X$, we define ${\rm
diameter}(A)=\sup\left\{ d(x,y): x,y \in A \right\}$. Clearly, for
set $A\subseteq X$, ${\rm diameter}(A)={\rm diameter}\left({\rm
cl}(A)\right)$.

A topological space is \underline{com}p\underline{letel}y
\underline{metrizable} if the topology is induced by a complete
metric. A separable, completely metrizable topological space is
called a \underline{Polish s}p\underline{ace}. Clearly, every
second countable, completely metrizable topological space is a
Polish space.

Any countable discrete space is Polish. $\mathbb{N}$ and
$\mathbf{2}$ with discrete topologies are Polish.
$\mathbb{R},\,\mathbb{R}^n,\,I=[0,1],\, I^n$ with the usual
topologies are Polish. Every $G_{\delta}$ subset of a Polish space
$X$ is Polish. The product of countably many Polish spaces is
Polish. $\mathbb{N}^{\mathbb{N}}$, the \underline{Hilbert}
\underline{Cube} $\mathbb{H}=[0,1]^{\mathbb{N}}$ and the
\underline{Cantor s}p\underline{ace} $\mathbf{2}^{\mathbb{N}}=
\mathcal{C}$ are Polish.

Note that the space $\mathbb{N}^{\mathbb{N}}$ is homeomorphic to
the space of positive irrational numbers in the open interval
$(0,1)$. The homeomorphism is achieved by associating an infinite
sequence $(n_o, n_1, n_2, ... )$ to a continued fraction
$1/(n_o+(1/n_1+(1/n_2+ ...))) \in (0,1)$. Therefore, we shall also
refer to $\Upsilon=\mathbb{N}^{\mathbb{N}}$ as the
\underline{s}p\underline{ace of irrational numbers}.

Notice that each of the topological spaces $\Upsilon \times
\Upsilon$, $\Upsilon^k,\,k=1,2,3,...$,
$\Upsilon^k\times\Upsilon^l$ ($k,l =1,2,3...$) and
$\Upsilon^{\mathbb{N}}$ are homeomorphic to
$\mathbb{N}^{\mathbb{N}}$.

Every $G_{\delta}$ subset of $\mathbb{N}^{\mathbb{N}}$ is
homeomorphic to a closed subset of $\mathbb{N}^{\mathbb{N}}$.
Every Polish space is a continuous image of
$\mathbb{N}^{\mathbb{N}}$.

For every Polish space $X$, there is a closed set $F\subseteq
\mathbb{N}^{\mathbb{N}}$ and a 1-1, continuous surjection $g:F\to
X$ such that $g\left(U\bigcap F\right)$ is an $F_{\sigma}$ set in
$X$ for every open set $U\in\mathbb{N}^{\mathbb{N}}$.

Every uncountable Polish space $X$ contains a homeomorph of the
Cantor Ternary Set and a homeomorph also of
$\mathbb{N}^{\mathbb{N}}$. Every uncountable Polish space is of
cardinality $\mathbf{c}$.

Every closed subspace of a Polish space is Polish. If $X_o$,
$X_1$, $X_2$, ... is a finite or infinite sequence of Polish
spaces then, so is their product $Y= \prod_{i=0}^{\infty}X_i$ a
Polish space. Every compact metric space is a Polish space.

Spaces $\mathbb{N}^{\mathbb{N}}$ and $\mathcal{C}$ are important
to us. A complete metric on $\mathbb{N}^{\mathbb{N}}$ compatible
with its topology is $\rho(\alpha, \beta)=1/\left(
\min\{n:\alpha(n)\neq\beta(n)\} + 1 \right)$ if $\alpha\neq\beta$
and $\rho(\alpha,\beta)=0$ otherwise. For $s\in
\mathbb{N}^{<\,\mathbb{N}}$ let $\Sigma(s)=\{ \alpha\in
\mathbb{N}^{\mathbb{N}}: s \prec \alpha\}$. The family of sets
$\{\Sigma(s): s\in\mathbb{N}^{<\,\mathbb{N}}\}$ is a {\em
clopen\/} base for $\mathbb{N}^{\mathbb{N}}$. Hence,
$\mathbb{N}^{\mathbb{N}}$ is a zero-dimensional Polish space.

For each $s\in \mathbb{N}^{<\, \mathbb{N}}$, $\Sigma(s)$ is
homeomorphic to $\mathbb{N}^{\mathbb{N}}$. Every $G_{\delta}$
subset of $\mathbb{N}^{\mathbb{N}}$ is homeomorphic to a closed
subset of $\mathbb{N}^{\mathbb{N}}$.

Also, for every Polish space $X$, there is a closed subset
$F\subseteq \mathbb{N}^{\mathbb{N}}$ and a 1-1, continuous
surjection $g:F\to X$ such that $g\left( U\bigcap F\right)$ is an
$F_{\sigma}$ set in $X$ for every open set $U$ in
$\mathbb{N}^{\mathbb{N}}$.

Let $A$ be a discrete topological space and the set
$X=A^{\mathbb{N}}$ be equipped with product topology. Then, $X$ is
a zero-dimensional completely metrizable space, it is Polish if
and only if $A$ is countable. The set $\left\{\alpha\in
A^{\mathbb{N}}: s\prec \alpha\right\}$ with $s\in
A^{<\,\mathbb{N}}$ is the base for its topology. We also denote it
by $\Sigma(s)$.

If $f$ is a continuous function on a subset of a metric space with
values in a Polish space $Y$ then, there exists a continuous
extension of $f$ to a $G_{\delta}$ set. \underline{Lavrentiev}'s
\underline{Theorem} proves further that: Let $A\subseteq X$ and
$B\subseteq Y$, $X$ and $Y$ being Polish spaces. Suppose $f:A\to
B$ is a homeomorphism. Then, there exist $G_{\delta}$ subsets
$A\subseteq A_1 \subseteq X$ and $B\subseteq B_1\subseteq Y$ and a
homeomorphism $f_1:A_1\to B_1$ which extends $f$.

A topological space $X$ is a Polish space if and only if it is
homeomorphic to a $G_{\delta}$ subset of the Hilbert cube. Every
$G_{\delta}$ subset $G$ of a completely metrizable (Polish) space
is completely metrizable (Polish). The converse of the above,
\underline{Alexandrov}'s \underline{Theorem}, is also true.

Let $X$ be a compact metrizable space and $Y$ a Polish space. Let
$C(X,Y)$ be the set of continuous functions from $X$ into $Y$. Let
a compatible complete on $Y$ be $\rho$ and define $\delta(f,g) =
\sup_{x\in X}\rho\left( f(x), g(x)\right), f,g\in C(X,Y)$. It is a
complete metric on $C(X,Y)$. The topology on $C(X,Y)$ induced by
the complete metric $\delta$ above is called the
\underline{to}p\underline{olo}gy \underline{of}
\underline{uniform} \underline{conver}g\underline{ence}.

If $(X,d)$ is a compact metric space and $(Y,\rho)$ is Polish,
then $C(X,Y)$ equipped with the topology of uniform convergence is
Polish.

For non-empty $A\subseteq X$ of $(X,d)$ and $\forall\,x\in X$,
define $d\left(x,A\right)=\min \left\{ d(x,a): a\in A\right\}$. It
is the \underline{distance} from the point $x$ to the set $A$.

Now, for a topological space $(X,\mathcal{T})$, consider the
family $\mathbb{K}(X)$ of all non-empty compact subsets of $X$.
The topology on $\mathbb{K}(X)$ generated by compact subsets of
$X$ of the form $ \{K \in \mathbb{K}(X):\,K \subseteq U \} $ and $
\{K \in \mathbb{K}(X):\,K\bigcap U \neq \emptyset\}$, $U$ open in
$X$, is called the \underline{Vietoris to}p\underline{olo}gy.

The sets $ [U_o, U_1, ..., U_n] = \left\{ K\in
\mathbb{K}(X):K\subseteq U_o\right.$ $\left. \&\;K\bigcap
U_i\neq\emptyset, 1\leq i \leq n\right\} $, with $U_o, U_1, ...,
U_n$ open in $X$, form a base for $\mathbb{K}(X)$. The set of all
finite, non-empty subsets of $X$ is dense in $\mathbb{K}(X)$.

On the family $\mathbb{K} (X)$, we define the
\underline{Hausdorff} \underline{metric} $\delta_H$ as \[
\delta_H(K,L) = \max \left(\max_{x\in K}\,d(x,L),\; \max_{y\in
L}\,d(y,K) \right)\]

The Hausdorff metric $\delta_H$ induces the Vietoris topology on
$\mathbb{K}(X)$. If $X$ is separable, so is $\mathbb{K}(X)$. If
$(X,d)$ is a complete metric space, so is $(\mathbb{K}(X),
\delta_H)$. If $(X,d)$ is Polish, so is $(\mathbb{K}(X),
\delta_H)$.

If $X$ is a metrizable space, then the set $K_f(X)=\left\{ L\in
K(X): L\;{\rm is\;finite} \right\}$ is an $F_{\sigma}$ set. Any
compact, dense-in-itself set is called p\underline{erfect}. Then,
for a separable and metrizable space $X$, the set $K_p(X)=\left\{
L\in K(X): L\;{\rm is\;perfect} \right\}$ is $G_{\delta}$.

For a locally compact Polish space $X$ and a base $B(X)$ for its
Polish topology, $B(X)$ can be given a topology generated by sets
of the type: $(S\in B(X):S\bigcap K = \emptyset\,\&\, S\bigcap
U_1\neq\emptyset\,\& \,S\bigcap U_2\neq\emptyset\,\&...\,\&\,
S\bigcap U_n\neq \emptyset)$ where $K$ ranges over the compact
subsets of $X$ and $U_1$, $U_2$, ..., $U_n$ range over open sets
in $X$. It is called the \underline{Fell to}p\underline{olo}gy,
and $B(X)$ with the Fell topology is Polish.

A subset $A$ of $(X,\mathcal{T})$ is
\underline{no}-\underline{where} \underline{dense} if ${\rm
cl}(A)$ has empty interior, {\em ie}, if $X\setminus {\rm cl}(A)$
is dense. For every closed set $A$, $A\setminus {\rm int}(A)$ is
nowhere dense.  Then, a set $A$ is nowhere dense iff every
nonempty open set $U$ contains another nonempty open set $V$ such
that $A\bigcap V=\emptyset$.

A set $A\subseteq X$ is \underline{mea}g\underline{er} or of
\underline{first cate}g\underline{or}y in $X$ if it is a countable
union of nowhere dense sets. Every meager set is contained in a
meager $F_{\sigma}$ set. A set which is not meager is of
\underline{second cate}g\underline{or}y in $X$. A subset $A$ is
\underline{co}-\underline{mea}g\underline{er} in $X$ if
$X\setminus A$ is meager in $X$. $A\subseteq X$ is co-meager in
$X$ iff it contains a countable intersection of dense open sets.

Let $(X,\,d)$ be Polish and $d$ a complete metric with ${\rm
diameter}(X) < 1$. Fix a nonempty set $A$. A \underline{Souslin
scheme} on $X$ is a system $\left\{ F_s: s\in A^{<\,\mathbb{N}}
\right\}$ of subsets of $X$ such that \begin{description}
\item{(i)} ${\rm cl}(F_{\widehat{s\,a}}) \subseteq F_s$ for all $s$
and $a$ \item{(ii)} for all $\alpha \in A^{\mathbb{N}}$, ${\rm
diameter}(F_{\alpha|n}) \to 0$ as $n\to \infty$.\end{description}

A Souslin scheme is a \underline{Lusin scheme} if in addition to
(i) and (ii) above the following is also satisfied:
\begin{description} \item{(iii)} for every $s, t \in A^{<\,\mathbb{N}}$, $s\bot
t\Rightarrow F_s\bigcap F_t= \emptyset$. \end{description} A
\underline{Cantor scheme} is a Lusin scheme with $A=\mathbf{2}$
and each $F_s$ is closed and nonempty.

For a Souslin scheme $\left\{ F_s: s\in A^{<\,\mathbb{N}}
\right\}$, equip $A^{\mathbb{N}}$ with the product of discrete
topologies on $A$. Then, \begin{description} \item{(a)} the set
$D=\left\{ \alpha\in A^{\mathbb{N}}: \forall\,n\,\left(
F_{\alpha|n} \neq 0\right) \right\}$ is a closed set. \item{(b)}
The set $\bigcap_nF_{\alpha|n}= \bigcap_n{\rm cl}\left(
F_{\alpha|n}\right)$ is a singleton for each $\alpha\in
D$.\end{description} Define $f: D\to X$ as $\left\{ f(\alpha)
\right\}=\bigcap_n=F_{\alpha|n}$ as the \underline{associated ma}p
of the scheme. The map $f$ is continuous.

Further, if $F_e=X$ and $\forall \, s\, \left(F_s=\bigcup_n
F_{\widehat{s\,n}}\right)$, the associated map $f$ is onto $X$.
For a Lusin scheme $f$ is one to one and, for a Cantor scheme
$\left\{ F_s: s\in 2^{<\,\mathbb{N}} \right\}$, $f$ is an
embedding in $X$.

The \underline{Cantor}-\underline{Bendixson Theorem} proves that
every separable space $X$ can be written as $X=Y\bigcup Z$ where
$Z$ is countable, $Y$ closed with no isolated point and $Y\bigcap
Z=\emptyset$. Also, every uncountable Polish space $X$ contains a
\underline{homeomor}p\underline{h} of $\mathcal{C}$ and of
$\mathbb{N}^{\mathbb{N}}$.

An \underline{e}q\underline{uivalence relation} $E\subseteq
X\times X$ on a Polish space is \underline{closed}
(\underline{o}p\underline{en}, $G_{\delta}, F_{\sigma}$ etc.) if
$E$ is a closed (open, $G_{\delta}, F_{\sigma}$ etc.) subset of
$X\times X$.

An \underline{al}g\underline{ebra} on a set $X$ is a collection
$\mathcal{A}$ of subsets of $X$ such that
\begin{description} \item{(${\rm \bf {a}}$)} $X\in \mathcal{A}$, \item{(${\rm
\bf {b}}$)} whenever $A\in \mathcal{A}$, $A^c\equiv X\setminus A
\in \mathcal{A}$, \item{(${\rm \bf {c}}$)} $\mathcal{A}$ is closed
under finite unions. As $X\in \mathcal{A}$, $\emptyset\equiv X^c
\in \mathcal{A}$. \end{description} An algebra closed under
countable unions is called as a
\underline{$\sigma$}-\underline{al}g\underline{ebra}.

Note that any $\sigma$-algebra is either finite or of cardinality
at least $\mathbf{c}$.

For any set $X$, $\mathcal{B}_1=\{\emptyset, X\}$ and
$\mathcal{B}_2=\mathcal{P}(X)$ are called the
\underline{indiscrete} and \underline{discrete} $\sigma$-algebras,
respectively. If the set $X$ is an uncountable set then,
$\mathcal{A}=\left\{ A\subseteq X: {\rm either\;}A\;{\rm
or\;}A^c\;{\rm is\;countable }\right\}$ is a
\underline{countable}-\underline{cocountable} $\sigma$-algebra.

A \underline{measurable s}p\underline{ace} is an ordered pair $(X,
\mathcal{A})$ with $\mathcal{A}$ being a $\sigma$-algebra of the
subsets of the set $X$. Members of the $\sigma$-algebra
$\mathcal{A}$ are called as the \underline{measurable sets}.

An intersection of a non-empty family of $\sigma$-algebras on a
set $X$ is a $\sigma$-algebra.

Let $\mathcal{S}$ be the family of all $\sigma$-algebras on $X$
containing a family $\mathcal{G}$ of subsets of $X$. Clearly,
$\mathcal{S}$ is always nonempty. Then, the intersection of all
the members of $\mathcal{S}$ is the smallest $\sigma$-algebra,
$\sigma(\mathcal{G})$, containing $\mathcal{G}$.
$\sigma(\mathcal{G})$ is said to be g\underline{enerated} by
$\mathcal{G}$. A $\sigma$-algebra $\mathcal{A}$ is said to be
\underline{countabl}y g\underline{enerated} if it has a countable
generator.

If $A_o, A_1, ..., A_n \in \mathcal{A}$, $\mathcal{A}$ is a
$\sigma$-algebra on set $X$, then the sets $\bigcap_nA_n=\left(
\bigcup_n A^c_n\right)^c$, $\limsup_nA_n
\equiv\bigcap_n\bigcup_{m\geq n}A_m$ and $\liminf_n A_n
\equiv\bigcup_n\bigcap_{m\geq n}A_m$ will always be some of the
measurable sets in the measurable space $(X,\mathcal{A})$.

Now, for $(X,\mathcal{A})$ with $\mathcal{A}=
\sigma(\mathcal{G})$, suppose $x,y\in X$ are such that for every
$G\in \mathcal{G}, x\in G$ if and only if $y\in G$. Then, for all
$A\in \mathcal{A}$, $x\in A$ if and only if $y\in A$ because
$\mathcal{B}=\left\{ A\subseteq X:x\in A \Longleftrightarrow y\in
A\right\}$ is a $\sigma$-algebra containing $\mathcal{G}$.

Next, if $(X,\mathcal{B})$ is a measurable space, $\mathcal{G}$ a
generator of $\mathcal{B}$, then there exists a countable
$\mathcal{G}' \subseteq\mathcal{G}$ such that $A\in
\sigma(\mathcal{G}')$.

Let $\mathcal{D}\subseteq \mathcal{P}(X)$ and $Y\subseteq X$. We
set $\mathcal{D}|Y=\left\{ B\bigcap Y:B\in \mathcal{D}\right\}$.
If $(X,\mathcal{B})$ is measurable space and $Y\subseteq X$ then,
$\mathcal{B}|Y$ is an \underline{induced}
\underline{$\sigma$}-\underline{al}g\underline{ebra} on $Y$, also
called the {\em trace of\/} $\mathcal{B}$. If $\mathcal{G}$
generates $\mathcal{B}$ then, $\mathcal{G}|Y$ generates
$\mathcal{B}|Y$.

{\em Unless stated otherwise, a subset of a measurable space will
be assumed to be equipped with the trace or the induced
$\sigma$-algebra}.

A collection $\mathcal{M}$ of subsets of a set $X$ is called
\underline{monotone} \underline{class} if it is closed under
countable non-increasing intersections and countable
non-decreasing unions. The \underline{Monotone} \underline{Class}
\underline{Theorem} states that the smallest monotone class
$\mathcal{M}$ containing an algebra $\mathcal{A}$ on a set $X$
equals $\sigma(A)$, the $\sigma$-algebra generated by
$\mathcal{A}$.

The $\sigma$-algebra generated by the topology on a measurable
space $X$ is called a \underline{Borel}
\underline{$\sigma$}-\underline{al}g\underline{ebra} and will be
denoted by $\mathcal{B}_X$. Sets in $\mathcal{B}_X$ will be called
\underline{Borel in $X$}. For an uncountable Polish space, the
Borel $\sigma$-algebra is of cardinality $\mathbf{c}$.

{\em Unless explicitly stated to the contrary, a metrizable space
will be assumed to be equipped with its Borel $\sigma$-algebra.}

Note that the Borel $\sigma$-algebra of a metrizable space $X$
equals the smallest family $\mathcal{B}_O$ ($\mathcal{B}_C$) of
subsets of $X$ that contains all open (closed) sets and that is
closed under countable intersections and countable unions, {\em
ie}, $\mathcal{B}_X=\mathcal{B}_O=\mathcal{B}_C$.

The Borel $\sigma$-algebra $\mathcal{B}_X$ of a metrizable space
$X$ can also be seen to equal the smallest family of subsets of
$X$ that contains all open (closed) subsets of $X$ and that is
closed under countable intersections and countable {\em
disjoint\/} unions.

A measurable set $A\neq\emptyset$ of a measurable space
$(X,\mathcal{A})$ is an \underline{$\mathcal{A}$}-\underline{atom}
if it has no non-empty measurable proper subset. No two distinct
atoms intersect. A measurable space is \underline{atomic} if $X$
is the union of its atoms. For metrizable $X$, $(X,\mathcal{B}_X)$
is atomic, singletons being $\mathcal{B}_X$-atoms.

A \underline{measurable ma}p is a map $f:(X,\mathcal{A}) \to (Y,
\mathcal{B})$ such that $f^{-\,1}(B)\in \mathcal{A}$ for every
$B\in \mathcal{B}$. A map $f$ is then measurable if and only if
$f^{-\,1}(B)\in \mathcal{A}$ for every $B\in \mathcal{G}$ where
$\mathcal{G}$ generates $\mathcal{B}$.

A measurable function $f:(X,\mathcal{B}_X)\to(Y,\mathcal{B}_Y)$ is
called \underline{Borel measurable} or simply \underline{Borel}.
If $X$ and $Y$ are metrizable spaces then, every continuous
function $f:X\to Y$ is Borel.

Let $(X_i, \mathcal{A}_i),\,i\in I,$ be a family of measurable
spaces and $X=\prod_iX_i$. The $\sigma$-algebra on $X$ generated
by $\left\{\pi^{-\,1}_i(B):B\in \mathcal{A}_i,\,i\in I\right\}$
where $\pi_i:X\to X_i$ are the projection maps, is called the
p\underline{roduct}
\underline{$\sigma$}-\underline{al}g\underline{ebra}. It is
denoted by $\bigotimes_i \mathcal{A}_i$. It is the smallest
$\sigma$-algebra such that each $\pi_i$ is measurable. The product
$\sigma$-algebra on $X\times Y$ where $(X,\mathcal{A})$ and
$(Y,\mathcal{B})$ are measurable spaces, will be denoted simply by
$\mathcal{A}\bigotimes \mathcal{B}$.

{\em Unless stated otherwise, we shall assume that the product of
measurable spaces is equipped with the product $\sigma$-algebra.}

Now, let $(f_n)$ be a sequence of measurable maps from a space $X$
to space $Y$, both measurable spaces, converging point-wise to
$f$. Then, $f:X\to Y$ is a measurable function.

If $X$ is a measurable space then, every Borel function $f:X\to
\mathbb{R}$ is the point-wise limit of a sequence of simple Borel
functions.

If $f:(X,\mathcal{A})\to (Y,\mathcal{B})$ and
$g:(Y,\mathcal{B})\to (Z, \mathcal{C})$ are measurable, then so is
$g\circ f:(X,\mathcal{A}) \to (Z,\mathcal{C})$ measurable. Also, a
map $f:(X, \mathcal{A}) \to \left(\prod X_i,
\bigotimes_i\mathcal{A}_i\right)$ is measurable if and only if its
composition with each projection map is measurable.

For metrizable spaces $X$ and $Y$, let $\mathcal{B}(X,Y)$ be the
smallest class of functions from $X$ to $Y$ containing all
continuous functions and closed under taking point-wise limits of
sequences of functions. Functions belonging to $\mathcal{B}(X,Y)$
are called the \underline{Baire functions}. Every Baire function
is Borel but the converse is not true.

However, for every metrizable $X$, every Borel $f:X\to \mathbb{R}$
is Baire. If $a,b\in\mathbb{R}$ and $f,g:X\to \mathbb{R}$ are
Baire, then so is $af+bg$ Baire. For $B\subseteq X$, with
metrizable $X$, the map $\chi_{_B}:X\to \mathbb{R}$ is a Baire
function. The
\underline{Lebes}g\underline{ue}-\underline{Hausdorff Theorem}
proves that every real-valued Borel function on a metrizable space
is a Baire function.

Some results that help reduce measurability problems to
corresponding topological problems are as follows.

Given a metrizable space $(X,\mathcal{T})$ and sequence $(B_n)$ of
its Borel subsets, there is a metrizable topology $\mathcal{T}'$
such that $\mathcal{T} \subseteq \mathcal{T}' \subseteq
\mathcal{B}_X$ and each $B_n\in \mathcal{T}'$. Topology generated
by $\mathcal{T}\bigcup\left\{B_n: n\in \right.$ $\left. \mathbb{N}
\right\} \bigcup\left\{B^c_n:n\in\mathbb{N}\right\}$ is such a
topology.

(If $\mathcal{T}$ and $\mathcal{T}'$ are topologies on $X$ with
$\mathcal{T}\subseteq\mathcal{T}'$, then $\mathcal{T}'$ is
\underline{finer} or \underline{stron}g\underline{er} or
\underline{lar}g\underline{er} than $\mathcal{T}$ and
$\mathcal{T}$ is \underline{coarser} or \underline{weaker} or
\underline{smaller} than $\mathcal{T}'$.\underline{Caution}:
coarser and finer or weaker and stronger are also used in the
sense opposite to the above.)

Next, if $(X,\mathcal{T})$ is Polish then, for every Borel set
$B\in X$ there is a finer Polish topology $\mathcal{T}_B$ on $X$
with $B$ clopen in $\mathcal{T}_B$ and $\sigma (\mathcal{T})=
\sigma (\mathcal{T}_B)$.

Also, for every sequence $(B_n)$ of Borel sets in a Polish space
$(X,\mathcal{T})$, there is a finer topology $\mathcal{T}'$ on $X$
generating the same Borel $\sigma$-algebra and making each $B_n$
clopen. Note also that every Borel subspace of a Polish space is
Polish.

Moreover, let $(X,\mathcal{T})$ be a Polish space, $Y$ a separable
metric space and $f:X\to Y$ be a Borel map. Then, there is a finer
Polish topology $\mathcal{T}'$ on $X$ generating the same Borel
$\sigma$-algebra such that $f:(X,\mathcal{T}') \to Y$ is
continuous.

A \underline{Bernstein Set} is a set $A$ of real numbers with both
$A\bigcap C$ and $(\mathbb{R}\setminus A)\bigcap C$ being {\em
uncountable\/} for any uncountable closed subset $C$ of
$\mathbb{R}$.

A map $f$ from a measurable space $X$ to a measurable space $Y$ is
\underline{bimeasurable} if it is measurable and $f(A)$ is
measurable for every measurable subset $A$ of $X$. A bimeasurable
bijection is an {\em isomorphism}. Thus, a bijection $f:X\to Y$ is
an isomorphism if and only if both $f$ and $f^{-\,1}$ are
measurable. If $X$ and $Y$ are measurable spaces and $f:X\to Y$,
$g:Y\to X$ are 1-1 bimeasurable maps then, $X$ and $Y$ are
isomorphic.

When $X$ and $Y$ are metrizable spaces equipped with Borel
$\sigma$-algebras and $f:X\to Y$ is an isomorphism, $f$ is called
a \underline{Borel isomor}p\underline{hism} and, $X$ and $Y$ as
\underline{Borel isomor}p\underline{hic}. Note that the Borel
$\sigma$-algebra of a countable metrizable space is discrete and,
hence, two countable metrizable spaces are Borel isomorphic if and
only if they are of the same cardinality.

A \underline{Standard Borel S}p\underline{ace} (SBS) is a
measurable space isomorphic to some Borel subset of a Polish
space. Then, a metrizable space $X$ is standard Borel if
$(X,\mathcal{B}_X)$ is standard Borel. A SBS equipped with a
probability measure will be called a \underline{Standard}
\underline{Probabilit}y \underline{S}p\underline{ace} (SPS).

For a compact metric space $X$, the space, $\mathcal{K}(X)$,  of
nonempty compact sets with Vietoris topology, being Polish, is a
standard Borel space. Interestingly, its Borel $\sigma$-algebra
$\mathcal{B}_{\mathcal{K}(X)}$ is generated by sets of the form
$\left\{ K\in \mathcal{K}(X): K\bigcap U\neq \emptyset\right\}$
where $U$ varies over open sets in $X$.

Now, let $X$ be a Polish space and $F(X)$ denote the set of all
nonempty closed subsets of $X$. Equip $F(X)$ with the
$\sigma$-algebra $\mathcal{E}(X)$ generated by sets of the form
$\left\{ F\in\mathcal{E}(X): F\bigcap U \neq \emptyset \right\}$,
where $U$ varies over open sets of $X$. The space $\left( F(X),
\mathcal{E}(X) \right)$ is called the \underline{Effros Borel
S}p\underline{ace} of $X$.

If $X$ is compact, $\mathcal{E}(X)=\mathcal{B}_{\mathcal{K}(X)}$,
{\em ie}, the Effros Borel space of a compact metrizable space is
standard Borel. The Effros Borel Space of a Polish space is
standard Borel.

For a Polish space $X$, the Borel space of $F(X)$ equipped with
the Fell topology is exactly the same as the Effros Borel Space as
a compact subset of a Polish space is closed and bounded.

Now, every standard Borel space $X$ is Borel isomorphic to a Borel
subset of the Cantor space $\mathcal{C}$. Next, for every Borel
subset $B$ of a Polish space $X$, there is a Polish space $Z$ and
a continuous bijection from $Z$ to $B$. The Borel isomorphism
theorem states that any two uncountable standard Borel spaces are
Borel isomorphic.

Note here that two Standard Borel Spaces can be Borel isomorphic
if and only if they are of the same cardinality.

Every Borel subset of a Polish space is a continuous image of
$\mathbb{N}^{\mathbb{N}}$ and a one-to-one, continuous image of a
closed subset of $\mathbb{N}^{\mathbb{N}}$. For every infinite
Borel subset $X$ of a Polish space, $|\mathcal{B}_X|=\mathbf{c}$.

The set of all Borel maps from $X$ to $Y$, these being uncountable
Polish spaces, is of cardinality $\mathbf{c}$. For $X$ a Polish
space, $A\subseteq X$, and $f:A\to A$ being a Borel isomorphism,
$f$ can be extended to a Borel isomorphism $g:X\to X$. We also
note that for an uncountable Polish space $X$ and a map
$f:X\to\mathbb{R}$, there is no Borel map $g:X\to\mathbb{R}$
satisfying $g(x)\leq f(x)$ for all $x$.

Now, for a nonempty set $X$ and an algebra $\mathcal{A}$ on $X$, a
\underline{measure} on $\mathcal{A}$ is a map $\mu:\mathcal{A}\to
[0, \infty]$ such that \begin{description} \item{(i)}
$\mu(\emptyset)=0$, \item{(ii)} $\mu$ is \underline{countabl}y
\underline{additive}, {\em ie}, if $A_o$, $A_1$, ... are pairwise
disjoint in $\mathcal{A}$ with $\bigcup_nA_n\in \mathcal{A}$ then,
$\mu\left(\bigcup_nA_n \right)=\sum_o^{\infty} \mu(A_n)$.
\end{description}

When $\mathcal{A}$ is understood from the context, we shall simply
say that $\mu$ is a measure on $X$. A measure $\mu$ is called
\underline{finite} if $\mu(X)< \infty$; it is
$\sigma$-\underline{finite} if $X$ can be written as a countable
union of sets in $\mathcal{A}$ of finite measure. It is called a
p\underline{robabilit}y \underline{measure} if $\mu(X)=1$.
Further, if all subsets of sets of measure zero are measurable, a
measure is said to be a \underline{com}p\underline{lete measure}.

If $m$ is a measure on $(X,\mathcal{A})$, then a set
$E\in\mathcal{A}$ is of {\em finite $m$-measure\/} if $m(E)<
\infty$; is of {\em $\sigma$-finite $m$-measure\/} if $\exists \;
\{E_i\},\;i\in\mathbb{N},\; E_i\in\mathcal{A}$ such that
$E\subseteq \bigcup_{i=1}^{\infty}E_i$ and
$m(E_i)<\infty,\;\forall\;i\in \mathbb{N}$. If
$m(A),\;A\in\mathcal{A}$ is finite ($\sigma$-finite) then the
measure $m$ is {\em finite ($\sigma$-finite) measure on
$\mathcal{A}$}. A measure is {\em totally finite or totally
$\sigma$-finite\/} if $m(X)$ is finite or $\sigma$-finite.

A \underline{measure s}p\underline{ace} is a triple $(X,
\mathcal{A}, \mu)$ where $\mathcal{A}$ is a $\sigma$-algebra on
$X$ and $\mu$ a measure. A measure space is called a
p\underline{robabilit}y \underline{s}p\underline{ace} if $\mu$ is
a probability measure on it.

For $(X,\mathcal{A})$ being a measurable space, $A\in \mathcal{A}$
and $x\in X$, let $\delta_x(A)=1$ if $x\in A$ and $\delta_x(A)=0$
otherwise. Then, $\delta_x$ is a measure on $\mathcal{A}$ and will
be called the \underline{Dirac measure} at $x$.

For a nonempty set $X$, $A\subseteq X$, let $\mu(A)$ denote the
number of elements in $A$, $\mu(A)=\infty$ if $A$ is infinite.
Then $\mu$ is a measure on $\mathcal{P}(X)$, called the
\underline{countin}g \underline{measure}.

Now, if $(X,\mathcal{A},\mu)$ is a measure space then, it is easy
to see that \begin{description} \item{(i)} $\mu$ is monotone,
\item{(ii)} $\mu$ is countably sub-additive, \item{(iii)} if the $A_n$'s are
measurable and nondecreasing then, $\mu\left( \bigcup_nA_n\right)=
\lim\,\mu\left(A_n\right)$, and \item{(iv)} if $\mu$ is finite and
$(A_n)$ is a non-increasing sequence of measurable sets then,
$\mu\left(\bigcap_nA_n\right)=\lim\,\mu(A_n)$. \end{description}

If $(X,\mathcal{B})$ is a measurable space, $\mathcal{A}$ an
algebra such that $\sigma(\mathcal{A})=\mathcal{B}$, and suppose
$\mu_1$ and $\mu_2$ are finite measures on $(X,\mathcal{B})$ such
that $\mu_1(A)=\mu_2(A)$ for every $A\in \mathcal{A}$, then
$\mu_1(A)=\mu_2(A)$ for every $A\in \mathcal{B}$. Furthermore, if
$\mathcal{A}$ is an algebra on $X$ and $\mu$ is a $\sigma$-finite
measure on $\mathcal{A}$ then, there exists a unique measure $\nu$
on $\sigma(\mathcal{A})$ that extends $\mu$.

Now, let $\mathcal{A}$ be the algebra on $\mathbb{R}$ consisting
of finite disjoint unions of non-degenerate intervals. For any
interval $I$, let $|I|$ denote the length of $I$. Let $I_o, I_1,
..., I_n$ be pairwise disjoint intervals and let
$A=\bigcup_{k=0}^nI_k$. Set $\lambda(A)=\sum_{k=0}^n|I_k|$. Then,
$\lambda$ is a $\sigma$-finite measure on $\mathcal{A}$. There is
then a unique measure on $\sigma(\mathcal{A})=
\mathcal{B}_{\mathbb{R}}$ extending $\lambda$. We call this
measure the \underline{Lebes}g\underline{ue measure} on
$\mathbb{R}$ and denote it by $\lambda$ itself.

Let $(X_n,\mathcal{A}_n,\mu_n)$, $n\in\mathbb{N}$, be a sequence
of probability spaces and $X=\prod_nX_n$. For any nonempty, finite
$F\subseteq \mathbb{N}$, let $\pi_F:X\to \prod_{n\in F}X_n$ be the
canonical projection map.

Define $\mathcal{A}=\left\{ \pi^{-\,1}_F (R): R\in
\bigotimes_{n\in F}\mathcal{A}_n,\,F\,{\rm finite} \right\}$.
Then, $\mathcal{A}$ is an algebra that generates the product
$\sigma$-algebra $\bigotimes_n\mathcal{A}_n$. Define further
$\prod_n\mu_n$ on $\mathcal{A}$ by $\prod_n\mu_n\left(
\pi^{-\,1}_F(R)\right)=\left(\times_{_{i\in F}}\mu_i\right)(R)$ as
a probability measure on $\mathcal{A}$. Then, there exists a
unique probability measure on $\bigotimes_n\mathcal{A}_n$ that
extends $\prod_n\mu_n$. We will call this extension the
p\underline{roduct of the} $\mu_n$'s and denote it by
$\prod_n\mu_n$. If $(X_n, \mathcal{A}_n, \mu_n)$ are the same,
say, $\mu_n=\mu$ for all $n$, then we denote the
p\underline{roduct} \underline{measure} by $\mu^{\mathbb{N}}$.

Let $X$ be a finite set with $n\, (n>0)$ elements and $\mathcal{A}
=\mathcal{P}(X)$. The \underline{uniform measure} on $X$ is the
measure $\mu$ on $\mathcal{A}$ such that $\mu\left(\{x\}\right) =
1/n$ for every $x\in X$. Let $\mu$ be the uniform probability
measure on the set $\mathbf{2}$. The product measure
$\mu^{\mathbb{N}}$ on $\mathcal{C}$ is a
\underline{Lebes}g\underline{ue measure} denoted also by
$\lambda$.

Let $(X,\mathcal{A}, \mu)$ be a measure space. A subset $A$ of $X$
will be called $\mu$-\underline{null} or simply a \underline{null
set} if there is a measurable set $B$ containing $A$ such that
$\mu(B)=0$. The measure space $(X,\mathcal{A},\mu)$ will be called
\underline{com}p\underline{lete} if every null set is measurable
in it. The counting measure and the uniform measure on a finite
set are complete.

If $(X,\mathcal{B},\mu)$ is a complete $\sigma$-finite measure
space, then $\mathcal{B}$ is closed under the Souslin operation.

An \underline{ideal} on a nonempty set $X$ is defined to be a
nonempty family $\mathcal{I}$ of subsets of $X$ such that we have
\begin{description} \item{(i)} $X \notin \mathcal{I}$,
\item{(ii)} whenever $A\in\mathcal{I}$, $\mathcal{P}(A) \in
\mathcal{I}$, and \item{(iii)} $\mathcal{I}$ is closed under
finite unions.\end{description} A $\sigma$-\underline{ideal} is an
ideal closed under countable unions. Notably, the family
$\mathcal{N}_{\mu}$ of all $\mu$-null sets is a $\sigma$-ideal.

If $\mathcal{E}$ is any collection of subsets of $X$, then there
exists a smallest $\sigma$-ideal containing $\mathcal{E}$, the
intersection of all $\sigma$-ideals containing $\mathcal{E}$. It
is called the $\sigma$-ideal g\underline{enerated b}y
$\mathcal{E}$ and is obtained by taking all sets of the form
$B\bigcap E$ with $B\in {\cal B}_X$, $E\in \mathcal{E}$ and taking
countable unions of such sets. Alternatively, the family
$\mathcal{I}=\left\{ A \subseteq X: A\subseteq
\bigcup_nB_n,\,B_n\in\mathcal{E} \right\}$ is the smallest
$\sigma$-ideal containing $\mathcal{E}$.

The $\sigma$-algebra generated by $\mathcal{A}\bigcap
\mathcal{N}_{\mu}$ is called the
$\mu$-\underline{com}p\underline{letion} or simply the
\underline{com}p\underline{letion} of the measure space $X$. We
shall denote the completion of the measure space as
$\bar{\mathcal{A}}^{\mu}$ and call the sets in
$\bar{\mathcal{A}}^{\mu}$ as $\mu$-\underline{measurable}.

Note that $\bar{\mathcal{A}}^{\mu}$ consists of all sets of the
form $A\triangle N$ where $A\in\mathcal{A}$, $N$ is null and
$\triangle$ denotes the \underline{s}y\underline{mmetric
difference} of sets. Further, $\bar{\mu}\left(A\triangle
N\right)=\mu(A)$ is a measure on the completion. It can also be
shown that the set $A$ is $\mu$-measurable if and only if there
exist measurable sets $B$ and $C$ such that $B\subseteq A\subseteq
C$ and $C\setminus B$ is null.

The set function $\mu^*:\mathcal{P}(X)\to [0,\infty]$ defined by
$\mu^*(A)=\inf\left\{ \mu(B):B\in \mathcal{A}\,\&\,A\subseteq B
\right\}$ is called the \underline{outer measure induced b}y
$\mu$. Clearly, for every set $A$ there is a set $B\in\mathcal{A}$
such that $A\subseteq B$ and $\mu(B)=\mu^*(A)$. Also, if $B'$ is
another measurable set containing $A$ then $B\setminus B'$ is
null.

Then, sets in $\bar{\mathcal{B}}_{_{\mathbb{R}}}^{\lambda}$,
$\lambda$ being the Lebesgue measure on the set of real numbers,
are the {\em Lebesgue measurable sets}. Note that $|\mathcal{B}| =
\mathbf{c}<2^{\mathbf{c}}$ and that there are Lebesgue measurable
sets which are not Borel. The Bernstein set, mentioned earlier, is
not Lebesgue measurable, for example. Note also that the Lebesgue
measure on $\mathbb{R}$ is \underline{translation}
\underline{invariant}, {\em ie}, for every Lebesgue measurable set
$E$ and every real number $x$, $\lambda(E) = \lambda(E+x)$ where
$E+x=\{y+x:y\in E\}$. Moreover, for every Lebesgue measurable set
$E$, the map $x\to \lambda \left( E\bigcap(E+x)\right)$ is
continuous. Then, if $E\subseteq \mathbb{R}$ is Lebesgue
measurable with positive Lebesgue measure, then the set
$E-E=\{x-y:x,y\in E\}$ can be shown to be a neighborhood of 0.

If $X$ is a metrizable topological space and $\mu$ is a finite
measure on $X$ then, for every Borel set $B$, we have
$\mu(B)=\sup\left\{ \mu(F): F\subseteq B,\,F\,{\rm
closed}\right\}$ $=\inf\left\{ \mu(U): U \supseteq B,\,U\,{\rm
open}\right\}$. We call $\mu$ a \underline{re}g\underline{ular}
\underline{measure} on $X$.

A \underline{si}g\underline{ned measure} is an extended,
real-valued, countably additive set function $\mu$ on the class,
$\mathcal{A}$, of all measurable sets of a measurable space
$(X,\mathcal{A})$ with $\mu(\emptyset)=0$, and $\mu$ assuming at
most one of the values $+\infty$ and $-\infty$.

If $\mu$ is a signed measure on a measurable space
$(X,\mathcal{A})$, we call a set $E$ $\mu$- p\underline{ositive}
(\underline{ne}g\underline{ative}) if, $\forall\;F\in\mathcal{A}$,
$E\bigcap F$ is measurable and $\mu(E\bigcap F)\geq 0$
($\mu(E\bigcap F)\leq 0$). The empty set is both $\mu$-positive
and $\mu$-negative in this sense.

If $\mu$ is a signed measure on $(X,\mathcal{A})$, then there
exist two disjoint sets $A,B\in \mathcal{A}$ such that $A\bigcup
B=X$ and $A$ is $\mu$-positive while $B$ is $\mu$-negative. The
sets $A$ and $B$ are said to form the \underline{Hahn
Decom}p\underline{osition} of $X$ relative to $\mu$. Note that the
Hahn decomposition is not unique.

For every $E\in \mathcal{A}$, we define $\mu^+(E)=\mu(E\bigcap
A)$, the \underline{u}pp\underline{er variation} of $\mu$, and
$\mu^-(E)=\mu(E\bigcap B)$, the \underline{lower variation} of
$\mu$, and $|\mu|\,(E)=\mu^+(E)+\mu^-(E)$, the \underline{total
variation} of $\mu$, where $A,B$ are as in the Hahn decomposition.
[Note that $|\mu(E)|$ and $|\mu|(E)$ are not the same.]

The upper, the lower and the total variations (of $\mu$) are
measures and $\mu(E)=\mu^+(E)-\mu^-(E)$
$\forall\;\,E\in\mathcal{A}$, the \underline{Jordon
decom}p\underline{osition}. If $\mu$ is finite or $\sigma$-finite,
then so are $\mu^+$ and $\mu^-$; at least one of $\mu^+$ and
$\mu^-$ is always finite.

A \underline{sim}p\underline{le function} on $(X,\mathcal{A})$ is
$f=\sum_{i=1}^n\alpha_i \chi_{_{E_i}}$ where $E_i\in \mathcal{A}$,
$\chi_{_{E_i}}$ is the {\em characteristic function of the set
$E_i$\/} and $\alpha_i\in\mathbb{R}$. This simple function $f$ is
$\mu$-\underline{inte}g\underline{rable} if $\mu(E_i)< \infty$
$\forall \;i$ for which $\alpha_i\neq 0$. The {\em $\mu$-integral
of $f$\/} is $\int f(x)d\mu(x)\;\mathrm{or}\; \int
f\,d\mu=\sum_{i=1}^n\alpha_i\mu(E_i)$.

If, $\forall\;\epsilon>0$, $\lim_{n\to\infty}$ $m\left( \{x\in X:
|f_n(x)\right.$$-f(x)|\geq \epsilon$$\left.\}\right)= 0$, a
sequence $\{f_n\}$ of a.e.\ finite-valued measurable functions is
said to \underline{conver}g\underline{e in measure} to a
measurable function $f$.

Given two integrable simple functions $f$ and $g$ on a measure
space $(X,\mathcal{A})$, define now a pseudo-metric
$\rho(f,g)=\int |f-g|\;d\mu$. A sequence $\{f_n\}$ of integrable
simple functions is \underline{mean fundamental} if
$\rho(f_n,g_m)\to 0$ if $n,m\to\infty$.

An a.e.\ finite-valued, measurable function $f$ on
$(X,\mathcal{A})$ is {\em $\mu$-integrable\/} if there is a mean
fundamental sequence $\{f_n\}$ of integrable simple functions
which converges in measure to $f$.

\underline{Lebes}g\underline{ue}-\underline{Radon}-\underline{Nikod}y\underline{m}
(LRN) Theorem \cite{measure-theory} states that: If
$(X,\mathcal{A},m)$ is a totally $\sigma$-finite measure space and
if a $\sigma$-finite measure $\nu$ on $\mathcal{A}$ is absolutely
continuous relative to $m$, then there exists a finite valued
measurable function $f$ on $X$ such that $\nu(E)=\int_{_E}f\,d\mu$
for every measurable set $E\in\mathcal{A}$. The function $f$ is
unique: if also $\nu(E)=\int_{_E}g\,d\mu$, then $f=g\;
(\mathrm{mod}\;\mu)$, {\it ie}, equality holding
\underline{modulo} a set of $\mu$-\underline{measure}
\underline{zero} or $\mu$-\underline{a}.\underline{e}.

If $\mu$ is a totally $\sigma$-finite measure and if
$\nu(E)=\int_{_E}f\,d\mu$ $\forall \;\;E\in\mathcal{A}$, we write
$f=\frac{d\nu}{d\mu}$ or $d\nu=f\,d\mu$. We call
$\frac{d\nu}{d\mu}$ the \underline{LRN}-\underline{derivative} and
all properties of ``differential'' hold for it $\mu$-a.e.

A measure on a Standard Borel Space is a \underline{Borel}
\underline{Measure}. A Borel measure $\mu$ on a SBS $X$ is
\underline{continuous} if $\mu\left(\{x\}\right)=0$ for every
$x\in X$.

Now, if $X$ is a Polish space, $\mu$ a finite Borel measure and
$\epsilon>0$ then, there exists a compact subset $K$ of $X$ such
that $\mu\left(X\setminus K\right)<\epsilon$.

To prove this above, we consider a compatible metric $d\leq 1$ on
$X$ and a regular system of sets $\{F_s:s\in
\mathbb{N}^{<\,\mathbb{N}}\}$ of nonempty closed sets such that
$F_e=X$, $F_s=\bigcup_nF_{\widehat{s\,n}}$ and ${\rm
diameter}(F_s)\leq 1/2^{|s|}$. The existence of such a system of
sets is provable by induction on $|s|$. Next, define positive
integers $n_o, n_1, ...$ such that for every $s=\left( m_o, m_1,
..., m_{k-1}\right)$ with $m_i\leq n_i$, $\mu\left(F_s\setminus
\bigcup_{j\leq n_k}F_{\widehat{s\,j}}\right) < \epsilon/(2^{k+1}.
n_o.n_1....n_{k-1})$. The set $K=\bigcap_k\bigcup_sF_s$ where the
union varies over all $s$, is the required, closed and totally
bounded, compact set for which $\mu\left(X\setminus
K\right)<\epsilon$.

Then, for a Polish space $X$, a finite Borel measure $\mu$ on $X$,
for every Borel set $B$ and for every $\epsilon>0$, there is a
compact set $K\subseteq B$ such that $\mu\left(B\setminus
K\right)<\epsilon$.

Let $\mu$ be a probability measure on $I=[0,1]$. Then, the
function $F(x)=\mu\left([0,1]\right),\,x\in I$ is called as a
\underline{distribution function} of $\mu$. It is a monotonically
increasing, right-continuous function such that $F(1)=1$.

Next, if $\mu$ is a continuous probability measure on a standard
Borel space $X$, then there is a Borel isomorphism $h:X \to I$
such that for every Borel subset $B$ of $I$, $\lambda(B)=
\lambda\left(h^{-\,1}(B)\right)$.

Let $(X,\mathcal{A})$ be a measurable space and $Y$ a second
countable metrizable space. A \underline{transition}
p\underline{robabilit}y on $X\times Y$ is a map $P:X\times
\mathcal{B}_Y \to [0,1]$ such that (i) for every $x\in X$,
$P(x,.)$ is a probability on $Y$ and (ii) for every
$B\in\mathcal{B}_Y$, the map $x\to P(x,B)$ is measurable. Then,
for every $A\in \mathcal{A} \bigotimes\mathcal{B}_Y$, the map
$x\to P(x, A_x)$ is measurable. In particular, for every
$A\in\mathcal{A}\bigotimes\mathcal{B}_Y$ such that $P(x,A_x)>0$,
$\pi_{_X}(A)$ is measurable.

Recall that a set of the first Baire category is a countable union
of nowhere dense sets. A subset $E$ of $X$ is said to have the
\underline{Baire} \underline{Pro}p\underline{ert}y
(\underline{BP}) if $E$ can be expressed as a symmetric difference
of an open set $G$ and a set $M$ of the first Baire category, {\em
ie}, expressible as the union $E\equiv G\triangle M=(G\setminus
M)\bigcup(M\setminus G)$. If $E$ has the property of Baire, then
so does its complement in $X$. Clearly, open sets and meagre sets
in $X$ have BP. Note also that every Borel subset of a metrizable
topological space has the Baire property.

The collection $\mathcal{D}$ of all subsets of a topological space
$X$ having the Baire property forms a $\sigma$-algebra to be
called the \underline{Baire}
\underline{$\sigma$}-\underline{al}g\underline{ebra}. Note that
the Baire $\sigma$-algebra of a topological space is closed under
the Souslin operation.

A space $X$ is called a \underline{Baire S}p\underline{ace} if no
nonempty open subset of $X$ is of first category in $X$ or
equivalently in itself. Every open subset of a Baire space is a
Baire space but a closed subset need not be. Every completely
metrizable topological space is a Baire space, the converse not
being true.

If $X$ is a standard Borel space, every Borel subset of $X$ has
the property of Baire since the $\sigma$-algebra of sets with the
property of Baire includes the Borel $\sigma$-algebra of $X$. The
collection of subsets of $X$ with the property of Baire is a
$\sigma$-algebra generated by open subsets together with the
subsets of the first Baire category. Subsets of first Baire
category in $X$ form a $\sigma$-ideal in the $\sigma$-algebra of
sets with the property of Baire.

For $A, B \in {\cal B}_X$, we \underline{write} $A=B\;({\rm
mod}\,\mathcal{N})$ if $A\setminus B$ and $B\setminus A$, both,
belong to $\mathcal{N}$.

A subset $B\subset X$, $B\in{\cal B}$, is said to be
\underline{decom}p\underline{osable} if it is expressible as a
union of two disjoint sets from ${\cal B}_X\setminus\mathcal{N}$.
Clearly, every such decomposable set belongs to ${\cal
B}_X\setminus\mathcal{N}$.

We say that the Borel $\sigma$-algebra ${\cal B}_X$ of subsets of
$X$ satisfies the \underline{countabilit}y \underline{condition}
if every collection of pairwise disjoint sets from ${\cal
B}_X\setminus\mathcal{N}$ is either finite or countably infinite.

Now, a homeomorphism of a topological space $X$ into the
topological space $X'$ is an isomorphism if it is 1-1 and if the
inverse mapping is also a homeomorphism.

The pivotal concept of the measure theory is, however, not an
isomorphism of measure spaces, but the concept of an
\underline{isomor}p\underline{hism} \underline{modulo}
\underline{zero}. Then, if upon removing from the corresponding
spaces appropriate sets of zero measure we obtain an isomorphism,
we say that the spaces are \underline{isomor}p\underline{hic}
\underline{modulo} \underline{zero}.

This above is achieved by the completion of the measure space $X$
with respect to the $\sigma$-ideal, $\mathcal{N}_{\mu}$, of
$\mu$-null sets.

In this mathematical framework, it is often enough to check a
result only for certain characteristic functions to conclude that
it holds for all measurable functions. For this purpose, we use
the concept of a semi-algebra.

A \underline{semi}-\underline{al}g\underline{ebra} on $X$ is a
collection $\mathcal{S}$ of subsets of $X$ which is closed under
finite intersections and such that the complement of any $S\in
\mathcal{S}$ is a finite disjoint union of members of
$\mathcal{S}$. A semi-algebra $\mathcal{S}$ generates a
$\sigma$-algebra $\mathcal{B}$ on $X$ if $\mathcal{B}$ is the
smallest $\sigma$-algebra containing $\mathcal{S}$.

Then, whenever results hold good for the characteristic functions
of the members of the semi-algebra $\mathcal{S}$, those results
hold good also for the members of the $\sigma$-algebra
$\mathcal{B}$.

For $(X,\mathcal{A},\mu)$ a measure space, let $\mathfrak{K}(\mu)$
be the set of all measurable sets with finite $\mu$-measure. For
any $E,F\in\mathfrak{K}(\mu)$, let $\rho(E,F)=\mu(E\triangle F)$.
The function $\rho$ so defined is a metric on $\mathfrak{K}(\mu)$
and the metric space $(\mathfrak{K}(\mu),\rho)$ is called the
\underline{metric s}p\underline{ace of} or \underline{associated
to} $(X,\mathcal{A},m)$.

Note that the metric space of a finite measure algebra
$(\mathcal{B},\mu)$ is complete. A measure algebra
$(\mathcal{B},\mu)$ is called as \underline{se}p\underline{arable}
if the metric space associated to it is separable.

\underline{Carath\'{e}odor}y's \underline{Theorem}: If
$(\mathcal{B},\mu)$ is a normalized, separable and non-atomic
measure algebra, then there is an isomorphism from $(\mathcal{B},
\mu)$ onto the measure algebra of the unit interval $(0,1)$.

Let $(X,\mathcal{A},\mu)$ be a space with complete and normalized
measure $\mu$. Let us denote by $F(\mathcal{F})$ the Borel
structure generated by a family $\mathcal{F}$ of measurable
subsets of $X$.

A countable collection, $\mathcal{F}=\{F_i:i\in I\}$, of
measurable subsets $F_i$ of a measure space $(X,\mathcal{A}, \mu)$
is said to be a \underline{basis} of the space $X$ if
\begin{description} \item{(1)} for any $A\in\mathcal{A}$ there is
a set $B\in F(\mathcal{F})$ such that $B\subset A$, $\mu\left( B
\setminus A\right)=0$, \item{(2)} for any $x_1,x_2\in X$, $x_1\neq
x_2$, there is an $i\in I$ such that either $x_1\in F_i$ \&
$x_2\notin F_i$ or $x_2\in F_i$ \& $x_1\notin F_i$.
\end{description}

Now, suppose $e_i=\pm 1$ and $F^{(e_i)}_i=F_i$ if $e_i=1$ and
$F^{(e_i)}=X\setminus F_i$ if $e_i=-\,1$. Then, to any sequence of
numbers $\{e_i:i\in I\}$ corresponds the intersection
$\bigcap_{i\in I}F^{(e_i)}_i$ with every such intersection
containing no more than one point of $X$.

Then, the space $(X,\mathcal{A},\mu)$ is called
\underline{com}p\underline{lete} with respect to the basis
$\mathcal{F}$ if all the intersections $\bigcap_{i\in
I}F^{(e_i)}_i$ are nonempty.

Moreover, the space $(X,\mathcal{A},\mu)$ will be called
\underline{com}p\underline{lete} ({\em \underline{mod}\/}
\underline{0}) with respect to the basis $\mathcal{F}$ if $X$ can
be included as a subset of full measure into a certain measure
space $(\bar{X}, \bar{\mathcal{A}},\bar{\mu})$ which is complete
with respect to its own basis $\bar{\mathcal{F}}=\{\bar{F}_i: i\in
I\}$ and satisfying $\bar{F}_i\bigcap X=F_i$ for all $i\in I$.

A space which is complete (mod 0) with respect to one of its basis
is also complete (mod 0) with respect to any other basis.

A measure space $(X,\mathcal{A},\mu)$ which is complete (mod 0)
with respect to one of its basis is called as a
\underline{Lebes}g\underline{ue} \underline{S}p\underline{ace}.

The notion of Lebesgue space is very wide. Still, Lebesgue spaces
possess many nice properties. To mention one here, any
automorphism $T$ of a measure space $(X,\mathcal{A},\mu)$ induces
an isomorphism $S$ of the $\sigma$-algebra $\mathcal{A}$ onto
itself as: $S(A)=TA, A\in \mathcal{A}$. For a Lebesgue space, the
converse that any isomorphism of the $\sigma$-algebra induces an
automorphism of the measure space is also \underline{true}.

A Lebesgue space is isomorphic (mod 0) to the ordinary Lebesgue
space of the unit interval. The unit interval is therefore a
representative object of the Lebesgue spaces.

Let $(X,\mathcal{A},\mu)$ and $(Y,\mathcal{B},\nu)$ be Lebesgue
spaces and $\Phi:(\bar{\mathcal{A}},\bar{\mu})\to
(\bar{\mathcal{B}},\bar{\nu})$ a homeomorphism of the associated
measure algebras. Then, there exists a set of measure zero
$A\subset X$ and a measurable function $\phi:X\setminus A \to Y$
such that $\phi^{-\,1}$ coincides with $\Phi$ as a map
$(\bar{\mathcal{A}},\bar{\mu})\to(\bar{\mathcal{B}},\bar{\nu})$.
In this case, we shall say that the map $\Phi$ \underline{arises}
\underline{from} \underline{a} p\underline{oint}
\underline{homeomor}p\underline{hism} (\underline{mod}
\underline{0}).

Therefore, for Lebesgue spaces, the notions of point homeomorphism
(mod 0) and homeomorphism of associated measure algebras of sets
of zero measure essentially coincide.

\underline{Theorem}: If $X$ is a complete separable metric space
and $\mathcal{B}$ is the completion of its Borel $\sigma$-algebra
with respect to a Borel probability measure $\mu$ on $X$, then
$(X,\mathcal{B},\mu)$ is a Lebesgue space.

A partition of a measure space $(X,\mathcal{A},\mu)$ is, by
definition, any family $\Xi=\{C_i:i\in I\}$ of nonempty disjoint
subsets of $X$ such that $\bigcup_iC_i=X$. Moreover, if
$\bigcup_iC_i=X\,(\mathrm{mod}\,0)$, then we call $\Xi$ as a
p\underline{artition} (\underline{mod} \underline{0}).

The sets $A\in \mathcal{A}$ which are the unions of the members of
$\Xi$ are called \underline{measurable} \underline{with}
\underline{res}p\underline{ect} \underline{to} $\Xi$ or simply
$\Xi$-sets.

A partition $\Xi$ is called \underline{measurable} if there is a
countable family, $\mathcal{G}=\{G_i:i\in I\}$,  of subsets of $X$
which are $\Xi$-sets and such that for all $C_1,C_2\in\Xi$ there
is an $i\in I$ such that either $C_1\subset G_i$ \& $C_2\nsubseteq
G_i$ or $C_2\subset G_i$ \& $C_1\nsubseteq G_i$.

The quotient space of a Lebesgue space by a measurable partition,
{\em ie}, $X\diagup\Xi$, is Lebesgue.

There of course exists an equivalence relation between the
measurable partitions of a Lebesgue space and the complete
$\sigma$-algebras on it.

The elements $C\in\Xi$ of a measurable partition can themselves be
transformed into spaces with measure $\mu_C$ and these measures
play the role of conditional probabilities. Thus, a system of
measures $\{\mu_C\}, C\in \Xi$ is said to be a
\underline{canonical} \underline{s}y\underline{stem}
\underline{of} \underline{conditional} \underline{measures}
belonging to the partition $\Xi$ if \begin{description} \item{(i)}
$\mu_C$ is defined on some $\sigma$-algebra $\mathcal{A}_C$ of
subsets of $C$, \item{(ii)} the space $(C, \mathcal{A}_C,\mu_C)$
is Lebesgue,
\item{(iii)} $\forall A\in\mathcal{A}$, the set $A\bigcap C\in
\mathcal{A}_C$ for almost all $C\in X\diagup\Xi$, the function
$\mu_C \left(A\bigcap C\right)$ is measurable on $X\diagup\Xi$ and
$\int_{X\diagup\Xi}\mu_C\left(A\bigcap
C\right)\,d\mu$.\end{description}

Every measurable partition possesses a canonical system of
conditional measures and this system is
\underline{uni}q\underline{ue} (\underline{mod} \underline{0}),
{\em ie}, any other system of conditional measures coincides with
it for almost all $C\in X\diagup \Xi$. Conversely, if some
partition of $X$ possesses a canonical system of conditional
measures then it is a measurable partition.

The forward image of a measurable subset of $X$ under a measurable
function $f$ {\em need not be\/} measurable, in general.

\underline{Lusin}'s \underline{Theorem}: If $f$ is a measurable
function from a Standard Borel Space into another Standard Borel
Space and if $f$ is \underline{countable to zero}, {\em ie}, if
the inverse image of every singleton set is at most countable,
then the forward image under $f$ of a Borel set is Borel.

Now, a one-one measurable map $T$ of a Borel space $(X, {\cal B})$
onto itself such that $T^{-\,1}$ is also measurable is called a
\underline{Borel automor}p\underline{hism}.

That is to say, a Borel automorphism of $(X, {\cal B})$ is a
one-one and onto map $T:X\to X$ such that $T(B)\in {\cal
B}\;\forall\; B \in \mathcal{B}$.

An automorphism of $X$ onto $X$ is, in general, not a Borel
automorphism. But, if $(X, {\cal B})$ is a Standard Borel Space
then a measurable one-one map of $X$ onto $X$ is a Borel
automorphism.

\underline{Ramsa}y-\underline{Macke}y \underline{Theorem}: If
$T:X\to X$ is a Borel automorphism of the standard Borel space
$(X,\mathcal{B}_X)$, then there exists a topology $\mathcal{T}$ on
$X$ such that
\begin{description} \item{(a)} $(X, \mathcal{T})$ is a complete,
separable, metric space \item{(b)} Borel sets of $(X,
\mathcal{T})$ are precisely those in $\mathcal{B}_X$ \item{(c)}
$T$ is a homeomorphism of $(X, \mathcal{T})$.\end{description}
Note that $X$ is {\em same\/} for $(X,\mathcal{B}_X)$ and
$(X,\mathcal{T})$.

Alternatively, if $X$ is the underlying set and if $T$ is a Borel
automorphism on a SBS $(X, \mathcal{B})$ and $\mathcal{C}\subseteq
\mathcal{B}$ is a countable collection, then there exists a
complete separable metric topology, {\em ie}, Polish topology,
$\mathcal{T}$, on $X$ such that \begin{description}
\item{(1)} $T$ generates the $\sigma$-algebra $\mathcal{B}$ \item{(2)} $T$ is
a homeomorphism of $(X, \mathcal{T})$ \item{(3)} $\mathcal{C}
\subseteq \mathcal{T}$, and lastly, \item{(4)} $\mathcal{T}$ has a
clopen base, {\em ie}, sets which are both open and closed are in
$\mathcal{T}$.
\end{description}

If $\mathcal{T}_o$ is a Polish topology on $X$ which generates
$\mathcal{B}$, $\{\mathcal{T}_i,\;i\in \mathbb{N}\}$ are Polish
topologies on $X$ with $\mathcal{T}_o \subseteq \mathcal{T}_i
\subseteq \mathcal{B}$ then $\exists$ a Polish topology
$\mathcal{T}_{\infty}\;\left( \subseteq \mathcal{B}\right)$ such
that $\bigcup_{i=1}^{\infty} \mathcal{T}_i \subseteq
\mathcal{T}_{\infty}$ and $\mathcal{T}_{\infty}$ is the topology
generated by all {\em finite intersections\/} of the form
$\bigcap_{i=1}^n\,G_i,\;G_i\in \mathcal{T}_i$ for $i, n \in
\mathbb{N}$.

Further, given $B\in\mathcal{B}$, there exists a Polish topology
$\bar{\mathcal{T}}$, $\mathcal{T}_o\subseteq\bar{\mathcal{T}
}\subseteq \mathcal{B}$ such that $B\in\bar{\mathcal{T}}$.
Moreover, $\bar{\mathcal{T}}$ can be chosen to have a clopen base.

We also note that, for any countable collection
$(B_j)_{j=1}^{\infty}\subseteq \mathcal{B}$, there exists a Polish
topology $\mathcal{T}$ (which can be chosen to have a clopen base)
such that $\mathcal{T}_o\subseteq \mathcal{T} \subseteq
\mathcal{B}$ and for all $j$, $B_j\in\mathcal{T}$.

Further, if $T$ is a homeomorphism of a Polish space $X$ then
there exists a compact metric space $Y$ and a homeomorphism $\tau$
of $Y$ such that $T$ is isomorphic as a homeomorphism to the
restriction of $\tau$ to a $\tau$-invariant $G_{\delta}$ subset of
$Y$. We can choose the $\tau$-invariant set to be dense in $Y$.
This result is due to N. Krylov and N. Bogoliouboff.

Combined with the theorem of Ramsay and Mackey, this shows that a
Borel automorphism on a Standard Borel Space can be viewed as a
restriction of a homeomorphism of a compact metric space to an
invariant $G_{\delta}$ subset.

\underline{To}p\underline{olo}g\underline{ical Grou}p is a triple
$(G, \diamond, \mathcal{T})$ where $G$ is a set, $\diamond$ is a
group multiplication on $G$ and $\mathcal{T}$ is a topology on $G$
such that \begin{description} \item{(i)} the
\underline{multi}p\underline{lication ma}p $m:G\times G\to G$ is
continuous relative to $\mathcal{T}$, \item{(ii)} the
\underline{inversion function} $\mathbf{i}:G\to G$ is continuous
relative to $\mathcal{T}$, and \item{(iii)} if $e$ is the group
identity, then the singleton set $\{e\}$ is closed in $G$, {\em
ie}, $G\setminus \{e\}\in\mathcal{T}$. \end{description} When
considering only the group properties of a topological group, we
refer to it as an \underline{al}g\underline{ebraic}
g\underline{rou}p and group properties as
\underline{al}g\underline{ebraic} properties. Also, $a\diamond
b\equiv ab$ for $a,b\in G$.

In what follows, we adopt two useful notations:
\begin{description} \item{(a)} if $A,B\subset G$, then $AB=\{ab:a\in A,
\,b\in B\}$ and \item{(b)} for $A\subset G$,
$A^{-\,1}=\{a^{-\,1}:a\in A\}$. \end{description} Then, a subset
$H\subset G, H\neq \emptyset$, is a subgroup of $G$ iff $HH\subset
H$ and $H^{-\,1}\subset H$, both.

A topological space $X$ is called
\underline{homo}g\underline{eneous} if for every $x,y\in X$, there
exists a group homoeomorphism $f:X\to X$ such that $f(x)=y$. Any
topological group is necessarily a homogeneous topological space.

A topological group whose underlying space is a
\underline{manifold} is a g\underline{rou}p \underline{manifold}.
Group $G$ is a group manifold if and only if $e\in G$ has a
neighborhood $U$ which is homeomorphic to $\mathbb{R}^n$.

Neighborhoods of the identity of a topological group are called as
\underline{nuclei}.

For $x\in X$, let $\mathfrak{N}$ be a family of neighborhoods of
$x$ such that every neighborhood of $x$ contains some member of
$\mathfrak{N}$. Then, $\mathfrak{N}$ is called as a
\underline{local} (\underline{nei}g\underline{hborhood})
\underline{base} at $x$.

In particular, a local base $\mathfrak{N}$ at $x$ has the
properties: \begin{description} \item{(i)} $M,N\in
\mathfrak{N}\Rightarrow M\bigcap N\in\mathfrak{N}$, \item{(ii)}
$M\subset N\subset X$ and $M\in\mathfrak{N}$ $\Rightarrow
N\in\mathfrak{N}$, \item{(iii)} $N\in\mathfrak{N}$ implies that
there always exists $M\in\mathfrak{N}$ such that
$M\,M^{-\,1}\subset N$, \item{(iv)} $N\in\mathfrak{N}$ implies
that for all $g\in G$, $g^{-\,1}Ng \in\mathfrak{N}$, and
\item{(v)} $\bigcap\mathfrak{N}=\{e\}$. \end{description}

Given an algebraic group $G$ and family $\mathfrak{N}$ of subsets
of $G$ with above properties, there exists a {\em unique\/}
topology $\mathcal{T}$ for $G$ making $G$ a topological group and
$\mathfrak{N}$ is exactly the family of nuclei.

If $X$ is a homogeneous space then, a local base at a single point
$x\in X$ determines a local base at {\em every other\/} point of
$X$ and, hence, also determines the entire topology $\mathcal{T}$
of $X$.

A neighborhood $S$ of $e\in G$ will be called as a
\underline{s}y\underline{mmetric nei}g\underline{hborhood} if
$S=S^{-\,1}$. If $N$ is any neighborhood of $e$, then $N\bigcap
N^{-\,1}\subset N$ is a symmetric neighborhood of $e$.
Consequently, if $N$ is any neighborhood of $e$ in a topological
group $G$, then there exists a symmetric neighborhood $S$ of $e$
with $SS=SS^{-\,1}\subset N$.

If $H$ is any algebraic subgroup of $G$ then the family $G\diagup
H$ is called a \underline{coset s}p\underline{ace} of $G$ and is a
quotient set of $(G,\mathcal{T})$ with quotient topology. The
quotient map $q:G\to G\diagup H$ is always open. The quotient of a
topological group modulo a closed normal subgroup is a topological
group under the quotient topology and the group product.

An open subgroup of a topological group must be closed as well.
The closure of a subgroup of a topological group is always a
subgroup, and the closure of a normal subgroup is always normal.
The interior of a subgroup need not be a subgroup. The product of
a closed set with a compact set is closed but not necessarily
compact. The quotient map $q:G\to G\diagup H$ is a closed map
whenever $H$ is a compact subset of $G$.

There can be subgroups of a topological group which are neither
open nor closed.

Given two topological groups $G$ and $G'$, and a continuous
function $f:G\to G'$, $f$ is a \underline{mor}p\underline{hism}
\underline{of to}p\underline{olo}g\underline{ical}
g\underline{rou}p\underline{s} if and only if $f$ is both a map of
on the underlying topological spaces and an algebraic morphism
(group homomorphism) on the underlying groups. An onto or
surjective morphism is called an
\underline{e}p\underline{imor}p\underline{hism} and a 1-1 or
injective morphism is called a
\underline{monomor}p\underline{hism}. An
\underline{isomor}p\underline{hism} of topological groups is an
algebraic morphism of underlying groups and a homeomorphism of
underlying topological spaces.

If function $f:G\to G'$ is an algebraic morphism from one
topological group to another and is continuous at the single point
$e$ of its domain, then $f$ is continuous at every other point of
$G$ and, hence, is also a topological morphism. Consequently, the
continuity of a morphism needs to be established by considering
only nuclei.

The group $\mathbb{R}\diagup \mathbb{Z}$, where $\mathbb{Z}$ is
the set of integers, of additive real numbers modulo integers is
topologically isomorphic to circle $S^1$.

If $p:X\to Y$ is a map, a \underline{cross section} to $p$ is a
map $s:Y\to X$ which is a right inverse for $p$, {\em ie}, $p\circ
s=\mathbf{1}_{_Y}$, $\mathbf{1}_{_Y}$ being the identity map of
$Y$. The cross section is said to be {\em at a point} $x\in X$ if
$x$ is a value of $s$, then $x=s\circ p(x)$.

Let $F$ be the set of functions from $X$ to $Y$. The
\underline{com}p\underline{act}-\underline{o}p\underline{en}
(\underline{CO}) \underline{to}p\underline{olo}gy on $F$ is a {\em
unique\/} topology on $F$ generated by the {\em subbase\/} of all
sets of the form $(K,S)=\left\{ f: f\in F\,
\mathrm{and}\,f(K)\subset S\right\}$, where $K$ is compact in $X$
and $S$ is open in $Y$.

The \underline{evaluation function}, $\mathbf{e}:F\times X\to Y$
carries each ordered pair $(f,x)$ to $\mathbf{e}(f,x)=f(x)\in Y$.
A topology for $F$ is called \underline{admissible} if and only if
the evaluation function $\mathbf{e}$ is continuous. The
CO-topology is the finest admissible topology.

A \underline{To}p\underline{olo}g\underline{ical Transformation
Grou}p (\underline{TTG}) or a group of transformations is an
admissible group $G$ of functions on a fixed Hausdorff topological
space $X$ with composition of functions as the group
multiplication. Members of $G$ must all be homeomorphisms of $X$
onto itself. More formally, a TTG is a pair $(G,X)$ where $G$ is a
topological group whose elements are permutations of $X$, $X$
being a Hausdorff space such that \begin{description} \item{(i)}
for all $f,g\in G$ and for every $x\in X$, $(fg)(x)=f[g(x)]$,
\item{(ii)} every $f\in G$ is a homeomorphism of $X$ onto itself,
and \item{(iii)} the evaluation function, $\mathbf{e}$, is
continuous on $G\times X\to X$. \end{description} Every TTG on $X$
must contain the identity map $\mathbf{1}_{_X}$, which is
necessarily also the identity element of the group $G$.

The group $G$ is said to \underline{act on $X$} and the evaluation
is called the \underline{action of $G$ on $X$}. A TTG on $X$ is
called as \underline{transitive} if for all $x,y\in X$ there
exists $g\in G$ such that $g(x)=y$. Note that any group of
matrices is non-transitive on $\mathbb{R}^n$ as the origin is left
fixed by linear transformations.

A topological group is a Baire space if and only if it is of
second category in itself. Every open subset of a Baire space is a
Baire space. Every completely metrizable topological space is a
Baire space, but the converse need not be true.

Let $X$ be a topological space, $A,U\subset X$ with $U$ open.
Then, $A$ is \underline{mea}g\underline{er}
(\underline{co}-\underline{mea}g\underline{er},
\underline{non}-\underline{mea}g\underline{er}) in $U$ if
$A\bigcap U$ is meager (co-meager, non-meager) in $U$.

Let $X$ and $Y$ be metrizable spaces. A function $f:X\to Y$ is
called \underline{Baire measurable} if for every open $U \subset
Y$, $f^{-\,1}(U)$ has BP. [\underline{Caution}: Baire measurable
functions are \underline{not} the same as Baire functions defined
earlier.] Clearly, every Borel function is Baire measurable.

Let $Y$ be second countable and $f:X\to Y$ be Baire measurable.
Then, thee exists a co-meager set $A\subset X$ such that $f|A$ is
continuous.

Let $G$ be a completely metrizable topological group and $H$ a
second countable topological group. Then every Baire measurable
morphism $\varphi:G\to H$ is continuous. In particular, every
Borel morphism $\varphi:G\to H$ is continuous.

\underline{Pettis}'s \underline{Theorem} \footnote{References to
works mentioned by name in this section can be found in
\cite{srivastava}.} proves that if $G$ is a Baire topological
group and $H$ a non-meager subset with BP, then there exists a
neighborhood $U$ of the group identity contained in $H^{-\,1}H$.
Consequently, it follows that every non-meager Borel subgroup $H$
of a Polish group $G$ is open.
\goodbreak

\underline{Notation}: For $E\subset X\times Y$, $x\in X$ and $y\in
Y$, we set $E_x=\{y\in Y:(x,y)\in E\}$ and $E^y=\{x\in X:(x,y)\in
E\}$.

If $X$ is a Baire space and $Y$ second countable and supposing
$A\subseteq X\times Y$ is closed as well as no-where dense, then
$\{x\in X: A_x\,\mathrm{is\,nowhere\,dense}\}$ is a dense
$G_{\delta}$ set.

Let $X$ be a non-empty set, $Y$ a topological space, $A\subset
X\times Y$, and $U$ nonempty, open subset in $Y$. We set
$A^{\triangle U}=\left\{ x\in X:A_x\,\mathrm{is\,nonmeager\,in\,}
U\right\}$ and $A^{*U}=\left\{ x\in X:A_x\,\mathrm{is\,
comeager\,in\,} U\right\}$.

\underline{Kuratowski}-\underline{Ulam} \underline{Theorem}: If
$X$ and $Y$ are second countable Baire spaces and $A\subseteq
X\times Y$ has BP, then the following are equivalent:
\begin{description} \item{(1)} $A$ is meager (co-meager), \item{(ii)}
 $\{x\in X: A_x\,\mathrm{is\,meager\,}$ $\mathrm{(co-meager)}\}$ is co-meager,
\item{(iii)} $\{y\in Y: A^y\,\mathrm{is}$
$\mathrm{meager\,(co-meager)}\}$ is co-meager. \end{description}

Let $(X,\mathcal{A})$ be a measurable space and $Y$ a Polish
space. For every $A\in\mathcal{A}\bigotimes\mathcal{B}_{_Y}$ and
$U$ open in $Y$, the sets $A^{\triangle U}$, $A^{*U}$ and $\{x\in
X:A_x\,\mathrm{is\,meager\,in\,} U\}$ are in $\mathcal{A}$.

Let $G$ be a Polish group that is acting continuously on a Polish
space $X$. For any $W\subseteq X$ and any nonempty open
$U\subseteq G$, define the \underline{Vau}g\underline{ht}
\underline{transforms} as: $W^{\triangle U}=\{ x\in X:\{g\in
U:g\cdot x\in W\}\, \mathrm{is\,nonmeager}\}$ and $W^{* U}=\{ x\in
X:\{g\in U:g\cdot x\in W\}\, \mathrm{is\,comeager}\}$.

Then, we have \begin{description} \item{(i)} $W^{\triangle U}$ is
invariant, \item{(ii)} $W$ is invariant implies that
$W=W^{\triangle U}$, \item{(iii)} $\left(\bigcup_n
W_n\right)^{\triangle U}$ $=\bigcup_n\left(W_n^{\triangle U}
\right)$, \item{(iv)} if $W\subseteq X$ is Borel and $U\subseteq
G$ is open, then $W^{\triangle U}$ and $W^{*U}$ are Borel.
\end{description}

We shall call a $\sigma$-algebra $\mathcal{B}$ on $X$
\underline{Marczewski} \underline{com}p\underline{lete} if for
every $A\subseteq X$ there exists $\hat{A}\in \mathcal{B}$
containing $A$ such that for every $B\in\mathcal{B}$ containing
$A$, every subset of $\hat{A}\setminus B$ is in $\mathcal{B}$.
Such a set $\hat{A}$ will be called a \underline{minimal}
$\mathcal{B}$-\underline{cover of} $A$.

Every $\sigma$-finite complete measure space is Marczewski
complete. Notably, Baire $\sigma$-algebra of any topological space
is Marczewski complete.

\underline{Marczewski}'s \underline{Theorem}: if $(X,\mathcal{B})$
is a measurable space with $\mathcal{B}$ Marczewski complete, then
$\mathcal{B}$ is closed under the Souslin operation.

We call a collection of point-sets (subsets of metrizable spaces)
as a p\underline{oint}-\underline{class}. For example, we shall
speak of point-classes of open sets, closed sets, Borel sets etc.

Let $X$ be a metrizable space. For ordinals $\alpha$, $1\leq
\alpha<\omega_1$, {\em ie}, for countable ordinals, define the
following point-classes by transfinite induction: \beq
\mathbf{\Sigma}^0_1(X)&=& \left\{
U\subseteq X:U\, {\rm open}\right\} \n \\
\mathbf{\Pi}^0_1(X)&=&\left\{ F\subseteq X: F \,{\rm
closed}\right\} \n \eeq for $1<\alpha<\omega_1$,
\[ \mathbf{\Sigma}^0_{\alpha}(X)= \left(\bigcup_{\beta<\alpha}
\mathbf{\Pi}^0_{\beta}(X)\right)_{\sigma} \] and
\[\mathbf{\Pi}^0_{\alpha}(X)= \left(\bigcup_{\beta<\alpha}
\mathbf{\Sigma}^0_{\beta}(X)\right)_{\delta} \] Finally, for every
$1\leq\alpha<\omega_1$, \[ \mathbf{\triangle}^0_{\alpha}(X)=
\mathbf{\Sigma}^0_{\alpha}(X)\bigcap\mathbf{\Pi}^0_{\alpha}(X)\]

Note that $\mathbf{\triangle}^0_1(X)$ is the family of all clopen
subsets, $\mathbf{\Sigma}^0_2(X)$ is the family of all
$F_{\sigma}$ subsets, and $\mathbf{\Sigma}^0_2(X)$ is the family
of all $G_{\delta}$ subsets of $X$. The families
$\mathbf{\Sigma}^0_{\alpha}(X)$, $\mathbf{\Pi}^0_{\alpha}(X)$, and
$\mathbf{\triangle}^0_{\alpha}(X)$ are called
\underline{additive}, \underline{multi}p\underline{licative}, and
\underline{ambi}g\underline{uous} classes respectively. If a
statement is true for all $X$, we shall omit the $X$ in the
brackets while stating the family of point-classes under
consideration.

A set in $\mathbf{\Sigma}^0_{\alpha}$ is called as an
\underline{Additive} \underline{Class} $\alpha$ \underline{Set},
that in $\mathbf{\Pi}^0_{\alpha}$ as a
\underline{Multi}p\underline{licative} \underline{Class} $\alpha$
\underline{Set} and that in $\mathbf{\triangle}^0_{\alpha}$ as an
\underline{Ambi}g\underline{uous} \underline{Class} $\alpha$
\underline{Set}.

Following elementary facts about these point-classes are easy to
establish: \begin{description} \item{(i)} Additive classes are
closed under countable unions, and multiplicative classes are
closed under countable intersections, \item{(ii)} All the classes
are closed under finite unions and finite intersections,
\item{(iii)} For all $1\leq \alpha<\omega_1$,
$\mathbf{\Sigma}^0_{\alpha}=\neg\mathbf{\Pi}^0_{\alpha}$ or
equivalently, $\mathbf{\Pi}^0_{\alpha}=\neg
\mathbf{\Sigma}^0_{\alpha}$, \item{(iv)} For $\alpha\geq 1$,
$\mathbf{\triangle}^0_{\alpha}$ is an algebra. \end{description}

The following results are also easy to establish:
\begin{description} \item{(i)} For every $1\leq \alpha<\omega_1$,
$\mathbf{\Sigma}^0_{\alpha}, \mathbf{\Pi}^0_{\alpha}
\,\subseteq\,\mathbf{\triangle}^0_{\alpha+1}$. Thus, the following
\underline{Hierarch}y \underline{of} \underline{Borel}
\underline{sets} in which any point-class is contained in every
point-class to its right is obtained: \bigskip

\begin{tabular}{ccccc} \phantom{m} & $\mathbf{\Sigma}^0_1$
&$\mathbf{\Sigma}^0_2$
&\hspace{.1in}$\mathbf{\Sigma}^0_3$ & ... \\ \phantom{m} \\
$\mathbf{\triangle}^0_1$ & \hspace{.4in} $\mathbf{\triangle}^0_2$
& \hspace{.6in}$\mathbf{\triangle}^0_3$ & ... \\  \phantom{m}
\\ \phantom{m} &$\mathbf{\Pi}^0_1$ &
$\mathbf{\Pi}^0_2$ & \hspace{.1in}$\mathbf{\Pi}^0_3$ & ...
\end{tabular}

\item{(ii)} For $\alpha>1$, $\mathbf{\Sigma}^0_{\alpha}=\left(
\mathbf{\triangle}^0_{\alpha}\right)_{\sigma}$ and
$\mathbf{\Pi}^0_{\alpha}=\left(\mathbf{\triangle}^0_{\alpha}
\right)_{\delta}$, it also being true for $\alpha=1$ when $X$ is a
zero-dimensional separable metric space, \item{(iii)} For metric
space $X$, we have the result that
$\mathcal{B}_X=\bigcup_{\alpha<\omega_1}
\mathbf{\Sigma}^0_{\alpha}(X)= \bigcup_{\alpha<\omega_1}
\mathbf{\Pi}^0_{\alpha}(X)$. \end{description} For any uncountable
Polish space, the inclusion in (i) is strict.

Let $X$ be an infinite separable metric space. Then,
$|\mathbf{\Sigma}^0_{\alpha}(X)|=|\mathbf{\Pi}^0_{\alpha}(X)|=\mathbf{c}$
and $|\mathcal{B}_X|=\mathbf{c}$.

Now, note that every set of additive class $\alpha>2$ is a
countable disjoint union of multiplicative class $<\alpha$ sets.

Let $X$ and $Y$ be metrizable spaces, $f:X\to Y$ a transformation,
and $1\leq\alpha<\omega_1$. We say that $f$ is \underline{Borel
measurable of class} $\alpha$, or simply \underline{of class}
$\alpha$, if $f^{-\,1}(U)\in \mathbf{\Sigma}^0_{\alpha}$ for every
open set $U$. The class 1 functions are the continuous functions.

A characteristic function $\chi_A, A\subseteq X$, is of class
$\alpha$ if and only if $A$ is of ambiguous class $\alpha$. Every
class $\alpha$ function is clearly Borel measurable.

Let $1\leq\alpha<\omega_1$. Let $\mathbf{\Gamma}_{\alpha}$ denote
one of the two point-classes of $\mathbf{\Pi}^0_{\alpha}$ or of
$\mathbf{\Sigma}^0_{\alpha}$ sets. For every second countable
metrizable space $Y$, there then exists a $U\in
\mathbf{\Gamma}_{\alpha}\left( \mathbb{N}^{\mathbb{N}} \times Y
\right)$ such that $A\in \mathbf{\Gamma}_{\alpha}(Y) \Rightarrow
\left(\exists\,x\in\mathbb{N}^{\mathbb{N}} \right)(A=U_x)$. We
shall call such a set $U$ a \underline{universal} for
$\mathbf{\Gamma}_{\alpha}$.

Let $1\leq \alpha<\omega_1$ and $\mathbf{\Gamma}_{\alpha}$ the
point-class of additive or multiplicative class $\alpha$ sets.
Then, for every uncountable Polish space $X$, there is a $U\in
\mathbf{\Gamma}_{\alpha}(X\times X)$ universal for
$\mathbf{\Gamma}_{\alpha}(X)$.

We also have that for $X$ being any uncountable Polish space and
$1\leq\alpha<\omega_1$, there exists an additive class $\alpha$
set that is not of multiplicative class $\alpha$. Hence, for every
uncountable Polish space $X$ and for any $\alpha$,
$\mathbf{\Sigma}^0_{\alpha}(X)\,\neq\,\mathbf{\Sigma}^0_{\alpha+1}(X)$.

Then, there does not exist a Borel set $U\subseteq X\times X$
universal for Borel subsets of $X$ for any Polish space $X$. A
fairly general conclusion is the following: Let a point-class
$\mathbf{\vartriangle}$ be closed under taking complements and
continuous pre-images. Then for no Polish space $X$ is there a set
in $\mathbf{\vartriangle}(X\times X)$ universal for
$\mathbf{\vartriangle}(X)$.

Now, let $X$ be a metrizable space and $1\leq\alpha<\omega_1$.
Suppose $(A_n)$ is a sequence of additive class $\alpha$ sets in
$X$. Then there exist $B_n\subseteq A_n$ such that (a) The $B_n$'s
are pairwise disjoint sets of additive class $\alpha$, (b)
$\bigcup_nA_n=\bigcup_nB_n$. Consequently, the $B_n$'s are of
ambiguous class $\alpha$ if $\bigcup_nA_n$ is so. The result also
holds for $\alpha=1$ if $X$ is zero-dimensional and second
countable. This above is known as the \underline{Reduction}
\underline{Theorem} \underline{for} \underline{Additive}
\underline{Classes}.

Let $X$ be metrizable and $1\leq\alpha<\omega_1$. Then for every
sequence $(A_n)$ of multiplicative class $\alpha$ sets with
$\bigcap_nA_n=\emptyset$, there exist ambiguous class $\alpha$
sets $B_n\supseteq A_n$ with $\bigcap_nB_n=\emptyset$. In
particular, if $A$ and $B$ are two disjoint subsets of $X$ of
multiplicative class $\alpha$, then there is an ambiguous class
$\alpha$ set $C$ such that $A\subseteq C$ and $B\bigcap
C=\emptyset$. This is also true for $\alpha=1$ if $X$ is
zero-dimensional and second countable. This above is known as the
\underline{Separation} \underline{Theorem} \underline{for}
\underline{Multi}p\underline{licative} \underline{Classes}.

Notably, the separation theorem does not hold for additive classes
and the reduction theorem does not hold for multiplicative
classes.

A sequence $(A_n)$ of sets is called
\underline{conver}g\underline{ent} if
$\liminf_nA_n=\limsup_nA_n=B$, say. In this case, we say that the
sequence $(A_n)$ converges to $B$ and write $\lim A_n=B$. Clearly,
when the sequence $(A_n)$ is convergent, we have that for every
$x\in X$, $x\in A_n$ for infinitely many $n$ if and only if $x\in
A_n$ for all but finitely many $n$.

Now, let $X$ be metrizable and $2<\alpha<\omega_1$. Suppose $A\in
\mathbf{\triangle}^0_{\alpha}(X)$. Then there is a sequence
$(A_n)$ of ambiguous class $<\alpha$ sets such that $A=\lim A_n$.
The result is also true for $\alpha=2$, provided that $X$ is
separable and zero-dimensional.

Let $2<\alpha<\omega_1$ and $X$ an uncountable Polish space. There
exists a sequence $A_n$ in $\mathbf{\Pi}^0_{\alpha}(X)$ with
$\limsup A_n=\emptyset$ such that there does not exist
$B_n\supseteq A_n$ in $\mathbf{\Sigma}^0_{\alpha}(X)$ with
$\limsup B_n=\emptyset$. This observation is due to A. Maitra, C A
Rogers and J E Jayne.

\underline{Theorem}: Suppose that $X$ and $Y$ are metrizable
topological spaces with $Y$ being second countable and
$2<\alpha<\omega_1$. Then for every Borel function $f:X\to Y$ of
class $\alpha$, there is a sequence $(f_n)$ of Borel maps from $X$
to $Y$ of class $<\alpha$ such that $f_n\to f$ point-wise.

To prove the above theorem, we use the following two lemmas:

\underline{Lemma 1} Suppose $Y$ is totally bounded. Then every
$f:X\to Y$ of class $\alpha$, $\alpha>1$, is the limit of a
uniformly convergent sequence of class $\alpha$ functions
$f_n:X\to Y$ of finite range.

\underline{Lemma 2}: Let $f:X\to Y$ be of class $\alpha>2$ with
range contained in a finite set $E=\{y_1, y_2, ..., y_n\}$. Then
$f$ is the limit of a sequence of functions of class $<\alpha$
with values in $E$.

Let $B\subseteq X\times Y$. For notational convenience, we shall
denote the p\underline{ro}j\underline{ection} $\pi_{_X}(B)$ of $B$
to $X$ by $\exists^YB$, {\em ie}, $\exists^YB=\{x\in X:(x,y)\in
B\,\mathrm{for\,some\,} y\in Y\}$. The
\underline{co}-p\underline{ro}j\underline{ection} of $B$ is
defined as: $\forall^YB=\{x\in X:(x,y)\in
B\,\mathrm{for\,all\,}y\in Y\}$. Then, clearly,
$\forall^YB=\left(\exists^YB^c\right)^c$.

For any point-class $\mathbf{\Gamma}$ and any Polish space $Y$, we
set $\exists^Y\mathbf{\Gamma}=\left\{ \exists^YB: B\in
\mathbf{\Gamma}(X\times Y),\,X\,\mathrm{is\,a}\right.$
$\left.\mathrm{Polish \,space}\right\}$, {\em ie}, $\exists^Y
\mathbf{\Gamma}$ is the family of sets of the form $\exists^YB$
where $B\in\mathbf{\Gamma}(X\times Y)$, $X$ being Polish. The
point-class $\forall^Y\mathbf{\Gamma}$ is similarly defined.

Let $X$ be a Polish space. We shall call a Borel subset of $X$ as
a \underline{Standard} \underline{Borel} \underline{Set}. Any
$A\subset X$ is called \underline{anal}y\underline{tic} if it is a
projection of a Borel subset $B$ of $X\times X$. The point-class
of analytic sets will be denoted by $\mathbf{\Sigma}^1_1$. A
subset $C$ of $X$ is called
\underline{co}-\underline{anal}y\underline{tic} if $X\setminus C$
is analytic. Then, a subset $A$ of $X$ is co-analytic if and only
if it is the co-projection of a Borel subset of $X\times X$. The
point-class of co-analytic sets will be denoted by
$\mathbf{\Pi}^1_1$. Clearly, we have $\mathbf{\Pi}^1_1=\neg
\mathbf{\Sigma}^1_1$. Finally, we define $\mathbf{\triangle}^1_1=
\mathbf{\Pi}^1_1\bigcap\mathbf{\Sigma}^1_1$.

All standard Borel sets are both analytic and co-analytic and,
hence, in point-class $\mathbf{\triangle}^1_1$. The converse that
every $\mathbf{\triangle}^1_1$ set is Borel was proved by Souslin.
This marked the recognition of descriptive set theory as an
independent subject.

The Theory of Analytic and Co-Analytic Sets is of fundamental
importance to the Theory of Borel Sets and Borel Functions. It
imparts the theory of Borel sets its deductive power.

\underline{Pro}p\underline{osition}: Let $X$ be a Polish space and
$A\subseteq X$. Then, the following are equivalent statements:
\begin{description} \item{(i)} $A$ is analytic, \item{(ii)} There is a Polish space $Y$ and a
Borel set $B \subseteq X\times Y$ whose projection is $A$,
\item{(iii)} There is a continuous map $f:\mathbb{N}^{\mathbb{N}}\to
X$ whose range is $A$, \item{(iv)} There is a closed subset $C$ of
$X\times \mathbb{N}^{\mathbb{N}}$ whose projection is $A$,
\item{(v)} For every uncountable Polish space $Y$ there is a
$G_{\delta}$ set $B$ in $X\times Y$ whose projection is $A$.
\end{description}

\underline{Pro}p\underline{osition}: (1) The point-class
$\mathbf{\Sigma}^1_1$ is closed under countable unions, countable
intersections and Borel pre-images. Consequently,
$\mathbf{\Pi}^1_1$ is also closed under these operations. (2) The
point-class $\mathbf{\Sigma}^1_1$ is closed under projection
$\exists^Y$, and $\mathbf{\Pi}^1_1$ is closed under co-projection
$\forall^Y$ for all Polish $Y$.

Let $B\subseteq X$ be analytic, in particular, Borel and $f:B\to
Y$ a Borel map. Then $f(B)$ is analytic.

\underline{Theorem}: For every Polish space $X$, there is an
analytic set $U\subseteq \mathbb{N}^{\mathbb{N}}\times X$ such
that $A\subseteq X$ is analytic if and only if $A=U_{\alpha}$ for
some $\alpha$, {\em ie}, $U$ is universal for
$\mathbf{\Sigma}^1_1(X)$.

\underline{Theorem}: Let $X$ be an uncountable Polish space. Then,
\begin{description} \item{(i)} There is an analytic set $U\subseteq X\times X$ such that
for every analytic set $A\subseteq X$, there is an $x\in X$ with
$A=U_x$, \item{(ii)} There is a subset of $X$ that is analytic but
not Borel.\end{description}

Thus, every uncountable standard Borel space contains an analytic
set that is not Borel.

Now, define for each $n\geq 1$, point-classes
$\mathbf{\Sigma}^1_n$, $\mathbf{\Pi}^1_n$ and
$\mathbf{\triangle}^1_n$ by induction on $n$ as follows. Let $n$
be any positive integer. Let $X$ be a Polish space. Take
\[\mathbf{\Sigma}^1_{n+1}(X)=\exists^X\mathbf{\Pi}^1_n( X\times
X)\] \[\mathbf{\Pi}^1_{n+1}(X)=\neg\mathbf{\Sigma}^1_{n+1}(X)\]
\[\mathbf{\triangle}^1_{n+1}(X)=\mathbf{\Sigma}^1_{n+1}(X)
\bigcap \mathbf{\Pi}^1_{n+1}(X)\] Sets thus defined are called the
p\underline{ro}j\underline{ective sets}.

\underline{Pro}p\underline{osition}: Let $n$ be a positive
integer. \begin{description} \item{(i)} The point-classes
$\mathbf{\Sigma}^1_n$ and $\mathbf{\Pi}^1_n$ are closed under
countable unions, countable intersections and Borel pre-images.
\item{(ii)} $\mathbf{\triangle}^1_n$ is a $\sigma$-algebra \item{(iii)} The
point-class $\mathbf{\Sigma}^1_n$ is closed under projections
$\exists^Y$ and the point-class $\mathbf{\Pi}^1_n$ is closed under
co-projections $\forall^Y$, when $Y$ is Polish. \end{description}

Let $B\subseteq X$ be $\mathbf{\Sigma}^1_n$ and $f:B\to Y$ be a
Borel map. Then, $f(B)\in\mathbf{\Sigma}^1_n$.

\underline{Pro}p\underline{osition}: For every $n\geq 1$,
$\mathbf{\Sigma}^1_n \bigcup \mathbf{\Pi}^1_n \subseteq
\mathbf{\triangle}^1_{n+1}$. Thus, we have the following
\underline{Hierarch}y \underline{of}
\underline{Pro}j\underline{ective} \underline{Sets} in which any
point-class is contained in every point-class to its
right:\bigskip

\begin{tabular}{ccccc} \phantom{m} & $\mathbf{\Sigma}^1_1$
&$\mathbf{\Sigma}^1_2$
&\hspace{.1in}$\mathbf{\Sigma}^1_3$ & ... \\ \phantom{m} \\
$\mathbf{\triangle}^1_1$ & \hspace{.4in} $\mathbf{\triangle}^1_2$
& \hspace{.6in}$\mathbf{\triangle}^1_3$ & ... \\  \phantom{m}
\\ \phantom{m} &$\mathbf{\Pi}^1_1$ &
$\mathbf{\Pi}^1_2$ & \hspace{.1in}$\mathbf{\Pi}^1_3$ & ...
\end{tabular}
\bigskip

Now, let $n\geq1$, $\mathbf{\Gamma}$ be either
$\mathbf{\Sigma}^1_n$ or $\mathbf{\Pi}^1_n$, and $X$ a Polish
space. There is a $U\subseteq \mathbb{N}^{\mathbb{N}}\times X$ in
$\mathbf{\Gamma}$ such that $A\subseteq X$ is in $\mathbf{\Gamma}$
if and only if $A=U_{\alpha}$ for some $\alpha$, {\em ie}, $U$ is
universal for $\mathbf{\Gamma}(X)$.

\underline{Theorem}: Let $X$ be an uncountable Polish space and
$n\geq1$. \begin{description} \item{(i)} There is a set $U\in
\mathbf{\Sigma}^1_n(X\times X)$ such that for every $A\in
\mathbf{\Sigma}^1_n(X)$, there is a $x$ with $A=U_x$, \item{(ii)}
There is a subset of $X$ that is in $\mathbf{\Sigma}^1_n(X)$ but
not in $\mathbf{\Pi}^1_n(X)$. \end{description}

For any Polish space $X$ and for any $n\geq1$, there is no set
$U\in \mathbf{\triangle}^1_n(X\times X)$ that is universal for
$\mathbf{\triangle}^1_n(X)$.

\underline{Theorem}: Let $X$ be a Polish space, $d$ a compatible
complete metric on $X$, and $A\subseteq X$. The following are
equivalent statements: \begin{description} \item{(i)} $A$ is
analytic, \item{(ii)} There is a regular scheme $\{F_s:s\in
\mathbb{N}^{<\,\mathbb{N}}\}$ of closed subsets of $X$ such that
for every $\alpha\in \mathbb{N}^{\mathbb{N}}$,
$\mathrm{diameter}\left(F_{\alpha|n} \right) \to 0$ and
$A=\mathcal{A}\left(\{F_s\}\right)$, \item{(iii)} There is a
system $\{F_s:s\in\mathbb{N}^{<\,\mathbb{N}}\}$ of closed subsets
of $X$ such that $A=\mathcal{A}\left( \{F_s\}\right)$.
\end{description}

Note that the point-class $\mathbf{\Sigma}^1_1$ is closed under
the Souslin operation. But, as there are analytic sets that are
not co-analytic, the point-class $\mathbf{\Pi}^1_1$ is not closed
under the Souslin operation. For an uncountable Polish space $X$
and $n\geq2$, all the point-classes $\mathbf{\Sigma}^1_n$,
$\mathbf{\Pi}^1_n$ and $\mathbf{\triangle}^1_n$ are closed under
the Souslin operation.

For every Polish space $X$, there is a pair of analytic sets $U_0,
U_1\subseteq \mathbb{N}^{\mathbb{N}}\times X$ such that for any
pair $A_o,A_1$ of analytic subsets of $X$ there is an $\alpha$
satisfying $A_i=(U_i)_{\alpha},\,i=0,1$.

For an uncountable Polish space $X$, there is a sequence $(U_n)$
of analytic subsets of $X\times X$ such that for every sequence
$(A_n)$ of analytic subsets of $X$ there is $x\in X$ with
$A_n=(U_n)_x$ for all $n$. Also, there is a set $U\in \mathcal{A}
\left( \mathbf{\Pi}^1_1(\mathbb{N}^{\mathbb{N}}\times X)\right)$
universal for $\mathcal{A}(\mathbf{\Pi}^1_1(X))$.

Note that for any uncountable Polish space $X$, it can be shown
that $\sigma\left( \mathbf{\Sigma}^1_1(X) \right)$ is
\underline{not} closed under the Souslin operation.

Now, a subset of $\mathbb{N}^{\mathbb{N}} \times
\mathbb{N}^{\mathbb{N}}$ is closed if and only if it is the body
of a tree $T$ on $\mathbb{N}\times\mathbb{N}$.  We therefore have
the following proposition:

\underline{Pro}p\underline{osition}: Let
$A\subseteq\mathbb{N}^{\mathbb{N}}$. Then, the following are
equivalent statements: \begin{description} \item{(i)} $A$ is
analytic, \item{(ii)} There is a tree $T$ on
$\mathbb{N}\times\mathbb{N}$ such that $\alpha\in
A\Longleftrightarrow T[\alpha]$ is well-founded,  as well as
$\alpha\in A\Longleftrightarrow T[\alpha]$ is well-ordered with
respect to $\leq_{_{KB}}$. \end{description}

Let $g:\mathbb{R}\times\mathbb{R}\to\mathbb{R}$ be a Borel
function. Define $f(x)=\sup_y g(x,y), x\in X$. If $f(x)<\infty$
for all $x$, the function $f$ need not be Borel.

We can characterize functions $f:\mathbb{R}\to\mathbb{R}$ of the
form $f(x)=\sup_yg(x,y)$, $g$ Borel. We call a function
$f:\mathbb{R}\to\mathbb{R}$ an $A$-\underline{function} if
$\{x:f(x)>t\}$ is analytic for every real number $t$.

Let $f(x)=\sup_yg(x,y), g:\mathbb{R}\times\mathbb{R}\to
\mathbb{R}$ be Borel. Assume $f(x)<\infty$. Then, for every real
number $t$, $f(x)>t\Longleftrightarrow (\exists y\in \mathbb{R}
(g(x,y)>t)$. So, $f$ is an A-function. Further, the function $f$
dominates a Borel function.

Moreover, the converse is also true. For every A-function
$f:\mathbb{R}\to\mathbb{R}$ dominating a Borel function there is a
Borel $g:\mathbb{R}\times\mathbb{R}\to\mathbb{R}$ such that $f(x)=
\sup_yg(x,y)$. Note that not all A-functions dominate a Borel
function.

Let $X$ be a Polish space and $A\subseteq X$. We say that $A$ is
$\mathbf{\Sigma}^1_1$-\underline{com}p\underline{lete} if $A$ is
analytic and for every Polish space $Y$ and for every analytic
$B\subseteq Y$, there is a Borel map $f:Y\to X$ such that
$f^{-\,1}(A)=B$. Notice that no $\mathbf{\Sigma}^1_1$-complete set
is Borel. This provides us a technique to establish the non-Borel
nature of analytic sets.

Let $X$ and $Y$ be Polish spaces and $A\subseteq X$, $B\subseteq
Y$. We say that $A$ is \underline{Borel} \underline{reducible} to
$B$ if there is a Borel function $f:X\to Y$ such that $f^{-\,1}
(B)=A$. Note that if an analytic set $A$ is Borel reducible to $B$
and $A$ is a $\mathbf{\Sigma}^1_1$-complete set then, $B$ is also
$\mathbf{\Sigma}^1_1$-complete.

Now, note that we can also define $\mathbf{\Pi}^1_1$-complete sets
analogously and the above results also hold for
$\mathbf{\Pi}^1_1$-complete sets.

\underline{Pro}p\underline{osition}: Let $X$ be an uncountable
Polish space and let $\mathbb{K}(X)$ be the family of all
non-empty compact subsets of $X$. Then
$U(X)=\{K\in\mathbb{K}(X):K\, \mathrm{is\,uncountable}\}$ is
$\mathbf{\Sigma}^1_1$-complete. Furthermore, the set
$\{K\in\mathbb{K}(X):K\, \mathrm{is\,countable}\}$ is
$\mathbf{\Pi}^1_1$-complete.

\underline{Pro}p\underline{osition}: (Marczewski) The set DIFF of
everywhere differentiable functions $f:[0,1]\to\mathbb{R}$ is
$\mathbf{\Pi}^1_1$-complete. In particular, it is a co-analytic,
non-Borel subset of the space of real-valued continuous functions
on $[0,1]$.

Let $\mu$ be a $\sigma$-finite measure on $(X,\mathcal{B}_X)$, $X$
Polish. Then every analytic subset of $X$ is $\mu$-measurable.
Further, every analytic subset of a Polish space has the Baire
Property.

We also note that if $X$ is an uncountable Polish space and
$\mathcal{B}$ be either the Baire $\sigma$-algebra or the
completion $\bar{\mathcal{B}}_X^{\mu}$ with $\mu$ being a
continuous probability on $X$, then no $\sigma$-algebra
$\mathcal{A}$ satisfying $\sigma(\mathbf{\Sigma}^1_1)\subseteq
\mathcal{A} \subseteq \mathcal{B}$ is countably generated.

Note that every uncountable analytic set contains a homeomorph of
the Cantor ternary set and, hence, is of cardinality $\mathbf{c}$.

Let $X$ and $Y$ be Polish spaces and $f:X\to Y$ a continuous map
with uncountable range. Then there is a homeomorph of the Cantor
set $C\subseteq X$ such that $f|C$ is 1-1.

\underline{Pro}p\underline{osition}: Let $X$ be Polish and
$A\subseteq X$. The following are equivalent:
\begin{description} \item{(i)} $A$ is analytic, \item{(ii)} There is a closed set
$C\subseteq X\times\mathbb{N}^{\mathbb{N}}$ such that $A=\{x\in
X:C_x\,\mathrm{is\,uncountable}\}$, \item{(iii)} There is a Polish
space $Y$ and an analytic set $B\subseteq X\times Y$ such that
$A=\{x\in X:B_x\,\mathrm{is\,uncountable}\}$. \end{description}

\underline{Sim}p\underline{son}'s \underline{Theorem}: If $X$ an
analytic subset of a Polish space, $Y$ a metrizable space, and
$f:X\to Y$ a Borel map, then $f(X)$ is separable.

Every Borel homomorphism $\varphi:G\to H$ from a completely
metrizable group $G$ to a metrizable group $H$ is continuous.

A set $A$ of real numbers has \underline{stron}g
\underline{measure} \underline{zero} if for every sequence $(a_n)$
of positive real numbers, there exists a sequence $(I_n)$ of open
intervals such that $|I_n|\leq a_n$ and $A\subseteq \bigcup_nI_n$.

Then, \begin{description} \item{(i)} Every countable set of real
numbers has strong measure zero, \item{(ii)} Every strong measure
zero set is of (Lebesgue) measure zero, \item{(iii)} Family of all
strong measure zero sets is a $\sigma$-ideal.\end{description}

Further, if $A\subseteq[0,1]$ is a strong measure zero set and
$f:[0,1]\to\mathbb{R}$ is a continuous map, then the set $f(A)$
has strong measure zero. Note that not all (Lebesgue) measure zero
sets of real numbers have strong measure zero. The Cantor ternary
set is not a strong measure zero set.

No set of real numbers containing a perfect set has strong measure
zero. The \underline{Borel} \underline{Con}j\underline{ecture} is
that no uncountable set of real numbers is a strong measure zero
set. Also, no uncountable analytic $A\subseteq \mathbb{R}$ has
strong measure zero.

There is a set $A$ of real numbers of cardinality $\mathbf{c}$
such that $A\bigcap C$ is countable for every closed, nowhere
dense set $C$. Such a set is a \underline{Lusin} \underline{Set}.
Every Lusin set has strong measure zero.

A co-analytic set is either countable or of cardinality
$\aleph_1$, {\em ie}, of cardinality $\mathbf{c}$.

The following separation theorems and the dual results - the
reduction theorems - are among some of the most important results
on analytic and co-analytic sets.

\underline{First Se}p\underline{aration Theorem}
(\underline{Anal}y\underline{tic Sets}): Let $A$ and $B$ be
disjoint analytic subsets of a Polish space $X$. Then there is a
Borel set $C$ such that $A\subseteq C$ and $B\bigcap C=\emptyset$.
In this case, we say that $C$ separates $A$ from $B$.

\underline{Theorem}: (Souslin) A subset $A$ of a Polish space $X$
is Borel if and only if it is both analytic and co-analytic, {\em
ie}, $\mathbf{\triangle}^1_1(X)=\mathcal{B}_X$.

Suppose $A_o,A_1, ...$ are pairwise disjoint analytic subsets of a
Polish space $X$. Then there exist pairwise disjoint Borel sets
$B_o,B_1,...$ such that $B_n\supseteq A_n$ for all $n$.

Let $E\subseteq X\times X$ be an analytic equivalence relation on
a Polish space $X$. Suppose $A$ and $B$ are disjoint analytic
subsets of $X$. Assume that $B$ is invariant with respect to $E$,
{\em ie}, $B$ is a union of $E$-equivalence classes. Then there is
an $E$-invariant Borel set $C$ separating $A$ from $B$.

Let $A$ be an analytic subset of a Polish space, $Y$ a Polish
space, and $f:A\to Y$ a 1-1 Borel map. Then $f:A\to f(A)$ is a
Borel isomorphism.

Let $X$ and $Y$ be two Polish spaces, $A\subseteq X$ be analytic,
and $f:A\to Y$ be any map. Then the following are equivalent
statements: \begin{description} \item{(i)} $f$ is Borel
measurable, \item{(ii)} $\mathrm{graph}(f)$ is Borel in $A\times
Y$, and \item{(iii)} $\mathrm{graph}(f)$ is
analytic.\end{description}

\underline{Solova}y \underline{Codin}g \underline{of}
\underline{Borel} \underline{Sets} - Let $(r_i)$ be an enumeration
of the rational numbers and let $J$ be the pairing function on
$\mathbb{N}\times\mathbb{N}$ defined as $J(m,n)=2^m(2n+1)$. We
define the Solovay coding recursively as follows:
\begin{description} \item{(s-i)} $\alpha\in\mathbb{N}^{\mathbb{N}}$ codes
$[r_i,r_j]$ if $\alpha(0)=0(\mathrm{mod}\,3)$, $\alpha(1)=i$, and
$\alpha(2)=j$, \item{(s-ii)} Suppose
$\alpha_i\in\mathbb{N}^{\mathbb{N}}$ codes $B_i\subseteq
\mathbb{R}, i=0,1,...$; then $\alpha\in \mathbb{N}^{\mathbb{N}}$
codes $\bigcup_iB_i$ if $\alpha(0)=1 (\mathrm{mod}\,3)$ and
$\alpha\left(J(m,n)\right)=\alpha_m(n)$, \item{(s-iii)} Suppose
$\beta\in\mathbb{N}^{\mathbb{N}}$ codes $B$, $\alpha(0)\equiv
2(\mathrm{mod}\,3)$, and $\alpha(n+1)=\beta(n)$. Then $\alpha$
codes $B^c$, \item{(s-iv)} $\alpha$ codes $B\subseteq \mathbb{R}$
only as per (s-i), (s-ii) and (s-iii) above. \end{description}

Then, we have the following: \begin{description} \item{(i)} Every
$\alpha\in \mathbb{N}^{\mathbb{N}}$ codes at most one subset of
$\mathbb{R}$, \item{(ii)} Every Borel subset of $\mathbb{R}$ is
coded by some $\alpha \in\mathbb{N}^{\mathbb{N}}$, \item{(iii)} If
a subset of $\mathbb{R}$ is coded by $\alpha$, it is Borel.
\end{description}

Next, we define a function $\Phi:\mathbb{N}^{\mathbb{N}}\times
\mathbb{N}\to\mathbb{N}^{\mathbb{N}}$ with the property that if
$\alpha$ codes a Borel set $B$, then $\Phi(\alpha,.)$ recovers the
Borel sets from which $B$ is constructed.

To achieve this above, we fix an enumeration $(s_n)$, without any
repetition, of $\mathbb{N}^{<\,\mathbb{N}}$ such that $s_n\prec
s_m \Rightarrow n\leq m$ with $s_0$ being the empty sequence.

Set $\Phi(\alpha,0)=\alpha, \alpha\in\mathbb{N}^{\mathbb{N}}$. Let
$n>0$ and suppose that $\Phi(\alpha,m)$ has been defined for all
$\alpha\in\mathbb{N}^{\mathbb{N}}$ and for all $m<n$. Let $m<n$
and $u$ be such that $s_n=\widehat{s_m\,u}$. Define for $i\in
\mathbb{N}$, $\Phi(\alpha,n)(i)=0$ if $\Phi(\alpha,m)(0)\equiv 0
(\mathrm{mod\,3})$, $=\Phi(\alpha,m)(J(u,i))$ if
$\Phi(\alpha,m)(0)\equiv 1(\mathrm{mod\,3})$,
$=\Phi(\alpha,m)(i+1)$ if $\Phi(\alpha,m)(0)\equiv 2
(\mathrm{mod\,3})$.

Then, the graph of $\Phi$ is Borel and, hence, $\Phi$ is Borel
measurable. Also, by induction on $n$, we see that if $\alpha$
codes a Borel set, then for all $n$, $\Phi(\alpha,n)$ codes a
Borel set.

Now, for $\beta\in\mathbb{N}^{\mathbb{N}}$, define $\bar{\beta}
\in \mathbb{N}^{\mathbb{N}}$ such that for every $n\in\mathbb{N}$,
$s_{\bar{\beta}(n)}=\left(\beta(0), \beta(1), ..., \beta(n-1)
\right)$. The map $\beta\to\bar{\beta}$ is continuous.

Now, define the co-analytic set $C=\left\{ \alpha\in
\mathbb{N}^{\mathbb{N}}:\right.$ $\left. (\forall \beta) (\exists
n)\Phi\left( \alpha, \bar{\beta(n)}\right)=0\right\}$. Then $C$ is
closed under Solovay's coding (s-i) - (s-iv).

Solovay then constructed an example of a non-Borel measurable
function $f:C\times\mathbb{R}\to\mathbf{2}^{\mathbb{N}}$ whose
graph is Borel in $C\times\mathbb{R}\times
\mathbf{2}^{\mathbb{N}}$.
\goodbreak

Next, let $X$ and $Y$ be Polish spaces, $A\subset X$ Borel, and
$f:X\to Y$ a 1-1 Borel map. Then, $f(A)$ is Borel.

Let $X$ be standard Borel and $Y$ be metrizable. Suppose there is
a 1-1 Borel map $f$ from $X$ onto $Y$. Then $Y$ is standard Borel
and $f$ a Borel isomorphism.

If $\mathcal{T}$ and $\mathcal{T}'$ be two Polish topologies on
$X$ such that $\mathcal{T}'\subseteq \sigma(\mathcal{T})$. Then,
$\sigma(\mathcal{T})=\sigma(\mathcal{T}')$.

\underline{Blackwell}-\underline{Macke}y \underline{Theorem}: Let
$X$ be an analytic subset of a Polish space and $\mathcal{A}$ be a
countably generated sub $\sigma$-algebra of the Borel
$\sigma$-algebra $\mathcal{B}_X$. Let $B\subseteq X$ be a Borel
set that is a union of atoms of $\mathcal{A}$. Then $B\in
\mathcal{A}$. (This result is not true if $X$ is co-analytic.)

Let $X$ be an analytic subset of a Polish space and
$\mathcal{A}_1$ and $\mathcal{A}_2$ be two countably generated sub
$\sigma$-algebras of the Borel $\sigma$-algebra $\mathcal{B}_X$
with the same set of atoms. Then $\mathcal{A}_1=\mathcal{A}_2$. In
particular, if $\mathcal{A}$ is a countably generated sub
$\sigma$-algebra containing all the singletons, then $\mathcal{A}
= \mathcal{B}_X$.

The \underline{Generalized} \underline{First}
\underline{Se}p\underline{aration} \underline{Theorem}: Let
$(A_n)$ be a sequence of analytic subsets of a Polish space $X$
such that $\bigcap_nA_n= \emptyset$. Then there exist Borel sets
$B_n\supseteq A_n$ such that $\bigcap_nB_n=\emptyset$. If $(A_n)$
satisfies the conclusion of this result, we call it
\underline{Borel se}p\underline{arated}.

Let $(E_n)$ be a sequence of subsets of $X$, $k\in\mathbb{N}$, and
$E_i=\bigcup_nE_{in}$ for $i\leq k$. Suppose $(E_n)$ is not Borel
separated. Then there exist $n_o, n_1, ..., n_k$ such that the
sequence $E_{on_o}$, $E_{1n_1}$, ..., $E_{kn_k}$, $E_{k+1}$,
$E_{k+2}$, ... is not Borel separated.

Let $(A_n)$ be a sequence of analytic subsets of a Polish space
$X$ such that $\limsup A_n=\emptyset$. Then there exist Borel sets
$B_n \supseteq A_n$ such that $\limsup B_n=\emptyset$. This result
is not true for co-analytic $A_n$'s.

\underline{Weak} \underline{Reduction}
\underline{Princi}p\underline{le} \underline{for}
\underline{Co}-\underline{Anal}y\underline{tic} \underline{Sets}:
Let $C_o$, $C_1$, $C_2$, ... be a sequence of co-analytic subsets
of a Polish space such that $\bigcup C_n$ is Borel. Then there
exist pairwise disjoint Borel sets $B_n\subseteq C_n$ such that
$\bigcup B_n=\bigcup C_n$.

Let $E$ be an analytic equivalence relation on a Polish space $X$.
Suppose $A_o$, $A_1$, $A_2$, ... are invariant analytic subsets of
$X$ such that $\bigcap A_n=\emptyset$. There then exist invariant
Borel sets $B_n\supseteq A_n$ with $\bigcap_nB_n=\emptyset$.
Hence, if $C_o$, $C_1$, $C_2$, ... is a sequence of invariant
co-analytic sets whose union is Borel, then there exist pairwise
disjoint invariant Borel sets $B_n\subseteq C_n$ with $\bigcup B_n
=\bigcup C_n$.

For the following considerations marked by *, let $X$ and $Y$ be
fixed Polish spaces and $(V_n)$ be a countable base for $Y$.

*\, Let $A_o$ and $A_1$ be disjoint analytic subsets of $X\times
Y$ with the sections $(A_o)_x, x\in X$, closed in $Y$. Then there
exists a sequence $(B_n)$ of Borel subsets of $X$ such that $A_1
\subseteq\bigcup_n\left(B_n\times V_n\right)$ and $A_o\bigcap
\bigcup_n\left(B_n\times V_n\right)=\emptyset$.

*\, \underline{Structure} \underline{Theorem} \underline{for}
\underline{Borel} \underline{Sets} \underline{with}
\underline{O}p\underline{en} \underline{Sections}: Suppose
$B\subseteq X\times Y$ is any Borel set with $B_x$ open, $x\in X$.
Then there is a sequence $(B_n)$ of Borel subsets of $X$ such that
$B=\bigcup\left(B_n\times V_n\right)$.

*\, Let $A_o$ and $A_1$ be disjoint analytic subsets of $X\times
Y$ with sections $(A_o)_x$ and $(A_1)_x$ closed for all $x\in X$.
Then there exist disjoint Borel sets $B_o$ and $B_1$ with closed
sections such that $A_o\subseteq B_o$ and $A_1\subseteq B_1$.

*\, Suppose $B\subseteq X\times Y$ is a Borel set with the
sections $B_x$ closed. Then there is a Polish topology
$\mathcal{T}$ finer than the given topology on $X$ generating the
same Borel $\sigma$-algebra such that $B$ is closed relative to
the product topology on $X\times Y$, $X$ being equipped with the
new topology $\mathcal{T}$.

*\, Let $A_o$ and $A_1$ be disjoint analytic subsets f $X\times Y$
with sections $(A_o)_x$ being compact. Then there exists a Borel
subset $B_o$ in $X\times Y$ with compact sections separating $A_o$
from $A_1$.

*\, Let $A_o, A_1\subseteq X\times Y$ be disjoint and analytic
with the sections $(A_o)_x$, $(A_1)_x$ closed. Then there exists a
Borel map $u:X\times Y\to [0,1]$ such that $y\to u(x,y)$ is
continuous for all $x$ and $u(x)=0$ if $x\in A_o$ and $u(x)=1$ if
$x\in A_1$. This result does not hold for co-analytic $A_o$,
$A_1$.

*\, Let $B\subseteq X\times Y$ be Borel with sections closed and
$f:B\to[0,1]$ a Borel map such that $y\to f(x,y)$ is continuous
for all $x$. Then there is a finer topology $\mathcal{T}$ on $X$
generating the same Borel $\sigma$-algebra such that when $X$ is
equipped with it, $B$ is closed and $f$ continuous. Hence, there
is a Borel extension $F:X\times Y\to[0,1]$ of $f$ such that $y\to
F(x,y)$ is continuous for all $x$. Notably, $[0,1]$ can be
replaced by any compact convex subset of $\mathbb{R}^n$ in this
result. However, this result does not hold for co-analytic $B$.

*\, In general, projection of a Borel set need not be Borel.
However, if $B\subseteq X\times Y$ is Borel and the sections $B_x$
are open (convex) (compact) in $Y$, then $\pi_X(B)$ is Borel in
$X$. The projection $\pi_X(B)$ is also Borel when for every $x\in
\pi_X(B)$, (i) the sections $B_x$ contains exactly one point, (ii)
$B_x$ is non-meager (iii) $P(x,B_x)>0$, where $P$ is any
transition probability on $X\times Y$. Furthermore, if $Y$ is
$\sigma$-compact (or, equivalently, locally compact) then the
projection of every Borel set $B$ in $X\times Y$ with $x$-sections
closed in $Y$ is Borel.

Now, let $(G,\diamond)$ be a Polish group and $H$ a closed
subgroup. Suppose $E=\{(x,y):xy^{-\,1}\in H\}$, {\em ie}, $E$ is
the equivalence relation induced by the right cosets of $H$. Then
the $\sigma$-algebra of invariant Borel sets is countably
generated. The converse of this result is also true.

Let $G$ be a Polish group and $H$ its Borel subgroup. Suppose that
the $\sigma$-algebra of invariant Borel sets is countably
generated. Then the subgroup $H$ is closed.

Let $X$ be a Polish space and $G$ a group of its homeomorphisms
such that for every pair $U,V$ of non-empty open sets there is
$g\in G$ with $g(U)\bigcap V\neq \emptyset$. Suppose $A$ is a
G-invariant Borel set, {\em ie}, $g(A)=A$ for all $g\in G$. Then
either $A$ or $A^c$ is meager in $X$.

Let $x\in X$. The set $G_x=\{g\in G:gx=x\}$ is called the
\underline{stabilizer} of $x$. Clearly, $G_x$ is a subgroup of the
group $G$.

Let $(G,\diamond)$ be a Polish group, $X$ a countably generated
measurable space with singletons as atoms and $(g,x)\to gx$ an
action of $G$ on $X$. Suppose that for a given $x$, the map $g\to
gx$ is Borel. Then the stabilizer $G_x$ is closed.

Let $G$ be a Polish group, $X$ a Polish space, and $a(g,x)=gx$ an
action of $G$ on $X$. Assume that $gx$ is continuous in $x$ for
all $g$ and Borel in $g$ for all $x$. Then the action is
continuous.

If $(G,\diamond)$ is a group with a Polish topology such that the
group operation $(g,h)\to gh$ is Borel, then $g\to g^{-\,1}$ is
continuous.

Note that if $(G,\diamond)$ is a group with a Polish topology such
that the group operation is separately continuous in each
variable, then $G$ is a topological group.

As a substantial generalization of the above, we have the result
that: If $(G,\diamond)$ is a group with a Polish topology such
that $h\to gh$ is continuous for every $g\in G$, and $g\to gh$ for
all $h$. Then $G$ is a topological group. This follows by showing
that the group operation $gh$ is jointly continuous. Also, for
every meager set $I$ and for every $g$, $Ig=\{hg: h\in I\}$ is
meager.

As a further generalization of the same result, we have that: If
$(G,\diamond)$ is a group with a topology that is metrizable,
separable, and Baire, and if the multiplication $gh$ is continuous
in $h$ for all $g$ and Baire measurable in $g$ for all $h$, then
$G$ is a topological group.

A \underline{norm on a set $S$} is a map $\varphi:S\to\mathrm{{\bf
ON}}$. Let $\varphi$ be a norm on a set $S$. Define
$\leq_{\varphi}$ as the binary relation $x\leq_{\varphi}y
\Leftrightarrow\varphi(x)\leq\varphi(y)$. Then $\leq_{\varphi}$ is
\begin{description} \item{(i)} reflexive, \item{(ii)} transitive,
\item{(iii)} \underline{connected}, {\em ie}, for every $x,y\in S$, at least one
of $x\leq_{\varphi}y$ or $y \leq_{\varphi}x$ holds, and
\item{(iv)} there is no sequence $(x_n)$ of elements in $S$ such
that $x_{n+1}<_{\varphi}x_n$ for all $n$, where
$x<_{\varphi}y\Leftrightarrow\varphi(x)<\varphi(y) \Leftrightarrow
x\leq_{\varphi}y\,\mathrm{and}\,\neg
y\leq_{\varphi}x$.\end{description} Such a binary relation,
satisfying (i)-(iv) above is called as a
p\underline{re}-\underline{well}-\underline{orderin}g on $S$.

Let $X$ be a Polish space and $A\subseteq X$ be co-analytic. A
norm $\varphi$ on $A$ is called a
$\mathbf{\Pi}^1_1$-\underline{norm} if there are binary relations
$\leq_{\varphi}^{\mathbf{\Pi}^1_1} \in \mathbf{\Pi}^1_1$ and
$\leq_{\varphi}^{\mathbf{\Sigma}^1_1} \in \mathbf{\Sigma}^1_1$ on
$X$ such that for $y\in A$, $x\in A\, \mathrm{and}\,
\varphi(x)\leq\varphi(y)\Leftrightarrow x
\leq_{\varphi}^{\mathbf{\Pi}^1_1} y \Leftrightarrow x
\leq_{\varphi}^{\mathbf{\Sigma}^1_1} y$.

Then, every $\mathbf{\Pi}^1_1$ set $A$ in a Polish space $X$
admits a $\mathbf{\Pi}^1_1$-norm $\varphi:A\to \omega_1$.

Let $X$ be a Polish space and $A\subseteq X$ co-analytic. A norm
$\varphi:A\to\mathrm{{\bf ON}}$ is a $\mathbf{\Pi}^1_1$-norm if
and only if there are binary relations
$\leq_{\varphi}^{\mathbf{\Sigma}^1_1}$ and
$<_{\varphi}^{\mathbf{\Sigma}^1_1}$, both in
$\mathbf{\Sigma}^1_1$, such that for every $y\in A$, $x\in A$ \&
$\varphi(x)\leq\varphi(y) \Leftrightarrow x
\leq_{\varphi}^{\mathbf{\Sigma}^1_1} y$ and $x\in A$ \&
$\varphi(x) <\varphi(y) \Leftrightarrow x
<_{\varphi}^{\mathbf{\Sigma}^1_1}$.

Let $A\subseteq X$ and $\varphi$ be a norm on $A$. Define
$\leq^*_{\varphi}$ and $<^*_{\varphi}$ on $X$ as: $x
\leq^*_{\varphi} y \Leftrightarrow x\in A\,\&\,(y\notin A\,
\mathrm{or}\,(y\in A\,\&\,\varphi(x)\leq\varphi(y)))$ and $x
<^*_{\varphi}y \Leftrightarrow x\in A\,\&\,(y\notin A\,
\mathrm{or}\,(y\in A\,\&\,\varphi(x)<\varphi(y)))$.

Let $X$ be a Polish space, $A\subseteq X$ co-analytic, and
$\varphi$ a norm on $A$. Then $\varphi$ is a
$\mathbf{\Pi}^1_1$-norm if and only if both $\leq_{\varphi}^*$ and
$<_{\varphi}^*$ are co-analytic.

Let us identify a tree $T$ on $\mathbb{N}$ with its characteristic
function $\chi_{_T}\in \mathbf{2}^{\mathbf{N}^{<\,\mathbf{N}}}$,
{\em ie}, $Tr=\{T\in\mathbf{2}^{\mathbf{N}^{<\,\mathbf{N}}}: T\,
\mathrm{is\,a\,tree\,on\,}\mathbb{N}\}$. Note that $Tr$ is a
$G_{\delta}$ set in $\mathbf{2}^{\mathbf{N}^{<\,\mathbf{N}}}$,
when $\mathbf{2}^{\mathbf{N}^{<\,\mathbf{N}}}$ is equipped with
the product of discrete topologies on $\mathbf{2}$ and, hence, is
a Polish space. Let $WF=\{T\in Tr: T\,\mathrm{is\,well\,founded}
\}$. $WF$ is $\mathbf{\Pi}^1_1$-complete and co-analytic.

Next, identify binary relations on $\mathbb{N}$ with points of
$\mathbf{2}^{\mathbb{N}\times\mathbb{N}}$ and equip
$\mathbf{2}^{\mathbb{N}\times\mathbb{N}}$ with the product of
discrete topologies on $\mathbf{2}$. Let $LO=\{ \alpha\in
\mathbf{2}^{\mathbb{N}\times\mathbb{N}}: \alpha\,\mathrm{is\,a\,
linear\,order}\}$. Then $LO$ is Borel. Define $WO=\{ \alpha\in
\mathbf{2}^{\mathbb{N}\times\mathbb{N}}:\alpha\,\mathrm{is\,a\,
well\,order}\}$. Then $WO$ is co-analytic and
$\mathbf{\Pi}^1_1$-complete since there exists a continuous map
$R:Tr\to\mathbf{2}^{\mathbb{N}\times\mathbb{N}}$ such that
$WF=R^{-\,1}(WO)$.

\underline{Boundedness} \underline{Theorem} \underline{for}
$\mathbf{\Pi}^1_1$-\underline{Norms}: Suppose $A$ is a
$\mathbf{\Pi}^1_1$ set in a Polish space $X$ and $\varphi:A\to
\omega_1$ a norm on $A$. Then for every $\mathbf{\Sigma}^1_1$ set
$B\subseteq A$, $\sup\{\varphi(x):x\in B\}<\omega_1$. Hence, $A$
is Borel if and only if $\sup\{\varphi(x):x\in A\}<\omega_1$.

\underline{Reduction} \underline{Princi}p\underline{le}
\underline{for} \underline{Co}-\underline{anal}y\underline{tic}
\underline{Sets}: Let $(A_n)$ be a sequence of $\mathbf{\Pi}^1_1$
sets in a Polish space $X$. Then there is a sequence $(A^*_n)$ of
$\mathbf{\Pi}^1_1$ sets such that they are pairwise disjoint,
$A^*_n \subseteq A_n$, and $\bigcup_nA^*_n=\bigcup_nA_n$.

Let $X$ be Polish and $A_0$, $A_1$ be co-analytic subsets of $X$.
Then there exist pairwise disjoint co-analytic sets $A^*_0$,
$A^*_1$ contained in $A_0$, $A_1$ respectively such that $A^*_0
\bigcup A^*_1=A_0\bigcup A_1$.

Note that analytic sets do not satisfy the reduction principle and
the co-analytic sets do not satisfy the separation theorems.

A very useful parametrization of Borel sets is provided by the
following: Let $X$ be a Polish space. Then there exist sets $C\in
\mathbf{\Pi}^1_1\left(\mathbb{N}^{\mathbb{N}}\right)$ and $V \in
\mathbf{\Pi}^1_1\left(\mathbb{N}^{\mathbb{N}}\times X\right)$, $U
\in \mathbf{\Sigma}^1_1\left(\mathbb{N}^{\mathbb{N}}\times X
\right)$ such that for every $\alpha\in C$, $U_{\alpha}=
V_{\alpha}$ and $\mathbf{\triangle}^1_1(X)=\{U_{\alpha}: \alpha
\in C\}$. In particular, there are a co-analytic set and an
analytic set contained in $\mathbb{N}^{\mathbb{N}}\times X$ that
are universal for $\mathbf{\triangle}^1_1(X)$.

Note that in the above we cannot replace $C\in \mathbf{\Pi}^1_1$
by $C\in\mathbf{\Sigma}^1_1$.

A \underline{Cho}q\underline{uet}-\underline{ca}p\underline{acit}y
on a Polish space $X$ is a set-map or a set-function
$I:\mathcal{P}(X)\to[0,\infty]$ such that \begin{description}
\item{(i)} $I$ is monotone, {\em ie}, $A\subseteq B \implies I(A)\leq
I(B)$, \item{(ii)} $A_0\subseteq A_1\subseteq A_2\subseteq ...
\implies \lim I(A_n)= I(A)$ where $A=\bigcup_nA_n$. We say that
$I$ {\em is going up}. \item{(iii)} $I(K)<\infty$ for every
compact $K\subseteq X$, and lastly, \item{(iv)} For every compact
$K$ and every $t>0$, $I(K) <t$ implies that there is an open set
$U\supseteq K$ such that $I(U)<t$. In this case, we say that $I$
{\em is right-continuous over compacta}. \end{description}

Let $\mu^*$ be the associated outer measure corresponding to a
finite Borel measure $\mu$ on a Polish space. Then, for any $A
\subseteq X$, $\mu^*=\inf \{ \mu(B):B\supseteq A, A\,
\mathrm{is\;Borel} \}$ is a capacity on $X$.

In general, if $I$ is a capacity on a Polish space $X$ and $I^*:
\mathcal{P}(X)\to[0,\infty]$ be defined as $I^*(A)=\inf \{ I(B):
B\supseteq A, B\,\mathrm{is\;Borel}\}$. Then $I^*$ is a capacity
on $X$.

Let $X$ be Polish and define $I:\mathcal{P}(X\times X)\to
\mathbf{2}$ by $I(A)=0$ if $\pi_1(A)\bigcap\pi_2(A)=\emptyset$,
and $I(A)=1$ otherwise, where $\pi_1$ and $\pi_2$ are two
projection maps on $X\times X$. For $A\subseteq X\times X$, let
$R[A]=\pi_1(A)\times\pi_2(A)$. Then $I$ is a capacity on $X\times
X$. This $I$ is the \underline{se}p\underline{aration}
\underline{ca}p\underline{acit}y on $X\times X$.

Let $X$, $Y$ be Polish spaces and $f:X\to Y$ be a continuous
function. Suppose that $I$ is a capacity on $Y$. Define $I_f(A)=
I(f(A)), A\subseteq X$. Then $I_f$ is a capacity on $X$.

Let $I$ be a capacity on a Polish space. Suppose $(K_n)$ is a
non-increasing sequence of compact subsets of $X$ decreasing to,
say, $K$. Then, $I(K_n)$ converges to $I(K)$.

Consider $I:\mathcal{P}\to\mathbf{2}$ defined by $I(A)=0$ if $A$
is contained in a $K_{\sigma}$ set and $I(A)=1$ otherwise. But $I$
is {\em not\/} a capacity since $I$ is not right-continuous over
compacta.

Let $X$ be a Polish space, $I$ a capacity on $X$, and $A\subseteq
X$. We say that $A$ is $I$-\underline{ca}p\underline{citable} if
$I(A)=\sup \{ I(K):K\subseteq A\,\mathrm{compact}\}$. A subset $A$
is called as \underline{universall}y
\underline{ca}p\underline{acitable} if it is $I$-capacitable with
respect to all capacities $I$ on $X$.

Let $X$, $Y$ be Polish spaces and $f:X\to Y$ a continuous map.
Assume $A\subseteq X$ is universally capacitable. Then $f(A)$ is
universally capacitable. (This is almost the only known stability
property of the class of universally capacitable sets.) Note that
the complement of a universally capacitable set need not be
universally capacitable.

Let $I$ be a capacity on a Polish space $X$ and $A\subseteq X$
universally capacitable. Then $I(A)=I^*(A)$ where $I^*(A)=\inf \{
I(B): B\supseteq A, B\,\mathrm{Borel}\}$, as defined earlier.

Note that the space $\mathbb{N}^{\mathbb{N}}$ of irrational
numbers is universally capacitable.

The \underline{Cho}q\underline{uet}
\underline{Ca}p\underline{acitabilit}y \underline{Theorem}: Every
analytic subset of a Polish space is universally capacitable.

Further, let $X$ be Polish and $I$ be the separation capacity on
$X\times X$. Assume that a rectangle $A_1\times A_2$ be
universally capacitable. If $I(A_1\times A_2)=0$ then there is a
Borel rectangle $B=B_1\times B_2$ containing $A_1\times A_2$ of
$I$-capacity $0$.

\underline{Second Se}p\underline{aration} \underline{Theorem}
\underline{for} \underline{Anal}y\underline{tic} \underline{Sets}:
Let $X$ be a Polish space and $A,B$ two analytic subsets. There
exist disjoint co-analytic sets $C$ and $D$ such that $A\setminus
B\subseteq C$ and $B \setminus A\subseteq D$.

Suppose $X$ is a Polish space and let $(A_n)$ be a sequence of
analytic subsets of $X$. Then there exists a sequence $(C_n)$ of
pairwise disjoint co-analytic subsets of $X$ such that $A_n
\setminus \bigcup_{m\neq n}A_m \subseteq C_n$.

Let $X$ be a Polish space and $(A_n)$ a sequence of analytic
subsets of $X$. Then there exists a sequence $(C_n)$ of
co-analytic subsets of $X$ such that we have $A_n\setminus \limsup
A_m \subseteq C_n$ and $\limsup C_n=\emptyset$.

Note that the Generalized First Separation Principle does not hold
for co-analytic sets.

Let $X$ be a Borel subset of a Polish space, $Y$ Polish and
$f:X\to Y$ Borel. Then $Z_f=\{y\in Y:f^{-\,1}(y)\, \mathrm{is\,
a\, singleton}\}$ is co-analytic.

Let $X$, $Y$ be Polish and $B\subseteq X\times Y$ a Borel set.
Then the set $Z=\{x\in X: B_x\,\mathrm{is\,a\,singleton}\}$ is
co-analytic.

If $X$, $Y$ are Polish and $B$ a Borel subset of $X\times Y$ such
that for every $x\in X$ the section $B_x$ is countable, then
$\pi_X(B)$ is Borel.

Let $X$, $Y$ be Polish and $f:X\times Y$ a countable-to-one Borel
map. Then $f(B)$ is Borel for every Borel set $B$ in $X$.

Let $X$ be Standard Borel, $Y$ polish, $A\subseteq X \times Y$
analytic with $\pi_X(A)$ uncountable and that $\forall\,
x\in\pi_X(a)$, the section $A_x$ is perfect. Then there is a
$C\subseteq \pi_X(A)$ homeomorphic to the Cantor ternary set and a
1-1 Borel map $f:C\times \mathbf{2}^{\mathbb{N}}\to A$ such that
$\pi_X\left(f(x,\alpha)\right)=x$, $\forall\,x$ and every
$\alpha$.

Now, the Axiom of Choice states that every family $\{A_i:i\in I\}$
of nonempty sets admits a choice function. It however does not
specify the procedure by which we can make the choice of such
sets. This situation leads to selection criteria or the selection
theorems.

A \underline{multifunction} $G:X\to Y$ is a map with domain $X$
and whose values are nonempty subsets of $Y$. For any $A\subset
Y$, we put $G^{-\,1}(A)=\{x\in X:G(x)\bigcap A\neq\emptyset\}$. We
call $\{(x,y)\in X\times Y:y\in G(x)\}$ the
g\underline{ra}p\underline{h} of the multifunction $G$ and will
denote it by $\mathrm{gr}(G)$. We have $G^{-\,1}(A)=\pi_X\left(
\mathrm{gr}(G)\bigcap(X\times A)\right)$.

A \underline{selection of a multifunction} $G:X\to Y$ is a point
map $s:X\to Y$ such that $s(x)\in G(x)$ for every $x\in X$.

Let $\mathcal{A}$ denote a class of subsets of $X$. We restrict
ourselves to cases where $\mathcal{A}$ is the $\sigma$-algebra or
$X$ being a Polish space and $\mathcal{A}$ being one of the
additive class $\mathbf{\Sigma}^0_{\alpha}(X)$.

For a Polish $Y$, a multifunction $G:X\to Y$ is called
\underline{$\mathcal{A}$}-\underline{measurable}
(\underline{stron}g\underline{l}y
\underline{$\mathcal{A}$}-\underline{measurable}) if
$G^{-\,1}(U)\in\mathcal{A}$ for every open (closed) set $U$ in
$Y$. We will often omit the prefix $\mathcal{A}$.

Suppose $X$ is a measurable space, $Y$ is a Polish space and
$F(Y)$ is the space of all nonempty closed sets in $Y$ with the
Effros Borel structure. Then a closed-valued multifunction $G:X\to
Y$ is measurable if and only if $G:X\to F(X)$ is measurable as a
point map.

A multifunction $G:X\to Y$ is called as \underline{lower}-
\underline{semicontinuous}
(\underline{u}pp\underline{er}-\underline{semicontinuous}) if
$G^{-\,1}(U)$ is open (closed) for every open (closed) set $U
\subseteq Y$. If $g:X\to Y$ be a continuous open (closed) onto map
then $G(x)=g^{-\,1}(x)$ is lower semicontinuous (upper
semicontinuous).

Let $Y$ be metrizable, $G:X\to Y$ strongly
$\mathcal{A}$-measurable, and $\mathcal{A}$ closed under countable
unions. Then $G$ is $\mathcal{A}$-measurable.

Let $(X,\mathcal{A})$ be a measurable space, $Y$ Polish and $G:X
\to Y$ a closed-valued measurable multifunction. Then $\mathrm{gr}
(G)\in \mathcal{A}\bigotimes\mathcal{B}_Y$. The converse of this
is, in general, not true.

If $X$, $Y$ are two Polish spaces and if $\mathcal{A}$ is a
sub-algebra of $\mathcal{B}_X$, then every compact-valued
multifunction $G: X\to Y$ whose graph is in
$\mathcal{A}\bigotimes\mathcal{B}_Y$ is seen to be
$\mathcal{A}$-measurable.

Now, let $B\subseteq X\times Y$. A set $C\subseteq B$ is called as
a \underline{uniformization of $B$} if for every $x\in X$, the
section $C_x$ contains at most one point and $\pi_X(C)=\pi_X(B)$.
That is to say, $C$ is a uniformization of $B$ if it is the graph
of a function $f:\pi_X(B)\to Y$. Such a map $f$ will be called
\underline{section of $B$}.

A Borel set $B\subseteq X\times Y$ admits a Borel uniformization
if and only if $\pi_X(B)$ is Borel and $B$ admits a Borel section.

If $C_1$ and $C_2$ are disjoint co-analytic subsets of $[0,1]$
that cannot be separated by Borel sets and if $B_s$ be a closed
subset of $[0,1]\times \Sigma(s)$ whose projection is $[0,1]
\setminus C_s,\,s=1$ or $2$, and if $B=B_1\bigcup B_2$, then $B$
is a closed subset of $[0,1]\times \mathbb{N}^{\mathbb{N}}$ whose
projection is $[0,1]$. Such a set $B$ does not admit a Borel
uniformization.

Let $\mathcal{D}$ be the partition of $X$ and $A\subset X$. We put
$A^*=\bigcup\{P\in\mathcal{D}:A\bigcap P\neq\emptyset\}$. Thus,
$A^*$ is the smallest invariant set containing $A$ and is called
the \underline{saturation of $A$}.

Let $X$ be a Polish space and $\mathcal{A}$ family of subsets of
$X$. A partition $\mathcal{D}$ will be called
\underline{$\mathcal{A}$}-\underline{measurable} if the saturation
of every open set is in $\mathcal{A}$. We then say that the
partition $\mathcal{D}$ of a Polish space $X$ as
\underline{closed}, \underline{Borel}, etc.\ if it is closed,
Borel etc.\ in $X\times X$. It is said to be \underline{lower}-
\underline{semicontinuous}
(\underline{u}pp\underline{er}-\underline{semicontinuous}) if the
saturation of every open (closed) set is open (closed).

A \underline{cross}-\underline{section} of $\mathcal{D}$ is a
subset $S$ of $X$ such that $S\bigcap A$ is a singleton for every
$A\in\mathcal{D}$. A \underline{section} of $\mathcal{D}$ is a map
$f:X\to X$ such that for any $x,y\in X$ (i) $x\mathcal{D}f(x)$ and
(ii) $x\mathcal{D}y\Rightarrow f(x)=f(y)$. To each section $f$ we
{\em canonically\/} associate a cross-section $S=\{x\in X: x=f(x)
\}$ of $\mathcal{D}$.

If $X$ is Polish and $\mathcal{D}$ is a Borel equivalence relation
on $X$, then the following are equivalent: \begin{description}
\item{(i)} $\mathcal{D}$ has a Borel section \item{(ii)} $\mathcal{D}$
admits a Borel cross section. \end{description} ({\em Notice that
we use here and elsewhere the same symbol $\mathcal{D}$ to denote
a partition of $X$ and a Borel equivalence relation on $X$.})

A partition $\mathcal{D}$ is said to be \underline{countabl}y
\underline{se}p\underline{arated} if there is a Polish $Y$ and a
Borel map $f:X\to Y$ such that $x\mathcal{D}x'\Leftrightarrow f(x)
=f(x')$.

For a partition $\mathcal{D}$ on a Polish space $X$, the following
are equivalent statements: \begin{description} \item{(i)}
$\mathcal{D}$ is countably-generated, \item{(ii)} There exists a
Polish space $Y$ as well as a sequence of Borel maps $f_n:X\to Y$
such that $\forall\,x,y\,\left(x\mathcal{D}y \Leftrightarrow
\forall\,n\,\left(f_n(x)=f_n(y)\right)\right)$, \item{(iii)} There
then exists a sequence $(B_n)$ of invariant Borel subsets of $X$
such that for all $\,x,y\,\left(x\mathcal{D}y \Leftrightarrow
\forall\, n\,\left(x\in B_n \Leftrightarrow y\in
B_n\right)\right)$, in short, $X\times Y \setminus \mathcal{D}
=\bigcup_n \left( B_n\times B^c_n\right)$.
\end{description}

Every closed equivalence relation $\mathcal{D}$ on a Polish space
$X$ is countably generated. Every Borel measurable partition of a
Polish space into $G_{\delta}$ sets is countably separated.

Let $\mathcal{D}$ be a partition of a Polish space $X$ and let
$X\diagup \mathcal{D}$ denote the set of all
$\mathcal{D}$-equivalence classes. Suppose $q:X\to X\diagup
\mathcal{D}$ be the canonical quotient map. Then $X\diagup
\mathcal{D}$ equipped with the largest $\sigma$-algebra making $q$
measurable is called the q\underline{uotient} \underline{Borel}
\underline{s}p\underline{ace}. The quotient $\sigma$-algebra then
consists of all subsets $E$ of $X\diagup \mathcal{D}$ such that
$q^{-\,1}(E)$ is a Borel subset in $X$.

If $\mathcal{D}$ is any countably separated partition of a Polish
space $X$, then the quotient Borel space $X\diagup \mathcal{D}$ is
seen to be Borel isomorphic to some analytic set in a Polish
space.

Importantly, note however that the quotient of a Standard Borel
space by an equivalence relation need not at all be isomorphic to
the Borel $\sigma$-algebra of a metric space.

Let $\mathcal{D}$ be a Borel partition of a Polish space $X$. Then
the following are equivalent: \begin{description} \item{(i)}
$\mathcal{D}$ is countably separated, \item{(ii)} The
$\sigma$-algebra $\mathcal{B}^*$ of $\mathcal{D}$-invariant Borel
sets is countably generated. \end{description}

Now, let $Y$ denote a Polish space, $d<1$ a compatible complete
metric on $Y$, $X$ a nonempty set, and $\mathcal{L}$ an algebra of
subsets for the results marked by * below.

*\,
\underline{Kuratowski}-\underline{R}y\underline{ll}-\underline{Nardzewski}
Theorem: Every $\mathcal{L}_{_{\sigma}}$-measurable, closed-valued
multifunction $F:X\to Y$ admits an
$\mathcal{L}_{_{\sigma}}$-measurable selection.

The proof for the above theorem rests on the following two lemmas.

*\, Suppose $A_n\in\mathcal{L}_{_{\sigma}}$. Then there exists
$B_n \subseteq A_n$ such that the $B_n$'s are pairwise disjoint
elements of $\mathcal{L}_{_{\sigma}}$ and $\bigcup_nA_n=\bigcup_n
B_n$.

*\, Let $f_n:X\to Y$ be a sequence of
$\mathcal{L}_{_{\sigma}}$-measurable functions converging
uniformly to $f:X\to Y$. Then $f$ is
$\mathcal{L}_{_{\sigma}}$-measurable.

For a Polish space $X$ and $F(X)$ being the space of nonempty
closed subsets of $X$ with Effros Borel structure, there is a
measurable $s:F(X)\to X$ such that $s(F)\in F$ for all $F\in
F(X)$.

Let $(T,\mathcal{T})$ be a measurable space and $Y$ a separable
metric space. Then every $\mathcal{T}$-measurable, compact-valued
multifunction $F:T\to Y$ admits a $\mathcal{T}$-measurable
selection.

If $Y$ is a compact metric space, $X$ a metric space and $f:Y\to
X$ a continuous onto map then, there is a Borel map $s:X\to Y$ of
class $2$ such that $f\circ s$ is the identity map on $X$.

Suppose $T$ is a nonempty set, $\mathcal{L}$ an algebra on $T$,
and $X$ a Polish space. Let $F:T\to X$ is a closed-valued
$\mathcal{L}_{_{\sigma}}$-measurable multifunction. Then there is
a sequence $(f_n)$ of $\mathcal{L}_{_{\sigma}}$-measurable
selections of $F$ such that $F(t)=\mathrm{cl}\left(\{ f_n(t): n\in
\mathbb{N}\}\right),\,t\in T$. Results of this kind are generally
also called by the name of \underline{Castain}g's
\underline{Theorems}.

Furthermore, let $T$ be a nonempty set, $\mathcal{L}$ be an
algebra on $T$, $X$ be a Polish space, and $F:T\to X$ be a
closed-valued $\mathcal{L}_{_{\sigma}}$-measurable multifunction.
Then there is a map $f:T\times\mathbb{N}^{\mathbb{N}}\to X$ such
that (i) for every $\alpha\in\mathbb{N}^{\mathbb{N}}$, $t\to
f(t,\alpha)$ is $\mathcal{L}_{_{\sigma}}$-measurable, and (ii) for
every $t\in T$, $f(t,.)$ is a continuous map from
$\mathbb{N}^{\mathbb{N}}$ onto $F(t)$. Moreover, suppose that
$s:T\to X$ is an $\mathcal{L}_{_{\sigma}}$-measurable selection
for $F$ and $\epsilon>0$. Then the multifunction $G:T\to X$
defined as $G(t)=\mathrm{cl}\left(F(t)\bigcap B(s(t),\epsilon)
\right), \, t\in T$ is $\mathcal{L}_{_{\sigma}}$-measurable.

\underline{The}
\underline{Bhattachar}y\underline{a}-\underline{Srivastava}
\underline{Theorem}: Let $F:X\to Y$ be closed-valued as well as
strongly $\mathcal{L}_{_{\sigma}}$-measurable. Suppose $Z$ is a
separable metric space and $g:Y\to Z$ a Borel map of class $2$.
Then there is an $\mathcal{L}_{_{\sigma}}$-measurable selection
$f$ of $F$ such that $g\circ f$ is
$\mathcal{L}_{_{\sigma}}$-measurable.

Now, let $X$, $Y$ be two compact metric spaces, $f:X\to Y$ a
continuous onto map. Suppose that $A\subseteq Y$ and $1\leq \alpha
<\omega_1$, {\em ie}, countable ordinals. Then $f^{-\,1}(A)\in
\mathbf{\Pi}^0_{\alpha}(X) \Leftrightarrow A \in
\mathbf{\Pi}^0_{\alpha}(Y)$. Moreover, for $1\leq\alpha<\omega_1$,
$Z$ a separable metric space, and $g:X\to Z$ being a Borel map of
class $\alpha$, there is a class $2$ map $s:Y\to X$ such that
$g\circ s$ is of class $\alpha$ and $f\left( s(y)\right)=y$ for
all $y$.

\underline{Sch\"{a}l}'s \underline{Selection Theorem}: Let $(T,
\mathcal{T})$ be a measurable space, $Y$ be a separable metric
space, $G:T\to Y$ is a $\mathcal{T}$-measurable compact-valued
multifunction, $v$ be a real-valued function on $\mathrm{gr}(G)$,
that is the point-wise limit of a non-increasing sequence $(v_n)$
of $\mathcal{T}\bigotimes\mathcal{B}_Y|\mathrm{gr}(G)$-measurable
functions on $\mathrm{gr}(G)$ such that for each $n$ and each
$t\in T$, $v_n(t,.)$ is continuous on $G(t)$. Let $v^*(t)=\sup
\left\{ v(t,y):y\in G(t)\right\}, t\in T$. Then there is a
$\mathcal{T}$-measurable selection $g:T\to Y$ for $G$ such that
$v^*(t)=v(t,g(t))$ for every $t\in T$.

Theorems of the above type are also known as
\underline{Dubins}-\underline{Sava}g\underline{e}
\underline{Selection} \underline{Theorems} in the dynamic
programming literature.

\underline{Theorem} (Effros): Every lower-semicontinuous or
upper-semicontinuous partition $\mathcal{D}$ of a Polish space $X$
into closed sets admits a Boreal measurable section $f:X\to X$ of
class $2$. In particular, they admit a $G_{\delta}$ cross section.

\underline{Effros}-\underline{Macke}y \underline{Cross}
\underline{Section} \underline{Theorem}: Suppose $H$ is a closed
subgroup of a Polish group $G$ and $\mathcal{D}$ be the partition
of $G$ consisting of all the right cosets of $H$. Then
$\mathcal{D}$ admits a Borel measurable section of class $2$. In
particular, it admits a $G_{\delta}$ cross section.

Every Borel measurable partition $\mathcal{D}$ of a Polish space
$X$ into closed sets admits a Borel measurable section $f:X\to X$.
In particular, it admits a Borel cross section. This is one of the
most frequently used cross section theorems.

\underline{Miller}'s \underline{Theorem}: Let $(G,\diamond)$ be a
Polish group, $X$ Polish, and $a(g,x)=g\diamond x$ an action of
$G$ on $X$. Suppose for a given $x\in X$ that $g\to g \diamond x$
is Borel. Then \underline{the orbit} $\{g\diamond x:g\in G\}$ of
$x$ is Borel.

If a section is measurable with respect to all continuous
probability measures then it is called \underline{universall}y
\underline{measurable}.

Let $X$, $Y$ be Polish spaces, $B\subseteq X\times Y$ Borel, and
$C$ an analytic uniformization of $B$. Then $C$ is Borel.

\underline{Von} \underline{Neumann}'s \underline{Theorem}: Let
$X$, $Y$ be Polish spaces, $A\subseteq X\times Y$ analytic, and
$\mathcal{A}=\sigma\left(\mathbf{\Sigma}^1_1(X)\right)$ - the
$\sigma$-algebra generated by the analytic subsets of $X$. Then
there exists an $\mathcal{A}$-measurable section $u:\pi_X(A)\to Y$
of $A$.

Every analytic subset $A$ of the product of Polish spaces $X$, $Y$
admits a section $u$ that is universally measurable as well as
Baire measurable. Furthermore, if $A$ is Borel, then the graph of
the section $u$ is co-analytic.

Note that a 1-1 Borel map defined on a co-analytic set need not be
a Borel isomorphism, although those with domain analytic are.

Let $X$, $Y$ be Polish spaces and $f:X\to Y$ Borel. Then there is
a co-analytic set $C\subseteq X$ such that $f|C$ is 1-1 and
$f(C)=f(X)$.

Let $(X,\mathcal{E})$ be a measurable space with $\mathcal{E}$
closed under the Souslin operation, $Y$ a Polish space, and $A\in
\mathcal{E}\bigotimes\mathcal{B}_Y$. Then $\pi_X(A)\in
\mathcal{E}$, and there is an $\mathcal{E}$-measurable section of
$A$.

If $(X,\mathcal{A},P)$ be a complete probability space, $Y$ a
Polish space, and $B\in\mathcal{A}\bigotimes\mathcal{B}_Y$, then
$\pi_X(B)\in\mathcal{A}$, and $B$ admits an
$\mathcal{A}$-measurable section. This is essentially the form in
which Von Neumann proved his theorem originally.

\underline{Bur}g\underline{ess}'s \underline{Theorem}: Let a
Polish group $G$ act continuously on a Polish space $X$, inducing
an equivalence relation $E_G$. Suppose $E_G$ is countably
separated. Then it admits a Borel cross section.

A \underline{Lar}g\underline{e} \underline{Section}
\underline{Condition} is the one where sections \underline{do}
\underline{not} \underline{belon}g \underline{to a
$\sigma$}-\underline{ideal} with appropriate computability
property, {\em eg}, the $\sigma$-ideal of meager sets or the
$\sigma$-ideal of null sets. A \underline{Small}
\underline{Section} \underline{Condition} is the one for which the
sections \underline{do} \underline{belon}g \underline{to a
$\sigma$}-\underline{ideal} with appropriate computability
property, {\em eg}, the $\sigma$-ideal of of countable sets or the
$\sigma$-ideal of $K_{\sigma}$ sets.

\underline{Theorem} (Novikov): Let $X$, $Y$ be Polish spaces and
$\mathcal{A}$ a countably generated sub $\sigma$-algebra of
$\mathcal{B}_X$. Suppose $B\in\mathcal{A}\bigotimes\mathcal{B}_Y$
is such that the sections $B_x$ are compact. Then $\pi_X(B)\in
\mathcal{A}$, and $B$ admits an $\mathcal{A}$-measurable section.

\underline{Theorem} (Lusin): Let $X$, $Y$ be Polish spaces and $B
\subseteq X\times Y$ Borel with sections $B_x$ countable. Then $B$
admits a Borel uniformization.

Let $X$ be Polish and $\mathcal{D}$ a countably separated
partition of $X$ with all equivalence classes countable. Then
$\mathcal{D}$ admits a Borel cross section.

Clearly, Novikov's theorem and Lusin's theorem are uniformization
theorems for Borel sets with small sections.

Let $X$, $Y$ be Polish. A map $\mathcal{I}:X\to \mathcal{P} \left(
\mathcal{P}(Y)\right)$ is called \underline{Borel} \underline{on}
\underline{Borel} if for every Borel $B\subseteq X\times Y$, the
set $\{x\in X:B_x\in \mathcal{I}(x)\}$ is Borel.

Some of the important Borel on Borel maps are:
\begin{description} \item{(i)} Let $P$ be a transition probability on
$X\times Y$ with $X$, $Y$ being Polish. Then the map
$\mathcal{I}(x):X\to\mathcal{P} \left( \mathcal{P} (Y) \right)$
defined by $\mathcal{I}(x)=\{N\subseteq Y:P(x,N)=0 \}$ is Borel on
Borel. \item{(ii)} If $X$, $Y$ are Polish and $\mathcal{I}(x)$ the
$\sigma$-ideal of all meager sets in $Y$, then $\mathcal{I}$ is
Borel on Borel. \item{(iii)} If $X$, $Y$ are Polish and $G:X\to Y$
is a closed-valued Borel measurable multifunction, then define
$\mathcal{I}:X\to \mathcal{P} \left( \mathcal{P}(Y)\right)$ as:
$\mathcal{I}(x)=\{I\subseteq Y: I\,
\mathrm{is\,meager\,in\,}G(x)\}$. Then $\mathcal{I}$ is Borel on
Borel. \end{description}

\underline{Theorem} (Kechris): Let $X$, $Y$ be Polish. Now, assume
that $x\to \mathcal{I}_x$ is a Borel on Borel map assigning to
each $x\in X$ a $\sigma$-ideal $\mathcal{I}_x$ of subsets of $Y$.
Suppose $B\subseteq X\times Y$ is a Borel set such that for every
$x\in\pi_X(B)$, $B_x\notin\mathcal{I}_x$. Then $\pi_X(B)$ is
Borel, and $B$ admits a Borel section.

\underline{Theorem} (Kechris-Sarbadhikari): If $B$ is a Borel
subset of the product of two Polish spaces $X$, $Y$ with $B_x$
non-meager in $Y$ for every $x\in\pi_X(B)$, then $B$ admits a
Borel uniformization.

Thus, every Borel set $B\subseteq X\times Y$ with $B_x$ a dense
$G_{\delta}$ set admits a Borel uniformization.

\underline{Theorem} (Blackwell and Ryll-Nardzewski): Let $X$, $Y$
be Polish spaces, $P$ a transition probability on $X\times Y$, and
$B\subseteq X\times Y$ Borel with $P(x,B_x)>0$ for all $x\in
\pi_X(B)$. Then $\pi_X(B)$ is Borel, and $B$ admits a Borel
uniformization.

Let $X$, $Y$ be Polish, $\mathcal{A}$ a countably generated sub
$\sigma$-algebra of $\mathcal{B}_X$, and $P$ a transition
probability on $X\times Y$ with $x\to P(x,B)$ as
$\mathcal{A}$-measurable for every $B\in\mathcal{B}_Y$. For every
$E\in\mathcal{A}\bigotimes\mathcal{B}_Y$ and every $\epsilon
>0$, there is an $F\in\mathcal{A}\bigotimes\mathcal{B}_Y$
contained in $E$ such that $F_x$ is compact and $P(x,F_x)\geq
\epsilon.P(x,E_x)$.

\underline{Theorem} (Blackwell and Ryll-Nardzewski): Let $X$, $Y$
be Polish, $\mathcal{A}$ a countably generated sub
$\sigma$-algebra of $\mathcal{B}_X$, $P$ a transition probability
on $X\times Y$ with $x\to P(x,B)$ as $\mathcal{A}$-measurable for
every $B\in\mathcal{B}_Y$. Suppose $B\in\mathcal{A}
\bigotimes\mathcal{B}_Y$ is such that $P(x,B_x)>0$ for all $x\in
\pi_X(B)$. Then $\pi_X(B)\in\mathcal{A}$, and $B$ admits an
$\mathcal{A}$-measurable section.

\underline{Theorem} (Lusin): If $X$ and $Y$ are Polish and $B$ a
Borel set with $B_x$ countable, then $B$ is a countable union of
Borel graphs.

A subset $A$ of $X$ is called a p\underline{artial}
\underline{cross} \underline{section} if $A\bigcap C$ is at most a
singleton for every member $C$ of the partition $\mathcal{D}$ of
$X$.

If $\mathcal{D}$ is a countably separated partition of Polish $X$
into countable sets, then there is a sequence $(G_n)$ of partial
Borel cross sections of $\mathcal{D}$ with $\bigcup_n G_n=X$ and
if $G_n$ and $G_m$ are distinct, then $G_n\bigcup G_m$ is not a
partial cross section.

Let $X$ be Polish and $G$ a group of Borel automorphisms on $X$,
{\em ie}, each member of $G$ is a Borel isomorphism of $X$ onto
itself and $G$ is a group under composition. Define $xE_G y
\Leftrightarrow (\exists g\in G)(y=g(x))$. Then $E_G$ is an
equivalence relation on $X$. $E_G$ is called the
\underline{e}q\underline{uivalence} \underline{relation}
\underline{induced} \underline{b}y $G$. It is clearly analytic,
and Borel if $G$ is countable. The converse of this result also
holds.

Every Borel equivalence relation on a Polish space $X$ with its
equivalence classes being countable is seen to be induced by a
countable group of Borel automorphisms.

\underline{Theorem} (Miller): Every partition $\mathcal{D}$ of a
Polish space $X$ into $G_{\delta}$ sets such that the saturation
of every basic open set is simultaneously $F_{\sigma}$ and
$G_{\delta}$ admits a section $s:X\to X$ that is Borel measurable
of class $2$. In particular, such partitions admit a $G_{\delta}$
cross section.

\underline{Theorem} (Srivastava): Every Borel measurable partition
$\mathcal{D}$ of a Polish space $X$ into $G_{\delta}$ sets admits
a Borel cross section.

Let $X$ be a Polish space and $\Phi\subseteq \mathcal{P}(X)$. We
then \underline{sa}y \underline{that} $\Phi$ is $\mathbf{\Pi}^1_1$
\underline{on} $\mathbf{\Pi}^1_1$ if for every Polish space $Y$
and every $\mathbf{\Pi}^1_1$ subset $D$ of $Y\times X$, $\{y\in
Y:D_y\in\Phi\}\in \mathbf{\Pi}^1_1$.

\underline{The} \underline{Reflection} \underline{Theorem}: Let
$X$ be Polish and $\Phi\subseteq\mathcal{P}(X)$ $\mathbf{\Pi}^1_1$
on $\mathbf{\Pi}^1_1$. For every $\mathbf{\Pi}^1_1$ set $A\in\Phi$
there is a Borel $B\subseteq A$ in $\Phi$.

If $X$, $Y$ are Polish and $A\subseteq X\times Y$ analytic with
sections $A_x$ countable, then every co-analytic set $B$
containing $A$ contains a Borel set $E\supseteq A$ with all
sections countable.

\underline{Theorem} (Lusin): Every analytic set with countable
sections, in the product of two Polish spaces, can be covered by
countably many Borel graphs.

Let $X$ be Polish, $E$ an analytic equivalence relation on $X$,
and $C\subseteq X\times X$ a co-analytic set containing $E$. Then
there is a Borel equivalence relation $B$ such that $E\subseteq B
\subseteq C$.

Let $X$ be Polish, $P$ analytic, $C$ co-analytic, and $\mathcal{E}
(P) \subseteq C$. Then there is a Borel set containing $P$ such
that $\mathcal{E}(B)\subseteq C$.

For every analytic equivalence relation $E$ on a Polish space $X$
there exist Borel equivalence relations $B_{\alpha}$, $\alpha<
\omega_1$, such that $E=\bigcap_{\alpha<\omega_1}B_{\alpha}$.

Let $X$ be Polish and $\mathcal{C}$ be a sub $\sigma$-algebra of
the Borel $\sigma$-algebra $\mathcal{B}_X$. A \underline{weak}
\underline{com}p\underline{lement} of $\mathcal{C}$ is another sub
$\sigma$-algebra $\mathcal{D}$ of $\mathcal{B}_X$ such that
$\mathcal{C}\bigvee\mathcal{D}=\mathcal{B}_X$ where $\mathcal{C}
\bigvee \mathcal{D}=\sigma\left(\mathcal{C}\bigcup\mathcal{D}
\right)$. A weak complement $\mathcal{D}$ is called as
\underline{minimal} if no proper sub $\sigma$-algebra is a weak
complement.

A \underline{com}p\underline{lement} of $\mathcal{C}$ is a sub
$\sigma$-algebra $\mathcal{D}$ such that $\mathcal{C}\bigvee
\mathcal{D}=\mathcal{B}_X$ and $\mathcal\bigcap\mathcal{D} =
\{\emptyset, X\}$.

If $X$ is Polish and $\mathcal{C}$ is a countably generated sub
$\sigma$-algebra of $\mathcal{B}_X$, then every weak complement of
$\mathcal{C}$ contains a countably generated weak complement.

If $X$ is Polish, $\mathcal{C}\subseteq \mathcal{B}_X$, and if
$\mathcal{D}$ is a minimal weak complement, then $\mathcal{C}
\bigcap\mathcal{D}=\{\emptyset,X\}$, {\em ie}, $\mathcal{D}$ is
also a complement.

Let $X$ be an uncountable Polish space, then the
countable-cocountable $\sigma$-algebra does not have a complement.

\underline{Theorem}: Every countably generated sub
$\sigma$-algebra of the Borel $\sigma$-algebra of a Polish space
has a minimal complement.

Let $X$ be Polish and $\mathcal{C}$ a countably generated sub
$\sigma$-algebra of $\mathcal{B}_X$. Suppose $\mathcal{D}$ is a
countably generated sub $\sigma$-algebra of $\mathcal{B}_X$ such
that every atom $A$ of $\mathcal{D}$ is a partial cross section of
the atoms of $\mathcal{C}$. Further, assume that for any two
distinct atoms $C_1, C_2$ of $\mathcal{D}$, $C_1\bigcup C_2$ is
not a partial cross section of the set of atoms of $\mathcal{C}$.
Then $\mathcal{D}$ is a minimal complement of $\mathcal{C}$.

\underline{Uniformization} \underline{Theorem} (Arsenin \&
Kunugui): Let $B\subseteq X\times Y$ be a Borel set, $X$ and $Y$
being Polish, such that $B_x$ is $\sigma$-compact for every $x$.
Then $\pi_X(B)$ is Borel, and $B$ admits a Borel uniformization.

\underline{Theorem} (Saint Raymond): Let $X$, $Y$ be Polish spaces
and $A,B\subseteq X\times Y$ be analytic sets. Assume that for
every $x$, there is a $\sigma$-compact set $K$ such that $A_x
\subseteq K \subseteq B^c_x$. Then there exists a sequence of
Borel sets $(B_n)$ such that the sections $(B_n)_x$ are compact,
$A\subseteq \bigcup_nB_n$ and $B\bigcap\bigcup_nB_n=\emptyset$.

Let $X$, $Y$ be Polish spaces and $A\subseteq X\times Y$ a Borel
set with sections $A_x$ $\sigma$-compact. Then $A=\bigcup_nB_n$,
where each $B_n$ is Borel with $(B_n)_x$ compact for all $x$ and
for all $n$.

Let $B\subseteq X\times Y$ be a Borel set with sections $B_x$ that
are $G_{\delta}$ sets in $Y$. Then there exist Borel sets $B_n$
with open sections such that $B=\bigcap_nB_n$.

Let $B\subseteq X\times Y$ be a Borel set with sections $B_x$ that
are $F_{\sigma}$ sets in $Y$. Then there exist Borel sets $B_n$
with closed sections such that $B=\bigcup_nB_n$.

Now, recall that the family $F(X)$ of all closed subsets of a
Polish space $X$ with the Effros Borel structure is a Standard
Borel Space. A family $\mathcal{B}\subseteq F(X)$ is called
\underline{hereditar}y if whenever $A\in\mathcal{B}$ and $B$ is a
closed subset of $A$, then $B\in\mathcal{B}$.

A \underline{derivative} on $X$ is a map $D:F(X)\to F(X)$ such
that for $A,B\in F(X)$ (i) $D(A)\subseteq A$, and (ii) $A\subseteq
B \Rightarrow D(A)\subseteq D(B)$.

Some examples of derivatives are: \begin{description} \item{(i)}
Let $\mathcal{B}\subseteq F(X)$ be hereditary. Define
$D_{\mathcal{B}}(A)=\{x\in X:(\forall\; \mathrm{open}\,U\ni
x)(\mathrm{cl}(A\bigcap U)\notin \mathcal{B} )\}$. Then
$D_{\mathcal{B}}$ is a derivative on $X$. Note that if
$\mathcal{B}$ consists of sets with at most one point,
$D_{\mathcal{B}}(A)$  is the usual derived set of $A$. \item{(ii)}
Another important example is obtained by considering $\mathcal{B}$
to be the family of all compact subsets of $X$. \end{description}

Note the following property of $D_{\mathcal{B}}$: Let
$\mathcal{B}$ be hereditary $\mathbf{\Pi}^1_1$. Then the set
$\{(A,B)\in F(X)\times F(X):A\subseteq D_{\mathcal{B}}(B)\}$ is
analytic.

Let $X$ be Polish, $\mathcal{D}:F(X)\to F(X)$ a derivative on $X$,
$A\subseteq X$ closed, and $\alpha$ any countable ordinal. Define
$D^{\alpha}(A)$ by induction on $\alpha$ as follows:

$\hspace{.3in}D^0(A)=A$

$\hspace{.3in}D^{\alpha}(A)=D(D^{\beta}(A))$, if $\alpha=\beta+1$,
and

$\hspace{.3in}D^{\alpha}(A)=\bigcap_{\beta<\alpha}D^{\beta}(A)$,
if $\alpha$ is limit.

Hence, $\{D^{\alpha}(A):\alpha<\omega_1\}$ is a non-decreasing
transfinite sequence of closed sets. Then there is an $\alpha<
\omega_1$ such that $D^{\alpha}(A)=D^{\alpha+1}(A)$. The least of
such $\alpha$ will be denoted by $|A|_D$. We then set

$\hspace{.3in}D^{\infty}(A)=D^{|A|_D}(A)$ and

$\hspace{.3in}\Omega_D=\{A\in F(X):D^{\infty}(A)=\emptyset\}$.

Now, let $X$ be a Polish space and $\mathcal{B}\subseteq F(X)$
hereditary. Then $\Omega_{D_{\mathcal{B}}}=\mathcal{B}_{\sigma}
\bigcap F(X)$.

Let $X$ be a Polish space and $D$ a derivative on $X$ such that
$\{(A,B) \in F(X)\times F(X):A\subseteq D(B)\}$ is analytic. Then
we have that \begin{description} \item{(i)} $\Omega_D$ is
co-analytic, and \item{(ii)} for all analytic
$\mathcal{A}\subseteq \Omega_D$, $\sup\left\{ |A|_D: A\in
\mathcal{A}\right\} < \omega_1$. \end{description}

Let $\mathcal{F}\subseteq F\left(\mathbb{N}^{\mathbb{N}}\right)$
be a hereditary $\mathbf{\Pi}^1_1$ family. Suppose $X$ is a Polish
space and $H\subseteq X\times\mathbb{N}^{\mathbb{N}}$ a closed set
such that $H_x\in\mathcal{F}_{\sigma}$. Then there exists a
sequence $(H_n)$ of Borel sets such that $H=\bigcup_nH_n$ and
$(H_n)_x\in\mathcal{F}$ for all $x$.
\goodbreak

Now, every countably separated partition of a Polish space into
$\sigma$-compact sets admits a Borel cross section.

Let $X$, $Y$ be two Polish spaces and $A,B$ two disjoint analytic
subsets of $X\times Y$ such that $A_x$ is closed and nowhere dense
for all $x$. Then there exists a Borel $C\subseteq X\times Y$ such
that the sections $C_x$ are closed and nowhere dense, and such
that $A\subseteq C$ and $C\bigcap B=\emptyset$.

Let $X$, $Y$ be Polish spaces and $A,B$ disjoint analytic subsets
of $X\times Y$. Assume that the sections $A_x$ are meager in $Y$.
Then there is a sequence $(C_n)$ of Borel sets with sections
nowhere dense such that $A\subseteq \bigcup_nC_n$ and $\left(
\bigcup_nC_n\right)\bigcap B =\emptyset$.

For every Borel set $B\subseteq X\times Y$ with sections $B_x$
co-meager in $Y$, there is a sequence $(B_n)$ of Borel sets such
that $(B_n)_x$ is dense and open for every $x$ and $\bigcap B_n
\subseteq B$.

Let $X$, $Y$ ne Polish spaces. For $1\leq\alpha<\omega_1$, let
$\mathcal{F}_{\alpha}$ denote the family of all Borel subsets of
$X\times Y$ with $x$-sections of multiplicative class $\alpha$ and
let $\mathcal{G}=\neg\mathcal{F}_{\alpha}$. By transfinite
induction, we define families $\mathbf{\Sigma}^*_{\alpha}$,
$\mathbf{\Pi}^*_{\alpha}$ of subsets of $X\times Y$ as follows.
Take $\mathbf{\Pi}^*_0$ to be the subsets of $X\times Y$ of the
form $B\times V$, $B$ Borel and $V$ open. For $\alpha>0$, set
$\mathbf{\Sigma}^*_{\alpha}=\left( \bigcup_{\beta<\alpha}
\mathbf{\Pi}^*_{\beta}\right)_{\sigma}$ and
$\mathbf{\Pi}^*_{\alpha}=\neg\mathbf{\Sigma}^*_{\alpha}$.

Clearly, $\mathbf{\Sigma}^*_{\alpha}\subseteq
\mathcal{G}_{\alpha}$ and $\mathbf{\Pi}^*_{\alpha} \subseteq
\mathcal{F}_{\alpha}$. Note that $\mathbf{\Pi}^*_2=\mathcal{G}_2$
and $\mathbf{\Sigma}^*_2=\mathcal{F}_2$. Furthermore,
$\mathbf{\Sigma}^*_1$ is precisely the family of all Borel sets
with sections that are open.

\underline{Louveau}'s \underline{Theorem}: For every $1\leq\alpha
<\omega_1$, $\mathbf{\Sigma}^*_{\alpha}=\mathcal{F}_{\alpha}$.

\underline{Theorem} (Becker-Kechris): Suppose a Polish group $G$
acts continuously on a Polish space $X$ and $A$ is an invariant
Borel subset of $X$. Then there is a finer Polish topology on $X$
making $A$ clopen with the action still continuous.

\underline{Theorem} (Becker-Kechris): Suppose a Polish group $G$
acts on a Polish space and the action is Borel. Then there is a
finer topology on $X$ making the action continuous.

\underline{Weak} \underline{To}p\underline{olo}g\underline{cal}
\underline{Vau}g\underline{ht} \underline{Con}j\underline{ecture}:
Suppose a Polish group $G$ acts continuously on a Polish space
$X$. Then, under Cantor's Continuum Hypothesis, the number of
orbits in $\leq \aleph_o$ or equals $2^{\aleph_o}$.

Let $E$ be an equivalence relation on a Polish space $X$. In this
case, we then say that $E$ \underline{has} p\underline{erfectl}y
\underline{man}y \underline{e}q\underline{uivalence}
\underline{classes} if there is a nonempty, perfect subset of $X$
consisting of pair-wise $E$-inequivalent elements.

\underline{To}p\underline{olo}g\underline{ical}
\underline{Vau}g\underline{ht} \underline{Con}j\underline{ecture}:
Suppose a Polish group $G$ acts continuously on a Polish space
$X$. Then the number of equivalence classes is countable or
perfectly many.

\underline{Theorem} (Burgess): Suppose $E$ is an analytic
equivalence relation on a Polish space $X$. Then the number of
equivalence classes is $\leq\aleph_1$ or perfectly many.

The topological Vaught conjecture is equivalent to the following
statement: Suppose $G$ is a Polish group acting on a Standard
Borel Space $X$ and the action is Borel. Then the number of orbits
is $\leq\aleph_0$ or perfectly many.

\underline{Theorem}: The topological Vaught Conjecture holds if
$G$ is a locally compact Polish group.

Suppose $X$ is a Polish space and $E$ an equivalence relation on
$X$ which is meager in $X\times X$, Then $E$ has perfectly many
equivalence classes.

\underline{Theorem} (Stern): Let $E$ be an analytic equivalence
relation on a Polish space $X$ with all equivalence classes
$F_{\sigma}$. Then the number of equivalence classes is
$\leq\aleph_0$ or perfectly many.

\underline{Silver}'s \underline{Theorem}: Suppose $E$ is a
co-analytic equivalence relation on a Polish space $X$. Then the
number of equivalence classes is countable or perfectly many.

Suppose $\{A_{\alpha}:\alpha<\omega_1\}$ is a family of Borel
subsets of a Polish space $X$ and $E$ is the equivalence relation
defined on $X$ as $xEy \Leftrightarrow \forall\alpha(x\in
A_{\alpha} \Leftrightarrow y\in A_{\alpha}),\,x,y\in X$. Then the
number of $E$-equivalence classes is $\leq \aleph_1$ or perfectly
many.

Suppose $Z$ is a subset of a Polish space $X$ of cardinality $>
\aleph_1$ such that no two distinct elements of $Z$ are
$E$-equivalent. Then there exists an $\alpha<\omega_1$ such that
both $Z\bigcap A_{\alpha}$ and $Z\bigcap A^c_{\alpha}$ are of
cardinality $>\aleph_1$.

Note that the orbit $\{gx:g\in G\}$ of every point $x$ of a Polish
space $X$ under a continuous action of a Polish group $G$ is
Borel. So the equivalence relation $E_a$ on $X$ induced by the
action is analytic with all equivalence classes Borel.

\underline{Theorem} (Stern): Let $E$ be an analytic equivalence
relation on a Polish space $X$ such that all but countably many
equivalence classes are $F_{\sigma}$ or $G_{\delta}$. Then the
number of equivalence classes is $\geq\aleph_0$ or perfectly many.

\underline{Theorem} (Stern): Assume analytic determinacy. Let $E$
be an analytic equivalence relation on a Polish space $X$ such
that all but countably many equivalence classes are of bounded
Borel rank. Then the number of equivalence classes is $\leq
\aleph_0$ or perfectly many.

In closing this rapid survey of some mathematical results, we also
note the following.

Let $A$ be a subset of a Polish space $X$. A \underline{scale}
\underline{on} \underline{$A$} is a sequence of norms $\varphi_n$
on $A$ such that $x_i\in A$, $x_i\to x$, and $\forall
n(\varphi_n(x_i) \to \mu_n)$, {\em ie}, $\varphi_n(x_i)$ is
eventually constant and equals $\mu_n$ after a certain stage,
imply that $x\in A$ and $\forall n$ $( \varphi_n(x)\leq\mu_n)$.

If for each $n$, $\varphi_n:A\to\kappa$, then we say that
$(\varphi_n)$ is a $\kappa$-scale.

Given some ordinal $\kappa$, let the
\underline{lexico}g\underline{ra}p\underline{hical}
\underline{orderin}g $<_{\mathrm{lex}}$ be defined on $\kappa^n$
by the following. Let $\left(\mu(0),\mu(1),...,\mu(n-1)\right)$
$<_{\mathrm{lex}}$ $\left( \lambda(0), \lambda(1),\right.$
$\left....,\lambda(n-1)\right)$ $\Leftrightarrow \exists$
$i<n\,\left[ \forall\, j<i\,\left(\mu(j)=\lambda(j)\right)\right.$
$\left.\,\&\,\left(\mu(i)<\lambda(i)\right)\right]$.

The lexicographical ordering is a well-order with order type
$\kappa^n$. Denote by $\langle\mu(0),\mu(1),...,$
$\mu(n-1)\rangle$ the ordinal $<\kappa^n$ corresponding to $\left(
\mu(0),\mu(1)\right.$ $\left.,...,\mu(n-1)\right)$ under the
isomorphism of $(\kappa^n,<_{\mathrm{lex}})$ with $\kappa^n$.

Now, note that given some scale $(\varphi_n)$ on $A \subseteq
\mathbb{N}^{\mathbb{N}}$, we can always define a new scale as
follows $\psi_n(\alpha)=$ $\langle\psi_0(\alpha), \alpha(0),
\varphi_1 (\alpha), \alpha(1), ..., \varphi_n(\alpha), \alpha(n)
\rangle$ The scale $(\psi_n)$ has the following additional
properties \begin{description} \item{(i)} $\psi_n(
\alpha)\leq\psi_n(\beta)\Rightarrow \forall\,m\leq\,n\,$ $\left(
\psi_m(\alpha)\leq \right.$ $\left. \psi_m(\beta)\right)$, and
\item{(ii)} If $\alpha_i \in A$ $\psi_n(\alpha_i)\to\mu_n$, then
$\alpha_i\to\alpha$ for some $\alpha\in A$. \end{description}

Let $A$ be a subset of a Polish space $X$. A scale $(\varphi_n)$
defined on $A$ is called as a \underline{ver}y g\underline{ood}
\underline{scale} if \begin{description} \item{(i)}
$\varphi_n(x)\leq\varphi_n(y) \Rightarrow \forall\,m\leq
n\,(\varphi_m(x)\leq\varphi_m(y))$, \item{(ii)} If $x_i\in A$ and
$\varphi_n(x_i)\to\mu_n$ for all $n$, then $x_i\to x$ for some
$x\in A$.\end{description}

Given a very good scale $(\varphi_n)$ on $A$, we can then
\underline{select} \underline{a} \underline{uni}q\underline{ue}
p\underline{oint} from $A$ as follows. Let $A_0=\{x\in A:
\varphi_0(x)\,\mathrm{is\,least, \, say\,}\mu_0\}$, $A_1=\{x\in A:
\varphi_1(x)\,\mathrm{is\,least, \, say\,}\mu_1\}$, $A_2=\{x\in A:
\varphi_2(x)\,\mathrm{is\,least, \, say\,}\mu_2\}$, and so on.
Thus, we have $A_0\supseteq A_1\supseteq A_2 \supseteq ...$ and if
$x_i\in A_i$, then $\varphi_n(x_i)=\mu_n$ for all $i>n$. Since
$(\varphi_n)$ is a very good scale, there is an $x\in A$ such that
$x_i\to x$. Moreover, $x\in A_n$ for all $n$.

Now, let $y$ be any other point in $\bigcap_n A_n$. Consider the
sequence $x,y,x,y,...$ which is convergent as $(\varphi_n)$ is a
very good scale. Therefore, $x=y$. Thus $\bigcap_n A_n$ is a
singleton. The above procedure then selects a unique point from
$A$, called as the \underline{canonical} \underline{element}
\underline{of} $A$ \underline{determined} \underline{b}y
$(\varphi_n)$.

A scale $(\varphi_n)$ on a co-analytic subset $A$ of a Polish
space $X$ is called as a $\mathbf{\Pi}^1_1$-\underline{scale} if
each $\varphi_n$ is a $\mathbf{\Pi}^1_1$-norm on $X$.

If $(\varphi_n)$ is a $\mathbf{\Pi}^1_1$-scale on a co-analytic
$A\subseteq \mathbb{N}^{\mathbb{N}}$, then $(\varphi_n)$ defined
as $\psi_n$ is also $\mathbf{\Pi}^1_1$-scale.

\underline{Theorem}: Every co-analytic subset of
$\mathbb{N}^{\mathbb{N}}$ admits a very good
$\mathbf{\Pi}^1_1$-scale.

As a corollary of the above, we have the result: Let $X$ be a
Polish space and $A\subseteq X$ co-analytic. Then $A$ admits a
very good $\mathbf{\Pi}^1_1$-scale.

\underline{Kondo}'s \underline{Theorem}: Let $X$, $Y$ be Polish
spaces. Every co-analytic set $C\subseteq X\times Y$ admits a
co-analytic uniformization.

\subsubsection*{Categorical Matters}

Now, a \underline{cate}g\underline{or}y $\mathfrak{C}$ is a
structure comprising the following mathematical data
\begin{description}
\item{(C-i)} a class whose members $A$, $B$, ... are called
\underline{ob}j\underline{ects} of $\mathfrak{C}$, \item{(C-ii)}
for each pair of objects $A$, $B$, there is given a set
$\mathfrak{C}(A,B)$, called the set of
\underline{mor}p\underline{hisms} \underline{from} \underline{$A$}
\underline{to} \underline{$B$}: we write $f:A\to B$ to indicate
that $f\in\mathfrak{C}(A,B)$, \item{(C-iii)} for each triple of
objects $A$, $B$, $C$, a law of composition $\mathfrak{C}(A,B)
\times \mathfrak{C}(B,C)\to\mathfrak{C}(A,C)$ is well defined for
morphisms \end{description} which is subject to the following two
axioms \begin{description}
\item{(C-iv)} \underline{Associativit}y:
If $f:A\to B$, $g:B\to C$ and $h:C\to D$, then $(fg)h=f(gh)$.
\item{(C-v)}  \underline{Identities}:
For each $A\in\mathfrak{C}$, there exists a morphism
$e_A\in \mathfrak{C}(A,A)$ such that for all $f:A\to B$, $e_Af=f$,
and for all $g:C\to A$, $ge_A=g$.
\end{description} [We have adopted a right-handed notation
for representing morphisms. Many authors use left-handed one or
some other notation.]

As can be established, the class of all topological spaces forms a
category with (categorical) morphisms as continuous maps between
them.

Then, consider a class of standard Borel spaces. Since each
standard Borel space is also a topological space, these spaces
form a \underline{cate}g\underline{or}y that satisfies conditions
(C-i) to (C-v) and an additional mathematical condition, that of
each of its objects being isomorphic to some Borel subset of a
Polish space. This additional condition is easily seen to be
compatible with the conditions (C-i) to (C-v) of this class being
a category.

We then have the Category of Standard Borel Spaces with continuous
maps being (categorical) morphisms between objects of this
category.

Then, each member of the category of standard Borel spaces is
``related'' to another of its members by a categorical morphism.
In particular, it will be related to $\mathbb{R}^3$ [See also
\cite{rylov}.].

We have summarized above relevant definitions and results (without
proofs). We note, in advance, that the {\em physical space\/} of
Universal Relativity will be taken as a specific Standard Borel
Space (of cardinality $\mathbf{c}$ or $\aleph_1$). Many of the
aforementioned results about measurable sets, Borel point-classes,
measurable partitions, group of Borel automorphisms etc.\ will
then be relevant.

Furthermore, results related to countable sets will be relevant to
us when we will define appropriate notion of a point-object. It is
then the reason why some of the results related to countable sets
have also been mentioned above.

\subsection*{Dynamical systems}
Differential equations of classical mechanics led to the study of
(the generalized concept of) dynamical systems.

There are the following three major situations for the
mathematical analysis of the evolution of the (topological) space
$X$, namely,
\begin{description}
\item{(i)} $X$ is a topological space and $T$ is a homeomorphism.
Such studies are called by the name of
\underline{To}p\underline{olo}g\underline{ical}
\underline{D}y\underline{namics}.
\item{(ii)} $X$ is a measure space and $T$ is
a measure-preserving transformation. Related studies are called
the \underline{Measurable} \underline{D}y\underline{namics}.
\item{(iii)} $X$ is a differentiable manifold and $T$ is a diffeomorphism.
Related studies are called by the name of
\underline{Differentiable} \underline{D}y\underline{namics}.
\end{description}

The above three cases of course overlap considerably and it is
possible to switch from one to another situation, as per
convenience. It is often rewarding to view the same example from
all the three perspectives, when permissible.

In the general context of dynamical systems, Poincar$\acute{e}$
pointed out that if a cross section existed for a continuous flow
on a compact manifold, one could, equivalently, study the complete
flow using a homeomorphism of the cross section onto itself
induced by the flow.

Further, Birkhoff, in particular, pointed out the equivalence of
the existence of a global cross section and a flow parameter that
increased along streamlines of the flow. He explicitly showed that
the original flow could be reconstructed with the knowledge of the
cross section, the induced homeomorphism and the value of flow
parameter for first return of points in the cross section.

The theory of dynamical systems is fundamental to the present
studies. Hence, to build the required mathematical vocabulary, we
provide below a rapid survey of its concepts. A knowledgeable
reader may wish to skip it.

If $X$ is a space, its evolution is a transformation $T_t:X\to X$
where $t$ is the parameter labelling the transformation. In
general, we shall be interested in a one-parameter family
$\{T_t:t\in\mathbb{R}\}$ of transformations of $X$ onto itself.

When the laws governing the space $X$ do not change with the
parameter $t$, we have $T_{s+t}=T_sT_t$, in which case, $T_t$ is
called a \underline{flow} or a g\underline{rou}p
\underline{action} \underline{of} \underline{$\mathbb{R}$}
\underline{on} \underline{$X$}, {\em ie}, action of the additive
group of the real line on $X$.

We could, sometimes, also confine ourselves to actions of the
additive group of $\mathbb{Z}$, the set of integers, {\em ie}, to
\underline{iterates} of a single Borel automorphism of a Standard
Borel Space.

Let $(X,\mathcal{B})$ be a Standard Borel Space. Then, $T_t,t\in
\mathbb{R}$ is called as a j\underline{ointl}y
\underline{measurable} \underline{flow} of a Borel automorphism on
$X$ if, for each $t\in\mathbb{R}$, $T_t$ is a Borel automorphism
of $X$ such that
\begin{description}
\item{(1)} the map $(T, x) \mapsto T_t x$ from $\mathbb{R}\times X \to
X$ is measurable, where $\mathbb{R}\times X$ is endowed with the
usual product Borel structure \item{(2)} $T_0x=x \;\forall\;x \in
X$ and, \item{(3)} $T_{t+s}x=T_t \circ T_sx$ for all $t,s \in
\mathbb{R}$ and for all $x\in X$. \end{description}

As noted before, we shall be dealing with a certain standard Borel
space (of cardinality $\mathbf{c}$). Therefore, for convenience, a
one-point compactification of space $X$ as $\hat{X}=X\bigcup\{
\infty \}$ and the extension of the dynamical system $T$ to
$\hat{T}$ in $\mathbb{R}\times \hat{X}$ with
$\hat{T}(t,\infty)=\infty$ for any $t\in \mathbb{R}$ will be
always assumed. We will, however, omit the overhead hat on
relevant quantities.

We shall, generally, refer to the pair $(X,T_t)$ as a
\underline{d}y\underline{namical} \underline{s}y\underline{stem}.

Let us define the following sets under the action of a dynamical
system $T_t$ on $X$:
\begin{itemize}
\item the future limit of a point $x\in X$ as: \\ $\Omega^+(x)= \{ y\in
X:T_tx\to y\; \mathrm{when}\; t\to\infty\}$ \item the past limit
of a point $x\in X$ as: \\ $\Omega^-(x)= \{ y\in X:T_tx\to y\;
\mathrm{when}\; t\to -\infty\}$\end{itemize} A point $x\in X$ is
said to be an \underline{invariant} p\underline{oint} of the
dynamical system if $T_t(\{x\})=\{x\}$ for all $t$. When we only
have $\Omega^+(x) = \Omega^-(x)$, we shall say that $y$ is a
\underline{limit} p\underline{oint} of $T$ and $x$ may ``wander''
in $X$ for the intermediate values of $t$. A point $x$ will be
called a p\underline{oint} of
\underline{as}y\underline{m}p\underline{totic} \underline{rest} of
the dynamical system $T$ if $\Omega^+(x) = \Omega^-(x)=\{x\}$.

In general, we shall also adopt the following notations: for $S
\subset X$ and $I\subset \mathbb{R}$,
\begin{description} \item{$\bullet$} $T(I,S)\equiv T_I(S)=\{T(t,x):t\in I, x\in S\}$,
\item{$\bullet$} $T(S)=T(\mathbb{R},S)$, \item{$\bullet$} $T^+(S)=T([0,\infty), S)$,
\item{$\bullet$} $T^-(S)=T((-\infty,0],S)$. \end{description} Then, $T(I,S)$
is the {\em history of set $S$\/} for some interval of the
parameter $t$, $T(S)$ is the {\em entire history of set $S$},
$T^+(S)$ is the {\em future history of set $S$}, and $T^-(S)$ is
the {\em past history of set $S$}.

Then, a set $S$ is said to be \underline{invariant} under the
dynamical system if $T_t(S)=S,\,\forall\, t\in \mathbb{R}$.

Now, let $(X, \mathcal{B}, \mu)$ be a complete probability space,
{\em ie}, a set $X$ with its $\sigma$-algebra $\mathcal{B}$ of
measurable subsets and a countably additive non-negative set
function $\mu$ on $\mathcal{B}$ with $\mu(X)=1$ and $\mathcal{B}$
containing all subsets of sets of measure zero.

Let $T:X\to X$ be a 1-1 and onto map such that $T$ and $T^{-\,1}$,
both, are measurable, {\em ie}, $T^{-\,1}\mathcal{B}=T\mathcal{B}
=\mathcal{B}$. [Notice that $T$ may be well-defined and one-one,
onto only after a set of measure zero is discarded from $X$.]

Now, let $\mu\left(T^{-\,1}A\right)=\mu\left(A\right)$ for all
$A\in \mathcal{B}$. Such a transformation is called as a
\underline{measure} p\underline{reservin}g
\underline{transformation} (\underline{MPT}).

For our studies related to dynamical systems, a fundamental system
will then be $(X,\mathcal{B},\mu,T)$ with $T$ being a MPT. For the
one-parameter case, we will assume that $T_t$ is a MPT for all
$t\in\mathbb{R}$, the map $(t,x)\mapsto T_tx$ is jointly
measurable from $\mathbb{R}\times X\to X$, $T_0$ being the
identity map.

Now, if $T:X\to X$ is a MPT, then the set $\{ T^n:n\in\mathbb{Z}
\}$ is the \underline{orbit} of a point $x\in X$ representing a
complete history of the system from infinite past to infinite
future values of $t$. The $\sigma$-algebra $\mathcal{B}$ is then
the family of {\em events\/} with $T$-invariant measure $\mu$
specifying the $t$-independent probabilities of the occurrence of
these history events.

A function $f:X\to \mathbb{R}$ on a measurable space $X$ is a
\underline{measurable function} if $\mathfrak{Support}\,(f)\bigcap
f^{-\,1}(M)$ is a measurable set where $M$ is any Borel subset of
the real line $\mathbb{R}$.

\underline{Basic} \underline{Er}g\underline{odic}
\underline{Theorem}: Given a measure preserving transformation $T$
of a probability space $(X,\mathcal{B},\mu)$, let $B\in
\mathcal{B}$ be any measurable set. Define $S_n(x)=\#\{i: 0\leq
i<n, T^ix\in B\}$ and $A_n(x)=S_n(x)/n, x\in X$. Then, for
$\mu$-almost every $x\in X$, there exists $A(x)=\lim_{n\to \infty}
A_n(x)$. Moreover, $\int_XA(x)\,d\mu(x)=\mu(B)$. [The symbol $\#$,
in general, indicates that the quantity under consideration is a
number.]

Consider now those functions $f:X\to\mathbb{R}$ for which
$\int_X|f|^p d\mu$ is defined and is finite. Define the
``distance'' between two functions as $d(f,g)=\left\{
\int_X|f-g|^p d\mu\right\}^{1/p}$. The resulting metric space is
known as the $L^p$-space associated to the measure space
$(X,\mathcal{A},\mu)$ and is denoted by $L^p(X,\mathcal{A},\mu)$.
A classical result, Riesz-Fisher Theorem, proves that $L^p$-spaces
are complete.

\underline{Birkhoff}'s \underline{Ergodic} \underline{Theorem}: If
$f\in L^1(X,\mathcal{A},\mu)$, then $\lim_{n\to\infty}\frac{1}{n}
\sum_{t=0}^{n-1}f\left( T^tx\right)$ exists $\mu$-a.e. and in
$L^1(X, \mathcal{A},\mu)$.

\underline{Kin}g\underline{man}'s \underline{Er}g\underline{odic}
\underline{Theorem}: Let $f_1$, $f_2$, ...$\in
L^1(X,\mathcal{A},\mu)$ be such that $\sup_n\int f_nd\mu> -\infty$
and $f_{n+m}(x)\leq f_n(x)+f_m(T^nx)$ for each $n,m>1$ and
$\mu$-a.e. $x\in X$. Then, $\lim_{n\to\infty}\frac{1}{n}f_n(x)$
exists $\mu$-a.e. and is in $L^1(X,\mathcal{A},\mu)$. [Birkhoff's
theorem treats the case $f_n(x)=\sum_{t=0}^{n-1}f_1(T^tx)$.]

The next important issue for us is that of the isomorphism of
dynamical systems. Let $(X,\mathcal{A},\mu,T)$ and $(X',
\mathcal{A}',\mu',T')$ be two dynamical systems. Then, they are
said to be isomorphic if there exists a map $\varphi:X\to X'$, an
isomorphism, such that \begin{description} \item{(i-1)} the map
$\varphi$ is measurable, \item{(i-2)} for each
$A'\in\mathcal{A}'$, $\mu(\varphi^{-\,1}A')=\mu'(A')$,
\item{(i-3)} for $\mu$-almost every $x\in X$,
$\varphi(Tx)=T'(\varphi x)$, \item{(i-4)} the map $\varphi$ is
invertible, {\em ie}, there exists a measurable and measure
preserving map $\psi:X'\to X$ such that $\psi(\varphi x)=x$ for
$\mu$-almost every $x\in X$ and $\varphi(\psi x')=x'$ for
$\mu'$-almost every $x'\in X'$. \end{description} If only
properties (i-1) to (i-3) hold, $\varphi$ will be called a
\underline{homomor}p\underline{hism} and
$(X',\mathcal{A}',\mu',T')$ is said to be a \underline{factor}
\underline{s}p\underline{ace} \underline{of}
$(X,\mathcal{A},\mu,T)$.

Now, a measure preserving transformation $T$ is said to be
\underline{er}g\underline{odic} if whenever $f:X\to\mathbb{R}$ is
a measurable function such that $f(Tx)=f(x)$ for $\mu$-almost
every $x\in X$, then $f$ is $\mu$-almost everywhere equal to a
constant.

Note that when $T$ is ergodic, $A(x)=\mu(B)$ for $\mu$-a.e. $x\in
X$ in the basic ergodic theorem.

It turns out that a system is ergodic if and only if the orbit of
\underline{almost} \underline{ever}y (\underline{a.e.}) point
$x\in X$ ``visits'' each set of positive measure, that is to say,
if $\mu(A)>0$ and $\mu(B)>0$ then $\mu\left(T^nA\bigcap
B\right)>0$ for some $n\in \mathbb{Z}$.

A \underline{recurrence} p\underline{ro}p\underline{ert}y is that
if $\mu(A)>0$ then $\mu\left(T^nA\bigcap A\right)>0$ for some
$n\in\mathbb{Z}$. A property which implies the ergodicity of a
system is that of \underline{stron}g \underline{mixin}g:
$\lim_{n\to\infty}\mu\left( T^nA\bigcap B \right)=\mu(A)\mu(B)$
for all $A,B\in\mathcal{B}$.

The question of characteristics that are identical for two
systems, the issue of {\em ergodic invariants}, leads us to the
problem of an appropriate \underline{classification}
\underline{of} \underline{s}y\underline{stems}.

Let $(X,\mathcal{B},\mu.T)$ be a system. Further, let $f:X\to (0,
\infty)$ be a measurable function on $X$. Construct a
one-parameter flow $\Upsilon=\{(x,t):0\leq t<f(x)\}$ under the
graph of $f$. Essentially, each point $x\in X$ flows such that we
identify the points $(x,f(x))$ and $(Tx,0)$. This flow $\Upsilon$
preserves the product of $\mu$ with Lebesgue measure and is called
as a \underline{flow} \underline{built} \underline{under}
\underline{the} \underline{function} $f$.

Under suitable conditions, every flow on the system
$(X,\mathcal{B},\mu, T)$ can be represented as flow built under a
function.

The Glimm-Effros Theorem \cite{trim6} states that: If $X$ is a
complete separable metric space and $G$ a group of homeomorphisms
of $X$ onto itself such that for some non-isolated point $x\in X$,
the set $Gx$, the orbit of $x$ under $G$, is dense in $X$, then
there is a continuous probability measure $\mu$ on Borel subsets
of $X$ such that every $G$-invariant Borel set has measure zero or
one.

A group $G$ of homeomorphisms of a Polish space $X$ admits a
\underline{recurrent} p\underline{oint} $x$ if there exists a
sequence $(g_n)_{n=1}^{\infty}$ of elements in $G$ such that
$g_nx\neq x\;\forall\;n$ and $g_nx\to x$ as $n\to\infty$. A
recurrent point $x$ is not isolated in the closure of $Gx$ and its
$G$-orbit is clearly dense in the closure of $Gx$.

A Borel set $W$ is said to be
\underline{$G$}-\underline{wanderin}g if the sets $gW,\;g\in G$
are pairwise disjoint. We \underline{write} $\mathcal{W}_{_G}$ for
the \underline{$\sigma$}-\underline{ideal} generated by
$G$-wandering Borel sets, it consists of {\em countable unions\/}
of $G$-wandering sets of $X$.

Now, a group action is called \underline{free} if, for each $x\in
X$, $g\mapsto gx$ is 1-1. Then, in this case, we have the result
that: If a group of homeomorphisms $G$ of a Polish space $X$ acts
\underline{freel}y and does not admit a recurrent point, then
$X\in \mathcal{W}_{_G}$.

Now, since a SBS is Borel-isomorphic to the Borel space of the
unit interval $X=[0,1]$ equipped with the $\sigma$-algebra
generated by its usual topology, we can restrict our discussion to
it as and when it is convenient.

Then, for any $x\in X$, let the {\em orbit of $x$ under $T$\/} be
the set $\{T^nx|n\in \mathbb{Z}\}$. We call a point $x\in X$ a
p\underline{eriodic} p\underline{oint of $X$} if $T^nx=x$ for some
integer $n$ and call the smallest such integer the
p\underline{eriod of $x$ under $T$}.

For $A\subseteq X$ and $x\in A$, we say that the point $x$ is {\em
recurrent in $A$\/} if $T^nx\in A$ for infinitely many positive
(\underline{fim}p) $n$ and for infinitely many negative
(\underline{fimn}) $n$ and we call the point $x$ a
\underline{recurrent} p\underline{oint}. For a metric space
$(X,d)$, a point $x\in X$ is recurrent if $\liminf_{n\to\infty}
d(x,T^nx)=0$.

Two Borel automorphisms, $T_1$ on a Borel space $(X_1,
\mathcal{B}_1)$ and $T_2$ on a Borel space $(X_2, \mathcal{B}_2)$,
are said to be \underline{isomor}p\underline{hic} if there exists
a Borel isomorphism $\phi: X_1\to X_2$ such that
$\phi\,T_1\,\phi^{-\,1}=T_2$.

We also say that Borel automorphisms $T_1$ and $T_2$ are
\underline{weakl}y \underline{e}q\underline{uivalent} or
\underline{orbit e}q\underline{uivalent} if there exists a Borel
automorphism $\phi: X_1\to X_2$ such that $\phi\left(
\mathrm{orb}(x,T_1)\right)=\mathrm{orb}\left( \phi(x),
T_2\right)$, $\forall$ $\;x\in X_1$.

If two Borel automorphisms are isomorphic then they are also
orbit-equivalent. However, the converse is, in general, not true.

Now, we say that a Borel automorphism $T$ is an
\underline{elementar}y \underline{Borel automor}p\underline{hism}
or that the {\em orbit space of $T$ admits a Borel
cross-section\/} or that \underline{$T$ admits a Borel
cross}-\underline{section} iff there exists a measurable set $B$
which intersects each orbit under $T$ in exactly one point.

Clearly, if $n$ is the period of $x$ under $T$, then the set $\{
x, Tx, T^2x, ..., T^{n-1}x\}$ consists of {\em distinct\/} points
of $X$. Now, for every positive integer $n$, let $E_n=\{ x\, |\,
Tx\neq x, ..., T^{n-1}x\neq x, T^nx=x\}$, and $E_{\infty}=\{x\,|\,
T^nx\neq x \;\mathrm{for \;all\;integers}\;n \}$. Then, each
$E_n,\;n<\infty,$ is Borel, $E_m\bigcap E_n=\emptyset$ if $m\neq
n$, and $\bigcup_{n=1}^{\infty}E_n=X$. Clearly, each $E_n$ is a
$T$-invariant Borel subset of $X$.

Now, if $y\in \{x, Tx, ..., T^{n-1}x\}$, then we clearly see that
$\{ x, Tx, ..., T^{n-1}x\}=\{y, Ty, ..., T^{n-1}y\}$. Moreover,
due to the natural order on $[0,1]$, if $y=\mathrm{min} \{ x, Tx,
..., T^{n-1}x\}$, then $y<Ty$, $y<T^2y$, ..., $y<T^{n-1}y$,
$y=T^ny$. Then, we can define $B_n=\{ y\in E_n \,|\, y<Ty,\,
...,\, y<T^{n-1}y\}$.

Then, for $n<\infty$, $B_n$ is a measurable subset of $E_n$ and it
contains exactly one point of the orbit of each $x\in E_n$. Note,
however, that $B_{\infty}$ need not be measurable.

Now, $X\setminus E_{\infty}=\bigcup_{n=1}^{\infty}
\bigcup_{k=0}^{n-1} T^kB_n$. The set $B=\bigcup_{k=1}^{\infty}
B_k$ is Borel and has the property that the orbit of any point in
$X\setminus E_{\infty}$ intersects $B$ in exactly one point. Let
$\mathbf{c}_n(T)$ denote the cardinality of $B_n$, $n<\infty$. The
sequence of integers $\{\mathbf{c}_{\infty}(T),\,
\mathbf{c}_1(T),\,\mathbf{c}_2(T),\, ...\}$ is called the
\underline{cardinalit}y \underline{se}q\underline{uence associated
to $T$}.

If $T_1$ and $T_2$ are orbit equivalent, then their associated
cardinality sequences are the same. Also, if $T_1$ and $T_2$ are
elementary and the associated cardinality sequences are the same,
then $T_1$ and $T_2$ are isomorphic and orbit equivalent.

A measurable subset $W\subset X$ is
\underline{$T$}-\underline{wanderin}g or \underline{wanderin}g
\underline{under} \underline{$T$} if $T^nW,\;n\in\mathbb{Z}$, are
pairwise disjoint. Clearly, a wandering set intersects the orbit
of any point in at most one point, it never intersects the orbit
of a periodic point.

The $\sigma$-ideal generated by all $T$-wandering sets in
$\mathcal{B}$ will be denoted by $\mathcal{W}_{_T}$ and will be
called a \underline{Shelah}-\underline{Weiss} ideal of $T$
\cite{weiss0}.

Note that if $T$ is a homeomorphism of a separable metric space
$(X,d)$ and $T$ has no recurrent points then $\mathcal{W}_{_T} =
\mathcal{B}$, {\it ie}, there is a wandering set $W$ such that
$X=\bigcup_{n=-\infty}^{\infty}T^nW$.

A subset $A\subset \mathrm{orb}(x,T)$ is called
\underline{bounded} \underline{below} (\underline{bounded}
\underline{above}) if the set of integers $n$ such that $T^nx\in
A$ is bounded below (bounded above). A subset $A\subset
\mathrm{orb}(x,T)$ is called \underline{bounded} iff it is both
bounded above and below. A set which is not bounded is called {\em
unbounded}.

A sufficient condition for a set $N\in \mathcal{B}$ to be a
$T$-wandering set, {\em ie}, a sufficient condition for
$N\in\mathbb{B}$ to belong to $\mathcal{W}_{_T}$, is that
$\forall\; x\in X$, $N\bigcap \mathrm{orb}(x,T)$ is either bounded
above or below.

One of the very basic results of the study of Borel automorphisms
is:

\underline{Poincar\'{e} Recurrence Lemma}: Let $T$ be a Borel
automorphism of a SBS $(X,\mathcal{B})$. Then, given
$A\in\mathcal{B}$ $\exists\;N\in \mathcal{W}_{_T}$ such that
$\forall\;x\in A_o=A\setminus N$ the points $T^nx$ return to $A$
fimp $n$ and fimn $n$.

Now, note also that if $x\in A_o=A\setminus N$ then $T^kx$ returns
to $A_o$ fimp $k$ and fimn $k$ because $N$ is $T$-invariant and
$x\notin N$.

Also, if $A\in \mathcal{B}$, and if $A_o=A\setminus N$ is as in
the Poincar\'{e} Recurrence Lemma, then
$\bigcup_{k=-\infty}^{\infty} T^kA=\bigcup_{k=0}^{\infty}T^kA \;
(\mathrm{mod}\; \mathcal{W}_{_T})$.

Now, suppose that $\mathcal{N}\subseteq \mathcal{B}$ is a
$\sigma$-ideal such that $T\mathcal{N}=T^{-\,1}\mathcal{N} =
\mathcal{N}$ and $\mathcal{W}_{_T}\subseteq \mathcal{N}$. Clearly,
given $A \in \mathcal{B}$, $\exists\;N\in \mathcal{N}$ such that
$\forall\; x\in A_o=A\setminus N$, $T^nx$ returns to $A_o$ fimp
$n$ and fimn $n$.

Of particular interest to us is a finite or $\sigma$-finite
measure $m$ on $\mathcal{B}$. The $\sigma$-ideal of $m$-null sets
in $\mathcal{B}$ will be \underline{denoted b}y $\mathcal{N}_m$.

A Borel automorphism $T$ is \underline{dissi}p\underline{ative}
relative to $m$ if there exists a $T$-wandering set $W$ in
$\mathcal{B}$ such that $m$ is supported on
$\bigcup_{n=-\infty}^{\infty}T^nW$.

On the other hand, a Borel automorphism $T$ is
\underline{conservative} with respect to $m$ or
\underline{$m$}-\underline{conservative} if $m(W)=0$ $\forall$
$T$-wandering sets $W\in\mathcal{B}$. Clearly, for any
$m$-conservative $T$, $\mathcal{W}_{_T}\subseteq \mathcal{N}_m$.

\underline{Poincar\'{e} Recurrence Lemma} \underline{for}
\underline{$m$}-\underline{conservative} \underline{$T$}: If $T$
is $m$-conservative and if $A\in \mathcal{B}$ is given, then
\underline{for} \underline{almost} \underline{ever}y
(\underline{f.a.e.}) $x\in A$ the points $T^nx$ return to $A$ fimp
$n$ and fimn $n$.

Further, if $m$ is a probability measure on $\mathcal{B}$, {\em
ie}, $m(X)=1$, and is $T$-invariant, {\em ie}, $m\circ
T^{-\,1}=m$, then $\mathcal{W}_{_T}\subseteq \mathcal{N}_m$,
$T\mathcal{N}_m =T^{-\,1}\mathcal{N}_m = \mathcal{N}_m$.

\underline{Poincar\'{e} Recurrence Lemma} (\underline{Measure}
\underline{Theor}y): If a Borel automorphism $T$ on
$(X,\mathcal{B})$ preserves a probability measure on
$\mathcal{B}$, and if $A\in \mathcal{B}$ is given, then f.a.e.
$x\in A$ the points $T^nx$ return to $A$ fimp $n$ and fimn $n$.

\underline{Poincar\'{e} Recurrence Lemma} (\underline{Baire
Cate}g\underline{or}y): If $T$ is a homeomorphism of a complete
separable metric space $X$ which has no $T$-wandering non-empty
open set, then for every $A\subseteq X$ with the property of Baire
(in particular, for any Borel set $A$) there exists a set $N$ of
the first Baire category (which is Borel if $A$ is Borel) such
that for each $x\in A\setminus N$, the points $T^nx$ return to
$A\setminus N$ fimp $n$ and fimn $n$.

Now, a measure-preserving automorphism $T$ on a Standard
Probability Space $(X, \mathcal{B}, \mu)$ is a
\underline{Bernoulli}-\underline{Shift} or
\underline{B}-\underline{shift} if there exists a finite or a
countably infinite partition $\mathcal{P}=\{P_1,\;P_2\;...\}$ of
$X$ into measurable sets such that
\begin{description} \item{(a)}
$\bigcup_{n=-\infty}^{\infty}T^n\mathcal{P}$ generates
$\mathcal{B}_X$ up to $\mu$-null sets \item{(b)} the family
$\{T^n\mathcal{P}\,|\, n\in \mathcal{Z}\}$ is independent in the
sense that for all $k$, for all distinct integers $n_1$, $n_2$,
..., $n_k$, and for all $P_1$, $P_2$, ..., $P_{i_{k}}$ $\in
\mathcal{P}$, the sets $T^{n_1}P_{i_{1}}$, $T^{n_2}P_{i_{2}}$,
..., $T^{n_k}P_{i_{k}}$ are independent, {\em ie}, $\mu\left(
T^{n_1}P_{i_{1}}\bigcap ... \bigcap T^{n_k}P_{i_{k}} \right) =
\prod_{j=1}^k \mu\left(T^{n_j}P_{i_{j}}\right)$ which in view of
the measure preserving character of $T$ is equal to
$\mu(P_{i_1})...\mu(P_{i_k})$.\end{description} We call the
partition $\mathcal{P}$ satisfying the above an
\underline{inde}p\underline{endent} g\underline{enerator of $T$}.

$T$ is \underline{$m$}-\underline{deterministic} (otherwise, {\em
non deterministic\/}) if $\forall \; n$,
$\mathcal{P}_n=\mathcal{P}_{n+1}\;(\mathrm{mod}\;m)$ in that,
given $A\in \mathcal{P}_n$, $\exists\;B\in \mathcal{P}_{n+1}$ such
that $m\left( A\triangle B\right)=0$. If $T$ is deterministic,
$\mathcal{P}_n = \mathcal{P}_k\,(\mathrm{mod}\,m),\;\forall\;n,k$.

A non-deterministic Borel automorphism $T$ is a
\underline{Kolmo}g\underline{orov} \underline{Shift} or
\underline{K}-\underline{shift} if
$\bigcap_{n=-\infty}^{\infty}\mathcal{P}_n$ consists of sets with
probability zero or one.

A B-shift is a K-shift and is of non-deterministic nature in the
above sense.

Now, a measure preserving Borel automorphism $T$ on a probability
space $(X, \mathcal{B}, m)$ is said to be
\underline{er}g\underline{odic} if for every $T$-invariant $A\in
\mathcal{B}$, $m(A)=0$ or $m(X\setminus A)=0$. Note that such a
$T$ is ergodic iff every real-valued measurable $T$-invariant
function $f$ is constant a.e.

Now, if $T$ is measure-preserving, ergodic and for some singleton
$\{x\}\in \mathcal{B}$, $m\left( \{x\}\right)>0$, then $x$ must be
a periodic point of $T$. A non-trivial measure-preserving ergodic
system is therefore the one for which $m$ is non-atomic.

The system $\left( X, {\cal B}, \mathcal{N}, T \right)$ is called
a \underline{Descri}p\underline{tive}
\underline{D}y\underline{namical} \underline{S}y\underline{stem}
\cite{trim6} \footnote{By descriptively ergodic it is implied that
the analysis will not involve measure theoretic considerations but
will only be based on the concepts involving Borel structure of
the space $X$.}.

Now, $T$ is said to be \underline{descri}p\underline{tivel}y
\underline{er}g\underline{odic} or that $T$ is said to {\em act in
a descriptively ergodic manner\/} if $T\mathcal{N}=\mathcal{N}$
and if $TA=A,\;A\in{\cal B}$ implies either $A\in \mathcal{N}$ or
$X\setminus A\in \mathcal{N}$.

\underline{Nadkarni}'s \underline{Theorem} \cite{trim6} states: if
$(X, {\cal B}, \mathcal{N}, T)$ is a descriptive dynamical system
such that
\begin{description} \item {(a)} every member of ${\cal B}\setminus\mathcal{N}$
is decomposable \item{(b)} ${\cal B}$ satisfies the countability
condition \item{(c)} $T$ is descriptively ergodic \item{(d)} $X$
is bounded,\end{description} then there exists a finite measure
$\mu$ on ${\cal B}$ such that
\begin{description}\item{(1)} $\mathcal{N}\;=\;\left\{ B\in{\cal B}: \mu(B)=0
\right\}$ \item{(2)} $\mu$ is continuous \item{(3)} $T$ is
$\mu$-measure preserving, and \item{(4)} $T$ is ergodic, {\em ie},
$TA=A,\; A\in {\cal B}$ implies that $\mu(A)=0$ or $\mu(X\setminus
A)=0$.
\end{description}

As a corollary of this theorem, we also have: Let $(X,\mathcal{B},
\mathcal{N},T)$ be a descriptive dynamical system such that
\begin{description} \item{(a)} every member of $\mathcal{B}\setminus
\mathcal{N}$ is decomposable, \item{(b)} $\mathcal{B}$ satisfies
the countability condition \item{(c)} $T$ is descriptively
ergodic, \item{(d)} $\exists\;B\in\mathcal{B}\setminus
\mathcal{N}$ which is bounded \end{description} Then, there exists
a unique continuous $\sigma$-finite measure $m$ on $\mathcal{B}$
such that its null sets in $\mathcal{B}$ form precisely the ideal
$\mathcal{N}$ and $T$ is ergodic and measure preserving with
respect to $m$.

Furthermore, it can also be shown \cite{trim6} that: for a SBS
$(X,\mathcal{B})$ and $T:X\to X$ a Borel automorphism, there
exists a finite continuous measure $m$ on $\mathcal{B}$ so that
$T$ is {\em non-singular\/} and ergodic iff there exists a
$\sigma$-ideal $\mathcal{N}\subseteq \mathcal{B}$ such that the
system $(X,\mathcal{B},\mathcal{N},T)$ has the following
properties
\begin{description} \item{(i)} every member of $\mathcal{B}\setminus
\mathcal{N}$ is decomposable \item{(ii)} $\mathcal{B}$ satisfies
the countability condition \item{(iii)} $T$ is descriptively
ergodic \item{(iv)} $\exists\;B\in \mathcal{B}\setminus
\mathcal{N}$ which is bounded \end{description}

Now, let $\mathcal{C}=\mathbf{2}^{\mathbb{N}}$, the countable
product of two point space $\mathbf{2}$ with product topology,
with the two point space $\mathbf{2}$ being given the discrete
topology and $\mathcal{B}$ its Borel $\sigma$-algebra.

If we drop from above $X$ the countable set of those sequences of
zeros and ones which have only finitely many zeros or finitely
many ones, then the remaining set, say, $Y$, can be mapped one-one
into $[0,1)$ by the map $\xi(x_1,x_2,...)=\sum_{i=1}^{\infty}
{x_i}/{2^i}$.  The image of $Y$ under this map is $[0,1)\setminus
D$ where $D$ is the set of rational numbers of the form $k/2^n$,
$0\leq k\leq 2^n$, $n\in\mathbb{N}$.

Now, if $x=(x_1, x_2, ...)\in Y$ and if $k$ is the first integer
such that $x_k=0$, then let us define the map, say,
$V=\xi^{-\,1}T\xi$ as $Vx=(0,0,...,0,1,x_{k+1},x_{k+2},...)$.
Then, $V$ replaces all the ones up to the first zero by zeros and
replaces the first zero by one, leaving all other coordinates of
$x$ unchanged.

We call the map $V$ on $Y$ the \underline{Diadic}
\underline{Addin}g \underline{Machine} (DAM) or the
\underline{Odometer}.

Now, a measure preserving automorphism $T$ on a probability space
$(X, \mathcal{B}, m)$ is ergodic iff $\forall\;A, B\in
\mathcal{B}$, $\frac{1}{n} \sum_{k=0}^{n-1}m\left( A\bigcap T^kB
\right) \to m\left( A\bigcap B\right)$ as $n\to\infty$. There are
two properties stronger than ergodicity which are also relevant to
us.

A measure preserving automorphism $T$ on a probability space $(X,
\mathcal{B}, m)$ is said to be \underline{weakl}y
\underline{mixin}g iff $\forall\;A, B\in \mathcal{B}$,
$\frac{1}{n} \sum_{k=0}^{n-1}|m\left( A\bigcap T^kB
\right)-m\left( A\bigcap B \right)| \to 0$ as $n\to\infty$. A
measuring preserving automorphism $T$ on $(X,\mathcal{B},m)$ is
said to be \underline{mixin}g if $\forall\;A, B \in \mathcal{B}$,
$m\left( A\bigcap T^kB \right) \to m\left(A\bigcap B\right)$ as
$n\to\infty$.

If a measure preserving Borel automorphism $T$ is mixing then it
is weakly mixing, and if $T$ is weakly mixing then it is ergodic.
However, a ergodic $T$ need not be weakly mixing and mixing. Also,
an ergodic and weakly mixing automorphism $T$ need not be mixing.

Let $T_1$ be a measure preserving Borel automorphisms on a
probability space $(X_1,\mathcal{B}_1, m_1)$ and $T_2$ be that on
$(X_2,\mathcal{B}_2,m_2)$. We say that $T_1$ and $T_2$ are
\underline{metricall}y \underline{isomor}p\underline{hic} if
$\exists$ $X'_1\subseteq X_1$ with $m_1\left(X_1\setminus
X_1'\right)=0$, $X_2'\subseteq X_2$ with $m_2\left( X_2\setminus
X_2'\right)=0$ and an invertible, {\em ie}, a one-one, onto,
measurable map with measurable inverse, measure preserving map
$\phi:X_1'\to X_2'$ such that $\phi T_1\phi^{-\,1}=T_2$.

A measure preserving automorphism $T$ gives rise to a
\underline{Unitar}y \underline{O}p\underline{erator}, $U_{_T}$,
as: $U_{_T}f=f\circ T,\;f\in L^2(X,\mathcal{B},m)$. The unitary
operator is linear, invertible with $U_{_T}^{-\,1}f=f\circ
T^{-\,1}$ and $L^2$-norm preserving, {\em ie}, $||U_{_T}f||_{_2}
=||f||_{_2}$.

We say that $\lambda$ is an \underline{ei}g\underline{envalue} of
$U_{_T}$ if $\exists$ a non-zero $f\in L^2(X,\mathcal{B},m)$, such
that $f\circ T=\lambda f$. Then, $f$ is an
\underline{ei}g\underline{enfunction} with eigenvalue $\lambda$.
An eigenvalue is \underline{sim}p\underline{le}, if up to a
multiplicative constant, it admits only one eigenfunction.

Let $L_o^2(X,\mathcal{B},m)\,=\,\{ f \in L^2(X, \mathcal{B},
m)\;|\; \int\,f dm$ $=0\}$, the subspace of functions orthogonal
to the constant functions. It is $U_{_T}$-invariant.

Now, $1$ is always an eigenvalue of $U_{_T}$ and that $1$ is a
simple eigenvalue of $U_{_T}$ iff $T$ is ergodic. Further, since
$U_{_T}$ is unitary, all eigenvalues of $U_{_T}$ are of absolute
value one.

Then, weakly mixing automorphisms $T$ are precisely those for
which $U_{_T}$ has no eigenvalue other than $1$. Also, $T$ is
ergodic iff $1$ is not an eigenvalue of $U_{_T}$ on $L_o^2(X,
\mathcal{B}, m)$.

If $U_{_T}$ and $U_{_{T'}}$ are unitarily equivalent, $T$ and $T'$
are \underline{s}p\underline{ectrall}y
\underline{isomor}p\underline{hic}. If measure preserving $T$ and
$T'$ are metrically isomorphic, then $U_{_T}$ and $U_{_{T'}}$ are
unitarily equivalent.

A measure preserving automorphism $T$ on a SPS $(X,
\mathcal{B},m)$ has \underline{discrete}
\underline{s}p\underline{ectrum} if $U_{_T}$ admits a complete set
of eigenfunctions. Then, if $T_1$ and $T_2$ are spectrally
isomorphic and $T_1$ has a discrete spectrum, then $T_2$ also has
a discrete spectrum and the corresponding unitary operators have
the same set of eigenvalues.

But, spectrally isomorphic measure preserving automorphisms are
not necessarily metrically isomorphic, in general. However, if the
measure preserving automorphisms defined on a SPS are ergodic with
discrete spectrum and are admitting the same set of eigenvalues,
then such spectrally isomorphic measure preserving automorphisms
are metrically isomorphic.

Note that in the case of a SPS, $U_{_T}$ can have at most a
countable number of eigenvalues, all of absolute value one.
Furthermore, in the same case, the eigenvalues of $U_{_T}$ form a
subgroup of the circle group $S^1$. Also, for each eigenvalue
$\lambda$ we can choose an eigenfunction $f_{\lambda}$ of absolute
value one so as to have $f_{\lambda}.f_{\nu}=f_{\lambda\nu}$ a.e.

Any two B-shifts are spectrally isomorphic but, in general, any
two B-shifts are not metrically isomorphic. Any two K-shifts are
spectrally isomorphic but, in general, any two K-shifts are not
metrically isomorphic.

For a finite partition $\mathcal{P}=\{P_1, P_2, ..., P_k\}$ of $X$
by members of $\mathcal{B}$, we define the \underline{entro}py of
$\mathcal{P}$ to be $\sum\,-\;m(P_i)\,\log_e{m(P_i)}$ and denote
it by $H(\mathcal{P})$. Then, we can define the
\underline{entro}py \underline{of} \underline{$\mathcal{P}$}
\underline{relative} \underline{to}
\underline{automor}p\underline{hism} \underline{$T$} defined as:
$h(\mathcal{P}, T) = \mathrm{lim\;sup}
\frac{1}{n}\,H\left(\bigvee_{k=0}^{n-1}T^{-\,1}
\mathcal{P}\right)$, where $\bigvee_{k=0}^{n-1}T^{-\,1}
\mathcal{P}$ is used to denote the partition generated by
$T^{-\,1} \mathcal{P}$, $k=0$, ..., $n-1$. Note that the
$\mathrm{lim\;sup}$ is indeed an increasing limit.

Then, we have the \underline{entro}py \underline{of}
\underline{the} \underline{automor}p\underline{hism} $T$, denoted
as $h(T)$, as: $h(T)=\mathrm{sup}\;h(\mathcal{P}, T)$, where the
supremum is taken over all finite partitions $\mathcal{P}$ of $X$.
Note that $h(T)$ is an invariant of the metric isomorphism.

Then, if $T$ is a B-shift with independent generating partition
$\mathcal{P}=\{P_1, P_2, ...\}$ then its entropy is
$h(T)=\sum\;-\;m(P_i)\log_e{m(P_i)}$. Now, any two B-shifts with
the same entropy can be shown to be metrically isomorphic.

For any set $A\in \mathcal{B}$, the set
$\bigcup_{k=-\infty}^{\infty} T^kA$ is called as the
\underline{saturation} of A with respect to $T$ or simply the
\underline{$T$}-\underline{saturation} \underline{of}
\underline{$A$}. We \underline{denote} it by {\it s}$_{_T}(A)$. A
point $x\in A$ is said to be a \underline{recurrent}
p\underline{oint} \underline{in} \underline{$A$} if $T^nx$ returns
to $A$ fimp $n$ and fimn $n$.

By Poincar\'{e} Recurrence Lemma, we can write $A$ as a disjoint
union of two measurable sets $B$ and $M$ such that every point of
$B$ is recurrent in $B$ (hence also in $A$) and no point of $M$ is
recurrent so that $M\in\mathcal{W}_{_T}$. Clearly, it follows that
$\bigcup_{n=0}^{\infty} T^nB=\bigcup_{n=-\infty}^{\infty}T^nB=$
{\it s}$_{_T}(B)$, since every point of $B$ is recurrent in $B$.

Now, given $x\in B$, let $n_{_B}(x)$ denote the smallest positive
integer such that $T^nx\in B$. Then, we can decompose $B$ into
pairwise disjoint sets $B_k,\;k\in \mathbb{N}$, where $B_k=\{x\in
B\; |\; n_{_B}(x)=k\}$ or, equivalently, $B_k=\{x\in B\;|\;
Tx\notin B, ..., T^{k-1}x \notin B, T^kx\in B\}$. Further, we have
$T^kB_k\subseteq B$ and that $B_k$, $TB_k$, ..., $T^{k-1}B_k$ are
pairwise disjoint.

Further, let $F_{\ell}=T^{\ell}\left( \bigcup_{k\,>\,\ell}B_k
\right)$ and note also that $F_{\ell} = TF_{\ell-1}\setminus B$,
where $F_o=B$. Now, we have $\bigcup_{k=0}^{\infty}T^kB =
\bigcup_{k=0}^{\infty}\bigcup_{i=0}^{k-1}T^kB_k =
\bigcup_{k=0}^{\infty}F_k = \bigcup_{k=-\infty}^{\infty}T^kB =$
{\it s}$_{_T}(B)$, with the middle two unions being pairwise
disjoint unions.

We call the set $B$ as the \underline{base} and the union
$\bigcup_{k=1}^{\infty}T^{k-1}B_k$ as the \underline{to}p
\underline{of} \underline{the} \underline{construction}. The above
construction is called as the \underline{Kakutani}
\underline{tower} \underline{over} \underline{base}
\underline{$B$}.

Now, if $m$ is any $T$-invariant probability measure on
$\mathcal{B}$ and if {\em we write
$B_{\star}=\bigcup_{k=0}^{\infty} T^kB$}, then we have $m\left(
B_{\star}\right) = \sum_{k=1}^{\infty} \sum_{i=0}^{k-1}m\left(
T^iB_k\right) = \sum_{k=1}^{\infty} k\, m\left(B_k\right) =
\int_B\,n_{_B}(x)\,dm$.

Let $m(B)\ge 0$. Then, we call the quantity
$\frac{1}{m(B)}\int_B\,n_{_B} (x)\,dm=m(B_{\star})/m(B)$ as the
\underline{mean} \underline{recurrence} \underline{time}
\underline{of} \underline{$B$}. Recall $A=B\bigcup M$, $M\in
\mathcal{W}_{_T}$. Then, $m(M)=m(M_{\star})=0$. Hence, $m(A)=m(B)$
and $m(A_{\star})=m(B_{\star})$. Thus, the above is also the {\em
mean recurrence time of $A$}.

Clearly, if $T$ is ergodic and $m(B)>0$ then we have
$B_{\star}=X\; (\mathrm{mod}\;m)$ since it is $T$-invariant and of
positive measure.

Now, consider the transformation $\mathcal{S}$ defined over
$B_{\star}$ as: \goodbreak \noindent
\begin{eqnarray} \mathcal{S}(x)= \left\{ \n
\begin{array}{cccc}
T(x) &{\rm if} \/x&\notin\bigcup_{k=1}^{\infty}T^{k-1}B_k = {\rm
Top}\\ \\ T^{-\/k+1}(x) &{\rm if} \,x&\in T^{k-1}B_k, \/k=1,2,...
\end{array}\right.
\end{eqnarray}

\noindent Then, $\mathcal{S}$ is periodic, the period being $k$
for points in $B_k$, and $\mathcal{S}$ agrees with $T$ everywhere
except at the top of the Kakutani tower. Further, if
$B_{\star}=X$, then $\mathcal{S}$ is defined on all of $X$.

Now, suppose $C_1\supseteq C_2\supseteq C_3\supseteq ...$ is a
sequence of sets in $\mathcal{B}$ decreasing to an empty set and
such that $\forall\;n$, we have \begin{description} \item{(i)}
every point of $C_n$ is recurrent, and that \item{(ii)}
$\bigcup_{k=0}^{\infty}T^kC_n=X$. \end{description} Let
$\mathcal{S}_n$ be the periodic automorphism as defined above with
$B=C_n$. Then, $\forall\;n$, $\mathcal{S}_n$ and
$\mathcal{S}_{n+1}$ agree except on the Top $T_{n+1}$ of the
Kakutani tower whose base is $C_{n+1}$. But $T_n\supseteq T_{n+1}$
and since $C_n$ decreases to $\emptyset$, $T_n$ also decreases to
$\emptyset$. Then, given any $x$, $\exists\;n(x)$ such that
$\forall\; k\geq n(x)$, $\mathcal{S}_k(x)$ are all the same and
equal to $T(x)$. Thus, $T$ is a limit in this sense of the
sequence of periodic automorphisms. Therefore, we obtain the
p\underline{eriodic} \underline{a}pp\underline{roximation}
\underline{of} \underline{automor}p\underline{hism}
\underline{$\;T$}.

\underline{Rohlin}'s \underline{Lemma} states that: If $T$ is
ergodic with respect to the $\sigma$-ideal of null sets of a
finite measure $m$, then given $\epsilon>0$ and $n\in\mathbb{N}$,
$\exists$ a set $C$ such that $C$, $TC$, ..., $T^{n-1}C$ are
pairwise disjoint and $m\left( X\setminus
\bigcup_{k=o}^{n-1}T^kC\right)\;<\; \epsilon$.

Now, let $B\in\mathcal{B}$ be such that every point of $B$ is
recurrent. Following Kakutani, the \underline{induced}
\underline{automor}p\underline{hism} \underline{on}
\underline{$B$} (mod $\mathcal{W}_{_T}$), denoted as $T_{_B}$, is
then defined as: $T_{_B}(x)=T^n(x), \;x\in B$, where $n=n_{_B}(x)$
is the smallest positive integer for which $T^n(x)\in B$. Note
that $T_{_B}(x)=T^k(x)$ if $x\in B_k$, $k=1,2,3....$. Then,
$T{_B}$ is one-one, measurable and invertible with
$T^{-\,1}(x)=T^n(x)$ where $n$ is the largest negative integer
such that $T^n(x)\in B$. Thus, $T_{_B}$ is a Borel automorphism on
$B$.

The induced Borel automorphism, $T_{_B}$, on $B$ has following
properties:
\begin{itemize}
\item $\mathrm{orb}(x,T_{_B})=B\bigcap\mathrm{orb}(x,T)$, $x\in B$
\item $T_{_B}$ is elementary iff $T$ restricted to {\it s}$_{_T}B$
is elementary
\item $W\subseteq B$ is $T_{_B}$-wandering if and only if $W$ is $T$-wandering
\item $\mathcal{W}_{_{T_{_B}}} =\mathcal{W}_{_T} \bigcap B$
\item if $T$ is ergodic and preserving a finite measure $m$ then
$T_{_B}$ is ergodic and preserves $m$ restricted to $B$,
\item If $\mathcal{N}$ is a $\sigma$-ideal in $\mathcal{B}$,
$\mathcal{W}_{_T}\subseteq \mathcal{B}$, and if $T$ is ergodic
with respect to $\mathcal{N}$, then $T_{_B}$ is ergodic with
respect to the restriction of $\mathcal{N}$ to $B$. In particular,
if $T$ is ergodic with respect to a finite continuous measure $m$
then $T_{_B}$ is ergodic with respect to the restriction of $m$ to
$B$
\item if $C\subseteq B$, then a point of $C$ is recurrent with respect to $T$
iff it is recurrent with respect to $T_{_B}$. If every point of
$C$ is recurrent then we have $T_{_C}=(T_{_B})_{_C}$.\end{itemize}

A broadened view of the induced automorphism defines it on a set
$A\in \mathcal{B}$ even if not every point of $A$ is recurrent.
For this, let us consider a set $B=\{x\in A\;|\; x
\;\mathrm{is\;recurrent \;in}\;A\}$. By Poincar\'{e} Recurrence
Lemma, $A\setminus B\in\mathcal{W}_{_T}$ and every point of $B$ is
recurrent in $B$. Then, the broadened induced automorphism
$T_{_A}$ is defined on all of $A$ iff every point of $A$ is
recurrent; otherwise $T_{_A}$ is defined on $A\;(\mathrm{mod}\;
\mathcal{W}_{_T})$. All the earlier properties of the induced
automorphism remain valid $(\mathrm{mod}\;\mathrm{W}_{_T})$ under
this broadened definition of $T_{_A}$. {\em Note however that the
stricter point of view is necessary for the descriptive aspects}.

Now, consider a Borel automorphism $T$ on $(X, \mathcal{B})$ and
let $f$ be a non-negative integer-valued measurable function on
$X$.

Let $B_{k+1}=\{x \;|\;f(x)=k\}$, $k=0,1,2,...$, $C_k=
\bigcup_{\ell\,>\,k}B_{\ell}$, $F_k=C_k\times\{k\}$, $Y=
\bigcup_{k=0}^{\infty}F_k$. If $Z=X\times\{0,1,2,...\}$, then
$Y\subseteq Z$ is the set $Y=\{(x,n)\;\;0\geq n\geq f(x)\}=$
Points in $Z$ below and including the graph of $f$.

Define $\Lambda$ on $Y$ as:
\begin{eqnarray} \Lambda(k,j)= \left\{ \n
\begin{array}{cccc}
(b,j+1)&{\rm if}\,b\in B_k \,{\rm and}\, 0\leq j\leq k-1\\
\\ (\Lambda(b), 0) &{\rm if}\,b\in B_k\, {\rm and}\, j=k-1
\end{array}\right.
\end{eqnarray}
This $\Lambda$ is a Borel automorphism on the space $Y$. We call
it the \underline{automor}p\underline{hism} \underline{built}
\underline{under} \underline{the} \underline{function} $f$ on the
space $X$. We call $X$ the \underline{base}
\underline{s}p\underline{ace} of $\;\Lambda$ and $f$ the
\underline{ceilin}g \underline{function} \underline{of}
\underline{$\Lambda$}. Note that if we identify $X$ with $X\times
\{0\}$, then $\Lambda_{_X}=T$ and we write $\Lambda=T^f$.

The automorphism built under a function has the following
properties: \begin{itemize}\item If $B\in\mathcal{B}$ with every
point of $B$ being recurrent and $B_{\star}=X$, then $T$ is
isomorphic to $(T_{_B})^f$, where $f(x)=n_{_B}(x)$, \item If
$A\subseteq Y$ is the graph of a measurable function $\xi$ on $X$,
then $(T^f)_{_A}$ and $T$ are isomorphic by $x\mapsto(x,\xi(x))$.
In particular, $(T^f)_{_A}$ and $T$ are isomorphic when $A=$ graph
of $f$, \item If $A\subseteq Y$ is measurable then we can find a
measurable $B$ with the same saturation as $A$ under $T^f$ and
such that $\forall\;x\in X$, $B\bigcap \{(x,i)\;|\;0\leq i \leq
f(x)\}$ is at most a singleton. Indeed, $B=\{(x,i)\in
A\,|\,(x,j)\notin A,\, 0\leq j<i\}$ can be chosen, \item Given
$T^f$ and $T^g$, they are isomorphic to automorphisms induced by
$T^{f+g}$ on suitable subsets. If $Y_1=\{(x,i)\,|\,0\leq i\leq
f(x)+g(x)\}$ on which $T^{f+g}$ is defined, then the sets
$\{(x,i)\,|\,0\leq i\leq f(x)\}$ and $\{(x,i)\,|\,0\leq i\leq
g(x)\}$ are subsets of $Y_1$ on which $T^{f+g}$ induces
automorphisms which are isomorphic to $T^f$ and $T^g$
respectively, \item If $m$ is a $\sigma$-finite $T$-invariant
measure on $X$, then $\exists$ a unique $\sigma$-finite
$T^f$-invariant measure $m_{_Y}$ on $Y$ such that $m_{_Y}$
restricted to $X\times\{0\}$ is $m$. The measure $m_{_Y}$ is
finite iff $m(X)$ is finite and $\int\,f\,dm$ is finite. Then, we
have $m_{_Y}(Y)=\sum_{k=1}^{\infty} k\,m(B_{k+1})=\int\, f\,dm <
\infty$. \item $T^f$ is elementary iff $T$ is elementary.
\end{itemize}

Now, given two Borel automorphisms $T_1$ and $T_2$, we say that
$T_1$ is a \underline{derivative} \underline{of} $T_2$, and write
$T_1 \prec T_2$, if $T_1$ is isomorphic to $(T_1)_{_A}$ for some
$A\in\mathcal{B}$ with $\bigcup_{k=0}^{\infty}T^k_1A=X$. If $T_1$
is a derivative of $T_2$, we call $T_2$ the p\underline{rimitive}
\underline{of} $T_1$. Two Borel automorphisms are said have a
\underline{common} \underline{derivative} if they admit
derivatives which are isomorphic. Similarly, two automorphisms are
said to have a \underline{common} p\underline{rimitive} if they
admit primitives which are isomorphic. If $T_1 \prec T_2$, then
clearly $T_2=T_1^f$ for some $f$.

Then, a lemma due to \underline{von} \underline{Neumann} states
that: Two Borel automorphisms have a common derivative iff they
have a common primitive.

Now, we say that two Borel automorphisms $T_1$ and $T_2$ are
\underline{Kakutani} \underline{e}q\underline{uivalent}, and we
write $T_1 \kakueq T_2$, if $T_1$ and $T_2$ have a common
primitive, or, equivalently the automorphisms T$_1$ and $T_2$ have
a common primitive. The Kakutani equivalence is reflexive,
symmetric and transitive. {\em Therefore, the Kakutani equivalence
is an equivalence relation for Borel automorphisms}.

Suppose $\mathcal{N}$ is a $\sigma$-ideal in $\mathcal{B}$. Then,
we say that $T_1$ and $T_2$ are \underline{Kakutani}
\underline{e}q\underline{uivalent} $(\mathrm{mod}\; \mathcal{N})$
if we can find two sets $M, N\in \mathcal{N}$, $M$ being
$T_1$-invariant and $N$ being $T_2$-invariant, such that
$T_1|_{_{X\setminus M}} \kakueq T_2|_{_{X\setminus N}}$. When
$\mathcal{N}$ is the $\sigma$-ideal of $m$-null sets of a
probability measure $m$ invariant under $T_1$ and $T_2$ both, we
get the measure theoretic Kakutani equivalence of Borel
automorphisms \cite{orw}.

Given a Borel automorphism $T$, a system of pairwise disjoint sets
$(C_o, C_1, ..., C_n) \in \mathcal{B}$ is called a
\underline{column} if $C_i=T^iC_o,\;0\leq i \leq n$. $C_o$ is
called the \underline{base} \underline{of} \underline{the}
\underline{column} and $C_n$ is called the \underline{to}p
\underline{of} \underline{the} \underline{column}. If
$D_o\subseteq C_o$, then $(D_o, TD_o, ..., T^nD_o)$ is called a
\underline{sub}-\underline{column} \underline{of} $(C_o, ...,
C_n)$.

Two columns $(C_o, ..., C_n)$ and $(B_o, ..., B_m)$ are said to be
\underline{dis}j\underline{oint} if $C_i\bigcap
B_j=\emptyset\;\forall \;i\neq j$. A finite or a countable system
of pairwise disjoint columns is called a
\underline{$T$}-\underline{tower}.

A $T$-tower with $r$ pairwise distinct columns may be written as
$\{C_{ij}\;|\; 0\leq i\leq n(j),\; 1\leq j\leq r\}$ where
$\{C_{0j}, ..., C_{n(j)j}\}$ is its j-th column.

Sets $C_{ij}$ are \underline{constituents} \underline{of}
\underline{the} \underline{$T$}-\underline{tower},
$\bigcup_kC_{0k}$ is a {\em base of the $T$-tower\/} and
$\bigcup_kC_{n(k)k}$ is a \underline{to}p \underline{of}
\underline{the} \underline{$T$}-\underline{tower}. The number of
distinct columns in a $T$-tower is a \underline{rank}
\underline{of} \underline{the} \underline{$T$}-\underline{tower}.

A $T$-tower is said to \underline{refine} a $S$-tower if every
constituent of $T$-tower is a subset of a constituent of the
$S$-tower.

$T$ has \underline{rank at most} $r$ if there is a sequence
$T_n,\;n\in \mathbb{N}$, of $T_n$-towers of rank $r$ or less such
that $T_{n+1}$ refines $T_n$ and the collection of sets in $T_n$,
taken over all $n$, generates $\mathcal{B}$. Then, $T$ has rank
$r$ if $T$ has rank at most $r$ but does not have rank at most
$r-1$. If $T$ does not have rank $r$ for any finite $r$, then $T$
has infinite rank.

Given a Borel automorphism $T$ on $(X,\mathcal{B})$, a partition
$\mathcal{P}$ of $X$, $\mathcal{P} \subseteq \mathcal{B}$, is a
g\underline{enerator} \underline{of} \underline{$T$} if
$\bigcup_{k=1}^{\infty} T^k\mathcal{P}$ generates $\mathcal{B}$. A
set $A\in\mathcal{B}$ is \underline{decom}p\underline{osable}
$(\mathrm{mod}\;\mathcal{W}_{_T})$ if we can write $A$ as a
disjoint union of two Borel sets $C$ and $D$ such that {\it
s}$_{_T}(C)=${\it s}$_{_T}(D)=${\it
s}$_{_T}(A\;(\mathrm{mod}\;\mathcal{W}_{_T}))$.

Let $\mathcal{P}=\{P_1, P_2,..., P_n\}\subseteq\mathcal{B}$ be a
partition of $X$ and let a measurable $C$ be such that
$\bigcup_{k=0}^{\infty}T^kC=X$. Then, on the basis of the first
return time $n(x)$ of each $x\in C$ and pairwise disjoint sets
$E_i=\{ x\in C\;|\; n(x)=i\}$ with union $\bigcup_iE_i=C$, there
exists a countable partition of $\{D_1,D_2,...\}$ of $C$ such that
each $P_i$ is a disjoint union of sets of the form $T^kD_i$,
$k=1,2,...$, $i=1,2,...$.

Now, a one-one and onto map $T:X\to X$ such that $T^kx\neq x$ for
all $k\neq 0$, and for all $x\in X$ is called a \underline{free
ma}p.

Every free Borel automorphism $T$ on a SBS $(X,\mathcal{B})$ is
\cite{djk} orbit equivalent to an induced automorphism by the DAM.

Further, every Borel set $A\in\mathcal{B}$ is clearly decomposable
$(\mathrm{mod}\;\mathcal{W}_{_T})$ for $T$ being a free Borel
automorphism on a countably generated and countably separated SBS.

Furthermore, given a free Borel automorphism $T$ on a countably
generated and countably separated SBS $(X,\mathcal{B})$, there
exists a sequence $C_n,\;n\in\mathbb{N}$, of Borel sets decreasing
to an empty set with {\it s}$_{_T}(C_n)=${\it s}$_{_T}(X\setminus
C_n)=X\; \forall\;n$, such that $\forall\;n$ the sets $C_n$,
$TC_n$, ..., $T^{n-1}C_n$ are pairwise disjoint, and such that
$\bigcap_{n=1}^{\infty}C_n=C_{\infty}$, say, is $T$-wandering.

Given a Borel automorphism $T$ on a countably generated and
countably separated SBS $(X,\mathcal{B})$, there exists a sequence
$T_n,\;n=1,2,...$ of periodic Borel automorphisms on $X$ such that
$\forall\;x$, $Tx=T_n x$ for all sufficiently large $n$.

Hence, the descriptive version of Rohlin's theorem \cite{rokhlin}
on generators is obtained \cite{weiss1} as: every free Borel
automorphism on a countably generated and countably separated SBS
admits a countable generator in a strict sense.

Note also that $T$ admits a countable generator iff $T$ admits at
most a countable number of periodic points \cite{kechris}.

Now, two subsets of $X$, $A, B \in {\cal B}$, are said to be
\underline{e}q\underline{uivalent} \underline{b}y
\underline{countable} \underline{decom}p\underline{osition}, and
we \underline{write} $A \sim B$, if
\begin{description}\item{(a)} $A=\bigcup_{i=1}^{\infty}A_i$, $A_i\bigcap
A_j=\emptyset$ for $i\neq j$, and $A_i \in {\cal B},\, i=1, 2,
...$ \item{(b)} $B=\bigcup_{i=1}^{\infty}B_i$, $B_i\bigcap
B_j=\emptyset$ for $i\neq j$, and $B_i \in {\cal B},\, i=1, 2,
...$ \item{(c)} there exist $n_1$, $n_2$, $... \in \mathbb{N}$
such that $\forall\;i\in\mathbb{N}, \;\;T^{n_i} A_i=B_i\;({\rm
mod}\,\mathcal{N})$.\end{description} The equivalence by countable
decomposition is an equivalence relation on ${\cal B}$.

Note that if $A_i\in {\cal B},\;i\in\mathbb{N}$ are pairwise
disjoint and $B_i\in {\cal B},\;i\in\mathbb{N}$ are pairwise
disjoint and if $\forall\; i\in\mathbb{N}, \; A_i\sim B_i$ then
$\bigcup_{i=1}^{\infty} A_i \;\sim\;\bigcup_{i=1}^{\infty}B_i$.

If $A\sim B$, then we say that $B$ is a \underline{co}py
\underline{of} \underline{$A$} and then, $A$ and $B$ have the same
measure with respect to a $T$-invariant $\sigma$-finite measure.

Further, we say that $A$ and $B$ are
\underline{e}q\underline{uivalent} \underline{b}y
\underline{countable} \underline{decom}p\underline{osition}
\underline{$(\mathrm{mod}\;m)$}, and we \underline{write} $A\sim
B\; (\mathrm{mod}\;m)$, if there exist sets $M$ and $N$ in
$\mathcal{B}$, of $m$-measure zero, such that $A\triangle M \sim
B\triangle N$.

A set $A \in {\cal B}$ is said to be
\underline{$T$}-\underline{com}p\underline{ressible}
\underline{in} \underline{the} \underline{sense} \underline{of}
\underline{Ho}p\underline{f} if there exists $B \subseteq A$ such
that $A\sim B$ and $m(A\setminus B)>0$. Clearly, if the set $X$ is
Hopf $T$-compressible then every of its subsets $B\in{\cal B}$ is
Hopf $T$-compressible.

If $\mu$ is a $T$-invariant finite measure on $\mathcal{B}$ and
having the same null sets as $m$, then $A\sim B\; (\mathrm{mod}
\;m)$ implies that $\mu(A)=\mu(B)$. Whenever such a $\mu$ exists,
no measurable sets of positive measure can be compressible in the
sense of Hopf and, in particular, $X$ is not Hopf
$T$-compressible.

In a descriptive setting, one can dispense with the measure and
consider only a SBS $(X,\mathcal{B})$ and a free Borel
automorphism $T$ on it.

Then, given $A, B\in \mathcal{B}$, we \underline{write}
$A\prec\prec B$ if there exists a measurable subset $C\subseteq B$
such that $A \sim C$ and {\it s}$_{_T}(B\setminus C)\;=\;${\it
s}$_{_T}B$, which is the smallest $T$-invariant set containing
$B$.

Now, we say that $A$ is
\underline{$T$}-\underline{com}p\underline{ressible} if
$A\prec\prec A$ or, equivalently, if we can write $A$ as a
disjoint union of two sets $C,\,D\in\mathcal{B}$ such that $A\sim
C$, and {\it s}$_{_T}(A)=$ {\it s}$_{_T}(C)=$ {\it s}$_{_T}(D)$.
The sets $C$ and $D$ together with the automorphism $T$ which
accomplishes $A\sim C$ is called a
\underline{com}p\underline{ression} \underline{of} $A$.

If $X$ is $T$-compressible, then we say that $T$ is
\underline{com}p\underline{ressible} or that {\em $T$ compresses
$X$}.

The above notion of compressibility has the following properties:
\begin{itemize} \item If $A\in\mathcal{B}$ is $T$-compressible then any
superset of $A$ in $\mathcal{B}$ having the same saturation as $A$
is compressible. In particular, {\it s}$_{_T}(A)$ is
$T$-compressible whenever $A$ is $T$-compressible, \item Since $T$
is a free automorphism, each orbit is infinite and
$T$-compressible as also the saturation of any $T$-wandering set.
However, every $T$-compressible $T$-invariant set in $\mathcal{B}$
is {\em not\/} the saturation of a $T$-wandering set in
$\mathcal{B}$ except in special cases, \item A finite non-empty
set is not $T$-compressible nor is a set $A$ $T$-compressible if
the orbit of some point intersects $A$ in a finite non-empty set.
Further, if there exists a $T$-invariant probability measure on
$\mathcal{B}$, then no set of positive measure is
$T$-compressible. In particular, $X$ is not $T$-compressible in
this case, \item Clearly, a subset of a $T$-compressible set need
not be $T$-compressible, \item If $E\in\mathcal{B}$ is
$T$-invariant, $T$-compressible, and if $F\in\mathcal{B}$ is a
$T$-invariant subset of $E$, then $F$ is $T$-compressible. The
countable pairwise disjoint union of $T$-invariant,
$T$-compressible sets in $\mathcal{B}$ is $T$-compressible.
Clearly, any countable union of $T$-invariant, $T$-compressible
sets in $\mathcal{B}$ is $T$-compressible, \item $T$-compressible
sets in $\mathcal{B}$ do not form a $\sigma$-ideal in
$\mathcal{B}$.

However, $T$-invariant, $T$-compressible sets in $\mathcal{B}$ are
closed under countable union and taking of $T$-invariant subsets
in $\mathcal{B}$. Hence, the collection $\mathcal{H}$ of subsets
in $\mathcal{B}$ whose saturations are $T$-compressible forms a
$\sigma$-ideal in $\mathcal{B}$ and we call $\mathcal{H}$ the
\underline{Ho}p\underline{f} \underline{ideal}. \item
$\mathcal{W}_{_T}=\mathcal{H}$ iff $X\in \mathcal{W}_{_T}$.
\end{itemize}

Note that the Hopf ideal is also equal to the $\sigma$-ideal
generated by $T$-compressible sets in $\mathcal{B}$. Note that
$\mathcal{W}_{_T}\subseteq \mathcal{H}$ since the saturation of
every $T$-wandering set in $\mathcal{W}_{_T}$ is $T$-compressible.

Let $\mathcal{N}\subseteq \mathcal{B}$ be a $\sigma$-ideal such
that
\begin{description} \item{(1)} $T\mathcal{N}=T^{-\,1}\mathcal{N}=\mathcal{N}$ and
\item{(2)} $\mathcal{W}_{_T}\subseteq\mathcal{N}$.\end{description} The Hopf ideal
$\mathcal{H}$; the $\sigma$-ideal of $m$-null sets in
$\mathcal{B}$ for any $T$-invariant $\sigma$-finite measure on
$\mathcal{B}$; and the $\sigma$-ideal of $m$-mull sets when $T$ is
$m$-conservative are few such ideals.

Then, two sets $A,\,B\in\mathcal{B}$ are said to be {\em
equivalent by countable decomposition $(\mathrm{mod}\;
\mathcal{N})$\/} if we can find sets $M,\,N\in\mathcal{N}$ such
that $A\triangle M\;\sim\;B\triangle N$. We then write $A\sim B\;
(\mathrm{mod}\; \mathcal{N})$. Note that if $A\sim B\;
(\mathrm{mod}\; \mathcal{N})$ then {\it s}$_{_T}(A)=$ {\it
s}$_{_T}(B)\;(\mathrm{mod}\; \mathcal{N})$. We write $A\prec\prec
B\;(\mathrm{mod}\; \mathcal{N})$ if there exists a set
$N\in\mathcal{N}$ such that $A\triangle N \prec\prec B\triangle
N$.

A set $A$ is {\em compressible $(\mathrm{mod}\; \mathcal{N})$\/}
if $\exists\;N\in \mathcal{N}$ such that $A\triangle N$ is
$T$-compressible. For a $T$-invariant set in $\mathcal{B}$ all the
three notions of compressibility, namely, $T$-compressibility,
compressibility $(\mathrm{mod}\;\mathcal{W}_{_T})$ and
compressibility $(\mathrm{mod}\;\mathcal{H})$, are equivalent.

Now, suppose that $A,B\in\mathcal{B}$ are equivalent by countable
decomposition. Let $A=\bigcup_{i=1}^{\infty}A_i$,
$B=\bigcup_{i=1}^{\infty}B_i$ be pairwise disjoint partitions of
$A$ and $B$ respectively, such that for suitable integers
$n_i,\;i\in \mathbb{N}$, $T^{n_i}A_i=B_i$.

The map $S:A\to B$ defined by $S(x)=T^{n_i}x$ if $x\in A_i$ is an
\underline{orbit} p\underline{reservin}g
\underline{isomor}p\underline{hism} between $A$ and $B$. In case
$A$ and $B$ are equivalent by countable decomposition
$(\mathrm{mod}\;\mathcal{N})$ then $S$ will be defined between $A
\triangle N$ and $B\triangle M$ for suitable sets $M,n\in
\mathcal{N}$. Such a $S$ is an {\em orbit preserving
isomorphism\/} between $A$ and $B$ $(\mathrm{mod}\;\mathcal{N})$.

The following results are then easily obtainable for $A,B,C, D \in
\mathcal{B}$: \begin{description}\item{(a)} If $A\supseteq
B\supseteq C$ and $A\sim C$ then $A\sim B$ \item{(b)} If $A\sim C
\subseteq B$ and $B\sim D\subseteq A$ then $A\sim B$, \item{(c)}
If $A\supseteq B\supseteq C \; (\mathrm{mod}\;\mathcal{N})$ and
$A\sim C\;(\mathrm{mod}\; \mathcal{N})$, then $A\sim
B\;(\mathrm{mod}\;\mathcal{N})$, \item{(d)} If $A\sim
C\;(\mathrm{mod}\;\mathcal{N})$, $C\subseteq B \;
(\mathrm{mod}\;\mathcal{N})$, and $B\sim D\; (\mathrm{mod}\;
\mathcal{N})$, $D\subseteq A\;(\mathrm{mod}\;\mathcal{N})$, then
$A\sim B\;(\mathrm{mod}\;\mathcal{N})$.\end{description} Note that
for (c) and (d) we remove suitable sets in $\mathcal{N}$ from
$A,B,C,D$.

Now, a set $A\in\mathcal{B}$ is {\em incompressible\/} if it is
not compressible and it is {\em incompressible
$(\mathrm{mod}\;\mathcal{N})$\/} if it is not compressible
$(\mathrm{mod}\;\mathcal{N})$. Note however that $A\in\mathcal{B}$
is incompressible $(\mathrm{mod}\;\mathcal{N})$ does not mean that
$A\triangle N$ is incompressible for a suitable set $N\in
\mathcal{N}$. Note also that for a set in $\mathcal{B}$ to be
incompressible $(\mathrm{mod}\;\mathcal{N})$ it is sufficient that
its saturation is incompressible $(\mathrm{mod}\;\mathcal{N})$.

Let $N$ be a positive integer. Then, it is easy to see that there
exists $B\in \mathcal{B}$ such that {\it s}$_{_T}(B)=X$ and
$\forall\;x\in B$, its first return time, $n_{_B}(x)$, is such
that $N\leq n_{_B}(x)\leq 2N$.

For any $F\in\mathcal{B}$ and $x\in X$, let us now define
$r_{\star}\left( x,F \right)=\liminf_{n\to\infty}\frac{1}{n}
\sum_{k=1}^n\mathbf{1}_F (T^kx)$ and $r^{\star}\left( x,F
\right)=\limsup_{n\to\infty} \frac{1}{n} \sum_{k=1}^n\mathbf{1}_F
(T^kx)$ where $\mathbf{1}_F$ is the identity function on set $F$.

Then, we note that given $0\leq b \leq 1$ and $\epsilon >0$, there
exists $F\in\mathcal{B}$ such that {\it s}$_{_T}(F)=X$ and
$b-\epsilon<\;r_{\star}(x,F)$, $r^{\star}(x,F)<b+\epsilon$.

We also note that, if $0<b<1$, then there exists $F\in\mathcal{B}$
such that {\it s}$_{_T}(F)=$ {\it s}$_{_T}(X-F)$ and for all $x\in
X$, $0< r^{\star}(x,F)<b$.

Further, for any $F\in\mathcal{B}$ and $\epsilon>0$, there exists
a measurable $G\subseteq F$ such that {\it s}$_{_T}(G)=$ {\it
s}$_{_T}(F-G)=$ {\it s}$_{_T}(F)$ and $r^{\star}(x,G)<\epsilon\;
(\mathrm{mod}\;\mathcal{W}_{_T})$.

Now, a key dichotomy: Let $E,F\in\mathcal{B}$ and let $f=
\mathbf{1}_{_E}-\mathbf{1}_{_F}$. Then, there exists a
$T$-invariant set $N\in\mathcal{W}_{_T}$ such that if $x\in
X\setminus N$, then either \begin{description}\item{(a)} for all
$y\in\,\mathrm{orb}(x,T)$, there exists $n\geq 0$ with
$\sum_{k=0}^nf(T^ky)\geq 0$, Or \item{(b)} the set of $y\in\;
\mathrm{orb}(x,T)$ such that $\forall\;n\geq 0$, $\sum_{k=0}^n
f(T^ky) < 0$ is unbounded to the left and right.
\end{description} These are {\em mutually exclusive\/} conditions.

Furthermore,  consider any decomposition of $X$ into pairwise
disjoint $T$-invariant sets $X_o$, $X_1$ $X_2$, $N$ with
$N\in\mathcal{N}$ and $X_o$, $X_1$ $X_2$ satisfying the properties
\begin{description}\item{(c)} $E\bigcap X_1 \prec\prec F\bigcap X_1$,
\item{(d)} $E\bigcap X_o\;\sim\; F\bigcap X_o$, \item{(e)} $E\bigcap X_2
\prec\prec F\bigcap X_2$.\end{description}

Such a decomposition will have the properties that \begin{itemize}
\item for $x\in X_1\;(\mathrm{mod}\;\mathcal{H})$ the set, say,
$A(x)=\{y\in\;\mathrm{orb}(x,T)\;|\; \sum_{k=0}^nf(T^ny)\,>
\,0\;\forall \;n\geq 0\}$ is unbounded to left and right, \item
for any $x\in X_o\;(\mathrm{mod}\;\mathcal{H})$, for all
$y\in\;\mathrm{orb}(x,T)$ $\exists\;n\geq 0$ such that
$\sum_{k=0}^nf(T^ky)=0$, \item for $x\in X_2(\mathrm{mod}\;
\mathcal{H})$, the set, say, $B(x)=\{y\in\;\mathrm{orb}(x,T)\;|\;
\sum_{k=0}^nf(T^ny)\,<\,0\;\forall \;n\geq 0\}$ is unbounded to
left and right. \end{itemize}

Moreover, $(\mathrm{mod}\;\mathcal{H})$, we have that \beq
\{x\;|\; r_{\star}(x,E)\,<\,r_{\star}(x,F)\} &\subseteq& X_2,\n \\
\{x\;|\; r^{\star}(x,E)\,<\,r^{\star}(x,F)\} &\subseteq& X_2,\n \\
\{x\;|\; r_{\star}(x,E)\,>\,r_{\star}(x,F)\} &\subseteq& X_1,\n \\
\{x\;|\; r^{\star}(x,E)\,>\,r^{\star}(x,F)\} &\subseteq& X_1\n
\eeq

Then, we have the following measure free version of the Birkhoff
point-wise Ergodic Theorem as: For any $E\in \mathcal{B}$, the set
of points $x$ for which limit $\lim_{n\to\infty}
\frac{1}{n}\sum_{k=0}^{n-1} \mathbf{1}_E(T^kx)$ does not exist
belongs to the Hopf ideal $\mathcal{H}$. That is to say, the set
$\{x\;|\;r_{\star}(x,E)\, <\,r^{\star}(x,E)\}$ is compressible.

For any $E\in \mathcal{B}$, let us now write $m(E,x)=\lim
\frac{1}{n} \sum_{k=0}^{n-1}\mathbf{1}_{_E}(T^kx)$. This $m$ is
countably additive $(\mathrm{mod}\;\mathcal{H})$ and
$T$-invariant. Moreover, we can show that
$m(E,x)=0\;(\mathrm{mod}\;\mathcal{H})$ iff $E\in \mathcal{H}$.

Now, let the Polish topology $\mathcal{T}$ on $X$ possess a
countable clopen base $\mathcal{U}$ that is closed under
complements, finite unions and applications of $T$. There then
exists a $T$-invariant set $N\in\mathcal{H}$ such that
$\forall\;x\in X\setminus N$, $m(A\bigcup B, x) = m(A,x)+m(B,x)$
whenever $A,B\in \mathcal{U}$ and $A\bigcap B = \emptyset$.

Fix $x\in X\setminus N$ and let us write $m(A,x)=m(A)$,
$A\in\mathcal{U}$. For any $B\subseteq X$, let us define
$m^{\star}(B)=\inf\left\{ \sum_{i=1}^{\infty}
m(U_i)\;|\;B\subseteq \bigcup_{i=1}^{\infty} U_i\right.$,
$\left.U_i\in\mathcal{U}\;\forall\;i\right\}$. This $m^{\star}$,
an outer measure on $\mathfrak{P}(X)$, is $T$-invariant, bounded
by one and $m^{\star}(X)=1$.

Recall \cite{measure-theory} that an outer measure $\mu^{\star}$
on the power set of a metric space $(X,d)$ is called a {\em metric
outer measure\/} if $\mu^{\star}(E\bigcup
F)=\mu^{\star}(E)+\mu^{\star}(F)$ whenever $d(E,F)>0$. If
$\mu^{\star}$ is a metric outer measure on $(X,d)$ then all open
sets, hence, all Borel sets, are $\mu^{\star}$-measurable. Then,
$m^{\star}$ defined above is a metric outer measure on $X$. The
restriction of $m^{\star}$ to $\mathcal{B}$ is a countably
additive $T$-invariant probability measure on $\mathcal{B}$.

Further, if $T$ is not free, then it has a periodic point on whose
orbit we can always put a $T$-invariant probability measure.

\underline{Ho}pf's \underline{Theorem}: if $T$ is a Borel
automorphism (free or not) of a Standard Borel Space
$(S,\mathcal{B})$ such that $X$ is $T$-incompressible, then there
exists a $T$-invariant probability measure on $\mathcal{B}$.

Now, a set $A\in\mathcal{B}$ is \underline{weakl}y
$T$-\underline{wanderin}g if $T^nA$ are pairwise disjoint for $n$
in some infinite subset of integers. Then, a non-singular
automorphism $T$ on a probability space $(X,\mathcal{B},m)$ admits
\cite{hakaku} an equivalent $T$-invariant probability measure if
and only if there does not exist any weakly $T$-wandering set of
positive measure.

But, $T$-compressibility of $X$ does not imply the existence of a
weakly $T$-wandering set $W\in \mathcal{B}$ such that {\it
s}$_{_T}(W)=X$ \cite{ehn}. If a measurable $A\in\mathcal{B}$ is
$T$-compressible then {\it s}$_{_T}(A)\prec\prec A$ and {\it
s}$_{_T}\sim A$.

Let $T_1$ and $T_2$ be Borel automorphisms on a Standard Borel
Space. Then, if $T_1$ and $T_2$ are orbit equivalent and if $T_1$
has an orbit of length $n$ then so has $T_2$. The cardinality of
the set of orbits of length $n$ for $T_1$ and $T_2$ is the same.
Further, if $\mathbf{c}_k(T_1)$ is the cardinality of the set of
orbits of length $k$, then for each $k\leq \aleph_o$,
$\mathbf{c}_k(T_1)=\mathbf{c}_k(T_2)$ whenever $T_1$ and $T_2$ are
orbit equivalent.

Dye's theorem \cite{dye} proves that: any two free ergodic measure
preserving Borel automorphisms on a Standard Probability Space
$(X,\mathcal{B},m)$ are orbit-equivalent $(\mathrm{mod}\;m)$.
Furthermore, we also note that if $T_1$ and $T_2$ are Borel
automorphisms both compressible and not admitting Borel
cross-sections, then $T_1$ and $T_2$ can be shown to be
orbit-equivalent \cite{chaubemgn}.

Let $M(X)=M(X,\mathcal{B},m)$ be the group of all measure
preserving automorphisms on the space $(X,\mathcal{B},m)$. Two
automorphisms in $M$ are identified if they agree a.e.

For a $T\in M$, let $[T]$ denote the \underline{full}
g\underline{rou}p \underline{of} $T$, {\em ie}, the collection of
all $\tau\in M$ such that f.a.e. $x\in X$, $\tau(x)=T^n(x)$ for
some integer $n=n(x)$. Note that $\tau\in [T]$ iff
$\mathrm{orb}(x,\tau)\subseteq \mathrm{orb}(x,T)$ f.a.e $x\in X$,
or equivalently, there exists a decomposition of $X=
\bigcup_{n\in\mathbb{Z}}A_n\;(\mathrm{mod}\;m)$ such that $X=
\bigcup_{n\in\mathbb{Z}}T^nA_n\;(\mathrm{mod}\;m)$, $T^nA_n$ being
pairwise disjoint, and $\tau(x)=T^n(x)$ for $x\in A_n,\;n\in
\mathbb{Z}$.

Let $A\in\mathcal{B}$ and $\tau\in[T]$. We shall write
$\tau\in[T]^+$ on $A$ in case $\tau(x)=T^n(x)$, where $n=n(x)>0$
a.e. on $A$.

An automorphism $T$ is called \underline{set} p\underline{eriodic}
\underline{with} p\underline{eriod} \underline{$k$}, for some
positive integer $k$, if there exists a partition
$\mathcal{P}=\{D_1, D_2, ..., D_k\}$ of $X$ associated with $T$
such that $D_i=T^{i-1}D_1$, for $1\leq i\leq k$ with each $D_i\in
\mathcal{B}$.

If every $x$ is $T$-periodic with period $k$, then it is clear
that $T$ is set periodic with period $k$. However, it should also
be noted that $T$ can be set periodic without having any periodic
points.

An automorphism $T\in M(X)$ is called as a \underline{weak}
\underline{von} \underline{Neumann}
\underline{automor}p\underline{hism} if
\begin{description}\item{(1)} $T$ is set periodic with period
$2^n$ for all $n\in\mathbb{N}$,
\item{(2)} A sequence $\{ \mathcal{D}_n(T)=\left( D_1^n, ...,
D_{2^n}^n\right)\}, \;n\in \mathbb{N}$, exists of partitions of
$X$ associated with $T$ satisfying
\begin{description}\item{(a)} $D_i^n=D_i^{n+1}\bigcup D_{i+2^n}^{n+1}$,
for $i=1,2$, ..., $2^n$, $n\in\mathbb{N}$ \item{(b)}
$D_i^n=T^{i-1}D_i^n$, for $i=1,2$, ..., $2^n$, with
$n\in\mathbb{N}$.\end{description} \end{description}

For $x\in D^n_1$, we shall call the finite sequence $(x, Tx, ...,
T^{2^n-1}x)$ a \underline{fiber} \underline{of}
\underline{len}g\underline{th} $2^n$. Two points $u,v\in X$ are
said to be {\em in the same fiber of length $2^n$\/} if for some
$x\in D_1^n$, $u=T^kx$, $v=T^{\ell}x$, where $0\leq k$, $\ell <
2^n-1$.

If, in addition to the above (1) and (2), we have
\begin{description}\item{(3)} the $\sigma$-field generated by
$\bigcup_{n=1}^{\infty} \mathcal{D}_n (T)$ is equal to
$\mathcal{B}\;(\mathrm{mod}\;m)$,\end{description} $T$ is called
as a \underline{von} \underline{Neumann}
\underline{automor}p\underline{hism}.

This above condition (3) means that there exists a $T$-invariant
set $N\in\mathcal{B}$ which is $m$-null and such that the
collection $\{ D\bigcap (X-n)\;|\;D\in
\bigcup_{n=1}^{\infty}\mathcal{D}_n(T)\}$ generates the
$\sigma$-algebra $\mathcal{B}$ restricted to $X-N$, equivalently,
the sets $D_k^n$ taken over all $n$ and all $k$ separate the
points of $X-N$.

For a weak von Neumann automorphism $T$, let $\mathcal{P}_n(T)$
denote the algebra generated by $\mathcal{D}_n(T)$. Then,
$\mathcal{P}_n(T)\subseteq \mathcal{P}_{n+1}(T)$ and the union
$\mathcal{P}(T)=\bigcup_{n=1}^{\infty}\mathcal{P}_n(T)$ is again
an algebra. For $A\in\mathcal{B}$, write $d(A)=\inf \{ m(A
\triangle B)\;|\;B\in \mathcal{P}(T)\}$. If $d(A)=0$ for every $A$
in a countable collection which generates $\mathcal{B}$ then $T$
is a von Neumann automorphism.

A DAM or Odometer $V$ on $\{0,1\}^{\mathbb{N}}$ is a von Neumann
automorphism. Furthermore, any two von Neumann automorphisms are
isomorphic modulo $m$-null sets.

Now, for ergodic $T\in M(X)$ and $\forall\;\;A, B\in \mathcal{B}$
with $0< m(A)=m(B)$, there exists a $J\in [T]$ such that $JB=A$
and $J\in[T]^+$ on $B$. Therefore, if $m(A)=m(B)$, then $T_{_A}$
and $T_{_B}$ are orbit equivalent. Indeed, $J$ when viewed as an
isomorphism from $A$ to $B$ establishes orbit equivalence
$(\mathrm{mod}\;m)$ between $T_{_A}$ and $T_{_B}$.

Moreover, let $T\in M(X)$ be ergodic and let $\epsilon >0$ be such
that $\epsilon < m(X)$. Then, there exists $A\in \mathcal{B}$ such
that $A\bigcap TA=\emptyset$ and $m(X-A\bigcup TA)=\epsilon$.
Also, there exists a weak von Neumann automorphism $\omega\in [T]$
such that $[\omega]=[T]$.

Further, if $\tau_1\in[T]$ is a set periodic automorphism with
period $2^{^K}$ such that $\mathcal{D}(\tau_1)=(D_1, ...,
D_{2^K})$ is a partition of $X$ associated with $\tau_1$, then,
for any $\epsilon>0$ and any set $A\in\mathcal{B}$, there exists a
weak von Neumann automorphism $\tau_1\in [T]$ and an integer $L>0$
that satisfy \begin{description} \item{(a)} $[\tau_1]=[\tau_2]$
\item{(b)} $\mathcal{D}(\tau_1)\subseteq \mathcal{D}_n(\tau_2)$ for
all $n\geq L$, where $\{\mathcal{D}_n
(\tau_2)\;|\;n\in\mathbb{N}\}$ are the partitions of $X$
associated with $\tau_2$ \item{(c)} $\{x\;|\; \tau_2(x)\neq
\tau_1(x)\} \subseteq D_{2^K}\in \mathcal{D}(\tau_1)$ \item{(d)}
for $n\geq L$, we have $m(A-A'_n)< \epsilon$, $m(A''-A)<
\epsilon$, where $A'_n= \bigcup D$ where union is over
$\mathcal{D}'_n= \{D\in \mathcal{D}_n(\tau_2)\;|\; D\subseteq A\}$
and $A''_n= \bigcup D$ where the union is over
$\mathcal{D}''_n=\{D\in\mathcal{D}_n (\tau_2)\;|\; m(A\bigcap
D)>0\}$. \end{description}

Under the same hypotheses as above, if we have in addition that
$\tau_1\in[T]^+$ on $X\setminus D_{2^K}$ for
$D_{2^K}\in\mathcal{D} (\tau_1)$, then the weak von Neumann
automorphism $\tau_2\in[T]$ and the positive integer $L>0$ chosen
above also satisfy
\begin{description}\item{(e)} $\tau_2\in[T]^+$ on $X\setminus D_{2^L}^L$ for $D_{2^L}^L \in
\mathcal{D}_L (\tau_2)$.\end{description}

Furthermore, there exists an integer $P>L$, and $C\in \mathcal{B}$
with $m(C)<\epsilon$ so that the following holds:
\begin{description}\item{(f)} $C(x,Tx)$ does not intersect $D_{2^P}^P\in
\mathcal{D}_{_P}$ for all $x\in X\setminus C$, where for $y\in
\mathrm{orb}(x,\tau_2)$ with $\tau_2^{n(x)}x=y$ and $C(x,y)=(x,
\tau_2x, ..., \tau_2^nx=y)$, if $n=n(x)\geq 0$ and also if
$C(x,y)=(x, \tau_2^{-\,1}x, ..., \tau_2^nx=y)$, if $n=n(x)<0$.
\end{description} In other words, $x$ and $Tx$ belong to the same
$\tau_2$-fiber of length $2^P$ for any $x\in X\setminus C$.

Then, given a free ergodic measure preserving automorphism $T$ on
a SPS $(X,\mathcal{B}, m)$, there exist two von Neumann
automorphisms $\tau_1$ and $\tau_2$ in $[T]$ such that (i)
$\tau_1\in [T]^+$ on $X$ and (ii) $[\tau_1]=[\tau_2]$.

Note that when two Borel automorphisms on $(X,\mathcal{B})$ are
free and uniquely ergodic, then the orbit equivalence holds
without discarding any set of measure zero. Moreover, any two free
Borel automorphisms on $(X,\mathcal{B})$, each admitting $n$
invariant ergodic probability measures, are orbit equivalent
whether we have $n$ as finite or countable or uncountable
\cite{djk}.

Now, we note that Krieger \cite{krieger} introduces an invariant
called the {\em ratio set, $r(T)$, of automorphism $T$\/} as a
closed subset of $[0,\infty)$ and $r(T)\bigcap (0,\infty)$ is a
closed multiplicative subgroup of $(0,\infty)$. Then, if
$r(T)=r(\tau)=[0,\infty)$ or if $r(T)= r(\tau)=\{0\}\bigcup
\{\alpha^k\;|\;k\in\mathbb{Z}\}$ for some $\alpha,\;0<\alpha<1$,
then $T$ and $\tau$ are orbit equivalent $(\mathrm{mod}\;m)$.

Extending these concepts to more general group actions is
possible. Then, let $G$ be Polish group of Borel automorphisms
acting in a jointly measurable manner on a SBS $(X,\mathcal{B})$.
Then, if $X$ is incompressible with respect to the $G$-action then
there exists a probability measure on $\mathcal{B}$ invariant
under the $G$-action \cite{trim6}.

However, note that further generalizations than above are limited
by counter examples.

For example, let $G$ now denote the group of {\em all\/} Borel
automorphisms of an uncountable Polish space with the property
that the set $\{ x\;|\; gx\neq x\}$ is of the first Baire
category. Then, $X$ is not compressible.

The $\sigma$-ideal $\mathcal{H}_{_G}$ generated by
$G$-compressible sets in $\mathcal{B}$ is the $\sigma$-ideal of
meagre Borel subsets of $X$. Hence, $X\notin \mathcal{H}_{_G}$.
However, every probability measure on $\mathcal{B}$ is supported
on a meagre set. Therefore, a $G$-invariant probability measure on
$\mathcal{B}$ does not exist.

A {\em flow\/} on a SBS $(X,\mathcal{B})$ is said to be
\underline{non}-\underline{sin}g\underline{ular} with respect to a
$\sigma$-finite measure $\mu$ on $\mathcal{B}$ if $\mu(A)=0$
implies that $\mu(T_t(A))=0$ for all $A\in\mathcal{B}$ and $t\in
\mathbb{R}$. In case, $\mu(T_t(A))=\mu(A)$ for all $t\in
\mathbb{R}$ and $A\in\mathcal{B}$, then we say that the
\underline{flow} p\underline{reserves} $\mu$.

Let $\sigma$ be a Borel automorphism on a SBS $(Y,\mathcal{C})$
and let $f$ be a positive Borel function on $Y$ such that
$\forall\;y \in Y$, the sums $\sum_{k=0}^{\infty}f(\sigma^ky)$,
$\sum_{k=0} ^{\infty} f(\sigma^{-\,k}y)$ are infinite. Let
$X=\{(y,t)\;|\; 0\leq t < f(y)\}$. Then, $X$ is the subset of
$Y\times\mathbb{R}$ strictly under the graph of $f$. Give
$Y\times\mathbb{R}$ the product Borel structure and restrict it to
$X$. We then obtain a new Borel space $(X,\mathcal{B})$.

A jointly measurable flow $T_t,\;t\in\mathbb{R}$, on $X$ can be
defined as follows: a point $(y,u)\in X$ moves vertically up with
``unit speed'' until it reaches the point $(y,f(y))$ when it goes
over to $(\sigma(y),0)$ and starts moving up again with unit
speed. The term \underline{unit} \underline{s}p\underline{eed}
means that the linear distance travelled in unphysical time $t$
equals $t$. The point thus reached at unphysical time $t>0$ is
defined to be $T_t(y,u)$. For $t<0$, $T_(y,u)$ is defined to be
the point $(y',u')$ such that $T_{-\\,t}(y',u')-(y,u)$. The point
$(y,0)$ is called the \underline{base} p\underline{oint}
\underline{of} $(y,u)$.

Analytically, the above is expressible as follows: Let $x=(y,u)\in
X$, and let $t\geq 0$. Then, $T_t(x)= T_t(y,u)=
\left(\sigma^ny,t+u-\sum_{k=0}^{n-1}f(\sigma^ky)\right)$ where $n$
is the unique integer such that $\sum_{k=0}^{n-1}f(\sigma^ky)\leq
t+u< \sum_{k=0}^nf(\sigma^ky)$. If $t<0$, the expression is
$T_t(x)=
\left(\sigma^{-\,n}y,t+u+\sum_{k=1}^{n}f(\sigma^{-\,k}y)\right)$
where $n$ is the unique integer such that $0\leq t+u+ \sum_{k=1}^n
f(\sigma^{-\,k}y)< f(\sigma^{-\,n}y)$. It is understood that
$\sum_{k=0}^{-\,1}$ and $\sum_{k=0}^0$ are equal to zero. It is
easy to verify that $T_t, \;t\in\mathbb{R}$ is indeed a flow on
$X$.

The flow $T_t, \;t\in\mathbb{R}$ as defined above is called the
{\em flow (or special flow) built under the function $f$ with base
automorphism $T$ and base space $(Y, \mathcal{C})$}.  Note that a
flow built under a function is a continuous version of
automorphism built under a positive integer-valued function. We
thus use the notation of $T^f$ for the continuous case also.

Let the base space $Y$ be Polish, the base automorphism $\sigma$ a
homeomorphism of $Y$ and $f$ continuous on $Y$. Let us give
$Y\times \mathbb{R}$ the product topology, where $\mathbb{R}$ has
the usual topology. Let $\bar{X}=\{(y,t)\;|\;0\leq t\leq f(y)\}$
be the closure of $X\subseteq Y\times\mathbb{R}$. Now, define $g:
\bar{X}\to X$ by $g(y,t)=(y,t)$ if $0\leq t\leq f(y)$ and $g(y,t)=
(\sigma y,o)$ if $t=f(t)$. The map $g$ identifies the point
$(f,f(y))$ with $(\sigma y, 0)$. Let $\mathcal{T}$ be the largest
topology on $X$ that makes $g$ continuous. Under this topology,
the flow  $\sigma^f$ is a jointly continuous flow of
homeomorphisms on $X$. The topology $\mathcal{T}$ can be shown to
be a Polish Topology.

Then, a jointly measurable flow is also jointly continuous with
respect to a suitable complete separable metric topology, the
Polish topology, on $X$ which also generates the $\sigma$-algebra
$\mathcal{B}$.

Let $T_t,\;t\in\mathbb{R}$ be a jointly measurable flow (without
fixed points) on a SBS $(X,\mathcal{B})$. Suppose we are able to
choose on each orbit of $T_t$ a non-empty discrete set of points
with all these points taken over all orbits forming a Borel set in
$\mathcal{B}$. Then, we suppose that there exists a Borel set
$Y\subseteq X$ such that $\forall\;x\in X$, the set $\{ t\;:\;
T_t(x)\in Y\}$ is a non-empty and discrete subset of $\mathbb{R}$.
Such a subset is called a \underline{countable}
\underline{cross}-\underline{section} \underline{of}
\underline{the} \underline{flow}.

Given a countable cross-section $Y$, we can write $X$ as the union
of three Borel sets $I$, $J$, $K$ as: $I=\{ x\in X\;|\; \{t\;|\;
T_t(x) \in Y\} \;\mathrm{is\;bounded\;below}\}$, $J=\{ x\in X\;|\;
\{t\;|\; T_t(x) \in Y\} \;\mathrm{is\;bounded\;above}\}$, $K=X-I
\bigcup J$. Let $i(x)=\inf \{t\;|\;T_t(x)\in Y\}$ and $j(x)=\sup
\{t\;|\; T_t(x) \in Y\}$. Then $i$ and $j$ are measurable
functions, so that $I$ and $J$, hence, also $K$, are measurable
sets.

Let us write $S(x)=T_{i(x)}(x),\;x\in I$. Then, $S(T_t(x))=S(x)$
$\forall\;t\in\mathbb{R}$ since $i(T_t(x))=i(x)-t$. The function
$S:I\to I$ is again measurable, and constant on orbits. Thus, if
we restrict the flow to $I$ then the orbit space admits a Borel
cross-section, the image of $I$ under $S$ being the required Borel
cross-section. Similarly for $J$. Therefore, in the set $I\bigcup
J$, the flow is isomorphic to a flow built under a function.

Then, there exists a Borel set $Y\subseteq X$ such that $\forall
\;x\in X$ the set $\{t\;|\;T_tx\in Y\}$ is non-empty, countable,
and discrete in $\mathbb{R}$, the flow is isomorphic to a flow
built under a function.

Thus, we note that every jointly measurable flow (without fixed
points) on a SBS admits a countable cross-section.

Further, for a jointly measurable $T_t$, it can be shown
\cite{vmw} that there exists a set $B\in\mathcal{B}$ such that
$\forall\;x\in X$ the sets $\{t\in\mathbb{R}\;|\; T_tx\in B\}$ and
$\{t\in\mathbb{R}\;|\;T_tx\notin B\}$ have positive Lebesgue
measure.

Then, it can further be shown \cite{vmw} that every jointly
measurable flow $T_t,\;t\in\mathbb{R}$ (without fixed points) on a
SBS $(X,\mathcal{B})$ admits a measurable subset $Y\subseteq X$
such that $\forall\;x\in X$ the set $\{ t\;|\;T_tx\in Y\}$ is
non-empty and discrete in $\mathbb{R}$. Therefore, we see that
every jointly measurable flow (without fixed points) on a SBS is
isomorphic to a flow built under a function.

For general finite measure preserving flows, this result was
proved in \cite{ambrose} while the refinement and adaptation of
that method to a descriptive setting can be found in \cite{vmw}.

As a corollary, every jointly measurable flow without fixed points
on a SBS $(X,\mathcal{B})$ is a flow of homeomorphisms under a
suitable Polish topology on $X$ which generates $\mathcal{B}$.

Furthermore, for a jointly measurable flow $T_t\;t\in\mathbb{R}$
(without fixed points) on a SBS $(X,\mathcal{B})$ and given $0\leq
\alpha\leq 1$, there exists $B\in\mathcal{B}$ such that
$\forall\;x\in X$ the orbit of $x$ spends the proportion $\alpha$
of time in $B$, that is, $\forall\;x\in X$,
$\frac{1}{N}\;\mathrm{Lebesgue\; measure}\;\{ t\;|\; T_tx\in
B,\,0\leq t < N\}\to \alpha$ as $N\to \infty$.

Note also that, under suitable modifications of the definition of
flow built under a function, these results hold for jointly
measurable flows with fixed points as well.

Now, consider the notion of a flow built under a function in a
measure theoretic setting. Let $(Y,\mathcal{B}_{_Y})$ be a SBS
equipped with a Borel automorphism $\tau:Y\to Y$ and a
$\sigma$-finite measure $n$ quasi-invariant for $\tau$.

[A measure $n$ on $\mathcal{B}$ is called
q\underline{uasi}-\underline{invariant} for $\tau$ if $n(B)=0$ iff
$n(\tau B)=0$ and is called \underline{conservative} for $\tau$ if
$n(W)=0$ for every $\tau$-wandering set $W$.]

Let $f$ be a positive Borel function on $Y$ such that
$\forall\;y$, the sums $\sum_{k=0}^{\infty} f(\tau^ky)$ and
$\sum_{k=1}^{\infty} f(\tau^{-\,1}y)$ are infinite. Let
$T_t,\;t\in\mathbb{R}$ be the flow $\tau^f$ built under $f$ with
base space $(Y,\mathcal{Y})$ and base automorphism $\tau$. It acts
on $Y^f=\{(y,t)\;|\; 0\leq t < f(y),\;y\in Y\}$.

Let $\ell$ denote the Lebesgue measure on $\mathbb{R}$ and let the
measure $n\time \ell$ on $Y\times \mathbb{R}$ be restricted to
Borel subsets of $Y^f$. Let us denote this measure on $Y^f$ by
$m=m_f$.

The flow $T_t,\;t\in\mathbb{R}$, when considered together with the
measure $m$ is called the {\em flow built under} $f$ in a
\underline{measure} \underline{theoretic} \underline{sense}. We
call the measure $n$ the \underline{base} \underline{measure}.

Now, \cite{jmmgn}, for any $t\in\mathbb{R}$,\[
\frac{dm_t}{dm}(y,u)=\frac{dn_{_{(t+u)y}}}{dn}(y)\;\;\;\;\mathrm{a.e.m.}\]
where $m_t$ and $n_k$ are the measures $m(T_t(.))$ and
$n(\sigma^k(.))$ respectively and $\frac{dm_t}{dm}$ denotes the
LRN derivative of a quasi-invariant measure \cite{measure-theory}.

Recall that the flow $T_t,\;t\in\mathbb{R}$, is the flow
$\sigma^f$. Then, as a corollary, we also see that $m$ is
quasi-invariant under the flow $\sigma^f$ iff $n$, the base
measure, is quasi-invariant under $\sigma$. $m$ is invariant under
$\sigma^f$ iff $n$ is invariant under $\sigma$.

Consider now a jointly measurable flow $\tau_t,\;t\in\mathbb{R}$,
on a SBS $(X,\mathcal{B})$ equipped with a probability measure $m$
quasi-invariant under the flow. Let us also assume, for
simplicity, that the flow $T_t,\;t\in\mathbb{R}$, is free. Then,
the map $t\to \tau_t x$ is one-one from $\mathbb{R}$ onto the
orbit $\{\tau_tx\;|\;t\in\mathbb{R}\}$. Thus, a Lebesgue measure
is definable on the orbit simply by transferring the Lebesgue
measure of $\mathbb{R}$ to it. Let us denote by $\ell_x$ this
Lebesgue measure on the orbit of $x$ under the flow.

Then, $m(A)=0$ iff $\ell(\{t\;|\;\tau_tx\in A\})=\ell_x(A)=0$ for
$m$-almost every $x$. [A property which holds for all $x\in X$
except for those $x$ in some $m$-null set is said to hold
$m$-almost everywhere.]

Let $\tau_t,\;t\in\mathbb{R}$, on $(X,\mathcal{B},m)$ and
$T_t,\;t\in\mathbb{R}$, on $(X',\mathcal{B}',m')$ be two
non-singular flows. We shall say that the two flows are
\underline{metricall}y \underline{isomor}p\underline{hic} if there
exist
\begin{description}
\item{(i)} $\tau_t$-invariant $m$-null set $M\in\mathcal{B}$ and
$T_t$-invariant $m'$-null set $M'\in\mathcal{B}'$, \item{(ii)} a
Borel automorphism $\phi$ of $X-M$ onto $X'-M'$
\end{description} such that $\forall\;t\in\mathbb{R}$, and $x'\in
X'-M'$ we have
\begin{description} \item{(a)} $\phi\circ\tau_t\circ\phi^{-\,1}(x')=T_tx'$,
\item{(b)} $m(\phi^{-\,1}(A'))=0\;\Longleftrightarrow m'(A')=0,\;\forall\;
A'\in\mathcal{B}'$, \item{(c)} in case the flows are measure
preserving we require $m\circ\phi^{-\,1}=m'$ in place of above
(b).\end{description}

Then, as shown in \cite{sgdani}, every non-singular free flow
$\tau_t,\;t\in\mathbb{R}$, on a SPS $(X,\mathcal{B},m)$ is
isomorphic to a flow built under a function in the measure
theoretic sense. The function which implements the isomorphism
preserves null sets.

On the other hand, the basic theorem of Ambrose \cite{ambrose}
states that: every free measure preserving flow on a SPS
$(X,\mathcal{B},m)$ is isomorphic to a flow built under a function
in the measure theoretic sense. The function which implements the
isomorphism preserves the measure. This holds also if $m$ is a
$\sigma$-finite measure \cite{krengel}.

We shall end our rapid survey of the basics and some results of
dynamical systems at this point. In surveying these mathematical
developments, our purpose was mainly to develop the required
mathematical vocabulary for the development of the Universal
Theory of Relativity. We therefore stated various definitions and
quoted results without any proofs. Details of proofs can be found
in the references provided.

However, we also note that the current mathematical apparatus of
the theory of dynamical systems will be found to be inadequate to
``visualize'' physical situations. (See later.) Certainly, some
``new'' conceptions will help here.

\subsection{Physical aspects of the mathematical formalism}

In the absence of any relevant motivation, we shall not consider
higher than four dimensions here, {\em ie}, we shall consider only
three spatial dimensions and one time dimension. Moreover, we
shall adopt the approach of dynamical systems and, hence, will
treat time as a parameter of the dynamical system.

Therefore, we assume that the physical world is describable using
a suitable 3-dimensional (topological) space, denoted by
$\mathfrak{S}$. We shall call $\mathfrak{S}$ the
p\underline{h}y\underline{sical} \underline{s}p\underline{ace}
\underline{of} \underline{Universal} \underline{Relativit}y.

Next, we will \underline{assume} that the cardinality of the
physical space $\mathfrak{S}$ is $\mathbf{c}$, {\em ie},
\underline{some} \underline{suitable} \underline{continuum}
\underline{underlies} \underline{the}
p\underline{h}y\underline{sical} \underline{world}. Further, {\em
we may expect the physical space $\mathfrak{S}$ to be Standard
Borel\/} and also to be a Lebesgue measure space. {\em Physical
objects\/} are then {\em Borel subsets\/} of $\mathfrak{S}$ with
{\em Borel measures\/} as their {\em physical properties}.

But, the 3-space $\mathbb{R}^3$ cannot be the {\em physical\/}
space  $\mathfrak{S}$ of Universal Theory of Relativity because
Newton's theory, which assumes $\mathbb{R}^3$ to be the underlying
physical space, does not describe the physical reality in its
totality.

However, $\mathbb{R}^3$ is a Standard Borel Space and, under our
assumptions, the space $\mathfrak{S}$ is related to $\mathbb{R}^3$
by a morphism, a continuous map, in the category of all standard
Borel spaces.

Next, let us recall that closed and bounded subsets of
$\mathbb{R}^3$ are compact. Consider the set
$\mathcal{K}(\mathbb{R}^3)$ of all non-empty compact subsets of
$\mathbb{R}^3$. This set can be equipped with the Vietoris
topology that is compatible with the Hausdorff metric $\delta_H$.
Since $\mathbb{R}^3$ is standard Borel with the usual metric, so
is $\left(\mathcal{K} (\mathbb{R}^3), \delta_H\right)$ standard
Borel with the topology induced by the Hausdorff metric.

The set $F(\mathbb{R}^3)$ of all nonempty closed subsets of
$\mathbb{R}^3$ can be equipped with the $\sigma$-algebra
$\mathcal{E}(\mathbb{R}^3)$ generated by sets of the form $\left\{
F\in\mathcal{E}(\mathbb{R}^3): F\bigcap U \right.$ $\left.\neq
\emptyset \right\}$, where $U$ varies over open sets of
$\mathbb{R}^3$. We then obtain the Effros Borel Space $\left(
F(\mathbb{R}^3), \mathcal{E}(\mathbb{R}^3) \right)$ of
$\mathbb{R}^3$. The Effros Borel Space of $\mathbb{R}^3$, being
Polish, is standard Borel.

Moreover, consider the base $B(\mathbb{R}^3)$ for the usual
topology of $\mathbb{R}^3$. Then, the Borel space of
$B(\mathbb{R}^3)$ equipped with the Fell topology, {\em ie}, the
pair of $B(\mathbb{R}^3)$ and the smallest $\sigma$-algebra
containing the Fell topology of $\mathbb{R}^3$, is exactly the
same as the Effros Borel Space of $\mathbb{R}^3$ because every
compact subset of $\mathbb{R}^3$ is closed and bounded.

The {\em physical space}, $\mathfrak{S}$, is then also related by
suitable morphism in the category of all standard Borel spaces to
the Polish space $\mathbb{K}(\mathbb{R}^3)$ equipped with the
Hausdorff metric $\delta_H$.

Essentially, the physical space $\mathfrak{S}$ of Universal
Relativity is some standard Borel space in the category of all
standard Borel spaces and we can utilize this fact to our
advantage.

As noted earlier, there exist hierarchies of Borel-point classes
as well as hierarchies of projective sets for a standard Borel
space. We are interested in these classes and in measures defined
on their member sets.

Now, the space $\mathfrak{S}$ is a Lebesgue Measure Space. Measure
preserving transformations of the physical measure space
$(\mathfrak{S}, \mathcal{B}_{\mathfrak{S}}, \mu)$ are therefore
natural for us to consider.

Let us call every member of a measurable partition (mod 0)
$\Upsilon$ of the p\underline{h}y\underline{sical}
\underline{measure} \underline{s}p\underline{ace} $(\mathfrak{S},
\mathcal{B}_{\mathfrak{S}}, \mu)$ as a \underline{basic}
(p\underline{h}y\underline{sical}) \underline{ob}j\underline{ect}.
Then, any $\Upsilon$-set, also a standard Borel set in
$\mathfrak{S}$, can be called as a
\underline{com}p\underline{ound}
(p\underline{h}y\underline{sical}) \underline{ob}j\underline{ect}.
Classes of measures defined on these physical objects,
mathematically well defined subsets of the space $\mathfrak{S}$,
are their physical properties then.

We could then define the distance between physical objects as the
Hausdorff distance between sets, using for this association the
standard Borel character of the space $\mathfrak{S}$. This is
clearly doable in a {\em continuous\/} manner using the morphism
in the category of all standard Borel spaces.

As any physical object moves, its {\em physical\/} movement is
then describable as the action of the transformation of the space
$\mathfrak{S}$ on it.

But, what distinguishes the physical space $\mathfrak{S}$ of
Universal Relativity from the other standard Borel spaces in the
category of all standard Borel spaces? Up to isometries, a metric
uniquely characterizes a metric space. We therefore look for some
property which helps us uniquely determine the metric for the
physical space $\mathfrak{S}$.

Now, according to Einstein's and Descartes's conceptions
\cite{ein-pop}, physical objects are the regions of space and vice
versa. Then, the physical space itself (and its unique
characteristic) must change as the physical objects change.

Then, the \underline{uni}q\underline{ue}
\underline{identif}y\underline{in}g \underline{characteristic} of
the physical space $\mathfrak{S}$ is provided by the following key
physical situation:
\begin{itemize}
\item {\em physical matter can be assembled (as well as reassembled) in any
arbitrary manner at any location in the Universe}.
\end{itemize}
But, this is equivalent to changing {\em continuously\/}
measurable partitions and Borel measures of the physical measure
space $\mathfrak{S}$. Perhaps \footnote{At the present stage, we
proceed here without the proof for this statement.}, for this, the
space $\mathfrak{S}$ needs to admit three, linearly independent,
homothetic Killing vectors which {\em uniquely determine\/} its
line element.

Now, a differentiable manifold $X$ admits a Riemannian metric, a
type $(0,2)$ tensor such that for all $p_1, p_2 \in X$,
$g(p_1,p_2)$ is symmetric and positive-definite with
$g(p_1,p_2)=0$ if and only if $p_1=p_2$. A Riemannian
pseudo-metric $\hat{g}$ is also a type $(0,2)$ tensor that is
symmetric and non-degenerate with $\hat{g}(p,p)=0$ for all $p\in
X$.

In general, a homothetic Killing vector captures the notion of the
scale-invariance of a differentiable manifold. A manifold that
conforms to some scale-invariance is then required to admit an
appropriate homothetic Killing vector ${\bf X}$ satisfying \be
{\cal L}_{\bf X} g_{ab}\;=\;2\,\Phi\,g_{ab} \label{hkveq} \ee
where $g_{ab}$ is the metric, ${\cal L}_{\bf X}$ is the Lie
derivative and $\Phi$ is an arbitrary constant.

This is also the broadest, Sophus Lie's, sense of the
scale-invariance leading not only to the reduction of the partial
differential equations to ordinary differential equations but
leading, simultaneously, also to their separation.

The Killing equation holds also for a Riemannian pseudo-metric
(Abraham \& Marsden, \cite{dyn-sys}, p.\ 144-157). In general, we
then demand that the space admitting no special symmetries, that
is no proper Killing vectors, admits {\em three\/} linearly
independent homothetic Killing vectors. Such a metric, from the
broadest (Lie) sense, admits three functions $P(x)$, $Q(y)$,
$R(z)$ of three space variables, conveniently called here, $x$,
$y$, $z$, each being a function of only one variable.

Based on the above considerations, we then demand that the space
$\mathfrak{S}$ admits three independent homothetic Killing vectors
\beq {\bf {\cal X}} = (f(x), 0, 0) \label{genhkv1} \\ {\bf {\cal
Y}} = (0, g(y), 0) \label{genhkv2} \\ {\bf {\cal Z}} = (0, 0,
h(z)) \label{genhkv3} \eeq for its line element \be
d\ell^2\;=\;g_{ab}dx^adx^b \ee Here the vectors ${\cal X}$, ${\cal
Y}$ and ${\cal Z}$ satisfy (\ref{hkveq}) with $\Phi_x$, $\Phi_y$,
$\Phi_z$ as corresponding constants.

As can be easily checked, the aforementioned demand, that the
continuum $\mathfrak{S}$ admits three linearly independent
homothetic Killing vectors, leads us, {\em uniquely}, to a
three-dimensional space, $\mathfrak{S}$, admitting the following
line element \cite{smw-issues} (after suitable redefinitions of
constants): \setcounter{equation}{0} \beq \label{3-metric}
d\ell^2= {P'}^2Q^2R^2\, dx^2 &+&P^2\bar{Q}^2 R^2\, dy^2 \n \\
&+&\;P^2Q^2\tilde{R}^2 \,dz^2 \eeq where we have $P\equiv P(x)$,
$Q\equiv Q(y)$, $R\equiv R(z)$ and $P'=dP/dx$, $\bar{Q}=dQ/dy$,
$\tilde{R}=dR/dz$. The vanishing of any of these spatial functions
is a {\em curvature singularity}, and constancy (over a range) is
a {\em degeneracy\/} of (\ref{3-metric}).

We will restrict to triplets of {\em nowhere-vanishing\/}
functions $P$, $Q$, $R$ and will also not consider any degenerate
situations for (\ref{3-metric}).

Now, with coordinates $x, y, z$:
\[\hat{g}_{ab}=\mathrm{diag}\,\left({P'}^2Q^2R^2, P^2\bar{Q}^2
R^2,P^2Q^2\tilde{R}^2 \right)\] Then, for some two
\underline{distinct} points $(x_1, y_1, z_1)$ and $(x_2,y_2,z_2)
\in \mathfrak{S}$ and for each of which $P'=\bar{Q}=\tilde{R}=0$,
the line element $d\ell^2=\hat{g}_{ab}dx^adx^b$ vanishes. The
$\hat{g}_{ab}$ is then a Riemannian {\em pseudo-metric\/} on the
space $\mathfrak{S}$.

Given $P$, $Q$, $R$, consider the equivalence class of $p\in
\mathfrak{S}$: $\hat{g}[p]=\{x:x\in \mathfrak{S}, \hat{g}(p,x) =
0\}$ and also the quotient space $\mathfrak{S}\diagup\hat{g}$. Let
$A,B\in \mathfrak{S}\diagup \hat{g}$. Define $g(A,B)=\hat{g}(p,q)$
with $p\in A$ and $q\in B$. Then, $g$ is a Riemannian metric on
$\mathfrak{S}\diagup \hat{g}$. On using $P,Q,R$ as coordinates on
$\mathfrak{S}\diagup \hat{g}$, we have
\[g_{ab} = \mathrm{diag}\, \left(Q^2R^2, P^2 R^2,P^2Q^2 \right)\]

Since $\hat{g}$ ``lives on'' $\mathfrak{S}$ while $g$ ``lives on''
$\mathfrak{S}\diagup \hat{g}$, we {\em distinguish\/} them. But,
$\mathfrak{S}$ and $\mathfrak{S}\diagup \hat{g}$ are homeomorphic
being of cardinality $\mathbf{c}$, both. Hence, we will, for
brevity, write $\mathfrak{S}$ for $\mathfrak{S}\diagup\hat{g}$.

If $\mathcal{T}$ denotes the metric topology induced by the metric
$g$ on $\mathfrak{S}$, more precisely on $\mathfrak{S}\diagup
\hat{g}$, then $(\mathfrak{S}, \mathcal{T})$ is an uncountable
Polish space and, hence, of cardinality $\mathbf{c}$. We then
obtain a Standard Borel Space $(\mathfrak{S},
\mathcal{B}_{\mathfrak{S}})$ where $\mathcal{B}_{\mathfrak{S}}$ is
the smallest $\sigma$-algebra of the subsets of $\mathfrak{S}$
containing $\mathcal{T}$.

We have therefore a unique characterization of the space
$\mathfrak{S}$ as a Standard Borel Space underlying Universal
Relativity.

Now, consider a subset of the space $\mathfrak{S}$ such that the
derivatives $P'$, $\bar{Q}$, $\tilde{R}$ occurring in the
pseudo-metric $\hat{g}_{ab}$ are of ``one fixed'' sign for all of
its points. We could then choose $P,Q,R$ as coordinates within
such a subset and could, then, write $\hat{g}_{ab}=
\mathrm{diag}\, \left(Q^2R^2, P^2 R^2,P^2Q^2 \right)$.

Hence, there are certain subsets of $\mathfrak{S}$, to be called
as the P-sets, on which the ``restriction'' of the function
$\hat{g}_{ab}$ is a function $g_{ab}$, {\em ie}, $g_{ab}=
\hat{g}_{ab}|P$. A P-set of $(\mathfrak{S},\hat{g})$ is never a
singleton subset of $\mathfrak{S}$ when the functions $P,Q, R$ are
as chosen. The P-sets are open sets of the Polish topology
$\mathcal{T}$ of $\mathfrak{S}$. Note also that every open set of
$(\mathfrak{S}, \mathcal{T})$ is {\em not\/} a P-set of
$(\mathfrak{S}, \hat{g})$. For example, a proper subset of a P-set
of $(\mathfrak{S}, \hat{g})$ will not be a P-set of
$(\mathfrak{S}, \hat{g})$ but it can be an open set of
$(\mathfrak{S}, \mathcal{T})$.

By construction, any two distinct P-sets, $P_i$ and $P_j$,
$i,j\;\in\;{\rm\bf N}$, $i\,\neq\,j$, are {\em pairwise disjoint
subsets\/} of $\mathfrak{S}$. Consequently, for a specific
pseudo-metric $\hat{g}_{ab}$, the collection of all the P-sets
provides us a {\em partition\/} of the space $\mathfrak{S}$.
Furthermore, each P-set is, in own right, an uncountable Polish
space with $g_{ab}=\hat{g}_{ab}|P$ as a metric compatible with its
(induced) Polish topology.

Being members of the Borel $\sigma$-algebra
$\mathcal{B}_{\mathfrak{S}}$ of the space $\mathfrak{S}$, the
P-sets are (Borel) measurable and, for specific $\hat{g}_{ab}$,
the collection of all the P-sets is a measurable (mod 0) partition
of the space $\mathfrak{S}$. Hence, {\em (Lebesgue) measures\/}
and {\em signed\/} measures on $\mathfrak{S}$ are natural for us
to consider.

Therefore, to every class of (Lebesgue) measures on such P-sets we
can associate an appropriate physical property of a {\em material
body}. A material body is always an {\em extended body}, since a
P-set cannot be a singleton subset of $\mathfrak{S}$.

The integration of measures on P-sets is always a well-defined
one, now for obvious mathematical reasons. A chosen measure can be
integrated over a P-set and an average value of the measure always
obtainable. This average value of measure provides then an
``averaged quantity'' characteristic of a P-set. Evidently, this
``average'' is a property of the entire P-set under consideration
and, therefore, of {\em every point\/} of that P-set.

A point of the P-set is then thinkable as having these averaged
properties of the P-set and, in this precise non-singular sense,
is thinkable as a (newtonian) {\em point-particle\/} possessing
those averaged properties. In this non-singular sense, points of
the space $\mathfrak{S}$ become point particles.

In essence, we have,  in a non-singular manner, then ``recovered''
the (newtonian) notion of a point particle from that of our notion
of a field - the underlying continuum $\mathfrak{S}$.

Further, the ``location'' of this point-particle will be {\em
indeterminate\/} over the {\em size\/} of that P-set because the
averaged property is also the property of every point of the set
under consideration. The individuality of a point particle is then
that of the corresponding P-set.

Now, for a specific pseudo-metric $\hat{g}_{ab}$, corresponding
collection $\Upsilon$ of all P-sets forms a measurable (mod 0)
partition of the space $\mathfrak{S}$. Recalling our earlier
terminology, a P-set is then a {\em basic (physical) object}.
Standard Borel Sets in $\mathcal{B}_{\mathfrak{S}}$ which are the
unions of the members of the measurable (mod 0) partition
$\Upsilon$, now the P-sets, are then the {\em compound (physical)
objects}.

Measures can also be integrated over compound (physical) objects
and a point of $\mathfrak{S}$ in the object is then also thinkable
as a (newtonian) point particle with these physical properties.
Location of such a point particle is then indeterminate over the
size of that object.

Then, the points of the underlying space $\mathfrak{S}$ can also
be attributed (physical) properties averaged over the size of an
object. Hence, we can also represent an object under consideration
as a (newtonian) point particle.

Therefore, we have the required characteristics of Descartes's and
Einstein's conceptions incorporated in the present formalism.
Clearly, we have then the non-singular notion of a point particle
as well as that of replacing any extended physical body by such a
non-singular point particle. Furthermore, physical bodies are also
represented as non-singular regions of the space $\mathfrak{S}$.
Thence, the union of the space and the physical objects is clearly
perceptible here.

Since $(\mathfrak{S},\mathcal{B}_{\mathfrak{S}})$ is a standard
Borel space, any measurable, one-one map of $\mathfrak{S}$ onto
itself is a Borel automorphism. Therefore, the Borel automorphisms
of $(\mathfrak{S},\mathcal{B}_{\mathfrak{S}})$, forming a group,
are natural for us to consider here. The Borel automorphisms of an
uncountable Polish space have been the subject of a recent study
\cite{clemens}.

As any Borel automorphism of the underlying space $\mathfrak{S}$
maps a chosen P-set or a $\Upsilon$-set to another, the integrated
properties may change and, consequently, the (initial)
characteristics of particle of that P-set/$\Upsilon$-set may
change.

[In this case, it is possible to adopt two views. In the first
one, the active point of view, we imagine that a chosen P-set or
an $\Upsilon$-set itself changes under the action of the Borel
transformation of $\mathfrak{S}$ and ``track'' these changes. In
the second, the passive point of view, we imagine that the action
of the Borel transformation of $\mathfrak{S}$ only maps a given
(measurable) set onto (measurable) another and track the changes
in measures.

We note that the standard methods of the theories of measures as
well as dynamical systems adopt the passive point of view and, it
seems that the active point of view is more closer to the
physicist's ways of thinking \footnote{It is, perhaps, for such
reasons that the current methods of dynamical systems appear, to
use too strong a word here, ``inadequate'' for the visualization
of physical processes in the Universal Theory of Relativity. The
author has not yet given any deeper attention to these issues.
However, he believes them to be interesting and certainly worthy
of investigation.}.]

Now, the {\em Hausdorff metric\/} provides the distance separating
P-sets and also the distance separating $\Upsilon$-sets. This
distance between sets will, henceforth, be called the {\em
physical distance\/} between P-sets or $\Upsilon$-sets (extended
physical bodies) because ``measurement in the physical sense'' can
be expected to yield only this quantity as distance separating
physical objects.

Measure-preserving Borel automorphisms of the space $\mathfrak{S}$
then ``transform'' a given set maintaining its characteristic
classes of (Lebesgue) measures, that is, its physical properties.

Non-measure-preserving Borel automorphisms change the
characteristic classes of Lebesgue measures (physical properties)
of a set while ``transforming'' it. Evidently, such considerations
also apply to even $\Upsilon$-sets.

At this point, we note that a general automorphism of the space
$\mathfrak{S}$ has two parts: one measure-preserving and one non
measure-preserving. This decomposition is maintained (mod 0).
Hence, only the measure-preserving transformations are studied in
the theory of dynamical systems.

Then, a {\em periodic\/} Borel automorphism or {\em periodic
component of\/} Borel automorphism \cite{trim6} of $\mathfrak{S}$
will lead to an {\em oscillatory motion\/} of a set while
preserving or not preserving its measures.

Therefore, a basic or compound (physical) object undergoing
periodic motion is a physical clock in the present framework. Such
an object undergoing oscillatory motion then ``displays'' the
time-parameter of the corresponding (periodic) Borel automorphism
since the period of the motion of such an object is precisely the
period of the corresponding Borel automorphism.

Then, within the present formalism, a {\em measuring clock\/} is
therefore any $\Upsilon$-set or an object undergoing {\em periodic
motion}. An $\Upsilon$-set or an object can also be used as a {\em
measuring rod}.

Therefore, in the present theoretical framework, measuring
apparatuses, measuring rods and measuring clocks, are on par with
every other thing that the formalism intends to treat.

A Borel automorphism of $(\mathfrak{S},
\mathcal{B}_{\mathfrak{S}})$ may change the physical distance
resulting into ``relative motion'' of objects. We also note here
that the sets invariant under the specific Borel automorphism are
characteristic of that automorphism. Hence, such sets will then
have their ``relative'' distance ``fixed'' under that Borel
automorphism and will be stationary relative to each other.

Now, in a precise sense, it follows that the position of the
point-particle (of averaged characteristics of its associated
$\Upsilon$-set) is ``determinable'' more and more accurately as
the size of that $\Upsilon$-set gets smaller and smaller. But,
complete localization of a point particle is not permissible here
since an $\Upsilon$-set is never a singleton subset of
$\mathfrak{S}$. The location of the point particle is then always
``indeterminate'' to the extent of the size of its $\Upsilon$-set.
Clearly, this is an intrinsic indeterminacy that cannot be
overcome in any manner.

Furthermore, a Borel automorphism of the space $\mathfrak{S}$
results in a ``relative motion'' of $\Upsilon$-sets and, hence, of
associated particles.

Clearly, therefore, a joint manifestation of Borel automorphisms
of the space $(\mathfrak{S}, \mathcal{B}_{\mathfrak{S}})$ and the
intrinsic indeterminacy in the location of a point particle, of
averaged measures on an $\Upsilon$-set, is a candidate reason
behind Heisenberg's indeterminacy relations in Universal
Relativity.

This is in complete contrast to their probabilistic origin as
advocated by the standard formalism of the quantum theory.

Notice now that, in the present considerations, we began with none
of the fundamental considerations of the concept of a quantum.
But, one of the basic characteristics of the conception of a
quantum, Heisenberg's indeterminacy relation, emerged out of the
present formalism.

Furthermore, in the present framework, we have also done away with
the ``singular nature'' of the particles and, hence, also with the
unsatisfactory dualism of the field (space) and the source
particle. We also have, simultaneously, well-defined laws of
motion (Borel automorphisms) for the field (space) and also for
the well-defined conception of a point particle (of averaged
measure characteristics of an basic or compound object). Then, the
present formalism is a {\em complete\/} field theory.

[At this point, we then also note that the Borel automorphisms of
$\mathfrak{S}$ need not be differentiable or, for that matter,
even continuous. Therefore, the present considerations also use,
for the most fundamental formalism of physics, a mathematical
structure different than that of the partial differential
equations. However, the question of the physical significance of
non-differentiable and non-continuous Borel automorphisms of
$\mathfrak{S}$ is a subject of independent detailed study.]

Now, any act of measurement is conceivable here only as a Borel
automorphism of the space $\mathfrak{S}$. Then, the measurability
of any characteristic of a point particle as defined in the
present framework is dependent on the ``Borel automorphism'' to be
used. But, from active point of view, that Borel automorphism may
change the very basic or compound object of measurement.

[This above situation could as well be an additional ``reason''
behind some indeterminacy relations. The demonstration of the
proposed origin(s) of indeterminacy relations will be the subject
of an independent study.]

The ``determined or observed'' characteristic of a particle is a
{\em different\/} conception here than its intrinsic
characteristics. The former notion clearly depends on the Borel
automorphism to be used for the measurement. For example, the
``observed velocity or momentum'' of a particle is a conception
dependent on the notion of the physical distance changing under
the action of a Borel automorphism of $\mathfrak{S}$. Clearly, the
coordination of the underlying continuum $\mathfrak{S}$ has
nothing whatsoever to do with the measurability here.

Importantly, dynamical systems on space $\mathfrak{S}$ follow the
``strict determinism'' in that the space coordinates map {\em
uniquely\/} under the action of a Borel automorphism of the space
$\mathfrak{S}$.

However, this is {\em not\/} the same ``causality'' as that of the
newtonian physical formulation.

The strict causality of newtonian conceptions implies that given
precise position and velocity of a particle at a given moment and
the total force acting on it, we can predict the precise position
and velocity of that particle at any later moment using
appropriate laws.

In the present context, strict newtonian causality would have
demanded that the position and the velocity of a point-particle
(definable in the present formalism as a point of the
$\Upsilon$-set with associated averaged measures) be exactly
determinable. This is of course not the situation for dynamical
systems on the space $\mathfrak{S}$.

But, {\em reality independent of any act of observation}, is then
ascribable to the phenomena as well as to the agencies of
observation in this formalism. Since physical objects are the
regions of the space and vice versa, the ``existence'' of the
$\Upsilon$-sets of the space $\mathfrak{S}$ is the ``existence''
of physical objects. This ``existence'' of physical bodies is,
obviously, independent of any act of observation.

Moreover, as we have seen earlier, the role of an observer in the
proposed Universal Theory of Relativity is very similar to that of
an observer in Newton's theory. This should not be construed in
any manner as an ``accidental'' situation. This is for the
following reasons.

The proposed Universal Theory of Relativity generalizes only the
notion of force in Newton's theory to that of transformations of
an appropriate standard Borel space $\mathfrak{S}$ and ``derives''
an appropriate concept of (newtonian) particle from the structure
of this space.

The Borel automorphism of the space $\mathfrak{S}$ is, obviously,
the {\em cause\/} behind an ``observable effect'' on the
$\Upsilon$-set or, equivalently, a physical object. Consequently,
to associate a definite cause to a definite effect has an
appropriate sense in the present formalism and this ``sense'' is
independent of an observer. Any intervention by any, conscious or
not, observer is therefore not needed to ``interpret'' the results
of observations. Newton's theory also had the same role for an
observer. It is a {\em passive role\/} for an observer.

{\em Objective reality\/} of physical phenomena, that the physical
phenomena are independent of the act of observation by an
observer, is then the underlying philosophical or conceptual basis
of the Universal Theory of Relativity. That the objective reality
of physical phenomena can be established in a mathematically and
physically consistent manner should now be evident.

In summary, the proposed generalization of the newtonian concept
of force by that of a transformation of the Standard Borel Space
$\mathfrak{S}$ has the ``genuine potential'' to provide us a
physical theory of everything. This generalization provides us,
essentially, a field theory that also contains a natural
non-singular notion of the (newtonian) particle to represent
physical bodies.

Moreover, measuring instruments and physical objects are then
treatable at par with each other as a result of this
generalization. This generalization is also in complete conformity
with the general principle of relativity.

A fundamental implication of the quantum conception is that of
indeterminacy, Heisenberg's indeterminacy. The proposed
generalization of the concept of force leads us to a mathematical
framework that explains indeterminacy as arising out of an
intrinsic fuzziness of the concept of a particle vis-\'{a}-vis the
field.

{\em The present formalism, of the dynamical systems of the
underlying continuum $\mathfrak{S}$, is then, already, a
unification of the ideas of the quantum theory and the general
principle of relativity.}

Surely, many details need to be worked out before we can test the
proposed Universal Relativity. However, some general observations
regarding experimental tests can also be reached at the present
stage of our theoretical developments. It is to such general
experimentally important observations that we now turn to.

\section{Experimental implications} \label{results}
The proposed Universal Relativity rests on the replacement of the
newtonian concept of force at a fundamental level with its
``natural'' generalization - a transformation (Borel automorphism)
of the underlying continuum (the space $\mathfrak{S}$). It should,
of course, be clear now that this replacement is that of the
``total force'' and not of its different, individual, components.

Another issue of special relevance to physical considerations is
that of the physical construction of reference frames.

Then, in ``fixing'' a physical reference frame, we are restricting
our attention to some specific $\Upsilon$-set of the space
$\mathfrak{S}$ and could be considering that it is unchanging.
Then, a subgroup of the full group of Borel automorphisms of
$\mathfrak{S}$ keeping that reference frame always invariant is
natural for us to consider. We could then be dealing with the
quotient of the space $\mathfrak{S}$ by the reference frame.

However, there certainly exist Borel automorphisms of
$\mathfrak{S}$ which affect the chosen $\Upsilon$-set. These Borel
automorphisms of this last type are precisely those which affect
the physical construction of the chosen reference frame.

Now, any member automorphism of the aforementioned subgroup could
have other invariant sets in addition to the $\Upsilon$-set of the
reference frame. All these invariant sets of an automorphism have
their Hausdorff distance ``fixed'' and, hence, are stationary
relative to each other. By our physical association, these
invariant sets of the space $\mathfrak{S}$ are the, basic or
compound, physical objects which are at ``rest'' relative to each
other.

Considerations of such type lead us, evidently, to a description
of the physical construction of a reference frame in its totality
and to a description of motion of other objects relative to it.

\begin{figure}
\includegraphics[width=2in,height=2in]{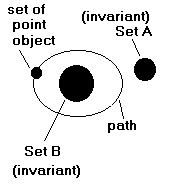}
\caption{\label{fig:rotmot} Circular Motion}
\end{figure}

For example, consider a point-object and its associated
$\Upsilon$-set. Let a Borel automorphism of the underlying
standard Borel space $\mathfrak{S}$ be such that it is measure
preserving and (its action) leading to a change in the Hausdorff
distance of the point object from that of a reference set (set
$A$) which we take to be an invariant set of that automorphism.
Also, let there be another set (set $B$), invariant under the same
automorphism, with respect to which the Hausdorff distance to the
chosen point-object is unchanging. (See Figure \ref{fig:rotmot}.)

Physically, this above situation could, as an example, represent
``revolution'' of the chosen point-object around the set $B$ that
is at rest relative to the reference set $A$.

The Borel automorphism in question describes the ``complete''
trajectory of the point object in this situation and, hence,
``encodes'' the {\em entire information about all the forces\/}
that are needed to determine the trajectory of the same point
object in theories using the notion of force.

[Because we limit ourselves here to only a general discussion of
physical issues arising in universal relativity, we shall not
explicitly display the Borel automorphism corresponding to the
motion described above. However, from the group properties of the
automorphisms of the standard Borel space, it is easy to see that
such an automorphism indeed exists.]

Clearly, this above is a consequence of the fact that the
universal relativity has only the law of motion, the Borel
automorphism of the space $\mathfrak{S}$, that is also the
``cause'' of motion in it.

Hence, as a result of the above, if some observation is
explainable in any theory using the concept of force then, the
same is explained in Universal Relativity by treating the {\em
involved total force(s)\/} as corresponding transformation(s). In
this way, Universal Theory of Relativity incorporates theories
that use the concept of force.

Essentially, a point object is a well-defined, non-singular,
notion in the mathematical framework of Universal Relativity. If
certain observation related to a physical body is ``explainable''
by representing the involved material body as a point particle
whose motion is describable by assuming a total force
(equivalently, potential field) then, in universal relativity, the
same ``total force'' is a Borel automorphism of the space
$\mathfrak{S}$ acting on the the $\Upsilon$-set of the imagined
point particle.

Of course, we can adopt this above procedure of ``realizing'' a
Borel automorphism of space $\mathfrak{S}$ for only the cases
which are ``explainable'' by theories using the concept of force.
This then also means that there will be situations in Universal
Relativity which will not be explainable by theories using the
concept of force.

As a consequence of the above, experimental situations can arise
for which Universal Relativity can be tested in the laboratory.

Experimental tests of the Universal Theory of Relativity will then
have to be developed keeping in mind this above. Specifically, we
should then look for or devise experimental situations for which
no ``natural'' explanations are offered by theories using the
concept of force \footnote{In this context, we also note the
so-called Pioneer Anomaly. The Doppler tracking of Pioneer 10 and
11 spacecrafts, both, apparently indicate an un-modelled
deceleration of order $10^{-\,9}\,m/s^2$ in the direction of the
inner solar system. Notably, the anomaly exceeds by five orders of
magnitude the corrections to newtonian motion predicted by
Einstein's theory of gravity. Any modification of gravitation
large enough to explain the anomaly is in contradiction with the
planetary ephemerides by nearly two orders of magnitude. However,
at the present moment, it cannot be concluded with certainty that
conventional effects do not cause the anomaly and new explanations
are required. See, for example, D Izzo and A Rathke (2005) {\it
Options for a non-dedicated test of the Pioneer Anomaly}, {\bf
Database: astro-ph/0504634}.}.

One such possibility is provided by the torsion balance experiment
that has been the basis of many important experimental results
related to gravity \cite{krishnan}. Therefore, as an example of
the situations that can arise in Universal Relativity, we discuss
below the ``explanations'' for the outcomes of the torsion balance
experiment.

\subsection*{The Torsion Balance Test}
Consider a torsion balance consisting of a thin wire, hanging from
one end at the roof, to which a dumbbell is hanging from its exact
middle at the other end of the wire. Let the masses of the
dumbbell be $m_1$ and $m_2$, $m_1=m_2=m$.

Next, consider ``external'' masses $m_3$ and $m_4$, $m_3=m_4=M$,
attached to the ground, $m_3$ placed on one side and $m_4$ on
another side of the dumbbell, with common center at the wire.

\begin{figure}
\includegraphics[width=2in,height=2in]{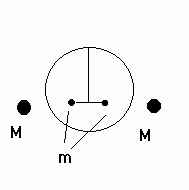}
\caption{\label{fig:tb} Torsion Balance}
\end{figure}

As per Newton's theory, there are two possible situations in which
{\em no torque\/} acts on the dumbbell of this assembly. The two
such arrangements of the masses are: 1] the external masses and
the dumbbell masses in one line and 2] line of the external masses
being perpendicular to the dumbbell. Then let the dumbbell be in
its natural equilibrium position in one of these states.

In Newton's theory, ``gravitational forces'' are supposed to act
between the ``external'' masses and the masses of the dumbbell,
and these forces ``cause'' the balance to ``torque'' when the
``external masses'' are {\em shifted}, say, from the situation 1]
to the situation 2]. This holds irrespective of other
``couplings'' of the masses, but these couplings can nullify the
``gravitational'' torque. Therefore, we then require and arrange
that these other possible couplings have been suitably eliminated
or minimized to adequate levels from the torsion balance assembly
of the above type.

Consequently, in Newton's theory, the balance will always be set
into ``oscillations'' when the positions of the external masses
are alternated between situations 1] and 2] above, say, by
rotating the external masses $m_3$ and $m_4$ about the common
center with the dumbbell.

In Universal Relativity, the balance setting into oscillations is
``explainable'' by treating the involved newtonian forces as
corresponding transformations acting on the space $\mathfrak{S}$.

But, as per Universal Relativity, we also have a transformation of
the ``external masses'' from the situation 1] to the situation 2]
{\em without\/} the balance ever getting torqued or set into
oscillations. Universal Theory of Relativity therefore
\underline{also} predicts that a ``null outcome'' is permissible
for the experiment of the above type.

Let us imagine a spherical shell whose outer surface is as
``frictionless'' as permissible. The torsion balance is situated
``inside'' the spherical shell while the ``external masses'' are
situated outside the spherical shell. [The reason why spherical
shell is mentioned here is, evidently, due to its being an
``invariant'' of the rotation map.]

Then, the motion of external masses can be such as to not affect
the interior of the spherical shell and, hence, not affecting the
state of the torsion balance located inside it.

To see this above, let us first note that the torsion balance is a
compound {\em object}, to be referred to as {\em object 1}, in the
sense described earlier. The external masses, together, are to be
considered also a compound {\em object}, {\em object 2}.

In our aforementioned experiment, we consider that the
``assembly'' of the torsion balance and the external masses is in
one of the two ``no torque states'' above at the beginning of the
experiment. Let $T_i$ be the (initial) transformation acting on
the assembly of object 1 and object 2, above, {\em ie}, let the
assembly be acted upon by an {\em initial\/} Borel automorphism
$T_i$ of the space $\mathfrak{S}$. The nature of the initial Borel
automorphism, although quite complicated, may be left unspecified
in our present considerations.

Let $T_r$ be the transformation of the rotational motion of only
the external masses and leaving the torsion balance unaffected or
the corresponding $\Upsilon$-sets invariant. That is, the map
$T_{r}$ acts on only the ``external masses or world'' to produce
their rotational motion.

The transformation acting on the assembly is then the mathematical
composition $T_{r}\circ T_i$ and its action is such that only the
external masses or world external to balance revolve around the
torsion balance with the $\Upsilon$-set of balance being an
invariant set of the composition $T_{r}\circ T_i$.

Clearly, such a transformation is possible when the external
masses are sufficiently away from the dumbbell masses so that the
effects of rotational motion of the external masses take time to
``propagate'' to the dumbbell masses or when the dumbbell is {\em
well-shielded\/} from {\em all\/} possible effects of the
rotational motion of external masses. Universal Theory of
Relativity then ``guarantees'' that the torsion balance will {\em
not\/} be torqued by the motion of the external masses in the
experimental situation imagined above.

Evidently, no ``explicit calculations'' using transformations
$T_i$ and $T_{r}$ are needed to reach this conclusion. Universal
Relativity always predicts a ``complete null effect'' in this
situation.

If, for the same experiment, we make the external masses move not
along a circle but along an {\em elliptical orbit}, there is to be
no change in the conclusion of Universal Relativity which predicts
a {\em complete null effect\/} even in this latter situation.
Highly eccentric elliptic orbit for external masses would imply
``enhancement'' of the newtonian non-null effect. An elliptical
orbit for the external masses could then be preferred over the
circular orbit for obvious reasons.

Then, in certain situations, the effect on torsion balance is
expected in Universal Relativity to be ``total null'' if shielding
of the balance is proper. Then, if the external masses ``rotate''
around the dumbbell masses sufficiently slowly, say, one
revolution in few minutes, and if we run the experiment for a
couple of weeks or months, so as to obtain good statistic, we can
test the Universal Theory of Relativity. Of course, this test can
only be considered successful if the ``null result'' is observed.
Else, in universal relativity, we shall be forced to conclude that
the ``isolation'' of the torsion balance is not achieved.

Thus, if a torsion-balance experiment is sufficiently carefully
performed in the aforementioned manner, we should be able to
verify the possible ``null effect'' prediction of the Universal
Theory of Relativity against that of the ``certainly non-null
effect'' of Newton's theory.

We then note that certain, extremely high precision, torsion
balance experiments are in use \footnote{For example, one such
torsion balance experiment is in operation at the Gauribidanur
Field Station of the Gravitation Group, Tata Institute of
Fundamental Research, Mumbai, India.}. Any of these torsion
balance assemblies can then be employed to verify the ``null
effect'' predicted by the Universal Theory of Relativity.

\section{Concluding Remarks} \label{conclude}

\noindent Arthur Schopenhauer: \\ \phantom{m}

$\clubsuit$ {\em All truth passes through three stages: First, it
is ridiculed. Second, it is vehemently opposed. Third, it is
accepted as being self-evident.\/} $\clubsuit$ \\  \phantom{m}

Over the centuries, some fundamental concepts that led Newton to
his theory acquired the status of being self-evident.
Specifically, Newton's concept of force is embedded so deep into
our thinking that we treat it as a self-evident concept. No doubt,
force is a very useful concept. But, while adopting this concept,
we are certainly required to attribute to physical matter source
properties that generate the forces in question.

However, we then choose to ignore one fundamental limitation of
this conception that the source properties so attributed to
physical matter cannot find any explanations with the theory that
uses the (newtonian) concept of force.

Developments in Physics that took place over the times since
Newton formulated his mechanics were, explicitly or implicitly,
based on the concept of force, developments in Quantum Theory not
being any exceptional.

[The concept of potential energy is a ``byproduct'' of the
newtonian concept of force.  From the very beginning, quantum
theory ``assumed'' this conception of potential energy to
formulate its various (Heisenberg) operators.]

Except, perhaps, for Hertz's attempt \cite{schlipp} [p. 31] when
he felt the need to replace the concept of potential energy by
some suitable other, no one attempted to generalize the concept of
force; such strong and gripping had been the influence of Newton's
thinking on the physicists.

In his attempts to incorporate the phenomenon of gravitation
within the overall framework of the theory of relativity, Einstein
reached the equivalence principle and, for the first time,
replaced the concept of force by that of the ``curvature'' of the
spacetime geometry.

Although Einstein replaced the concept of gravitational force by
that of the curvature of spacetime geometry, his formulation of
general theory of relativity did not live up to his intentions.
The same conception did not work for other forces of Nature, {\em
eg}, for Coulomb's force. Consequently, Einstein's mathematical
formulation of a theory (of gravitation) based on the principle of
general relativity is logically completely inappropriate.

Einstein had ``realized'' such problems with his formulation of 
general relativity. That is perhaps why he dubbed his equations of
general relativity as ``preliminary'' equations. This obvious
``failure'' led him to his numerous attempts at the Unified Field
Theory. Even these attempts failed.

Many others following the methods of Quantum Theory systematically
developed theoretical foundations for the successful description
of the micro-physical world. For this purpose, definite but ad-hoc
rules of obtaining the $\Psi$-function were first adopted and the
predictions checked by ingenious experimentations.

However, Einstein had, in his ways, also realized that these
developments in quantum theory too were not entirely satisfactory
from the perspective of some fundamental physical issues. He had,
time and again, warned \cite{schlipp, subtle} us against the
pitfalls of the conceptions behind these theories - his own
formulation of general relativity as well as the (probabilistic)
quantum theory.

Here, we systematically built on Einstein's aforementioned
intuition and replaced the newtonian concept of force with its
natural generalization as a transformation of the underlying
continuum. In doing so, we were led to a natural unification of
the ideas of the quantum theory and those of the general principle
of relativity.

The formalism we developed satisfies one of the foremost of the
requirements of a genuinely universal theory of physics: that the
constants of physics ``arise'' in it through mutual relationships
of physical objects and their values cannot be changed without
essentially destroying the underlying theoretical framework. It is
also the Universal Theory of Relativity that is in complete
agreement with the general principle of relativity.

In the last section, \S\,\ref{results}, we discussed an
experimental test, the torsion balance test, of the proposed
Universal Relativity. In accordance with the scientific
methodology, it would then be worthwhile to let this and such
tests decide whether the proposed generalization of the newtonian
concept of force is really what is appropriate one to describe the
workings of the physical world.

\acknowledgments I am indebted to S G Dani, M G Nadkarni, R V
Saraykar and V M Wagh for drawing my attention to literature on
theories of measures \& dynamical systems and, to N Krishnan as
well, for many helpful and encouraging discussions.

This work is dedicated to the memory of (Late) Professor L K Patel
who was a gifted relativist, mathematician and, above all, a
cheerful as well as a kindhearted person.

\end{document}